\documentclass[11pt,a4paper]{article}
\pdfoutput=1
\usepackage{jheppub}                

\usepackage{graphicx}
\usepackage{epsfig,multicol,bbm}
\usepackage{amsmath,amssymb,bbold,bm}

\title{Why is the mission impossible? -- Decoupling the mirror Ginsparg-Wilson fermions in the lattice models for two-dimensional abelian chiral gauge theories}

\author[]{Y. Kikukawa}
\affiliation[a]{Institute of Physics, the University of Tokyo, Tokyo 153-8902, Japan }
\emailAdd{kikukawa@hep1.c.u-tokyo.ac.jp}
\date{July 21, 2016}                                 

\subheader{
\begin{flushright}
\rm 
UT-Komaba/17-10
\end{flushright}
}

\abstract{
It has been known that the four-dimensional abelian chiral gauge theories of an anomaly-free set of Wely fermions can be formulated on the lattice preserving the exact gauge invariance and the required locality property in the framework of the Ginsparg-Wilson relation. This holds true in two dimensions. However, in the related formulation including the mirror Ginsparg-Wilson fermions and therefore having the simpler fermion path-integral measure, it has been argued that the mirror fermions do not decouple: in the 345 model with Dirac- and Majorana-Yukawa couplings to XY-spin field, the two-point vertex function of the (external) gauge field in the mirror sector shows a singular non-local behavior in the PMS phase. We re-examine why the attempt seems a ``Mission: Impossible'' in the 345 model. We point out that the effective operators to break the fermion number symmetries ('t Hooft operators plus others) in the mirror sector do not have sufficiently strong couplings even in the limit of large Majorana-Yukawa couplings. We also observe that the type of Majorana-Yukawa term considered there is singular in the large limit due to the nature of the chiral projection of the Ginsparg-Wilson fermions,  but a slight modification without such singularity is allowed by virtue of the very nature. We then consider a simpler four-flavor axial gauge model, the 1$^4$(-1)$^4$ model, in which the U(1)$_A$ gauge and Spin(6)($\cong$ SU(4)) global symmetries prohibit the bilinear terms, but allow the quartic terms to break all the other continuous mirror-fermion symmetries. We formulate the model so that 
it is well-behaved and simplified
in the strong-coupling limit of the quartic operators.
Through Monte-Carlo simulations in the weak gauge coupling limit, we show a numerical evidence that the two-point vertex function of the gauge field in the mirror sector shows a regular local behavior, and we still argue that  
all you need is killing the continuous mirror-fermion symmetries with would-be gauge anomalies non-matched. Finally, by gauging a U(1) subgroup of the U(1)$_A$$\times$ Spin(6)(SU(4)) of the previous model, we formulate the $2 1 (-1)^3$ chiral gauge model and argue that the induced fermion measure term satisfies the required locality property and provides a solution to  the reconstruction theorem. This gives us  ``A New Hope''  for the mission to be accomplished.

}


\begin{document}
\maketitle

\section{Introduction}

Chiral gauge theories have several interesting possibilities in their own dynamics: 
fermion number non-conservation due to chiral anomaly\cite{'tHooft:1976up, 'tHooft:1976fv}, 
various realizations of the gauge symmetry and global flavor symmetry\cite{Raby:1979my, Dimopoulos:1980hn},  
the existence of massless composite fermions suggested by 't Hooft's anomaly matching condition\cite{'tHooft:1979bh},
the classical scale invariance and the vanishing vacuum energy\cite{Holdom:2007gg,Holdom:2009ma} 
and so on. 
Unfortunately,  
little is known so far about the actual behavior of chiral gauge theories beyond perturbation theory.  
It is desirable to develop a formulation to study  
the non-perturbative dynamics of chiral gauge theories. 


Lattice gauge theory can 
provide a framework for non-perturbative formulation of chiral gauge theories, despite the well-known problem of the species 
doubling \cite{Karsten:1980wd,Nielsen:1980rz,Nielsen:1981xu,Friedan:1982nk}.\footnote{See \cite{Luscher:1999mt,Golterman:2000hr,Luscher:2000hn,Kaplan:2009yg} for the recent  reviews on this subject.}
A clue to this development is 
the construction of {\em local} and {\em gauge-covariant} lattice Dirac operators 
satisfying the Ginsparg-Wilson relation\cite{Ginsparg:1981bj,
Neuberger:1997fp,Hasenfratz:1998ri,Neuberger:1998wv,
Hasenfratz:1998jp,Hernandez:1998et}.
\begin{equation}
\gamma_5 D + D \gamma_5 = 2 a D \gamma_5 D .
\end{equation}
%
An explicit example of such lattice Dirac operator is given by the overlap Dirac operator \cite{Neuberger:1997fp,Neuberger:1998wv}, 
which was derived by Neuberger from the overlap formalism 
\cite{
Narayanan:wx,Narayanan:sk,Narayanan:ss,Narayanan:1994gw,Narayanan:1993gq,
Neuberger:1999ry,Narayanan:1996cu,Huet:1996pw,
Narayanan:1997by,Kikukawa:1997qh,Neuberger:1998xn,
Narayanan:1996kz,Kikukawa:1997md,Kikukawa:1997dv}.\footnote{
The overlap formula was derived from the five-dimensional approach of domain wall fermion proposed by 
Kaplan\cite{Kaplan:1992bt}.
In the vector-like formalism of domain wall fermion by Shamir\cite{Shamir:1993zy, Furman:ky,
Blum:1996jf, Blum:1997mz}, 
the {\em local} low energy effective action of the chiral mode 
is precisely given by the overlap Dirac 
operator \cite{Vranas:1997da,Neuberger:1997bg,Kikukawa:1999sy}.
}
By the Ginsparg-Wilson relation, it is possible to realize an exact chiral symmetry on the lattice\cite{Luscher:1998pq}
in the manner consistent 
with the no-go theorem.\cite{Kikukawa:1998pd, Luscher:1998kn, Fujikawa:1998if, Adams:1998eg, Suzuki:1998yz}
It is also possible to introduce Weyl fermions on the lattice
and 
this opens the possibility
to formulate anomaly-free chiral gauge theories on the 
lattice\cite{Luscher:1998du,Luscher:1999un,Luscher:1999mt,Luscher:2000hn,Suzuki:1999qw,
Neuberger:2000wq,Adams:2000yi,Suzuki:2000ii,Igarashi:2000zi,Luscher:2000zd,Aoyama:1999hg,Kikukawa:2000kd,Kikukawa:2001jm,Kikukawa:2001mw,Kadoh:2003ii,Kadoh:2004uu,Kikukukawa:2005ILFTN,Kadoh:2007wz,Kadoh:2007xb}.
In the case of U(1) chiral gauge theories,  
L\"uscher\cite{Luscher:1998du}  proved rigorously that
it is possible to construct the fermion path-integral measure 
which depends smoothly on the gauge field  and
fulfills the fundamental requirements such as 
locality, gauge-invariance and lattice symmetries.\footnote{For generic non-abelian chiral gauge theories, 
the construction in all orders of the weak gauge-coupling expansion was given by Suzuki\cite{Suzuki:2000ii,Igarashi:2000zi} and 
by L\"uscher\cite{Luscher:2000zd}.
}\footnote{In this formulation 
of U(1) chiral lattice gauge theories\cite{Luscher:1998du}, 
although the proof of the existence of the fermion measure is constructive,  the resulted formula 
of the fermion measure turns out to be rather complicated for the case of the finite-volume lattice. 
It also relies on the results obtained in the infinite lattice.  Therefore it does not provide a formulation 
which is immediately usable for numerical applications. See 
\cite{Kadoh:2003ii,Kadoh:2004uu,Kikukukawa:2005ILFTN,Kadoh:2007wz} for a simplified formulation toward a practical implementation.
}\footnote{This construction was extened to 
the SU(2)$\times$U(1) chiral gauge theory of the Glashow-Weinberg-Salam model\cite{Glashow:1961tr,Weinberg:1967tq,Salam:1968rm} based on the pseudo reality and anomaly-free condition
of SU(2) by Kadoh and the author\cite{Kadoh:2007xb}.
}
This gauge-invariant construction 
holds true in two dimensions.

In two dimensions, the target theories
are nothing but chiral Schwinger models of the 
sets of left- and right-handed Weyl fermions 
satisfying the anomaly-free conditions on the U(1) charges,
$\sum_\alpha(q^\alpha_L)^2 =\sum_{\alpha'}(q^{\alpha'}_R )^2$. 
The models are super-renormalizable and essentially solvable in the continuum limit\cite{Jackiw:1984zi}.
The effective action induced by the Weyl fermions can be obtained exactly in the continuum limit (i.e. 
taking the infinite UV cutoff limit
$\Lambda \rightarrow \infty$
and neglecting the irrelevant higher-order terms) up to a reguralization-dependent and gauge-noninvariant 
relevant term.
The total effective action of the U(1) gauge field is obtained 
exactly (in the Euclidean spacetime) as 
\begin{eqnarray}
\label{eq:chiral-Schwinger-model-one-loop}
S_{\rm eff} &=&
 \int d^2 x 
\left\{ 
\frac{1}{4} F_{\mu\nu} F_{\mu\nu} 
\right.
\nonumber\\
&&\qquad\quad
+
\frac{1}{2}
\frac{\sum_\alpha \, (q^\alpha_L)^2 e^2}{4\pi} \,\,\,
A_\mu 
\Big[
c \, \delta_{\mu\nu'} -
(\delta_{\mu\nu'}+ i \epsilon_{\mu\nu'}) 
\frac{\partial_{\nu'} \partial_{\mu'}}{\square} 
(\delta_{\mu'\nu}- i \epsilon_{\mu'\nu}) 
\Big]
A_\nu
\nonumber\\
&&\qquad\quad
\left.
+ 
\frac{1}{2}
\frac{\sum_{\alpha'}(q^{\alpha'}_R )^2 e^2}{4\pi}
A_\mu 
\Big[
c \, \delta_{\mu\nu'} -
(\delta_{\mu\nu'} -i \epsilon_{\mu\nu'}) 
\frac{\partial_{\nu'} \partial_{\mu'}}{\square} 
(\delta_{\mu'\nu}+ i \epsilon_{\mu'\nu}) 
\Big]
A_\nu
\right\} , 
\end{eqnarray}
where $c$ is the reguralization-dependent constant
and, 
when $c=1$, the effective action is gauge-invariant.
Due to the one-loop correction of the massless fermions, 
the U(1) gauge boson acquires the mass (square) as
$m_\gamma^2 = 
\big[
\sum_\alpha(q^\alpha_L)^2 +\sum_{\alpha'}(q^{\alpha'}_R )^2
\big] \frac{e^2}{2\pi}$.
%
%
%
In this respect, it is known in the continuum theory
that there is no
gauge-invariant
reguralization method for chiral gauge theories in general
and even in two-dimensions.
Then,
these two-dimensional theories are nice testing grounds for
the attempts/approaches to seek 
the exactly gauge-invariant and nonperturbative 
formulation of
chiral gauge theories, 
where one can compare the outcomes of the attempts/approaches 
with the exact results of the target continuum theories.

The above lattice construction\cite{Luscher:1998du} gives a gauge-invariant non-pertutbative
regurarization of the chiral Schwinger models in the framework of lattice gauge
theory 
for all possible topological sectors of 
the U(1) gauge field in two-dimensions.
%
One can verify in the (classical) continuum limit
$ a (\equiv \pi /\Lambda) \rightarrow 0$  that this lattice formulation reproduces
the exact results of the target continuum theories, eq.~(\ref{eq:chiral-Schwinger-model-one-loop}), with $c=1$.
And there are good reasons to believe that the lattice model (i.e. keeping the lattice spacing $a$ finite) 
has a simple phase-structure of the single phase with the massive U(1) gauge boson,
where the second-order critical point is given at the vanishing gauge-coupling constant, $g  a = 0$ (in unit of the lattice spacing $a$).

However, in the related formulation by Poppitz and his collaborators\cite{Bhattacharya:2006dc,Giedt:2007qg,Poppitz:2007tu,Poppitz:2008au,Poppitz:2009gt,Poppitz:2010at,Chen:2012di,Giedt:2014pha}
which includes the mirror degrees of freedom\cite{Montvay:1987ys, Montvay:1988av, Farakos:1990ex, Lin:1990ue, Lin:1990vi, Munster:1991xs, Lin:1991csa, Montvay:1992mv, Montvay:1992eg, Lin:1992qb, Lin:1993hp} 
in terms of Ginsparg-Wilson fermions and therefore has the simpler fermion path-integral measure, it was argued 
that the mirror fermions do not decouple: in the two-dimensional 345 model with Dirac- and Majorana-Yukawa couplings to XY-spin field, the two-point vertex function of the (external) U(1) gauge field in the mirror sector shows a singular non-local behavior in the paramagnetic strong-coupling phase(PMS)\cite{Giedt:2007qg,Gerhold:2007yb,Gerhold:2007gx}.\footnote{See \cite{Eichten:1985ft,Swift:1984dg,Smit:1985nu,Aoki:1987cb,Aoki:1988cu,Funakubo:1987zt,Funakubo:1988tg,
Golterman:1992yha,
Golterman:1990zu,Golterman:1991re,
Bock:1991bu,Bock:1992gp,Aoki:1991es} for the former attempts to decouple the species-doublers/mirror-fermions
by strong Yukawa and multi-fermion interactions.}
The singular non-local term turns out to be 
same as
the contribution of the massless Weyl fermions of the target sector.
%
It implies that
the U(1) gauge boson acquires twice as large as the mass square
expected in the target chiral Schwinger model.
This result seems puzzling because
the Dirac- and Majorana-Yukawa couplings can break 
two ``would-be anomalous'' global U(1) symmetries in the mirror sector, that is the
required condition for the decoupling of the mirror fermions,
as claimed 
by Eichten and Preskill
\cite{Eichten:1985ft,Banks:1991sh,Banks:1992af,Bhattacharya:2006dc}.
In their numerical simulations, though, a specific limit
of large Majorana-Yukawa couplings was taken to tame the sign problem of their lattice model.


On the other hand, 
this question of decoupling 
the mirror degrees of freedoms in the 345 model
was also
studied by Wang and Wen\cite{Wang:2013yta}
from the point of view of the Hamiltonian construction
based on Topological Insulators/Superconductors\cite{Wen:2013ppa,Wang:2013yta,You:2014oaa,You:2014vea,DeMarco:2017gcb}: 
based on the effective bosonic (bosonized) description
of the 2D Chern Insulator by the bulk Chern-Simon gauge theory
and the boundary chiral-boson theory with sine-Goldon couplings,
it was shown that the boundary phase can be fully gapped
in the 345(0) model 
by the two sine-Goldon couplings required precisely to break
the two ``would-be anomalous'' global U(1) symmetries in the mirror sector.
This result 
suggests that 
the mirror fermions can be decoupled by the suitable choice of the coupling strengths of the symmetry-breaking interactions.

In this paper, 
we 
re-examine why the attempt seems a ``mission impossible'' 
for the 345 model in the mirror-fermion approach with Ginsparg-Wilson fermions\cite{Bhattacharya:2006dc,Giedt:2007qg,Poppitz:2007tu,Poppitz:2008au,Poppitz:2009gt,Poppitz:2010at,Chen:2012di,Giedt:2014pha}.
We point out that the effective operators to break the fermion number symmetries ('t Hooft operators plus others) in the mirror sector do not have sufficiently strong couplings even in the limit of large Majorana-Yukawa couplings. We also observe that the type of Majorana-Yukawa term considered there is singular in the large limit due to the nature of the chiral projection of the Ginsparg-Wilson fermions, but a slight modification without such singularity is allowed by virtue of the very nature. 
Based on these results, we argue that
one can attribute the failure of decoupling
to the singular Majorana-Yukawa terms of the lattice model
and may expect a better result by modifying
the Majorana-Yukawa terms so that they are  well-behaved in the strong-coupling limit.

We then consider a simpler four-flavor axial gauge model, the 1$^4$(-1)$^4$ model, in which the U(1)$_A$ gauge and Spin(6)(SU(4)) global symmetries prohibit bilinear mass terms, but allow the quartic terms to break the other continuous mirror-fermion symmetry U(1)$_V$. We formulate the model so that 
it is well-behaved and simplified in the strong-coupling limit of the quartic operators. Through Monte-Carlo simulations in the weak gauge coupling limit, we show a numerical evidence
that the two-point vertex function of the U(1)$_A$ gauge field in the mirror sector shows a regular {\em local} behavior,
consistently with the decoupling of the mirror-fermions, and we argue that still all you need is killing the (continuous) mirror-fermion symmetries with would-be gauge anomalies non-matched,
as originally claimed 
by Eichten and Preskill\cite{Eichten:1985ft,Bhattacharya:2006dc,Wang:2013yta}.

Finally, 
we formulate the $2 1 (-1)^3$ chiral gauge model 
by gauging a U(1) subgroup of the U(1)$_A$$\times$ Spin(6)(SU(4))
of the previous model.
We show again a numerical evidence
that the two-point vertex function of the U(1) gauge field in the mirror sector shows a regular {\em local} behavior
through Monte-Carlo simulations in the weak gauge coupling limit.
We then deduce a definition of the (target) Weyl-field
measure of the $21(-1)^3$ chiral gauge model,
where 
the mirror-fermion part of the Dirac-field measure
is just saturated 
by the suitable products of the 't Hooft vertices
in terms of the mirror-fermion fields.
Based on the results of Monte-Carlo simulations,
we argue that the induced fermion measure term satisfies the required {\em locality} property 
and provides a solution to the reconstruction theorem
of the Weyl field measure 
in the framework of the Ginsparg-Wilson relation\cite{Luscher:1998du}. 
This result gives us a new hope for the mission to be accomplished.

The mirror-fermion models formulated with overlap fermions in this paper, the $1^4(-1)^4$- and $21(-1)^3$- models, 
can be also constructed through the 2+1D vector-like domain wall fermion
by adding suitable boundary interaction terms.
We give the explicit form of the boundary terms which precisely reproduce
the U(1)$_A$ $\times$ Spin(6)(SU(4))-invariant multi-fermion interaction in the mirror sector
without the singularity in the large-coupling limit (cf. \cite{Creutz:1996xc}).

The four-flavor model 
with the U(1)$_A$ $\times$ Spin(6)(SU(4))-invariant 
multi-fermion interaction, 
which we adopt for the mirror-fermion sector, 
is closely related to the eight-flavor 1D Majorana chain 
with the SO(7)-invariant quartic interaction (1D TSC with time-reverasl symmetry; class BDI in 1D classified by
$\mathbb{Z}_8$($\leftarrow\mathbb{Z}$))\cite{Fidkowski:2009dba,Fidkowski:2011,BenTov:2014eea}.
It also resembles the SU(4)/SO(4)-invariant 
reduced-staggered-fermion models in 3+1, 2+1, 1+1D, which are
used in the recent studies of ``mass without symmetry breaking''
\cite{BenTov:2015gra,Ayyar:2014eua,Ayyar:2015lrd,Ayyar:2016lxq,Catterall:2015zua,Catterall:2016dzf,Catterall:2017ogi,Schaich:2017czc}.\footnote{
We hope that
the formulation of the four-flavor model 
with the U(1)$_A$ $\times$ Spin(6)(SU(4))-invariant 
multi-fermion interaction in terms of overlap fermions
and its use for the construction of chiral lattice gauge theories
given in this paper provides an answer to the question raised by Catterall and his collaborators\cite{Catterall:2016dzf,Schaich:2017czc}.}
We clarify this relation by formulating the quantum 1D Majorana chain as the classical 1+1D Majorana-fermion model in Euclidean metric through the path-integral quantization.
By this relation, 
the rigorous result
about 
the mass gap of the eight-flavor 1D Majorana chain
with the SO(7)-invariant quartic interaction
by Fidkowski and Kitaev\cite{Fidkowski:2009dba} 
and its extension to the model with the reduced SO(6) symmetry
by Y.-Z. You and C. Xu \cite{You:2014vea}
suggest strongly 
that
the four-flavor axial model with U(1)$_A$ $\times$ Spin(6)(SU(4)) symmetry is indeed gapped. 
And vise versa: 
%
our numerical-simulation results that
the correlation lengths of the mirror-sector fields are
of order multiple lattice spacings
provide a numerical evidence for the mass gap
of the eight-flavor 1D Majorana chain
based on 
the framework of  1+1D Euclidean 
path-integral quantization.

Since the 2+1D domain wall fermion is nothing but
2+1D classical formulation of a 2D Topological 
Insulator\cite{Qi:2006xx,Creutz:1994ny,Qi:2008ew}
(Chern Insulator/IQHE without time-reversal symmetry;
class A in 2D classified by $\mathbb{Z}$), 
our result here 
provides the explicit procedure to bridge
between the two constructions
for 1+1D chiral gauge theories,
the 2+1D classical construction of
domain wall fermion with boundary interactions
to decouple the mirror-modes\cite{Creutz:1996xc}
and 
the 2D quantum Hamiltonian construction
of TI/TSC with gapped boundary phases\cite{Wen:2013ppa,Wang:2013yta,You:2014oaa,You:2014vea,DeMarco:2017gcb}.
And 
the mirror-fermion model in terms of overlap fermions
is obtained precisely as the 1+1D low-energy effective 
{\em local} lattice theory, 
and it can describe directly the gapless/gapped boundary phases.
We illustrate this relation for the case of
the eight-flavor 2D chiral p-wave TSC with time-reversal and Z$_2$ symmetries
(class D'/DIII+R in 2D classified 
by $\mathbb{Z}_8$($\leftarrow\mathbb{Z}$))\cite{Qi:2013dsa,Yao-Ryu:2013,Ryu-Zhang:2012,Gu-Levin:2014}.
This connection should hold true in lower and higher dimensions.
In particular, 
it would be useful
to examine
the Hamiltonian constructions
of 3+1D chiral gauge theories
based on the 4D TI/TSC with the ``proposed'' gapped boundary phases\cite{Wen:2013ppa,You:2014oaa,You:2014vea}
from the point of view of the 3+1D/4+1D Euclidean 
construction based on
the overlap/domain wall fermions.

This paper is organized as follows.
In section~\ref{sec:review-abelian-chiral-gauge-model-two-dim}, we first review the construction of U(1) chiral lattice gauge theories 
based on the Ginsparg-Wilson relation, 
adapted for two-dimensional theories.
In section~\ref{sec:mirror-fermion-approach-in-general},
we next review 
the mirror-fermion approach
with Ginsparg-Wilson fermions
to 
two-dimensional
U(1) chiral gauge theories. 
Section~\ref{sec:3450-model} is devoted to the re-examination of the 345 model with Dirac- and Majorana-Yukawa couplings to XY-spin field.
In section~\ref{sec:1^4(-1)^4-model}, we introduce the $1^4(-1)^4$ model and discuss its properties in detail. 
In section~\ref{sec:21(-1)^3-model}, 
we formulate the $21(-1)^3$ model and discuss its properties
in relation to the reconstruction theorem reviewed in 
section~\ref{sec:review-abelian-chiral-gauge-model-two-dim}.
In section~\ref{sec:DWF-to-Mirror}, 
the mirror-fermion models introduced 
in sections~\ref{sec:1^4(-1)^4-model} 
and \ref{sec:21(-1)^3-model}, 
the $1^4(-1)^4$- and $21(-1)^3$- models, 
are constructed through the 2+1D vector-like domain wall fermion
by adding the suitable boundary interaction terms.
In section~\ref{sec:relation-1D-2D-TI-TSC}, 
we discuss the relations of
the 1+1D/2+1D Euclidean formulation of
mirror-fermion/domain wall-fermion models
to the 1D/2D quantum Hamiltonian construction of
TI/TSC with gapped boundary phases.
In the final section~\ref{sec:summary-discussion}, we conclude with a few discussions.
 
\section{Two-dimensional abelian chiral gauge theories on the lattice in the framework of the Ginsparg-Wilson relation
}
\label{sec:review-abelian-chiral-gauge-model-two-dim}

%
%
%
%
%
%
%
%


In this section, we review the construction of U(1) chiral lattice gauge theories 
based on the Ginsparg-Wilson relation \cite{Luscher:1998du}, adapted for the two-dimensional theories.

We consider U(1) gauge theories where the gauge field 
couples to $N$ left-handed Weyl fermions with charges ${q}_\alpha$
and $N'$ right-handed Weyl fermions with charges ${q}'_{\alpha'}$
satisfying the anomaly cancellation condition, 
\begin{equation}
\label{eq:anomaly-cancellation-condition}
\sum_{\alpha=1}^N ({q}_\alpha)^2  - \sum_{\alpha'=1}^{N'} ({q}'_{\alpha'})^2 = 0.
\end{equation}
We assume the two-dimensional lattice of the finite size $L$ and choose lattice units, 
\begin{equation}
\Gamma =
 \left\{x=(x_1, x_2)  \in \mathbb{Z}^2 \, \vert \, \, 0 \le x_\mu < L \,  (\mu = 1,2) \right\} , 
\end{equation}
and adopt the periodic boundary condition for both boson fields and fermion fields. 

\subsection{Gauge fields}
We adopt the  compact formulation of U(1) gauge theory on the lattice.  
U(1) gauge fields on $\Gamma$ then are represented by link fields,  $U(x,\mu) \, \in \text{U(1)}$. 
We require the so-called admissibility condition on the gauge fields:
\begin{equation}
\vert F_{\mu\nu}(x) \vert  < \epsilon \qquad {\rm for \ all} \ \ 
x,\mu,\nu, 
\end{equation}
where the field tensor $F_{\mu\nu}(x)$ is defined from the plaquette variables, 
\begin{eqnarray}
F_{\mu\nu}(x) &=& \frac{1}{i} {\rm ln} P_{\mu\nu}(x), 
\quad - \pi < F_{\mu\nu}(x) \le \pi , \\
P_{\mu\nu}(x)&=&U(x,\mu)U(x+\hat\mu,\nu)U(x+\hat\nu,\mu)^{-1}
U(x,\nu)^{-1}, 
\end{eqnarray}
and $\epsilon$ is a fix number in the range $0 < \epsilon < \pi$. 
This condition ensures that 
the overlap Dirac operator\cite{Neuberger:1997fp,Neuberger:1998wv} is a
smooth and local function of  the gauge field if $| {\rm e}_\alpha | \epsilon <2/5$ for all $\alpha$
and
$| {\rm e}'_{\alpha'} | \epsilon <2/5$
for all $\alpha'$ \cite{Hernandez:1998et}.
The admissibility condition may be imposed dynamically by choosing the following action, 
\begin{equation}
S_G = \frac{1}{4g_0^2} \sum_{x\in \Gamma} \sum_{\mu,\nu}  L_{\mu\nu}(x), 
\end{equation}
where 
\begin{equation}
L_{\mu\nu}(x)= \left\{ \begin{array}{ll}
\left[ F_{\mu\nu}(x) \right]^2 
  \left\{ 1 -  \left[ F_{\mu\nu}(x) \right]^2 / \epsilon^2 \right\}^{-1}  
  & \quad \text{if } \vert F_{\mu\nu}(x) \vert < \epsilon , \\
  \infty & \quad \text{otherwise} .
  \end{array}
  \right.
\end{equation}

The admissible U(1) gauge fields can be classified by the magnetic fluxes, 
\begin{equation}
m_{\mu\nu} 
=
\frac{1}{2\pi}\sum_{s,t=0}^{L-1}F_{\mu\nu}(x+s\hat\mu+t\hat\nu) , 
\end{equation}
which are integers independent of $x$.  We denote the space of the admissible gauge fields with a given magnetic flux $m_{\mu\nu}$ by $\mathfrak{U}[m]$. As a reference point in the given 
topological sector $\mathfrak{U}[m]$, one may introduce the gauge field 
which has the constant field tensor equal to $2\pi m_{\mu\nu}/L^2 (< \epsilon)$ by
\begin{eqnarray}
V_{[m]}(x,\mu) 
&=&{\text{e}}^{-\frac{2\pi i}{L^2}\left[
L \delta_{ \tilde x_\mu,L-1} \sum_{\nu > \mu} m_{\mu\nu}
 \tilde x_\nu +\sum_{\nu < \mu} m_{\mu\nu} \tilde x_\nu
\right]}    \quad (\tilde x_\mu = x_\mu \text{ mod } L ).  
\end{eqnarray}
Then any admissible U(1) gauge field in $\mathfrak{U}[m]$ 
may be expressed as 
\begin{equation}
\label{eq:U-tilde}
U(x,\mu)=\tilde U(x,\mu) \, V_{[m]}(x,\mu) , 
\end{equation}
where  $\tilde U(x,\mu)$ stands for the dynamical degrees of freedom.



\subsection{Weyl fields}
Weyl fermions are introduced  based on the Ginsparg-Wilson relation.
We first consider  Dirac fields $\psi(x)$ 
which carry a Dirac index $s=1,2$ and a flavor index $\alpha = 1, \cdots, N$
and 
$\psi'(x)$ which carry a Dirac index $s' =1,2$ and a flavor index $\alpha' = 1, \cdots, N'$. 
Each components $\psi_\alpha(x)$ and ${\psi'}_{\alpha'}(x)$ couple to the link fields,  $U(x,\mu)^{q_\alpha}$ and $U(x,\mu)^{{q'}_{\alpha'}}$, 
respectively. 
We assume that the lattice Dirac operators $D$ and $D'$ acting on $\psi(x)$ and ${\psi}'(x)$, respectively satisfy the Ginsparg-Wilson 
relation\footnote{In this paper, we adopt the normalization of the lattice Dirac operator so that the factor $2$ appears in the right-hand-side of the Ginsparg-Wilson relation: 
$ \gamma_3 D + D \gamma_3 = 2 D \gamma_3 D $. }, 
\begin{equation}
\label{eq:GW-rel}
\gamma_3 D + D \hat \gamma_3 = 0, \qquad \hat \gamma_3 \equiv \gamma_5(1- 2 D ) , 
\end{equation}
\begin{equation}
\label{eq:GW-rel-prime}
\gamma_3 D' + D' {\hat \gamma}'_3  = 0, \qquad {\hat \gamma}'_3 \equiv \gamma_3(1- 2 D' ) , 
\end{equation}
and we define the projection operators as 
\begin{equation}
       P_{\pm} = \left( \frac{1\pm  \gamma_3}{2} \right) ,  \quad
\hat P_{\pm} = \left( \frac{1\pm \hat \gamma_3}{2} \right) , \quad
\hat P'_{\pm} = \left( \frac{1\pm \hat \gamma'_3}{2} \right) . 
 \end{equation}
The left-handed and right-handed Weyl fermions can be defined by imposing the
constraints, 
\begin{equation}
\psi_{-}(x) = \hat P_{-} \psi(x) ,  \quad \bar \psi_-(x) = \bar \psi(x) P_{+} , 
\end{equation}
\begin{equation}
\psi'_{+}(x) = \hat P'_{+} \psi'(x) ,  \quad \bar \psi'_+(x) = \bar \psi'(x) P_{-} .
\end{equation}
The action of the left-handed Weyl fermions is then given by 
\begin{equation}
S_W = \sum_{x\in\Gamma} \bar \psi_- (x) D  \psi_-(x)  + \sum_{x\in\Gamma} \bar \psi'_+ (x) D'  \psi'_+(x) . 
\end{equation}

The kernel of the lattice Dirac operator in finite volume, $D$ ($D'$),  may be represented 
through the kernel of the lattice Dirac operator in infinite volume, $D_\infty$, as follows:
\begin{equation}
\label{eq:DL-in-D-infity}
D(x,y) = D(x,y)_\infty + \sum_{n \in \mathbb{Z}^4, n \not = 0} D_\infty(x,y+n L) ,
\end{equation}
where $D_\infty(x,y)$ is defined with a periodic link field in  infinite volume. We assume that 
$D_\infty(x,y)$ posseses the locality property given by 
\begin{equation}
\label{eq:locality-D}
\| D_\infty(x,y) \|  \le C ( 1+\| x-y \|^p ) \, {\rm e}^{-\| x-y \| / \varrho}
\end{equation}
for some constants $\varrho > 0$, $C >0$ , $p \ge 0$, 
where  $\varrho$ is the localization range of the lattice Dirac operator. 

\subsection{Path-integral measure of the Weyl fermions}
\label{subsec:properties_of_measure_term}

The path-integral measure of the Weyl fermions may be defined by
the Grassmann integrations, 
\begin{equation}
{\cal D}[\psi_-] {\cal D}[\bar \psi_-]{\cal D}[\psi'_+] {\cal D}[\bar \psi'_+]  = \prod_j d c_j  \prod_k d \bar c_k  \prod_j d c'_j  \prod_k d \bar c'_k,   
\end{equation}
where $\{ c_j \}$, $\{ \bar c_k \}$ and  $\{ c'_j \}$, $\{ \bar c'_k \}$ are the Grassmann coefficients
in the expansion of the Weyl fields, 
\begin{equation}
\psi_-(x) = \sum_j  v_j(x) c_j , \quad  \bar \psi_-(x) = \sum_k \bar c_k \bar v_k(x) 
\end{equation}
\begin{equation}
\psi'_+(x) = \sum_j  u'_j(x) c'_j , \quad  \bar \psi'_+(x) = \sum_k \bar c'_k \bar u'_k(x) 
\end{equation}
 in terms of the chiral (orthonormal) bases defined by 
\begin{equation}
\hat P_{-} v_j(x) = v_j(x) ,    \quad 
\bar v_k (x) P_{+}  = \bar v_k (x) , 
\end{equation}
\begin{equation}
\hat P'_{+} u'_j(x) = u'_j(x) ,    \quad 
\bar u'_k (x) P_{-}  = \bar u'_k (x) .  
\end{equation}
Since the projection operators $\hat P_{-} $ and $\hat P'_{+} $ depend on the gauge field through $D$ and $D'$, respectively, 
the fermion measure also depends on the gauge field.  In this gauge-field dependence
of the fermion measure, there is  an ambiguity by a pure phase factor, because
any unitary transformations of the bases, 
\begin{equation}
\tilde v_j(x) = \sum_l v_l (x) \left(  {\cal Q}^{-1} \right)_{lj},  \qquad 
\tilde c_j = \sum_l   {\cal Q}_{jl}  c_l , 
\end{equation}
\begin{equation}
\tilde u'_j(x) = \sum_l u'_l (x) \left(  {\cal Q'}^{-1} \right)_{lj},  \qquad 
\tilde c'_j = \sum_l   {\cal Q'}_{jl}  c'_l , 
\end{equation}
induces a change of the measure by the pure phase factor $\det {\cal Q} \times \det {\cal Q'}$.
This ambiguity should be fixed
so that
it fulfills the fundamental requirements such as 
locality, 
gauge-invariance, integrability and lattice symmetries. 


\subsection{Reconstruction theorem of the fermion measure}

The effective action induced by the path-integration of the Weyl fermions  is given by
\begin{equation}
\label{eq:effective-action-W}
\Gamma [ U ] = \ln \{ \det  ( \bar v_k D v_j ) \, \det  ( \bar u'_k D' u'_j ) \}.
\end{equation}
Its variation with respect to the gauge field,
$\delta_\eta U(x,\mu) = i \eta_\mu(x) U(x,\mu)$, reads
\begin{eqnarray}
\label{eq:variation-effective-action-W}
\delta_\eta \Gamma [ U ] 
&=& {\rm Tr}\{ P_+ \delta_\eta D D^{-1}  \} + \sum_j  ( v_j , \delta_\eta v_j ) \nonumber\\
&+& {\rm Tr}\{ P_- \delta_\eta D' {D'}^{-1}  \}  + \sum_j  ( u'_j , \delta_\eta u'_j ) . 
\end{eqnarray}
Then the properties of the fermion measure can be characterized by the so-called measure term
which is given in terms of the chiral basis and its variation with respect to the gauge field as 
\begin{equation}
\label{eq:measure-term}
\mathfrak{L}_\eta = i \sum_j  ( v_j , \delta_\eta v_j ) + i \sum_j  ( u'_j , \delta_\eta u'_j ) . 
\end{equation}

The reconstruction theorem given in \cite{Luscher:1998du} asserts that 
if there exists a local current $j_\mu(x)$ which satisfies the following four 
properties, 
it is possible to reconstruct  the 
fermion measure (the bases $\{ v_j(x) \}$, $\{ u'_j(x) \}$) which depends smoothly on the gauge field  and
fulfills the fundamental requirements such as 
locality\footnote{We adopt the generalized notion of locality on the lattice given in \cite{Hernandez:1998et, Luscher:1998kn, Luscher:1998du} 
for Dirac operators and 
composite fields.  See also \cite{Kadoh:2003ii} for the case of the finite volume lattice.}, 
gauge-invariance, integrability and lattice symmetries\footnote{The lattice symmetries mean
translations, rotations, reflections and charge conjugation.}: 

\vspace{1em}
\noindent{\bf Theorem} \ \  {\sl Suppose $j_\mu(x)$ is a given current with the following 
properties\footnote{Throughout this paper, 
${\text Tr}\{\cdots\}$ stands for the trace over the lattice index $x$, 
the flavor index $\alpha (=1,\cdots,N)$ and the spinor index, while  
${\text tr}$ stands for the trace over the flavor and spinor indices only. 
${\text Tr}_L\{\cdots\}$ stands for the trace over the finite lattice, $x \in \Gamma$.}:

\begin{enumerate}
\item $j_\mu(x)$  is defined for all admissible gauge fields
and depends smoothly on the link variables.

\item $j_\mu(x)$ is gauge-invariant and transforms as an axial vector
  current under the lattice symmetries. 

\item The linear functional $\mathfrak{L}_\eta= \sum_{x\in \Gamma} \eta_\mu(x) j_\mu(x)$
is a solution of the integrability condition
\begin{equation}
\label{eq:integrability-condition}
\delta_\eta \mathfrak{L}_\zeta - \delta_\zeta \mathfrak{L}_\eta
= i {\rm Tr} \left\{ \hat P_- [ \delta_\eta \hat P_-, \delta_\zeta
\hat P_- ] \right\} 
+
i {\rm Tr} \left\{ \hat P'_+ [ \delta_\eta \hat P'_+, \delta_\zeta
\hat P'_+ ] \right\} 
\end{equation}
for all periodic variations $\eta_\mu(x)$ and $\zeta_\mu(x)$.

\item The anomalous conservation law holds: 
\begin{equation}
\label{eq:anomalous-conservation-law}
\partial_\mu^\ast j_\mu(x) 
=   {\rm tr}\{ Q \gamma_5(1-D)(x,x) \} - {\rm tr}\{ Q' \gamma_5(1-D')(x,x) \}, 
\end{equation}
where $Q=\text{diag}(q_1, \cdots, q_N)$ and $Q'=\text{diag}(q'_1, \cdots, q'_{N'})$. 

\end{enumerate}
Then there exists a smooth fermion integration measure in the vacuum sector such that
the associated current coincides with $j_\mu(x)$. The same is true in all other sectors
if the number of fermion flavors with $\vert q_\alpha \vert = q $ (or $\vert q'_{\alpha'} \vert = q $) is even for all odd $ q$. 
In each case the measure is uniquely determined up to a constant phase factor. 
}

In \cite{Luscher:1998du}, it is proved constructively that there exists 
a {\it local} current $j_\mu(x)$ which satisfies the properties required in the reconstruction theorem.  
In fact, the construction of the current  is not straightforward by two reasons. 
The first reason is that 
the locality property of the current must be maintained. 
The second reason is that 
the measure term must be smooth w.r.t. the gauge field, but 
the topology of the space of the admissible gauge fields in finite volume is not trivial. 
To take these points into account, the construction in \cite{Luscher:1998du} is made 
by sperating the part definable in infinite volume from the part of the finite volume corrections. 
Then, 
the explicit formula of the measure term turns out to be complicated.  
Therefore it does not provide a formulation which is immediately usable for practical numerical applications. 

In \cite{Kadoh:2007wz}, 
by formulating the procedure to solve the local cohomology problem of the U(1) gauge (chiral) anomaly within finite volume, 
a rather explicit formula of the local current $j_\mu(x)$ is derived as 
\begin{eqnarray}
\label{eq:average-L-diamond}
\sum_{x\in \Gamma} \eta_\mu(x) j_\mu(x)
&\equiv& \frac{1}{2^2 2!} \sum_{R \in  O(2,\mathbb{Z})} \det R \, \, 
 \mathfrak{L}^\diamond_{\eta}\vert_{U\rightarrow\{U\}^{R^{-1}},  
         \eta_\mu\rightarrow\{\eta_\mu \}^{R^{-1}} } , 
\end{eqnarray}
where
\begin{eqnarray}
\label{eq:measure-term-finite-volume}
\mathfrak{L}_\eta^\diamond &=& i \int_0^1 ds \, 
{\rm Tr} \left\{ \hat P_- [ \partial_s \hat P_-, 
\delta_\eta \hat P_- ] \right\} \Big\vert_{\tilde A_\mu \rightarrow s \tilde A_\mu}
\nonumber\\
&+& 
\delta_\eta \, 
\int_0^1 ds \, \sum_{x \in \Gamma} 
 \left\{   \tilde A_\mu^\prime(x)  \,   k_\mu(x) \right\}   
+{\mathfrak{W}}_\eta \vert_{U=U_{[w]} V_{[m]}, \eta=\eta_{[w]}}.
\end{eqnarray}
In this formula, 
the link field $U(x,\mu)$ in $\mathfrak{U}[m]$ is represented as
\begin{equation}
\label{eq:U-in-AT}
U(x,\mu) =
 {\rm e}^{i A^T_\mu(x)} \, \, 
U_{[w]}(x,\mu) 
 \, \Lambda(x)  \, \Lambda(x+\hat\mu)^{-1} \,\,
V_{[m]}(x,\mu)
, 
\end{equation}
where 
$A^T_\mu(x)$ is the transverse vector potential in satisfying 
\begin{eqnarray}
&&\partial_\mu^\ast  A^T_\mu(x) = 0, \qquad
     \sum_{x\in \Gamma} A^T_\mu(x) = 0, \\
&& 
\partial_\mu A^T_\nu(x)-\partial_\nu A^T_\mu(x) + 2\pi m_{\mu\nu}/L^2 
= F_{\mu\nu}(x) ,  
\end{eqnarray}
$U_{[w]}(x,\mu)$ represents the degrees of freedom of  the Wilson lines, 
\begin{equation}
\label{eq:link-field-wilson-lines}
U_{[w]}(x,\mu) = \left\{ 
\begin{array}{cl}
w_\mu
&\quad  \text{if $x_\mu = L-1$},  \\
1           &\quad \text{otherwise,  }  
\end{array} \right.
\end{equation}
with the phase factor $w_\mu \in U(1)$ and 
$\Lambda(x)$ is the gauge function satisfying $\Lambda(0)=1$. 
$\tilde A_\mu(x)$ is then defined by 
\begin{eqnarray}
\tilde A_\mu(x) &=& A_\mu^T(x) - \frac{1}{i}\partial_\mu \Big[ \ln \Lambda(x) \Big] ; \qquad
\frac{1}{i} \ln \Lambda(x) \in (-\pi, \pi] .
\end{eqnarray}
$k_\mu(x)$ is the gauge-invariant local current which satisfies
\begin{equation}
\partial_\mu^\ast k^{}_\mu(x) =  {\rm tr}\{ Q \gamma_5(1-D)(x,x) \} - {\rm tr}\{ Q' \gamma_5(1-D')(x,x) \} 
\end{equation}
and 
transforms as an axial vector field under the lattice symmetries. 
${\mathfrak{W}}_\eta \vert_{U=U_{[w]} V_{[m]}, \eta=\eta_{[w]} }$ is the additional measure term 
at the gauge field $U(x,\mu)=U_{[w]}(x,\mu) V_{[m]}(x,\mu)$
with the variational parameters in the directions of the Wilson lines, 
$\eta_{\mu [w]}(x)=\sum_{\nu} \delta_{\mu\nu} \delta_{x_\nu, L-1} \eta_{(\nu)}$ 
%
%
Using these formula, it is indeed feasible to compute numerically  the gauge-field dependence of the Weyl fermion measure in two-dimensions.\cite{Kikukukawa:2005ILFTN}

\section{Mirror-fermion approach with the Ginsperg-Wilson fermions}
\label{sec:mirror-fermion-approach-in-general}

In this section, 
we review 
the mirror-fermion approach\cite{Montvay:1987ys, Montvay:1988av, Farakos:1990ex, Lin:1990ue, Lin:1990vi, Munster:1991xs, Lin:1991csa, Montvay:1992mv, Montvay:1992eg, Lin:1992qb, Lin:1993hp} with the Ginsparg-Wilson fermions\cite{Bhattacharya:2006dc,Giedt:2007qg,Poppitz:2007tu,Poppitz:2008au,Poppitz:2009gt,Poppitz:2010at,Chen:2012di,Giedt:2014pha} to lattice models of two-dimensional
abelian chiral gauge theories. 
%

In the mirror fermion approach 
in the framework of the Ginsparg-Wilson fermions,
the opposite chirality fields $\psi_+(x)$ and $\psi'_-(x)$ are also considered:
\begin{equation}
\psi_{+}(x) = \hat P_{+} \psi(x) ,  \quad \bar \psi_+(x) = \bar \psi(x) P_{-} , 
\end{equation}
\begin{equation}
\psi'_{-}(x) = \hat P'_{-} \psi'(x) ,  \quad \bar \psi'_-(x) = \bar \psi'(x) P_{+} .
\end{equation}
These fields, which are referred as mirror fermions,  are assumed to be dynamical, but to be decoupled
by acquiring the masses of order the inverse lattice spacing $1/a$ through the dynamical effect of certain (gauge-invariant) local interactions
among the mirror fermion fields and  additional auxiliary boson fields.
The action of the mirror sector is then given in the form 
\begin{eqnarray}
S_M &=& \sum_{x\in\Gamma}
\big\{ \bar \psi_+ (x) D  \psi_+(x) 
      + \bar \psi'_- (x) D'  \psi'_-(x)  \big\}
\nonumber\\
&+&
\sum_{x \in \Gamma} {\cal V}\big( \psi_+(x), \bar \psi_+(x), \psi'_-(x), \bar \psi'_-(x), \Phi(x), U(x,\mu) \big)
+
\sum_{x \in \Gamma} \kappa \, \big\vert \nabla \Phi(x) \big\vert^2,
\end{eqnarray}
where $\Phi(x)$ stands for the additional boson fields collectively,
and the total action of the lattice model is assumed to be
\begin{equation}
S_{\rm mirror} = S_G  + S_W + S_M .
\end{equation}

The path-integral measures of the mirror fermion fields  may be defined by
\begin{equation}
{\cal D}[\psi_+] {\cal D}[\bar \psi_+]{\cal D}[\psi'_-] {\cal D}[\bar \psi'_-] 
= \prod_j d b_j  \prod_k d \bar b_k  \prod_j d b'_j  \prod_k d \bar b'_k   , 
\end{equation}
where $\{ b_j \}$, $\{ \bar b_k \}$ and  $\{ b'_j \}$, $\{ \bar b'_k \}$ are the grassman coefficients
in the expansion of the mirror fermion fields, 
\begin{equation}
\psi_+(x) = \sum_j  u_j(x) b_j , \quad  \bar \psi_+(x) = \sum_k \bar b_k \bar u_k(x) 
\end{equation}
\begin{equation}
\psi'_-(x) = \sum_j  v'_j(x) b'_j , \quad  \bar \psi'_-(x) = \sum_k \bar b'_k \bar v'_k(x) 
\end{equation}
 in terms of the chiral (orthonormal) bases defined by 
\begin{equation}
\hat P_{+} u_j(x) = u_j(x) ,    \quad 
\bar u_k (x) P_{-}  = \bar u_k (x) , 
\end{equation}
\begin{equation}
\hat P'_{-} v'_j(x) = v'_j(x) ,    \quad 
\bar v'_k (x) P_{+}  = \bar v'_k (x) .  
\end{equation}
On the other hand, since the target Weyl fermions and the mirror fermions
now consist the Dirac pairs (in the sense of the Ginsparg-Wilson fermions) as $\psi = \psi_- + \psi_+$, $\psi' = \psi'_+ + \psi'_-$,
the path-integral measures of the fermion fields can be defined simply by 
\begin{equation}
{\cal D}[\psi] {\cal D}[\bar \psi]{\cal D}[\psi'] {\cal D}[\bar \psi']  = 
\prod_{x,s,\alpha}  \,  d \psi_{s \alpha } (x)  d \bar \psi_{s \alpha } (x)  
\prod_{x,s',\alpha'}  \, d \psi'_{s' \alpha'}(x)  d \bar \psi'_{s' \alpha'}(x) , 
\end{equation}
which are independent of the gauge fields and are manifestly gauge invariant.
This fact implies that one can always choose the bases of the Dirac fields,  $ \{ u_j(x), v_j(x) \}$ and $\{ u'_j(x), v'_j(x) \}$, 
so that the Jacobian factors,  $\det (u_j(x), v_j(x) )$ and $\det (u'_j(x), v'_j(x) )$, are independent of the gauge fields: 
\begin{eqnarray}
\label{eq:uv-u'v'-relations}
&&  \sum_j ( u_j, \delta_\eta u_j) + \sum_j ( v_j, \delta_\eta v_j)  = 0 ,  \nonumber\\
&&  \sum_j ( u'_j, \delta_\eta u'_j) + \sum_j ( v'_j, \delta_\eta v'_j)  = 0 .
\end{eqnarray}
Adjusting the overall constant phase factors of the Jacobians,
one obtain
\begin{equation}
{\cal D}[\psi_-] {\cal D}[\bar \psi_-]{\cal D}[\psi'_+] {\cal D}[\bar \psi'_+] \times 
{\cal D}[\psi_+] {\cal D}[\bar \psi_+]{\cal D}[\psi'_-] {\cal D}[\bar \psi'_-] 
=
{\cal D}[\psi] {\cal D}[\bar \psi]
{\cal D}[\psi'] {\cal D}[\bar \psi'] .
\end{equation}
This factorization of the path-integral measure, as well as the action, into the target and mirror sectors is the characteristic feature of the mirror Ginsperg-Wilson fermion approach.\cite{Poppitz:2007tu}\footnote{These sectors are not completely independent each other with respect to the coupling to the gauge link fields because of the constraints, eqs.~(\ref{eq:uv-u'v'-relations}). }
For later convenience, we introduce the abbreviations for the path-integrations of the parts of 
the target-sector and the mirror-sector fields as follows:
\begin{eqnarray}
 \big\langle  {\cal O}_{W}  \big\rangle_{W}  
 &\equiv& 
 \int 
{\cal D}[\psi_-] {\cal D}[\bar \psi_-]{\cal D}[\psi'_+] {\cal D}[\bar \psi'_+] \, 
 {\rm e}^{-S_W}  \, {\cal O}_{W} , 
\\
 \big\langle  {\cal O}_{M} \big\rangle_{M}  
 &\equiv& 
 \int 
{\cal D}[\psi_+] {\cal D}[\bar \psi_+]{\cal D}[\psi'_-] {\cal D}[\bar \psi'_-] 
{\cal D}[\Phi] \,
 {\rm e}^{-S_M}  \, {\cal O}_{M} ,
\end{eqnarray}
and
\begin{eqnarray}
 \big\langle  {\cal O}_{WM}  \big\rangle_{WM}  
 &\equiv& 
 \int {\cal D}[\psi] {\cal D}[\bar \psi] {\cal D}[\psi'] {\cal D}[\bar \psi']  {\cal D}[\Phi] \, {\rm e}^{-S_W-S_M}  \,
 {\cal O}_{WM}. 
\end{eqnarray}
In the last formula, 
the result of the path-integration is independent of the choice of the chiral bases. To make this fact clear, 
$S_W$ and $S_M$ may be represented simply with the Dirac fields $\psi(x)$, $\psi'(x)$ as 
\begin{eqnarray}
\label{eq:S_W}
S_W &=& \sum_{x\in\Gamma} \bar \psi (x) P_+ D  \psi(x)  + \sum_{x\in\Gamma} \bar \psi' (x) P_- D'  \psi'(x) ,  \\
S_M &=& \sum_{x\in\Gamma}
\big\{ \bar \psi (x) P_- D  \psi (x) 
      + \bar \psi' (x) P_+ D'  \psi'(x)  \big\}
\nonumber\\
&+&
\sum_{x \in \Gamma} {\cal V}\big( \hat P_+  \psi(x), \bar \psi(x) P_- , \hat P'_- \psi'(x), \bar \psi'(x) P_+, \Phi(x), U(x,\mu) \big)
+
\sum_{x \in \Gamma} \kappa \, \big\vert \nabla \Phi(x) \big\vert^2. \nonumber\\
\end{eqnarray}
With these abbreviations, the factorization means 
$
 \big\langle  {\cal O}_{W} {\cal O}_{M}  \big\rangle_{WM} 
=
\big\langle  {\cal O}_{W}  \big\rangle_{W}
\,
\big\langle  {\cal O}_{M}  \big\rangle_{M}
$.

The effective action induced by the path-integration of the mirror-sector fields as well as the target Weyl fermions is then represented by 
\begin{eqnarray}
\Gamma_{\rm mirror} \big[ U \big]  &=&  \ln \big\{ \big\langle  1 \big\rangle_{WM} \big\}, 
\end{eqnarray}
Its variation with respect to the gauge field reads
\begin{eqnarray}
\delta_\eta \Gamma_{\rm mirror} [ U ] 
&=&  \left\{  \big\langle - \delta_\eta S_W  \big\rangle_{WM}   +  \big\langle - \delta_\eta S_M  \big\rangle_{WM}  \right\} 
\slash  \big\langle  1  \big\rangle_{WM}
\nonumber\\
&=&
\big\langle - \delta_\eta S_W  \big\rangle_{W} 
\slash  \big\langle  1  \big\rangle_{W}
+  
\big\langle - \delta_\eta S_M  \big\rangle_{M} 
\slash  \big\langle  1  \big\rangle_{M}
\nonumber\\
&=& {\rm Tr}\{ P_+ \delta_\eta D D^{-1}  \}     
+ {\rm Tr}\{ P_- \delta_\eta D' {D'}^{-1}  \}  
+  \big\langle  - \delta_\eta S_M  \big\rangle_{M} \slash  \big\langle  1  \big\rangle_{M} . 
\end{eqnarray}
By comparing this result 
with Eqs.~(\ref{eq:variation-effective-action-W}) and (\ref{eq:measure-term}), one can see that the contribution of the mirror sector, 
$\big\langle  - \delta_\eta S_M  \big\rangle_{M} \slash  \big\langle  1  \big\rangle_{M}$, 
should play the role of the measure term $\mathfrak{L}_\eta$. 

If the mirror-sector fields could successfully decouple
by acquiring the masses of order $1/a$, 
all these fields should have the short range correlation lengths
of order multiple lattice spacings. Moreover,
these fields should leave only local terms of the gauge fields
in the induced effective action, 
according to the decoupling theorem (and from the more general point of view of the Wilsonian renormalization group).
This implies that the contribution of the mirror sector, 
$ \big\langle - \delta_\eta S_M  \big\rangle_{M} \slash  \big\langle  1  \big\rangle_{M} $, should be a local function of the gauge fields.
In the weak gauge-coupling expansion, the vertex functions are derived from this contribution as
\begin{eqnarray}
&& \big\langle - \delta_\eta S_M  \big\rangle_{M} \slash  \big\langle  1  \big\rangle_{M} 
\nonumber\\
&&=
\sum_{m=0}^\infty
\frac{1}{L^{2+2m}} \, \frac{1}{m!}\, 
\sum_{k,p_1,\cdots,p_m} \tilde \eta_\mu(-k)\,
 \tilde \Gamma'_{\mu \nu_1, \cdots, \nu_m} (k,p_1,\cdots,p_m ) \,
 \tilde A_{\nu_1}(p_1)\cdots  \tilde A_{\nu_m}(p_m) 
\nonumber\\
\end{eqnarray}
and they should be reguler(analytic) w.r.t. the external momenta.
These conditions are indeed consistent with the requirement of the locality properties of the measure-term in the reconstruction theorem.

In order to  achieve the above situation, one important requirement about the fermion symmetries of the mirror-sector action  
follows from the consideration of 't Hooft anomaly matching condition.\cite{Eichten:1985ft,Bhattacharya:2006dc,Wang:2013yta}
{\em If there exists a global continuous fermion symmetry in $S_M$, 
it must be free from the ``would-be gauge anomaly'', i.e. 
that global symmetry can be gauged successfully without encountering  gauge anomalies.}
This is because the ``would-be gauge anomaly'' implies an IR singularity in the  (gauge-invariant)
correlation function of the  symmetry currents and it in turn
implies certain massless states in the spectrum of the model, so that they can saturate the IR  singularity. 
This contradicts the required situation.

\section{345 model in the mirror-fermion approach
}
\label{sec:3450-model}

In this section, we first review the 345 model in the
mirror fermion approach with the Ginsparg-Wilson fermions\cite{Bhattacharya:2006dc,Giedt:2007qg,Poppitz:2007tu,Poppitz:2008au,Poppitz:2009gt,Poppitz:2010at,Chen:2012di,Giedt:2014pha},
which is formulated by introducing all possible
Dirac- and Majorana-Yukawa couplings with XY-spin field
to break the global continuous symmetries of the mirror sector. 
We next examine why the attempt seems a ``Mission impossible'' in the 345 model.
We point out that the effective fermionic operators to break 
the 
symmetries U(1)$_f$ and U(1)$_{f'}$ 
in the mirror sector do not have sufficiently strong couplings even in the limit of large Majorana-Yukawa couplings. 
We observe also that the type of Majorana mass term considered there is singular in the large limit due to the nature of the chiral projection of the Ginsparg-Wilson fermions,  but a slight modification without such singularity is allowed by virtue of the very nature.


\subsection{345 model}
The 345 model is defined by the charge assignment of the U(1) gauge symmetry as\footnote{One should note that
this charge assignment does not satisfy
the assumption of the reconstruction theorem
that
the number of fermion flavors with $\vert q_\alpha \vert = q $ (or $\vert q'_{\alpha'} \vert = q $) is even for all odd $ q$. 
}
\begin{equation}
Q={\rm diag}(q_1, q_2) ={\rm diag}(3, 4) ,\quad
Q'={\rm diag}(q'_1, q'_2) ={\rm diag}(5, 0) .
\end{equation}
The neutral fermion is introduced as a spectator.
Let us index the components of the Weyl fields by their charges, $q$, $q'$ as 
\begin{eqnarray}
&& 
\psi_- = ( \psi_{3 -}, \psi_{4 -} ) , \quad 
\bar \psi_- = ( \bar\psi_{3 -}, \bar\psi_{4 -} ) , \\
&& 
\psi'_+
= ( \psi_{5 +}, \psi_{0 +}) , \quad
\bar\psi'_+ = ( \bar\psi_{5 +}, \bar\psi_{0 +}).
\end{eqnarray}
We also specify the representations of the gamma matrices by the Pauli matrices as
$\gamma_1=\sigma_1, \gamma_2=\sigma_2, \gamma_3 =\sigma_3$,
and of the charge conjugation operator as $c_D = i \gamma_2$.

Accordingly, 
let us index the components of 
the Mirror fermion fields as
\begin{eqnarray}
&& 
\psi_+ = ( \psi_{3 +}, \psi_{4 +} ) , \quad 
\bar \psi_+ = ( \bar\psi_{3 +}, \bar\psi_{4 +} ) , \\
&& 
\psi'_- = ( \psi_{5 -}, \psi_{0 -}) , \quad
\bar\psi'_- = ( \bar\psi_{5 -}, \bar\psi_{0 -}).
\end{eqnarray}
Without interaction, the fermionic symmetries of the mirror-sector 
are as listed in the table \ref{table:345-anomaly}.
\begin{table}[bht]
\begin{center}
\begin{tabular}{|c||c|c|c|c|l|l|}
\hline
 & $+$ & $+$ & $-$ & $-$ & gauge anomaly & chiral anomaly \\ \hline
U(1)$_g$ & 3 & 4 & 5 & 0 & matched (gauged)& ---\\
U(1)$_b$ & 2 & 1 & 2 & 1 & matched (can be gauged) & anomaly free \\
U(1)$_f$& 1 & 1 & 1 & 0 & not matched & anomalous \\
U(1)$_{f'}$& 0 & 0 & 0 & 1 & not matched & anomaly free (can be anomalous) \\ \hline
\end{tabular}
\end{center}
\caption{Fermionic continuous symmetries in the mirror sector
of the 345 model and their would-be gauge anomalies}
\label{table:345-anomaly}
\end{table}

\vspace{1em}
\noindent
The two types of gauge-invariant local operators 
\begin{equation}
O_{f^{}} = 
\big( \psi_{3+} \big)^1  \, 
\big(\psi_{4+} \big)^3 \, 
\big( \bar \psi_{5-} \big)^3  \, 
\psi_{0-}, \qquad
O_{f'} = 
\big( \psi_{3+} \big)^2  
\psi_{4+} 
\big( \bar \psi_{5-} \big)^2 
\bar \psi_{0-}
\end{equation}
can break the symmetries, U(1)$_f$ and U(1)$_{f'}$.  
The product of these operators,
\begin{eqnarray}
O_{T} &=& O_{f^{}} \, O_{f'} 
=
\big( \psi_{3+} \big)^3  \, \big(\psi_{4+} \big)^4 \, 
\big( \bar \psi_{5-} \big)^5  \, \bar \psi_{0-}  \psi_{0-}, 
\end{eqnarray}
involves the 't Hooft vertex which can induced by the U(1) instantons in two-dimensions.
\footnote{One can also include
another operator, 
$O_{f^{''}} = 
\big( \psi_{3+} \big)^1  \, 
\big(\bar \psi_{4+} \big)^2 \, 
\big( \bar \psi_{5-} \big)^1 \, 
\big( \bar \psi_{0-} \big)^2 $,
which breaks only U(1)$_{f'}$.
}
To define the actual operators, one needs the point-splitting procedure because of the fermi statistics. 
A possible choice is the following: 
\begin{eqnarray}
O_{f}(x)&=& 
 \psi_{0-}(x) c_D \psi_{3+}(x) \, 
\bar \psi_{5-}(x) \psi_{4+}(x) \,
\big\{ \square \, \big( \bar \psi_{5-}(x) \psi_{4+}(x) \big)
\big\} ^{2} , 
\\
O_{f'}(x) &=&
\bar \psi_{0-}(x) \psi_{3+}(x) \, 
\bar \psi_{5-}(x) \psi_{4+}(x) \,
\square \big( \bar \psi_{5-}(x) \psi_{3+}(x) \big), 
\\
O_{T}(x)&=& 
\bar \psi_{0-}(x) \psi_{3+}(x) \, 
\bar \psi_{5-}(x) \psi_{4+}(x)  \times \nonumber \\
&& 
\square \, \big( \bar \psi_{5-}(x) \psi_{3+}(x) \big) \, 
\big\{ \square \, \big( \bar \psi_{5-}(x) \psi_{4+}(x) \big)
\big\} ^{3} \, 
\square \, \big( \psi_{0-}(x) c_D \psi_{3+}(x) \big) ,
\end{eqnarray}
where
\begin{equation}
\square \, O_q (x) \equiv 
\sum_\mu
\big( 
U(x,\mu)^{q} O_q(x+\hat\mu) + 
U(x-\hat\mu,\mu)^{-q} O_q(x-\hat\mu)
- 2  O_q(x)
\big).
\end{equation}


\subsection{Mirror sector of the 345 model with Dirac- and Majorana-type Yukawa couplings to XY spin field}

In \cite{Bhattacharya:2006dc,Giedt:2007qg,Poppitz:2007tu,Poppitz:2008au,Poppitz:2009gt,Poppitz:2010at,Chen:2012di,Giedt:2014pha}, the 345 model is formulated by introducing 
all possible Dirac- and Majorana-type Yukawa couplings to the XY spin field in order to break the global continuous symmetries of the mirror sector. 
\begin{eqnarray}
S_M &=& \sum_{x\in\Gamma}\, z
\big\{ \bar \psi_+ (x) D  \psi_+(x) 
      + \bar \psi'_- (x) D'  \psi'_-(x)  \big\}
\nonumber\\
&+&
\sum_{x \in \Gamma}
\sum_{q,q'}
\big\{
y_{q q'} \,
\bar \psi_{q +}(x) \psi'_{q' -}(x) \, 
\phi(x)^{q - q'}
+
y_{q' q} \,
\bar \psi'_{q' -}(x) \psi_{q +}(x) \, 
\phi(x)^{q'- q }
\big\}
\nonumber\\
&+&
\sum_{x \in \Gamma}
\sum_{q,q'}
\big\{
\bar h_{q q'} \,\bar \psi_{q +}(x) \,c_D\, \bar\psi'_{q' -}(x) \, 
\phi(x)^{q + q'}
+
h_{q q'} \,\psi_{q +}(x) \,c_D\, \psi'_{q' -}(x) \, 
\phi(x)^{-q - q'}
\big\}
\nonumber\\
&+&
 \sum_{x \in \Gamma, \mu} \frac{\kappa}{2}
\big\{\,
2 
-\phi(x)^{-1} U(x,\mu) \phi(x+\hat \mu) 
-\phi(x+\hat \mu)^{-1} U(x,\mu)^{-1} \phi(x)
\big\} ,
\end{eqnarray}
where $y_{q q'}^\ast = y_{q' q}$ and $h_{q q'}^\ast = \bar h_{q q'}$ for hermeticity. 
The path-integral measure of the XY spin field is defined by
\begin{eqnarray}
{\cal D}[\phi]
&=&
\prod_{x}\delta( |\phi(x)|-1) d\phi(x) d\phi^*(x) / 2i  .
\end{eqnarray}
A comment is in order about our conventions. 
There are several differences in the conventions from those in the original works\cite{Bhattacharya:2006dc,Giedt:2007qg,Poppitz:2007tu,Poppitz:2008au,Poppitz:2009gt,Poppitz:2010at,
Chen:2012di,Giedt:2014pha}. 
First of all, the definition of the chiral projection of the Ginsparg-Wilson fermions is opposite: here $\hat \gamma_3 =\gamma_3(1-2D)$ is used for the field and  $\gamma_3$ for the anti-field (as usual).
Secondly, the Majorana-Yukawa couplings are defined here with $c_D = i \gamma_2$, but not $\gamma_2$. 
(Our choice of
the representation of the Dirac gamma matrices
in the Euclidean metric 
is specified 
as $\gamma_0 = \sigma_1, \gamma_1 = \sigma_2$, $\gamma_3=\sigma_3$.)
Therefore, it is $i h_{q q'} (-i \bar h_{q q'})$ which corresponds to the coupling 
$h_{q q'}$ in the original 
works.
Thirdly, 
the chirality assignments of the target Weyl fermions and the mirror fermions is opposite: here $(3-,4-,5+,0+)$ for the target Weyl fermions, while $(3+,4+,5-,0-)$ for the mirror fermions. 

The model was studied in the parameter regions where $\kappa$ is small and $y_{q q'}$, $h_{q q'}$ are large compared to $z$, so that the model is within the so-called PMS (paramagnetic strong-coupling) phase where the XY spin field is disordered,
and the fermion fields form certain bound states (with the XY spin field or among the fermion fields) and acquire masses in the manner consistent with the chiral gauge invariance. The typical values of the coupling constants
are listed in table~\ref{fig:345-DMY-parameters}, which are the values used in the latest numerical study 
in \cite{Chen:2012di,Giedt:2014pha}.
\begin{table}[bth]
\begin{center}
\begin{tabular}{|c|c|cccc|cccc|}
\hline 
 $\kappa$ & $z$ & $y_{35}$ & $y_{30}$ & $y_{45}$ & $y_{40}$ & 
          $ i h_{35}$ & $i h_{30}$ & $i h_{45}$ & $i h_{40}$ \\ \hline
0.5 & 0.0 & 1.0 & 1.0 & 1.0 & 1.0 & 3.00278 & 30.3214 & 23.7109 & 3.08123
\\ \hline
\end{tabular}
\end{center}
\caption{The values of the coupling constants}
\label{fig:345-DMY-parameters}
\end{table}
It was then claimed through the Monte-Carlo studies that, although the XY spin field and the mirror fermion fields both have short correlation lengths indeed in the parameter 
region of their choice, 
the two-point vertex function of the (external) gauge field in the mirror sector $ \tilde \Pi'_{\mu \nu} (k)$, which is defined by
\begin{eqnarray}
\label{eq:two-point-vertex-U1-gauge-field}
\frac{1}{L^2}\sum_{k} \tilde \eta_\mu(-k)\, \tilde \Pi'_{\mu \nu} (k) \, \tilde \zeta_\nu(k) &=&
\delta_\zeta 
\left[\big\langle  - \delta_\eta S_M  \big\rangle_{M} \slash  \big\langle  1  \big\rangle_{M} 
\right]\, \Big\vert_{U(x,\mu) \rightarrow 1} , 
\end{eqnarray}
shows a singular non-local behavior there. It was also shown that the normalization of the singular term matches well with that of the target Weyl fermion fields. This singularity implies that there remains certain massless states in the mirror sector which are charged under the gauged U(1), and
the model 
looks vector-like, where both the target Weyl fermions and the mirror fermions remain massless and couples to the U(1) gauge field.

\subsection{Why is the mission impossible in the 345 model with Dirac- and Majorana-type Yukawa couplings to XY spin field ?}

We now examine why the attempt seems a ``Mission impossible'' in the 345 model with the Dirac- and Majorana-Yukawa couplings with XY-spin field. 
We will point out that 
the effective operators to break the fermion number symmetries U(1)$_f$ and U(1)$_{f'}$ 
in the mirror sector do not have sufficiently strong couplings even in the limit of large Majorana-Yukawa couplings. 
We will observe also that the type of Majorana mass term considered there is singular in the large limit due to the nature of the chiral projection of the Ginsparg-Wilson fermions,  but a slight modification without such singularity is allowed by virtue of the very nature.

\subsubsection{Strength of the effective fermionic operators
to break U(1)$_f$ and U(1)$_{f'}$}

Let us consider to evaluate the partition function of the mirror fermion sector
by 
the weak-coupling expansion w.r.t. to
$\kappa$, assuming $\kappa \ll 1$,
\begin{eqnarray}
 \big\langle  1 \big\rangle_{M}  
 &\equiv& 
 \int 
{\cal D}[\psi_+] {\cal D}[\bar \psi_+]{\cal D}[\psi'_-] {\cal D}[\bar \psi'_-] 
{\cal D}[\phi] \,
 {\rm e}^{-S_M} , 
\end{eqnarray}
where the original action of the mirror fermion sector, $S_M$, 
is given by
\begin{eqnarray}
S_{M} &=& \sum_{x} {\cal L}(x) + \kappa S_{B} ,
%
\end{eqnarray}
\begin{eqnarray}
{\cal L}(x) &=& 
z
\left\{  
\bar \psi_{3 +}(x) D_3 \psi_{3 +}(x)
+
\bar \psi_{4 +}(x) D_4 \psi_{4 +}(x)
+
\bar \psi_{5 -}(x) D_5 \psi_{5 -}(x)
+
\bar \psi_{0 -}(x) D_0 \psi_{0 -}(x)
\right\} 
\nonumber\\
&+&
\left\{\,\,\,\,
y_{35} \,\bar \psi_{3 +}(x) \psi_{5 -}(x) \, \phi(x)^{-2}
+
y_{53} \,\bar \psi_{5 -}(x) \psi_{3 +}(x) \, \phi(x)^{2}
\right. \nonumber\\
&& \left.\qquad
+
y_{30} \,\bar \psi_{3 +}(x) \psi_{0 -}(x) \, \phi^{}(x)^{3}\,\,
+
y_{03} \,\bar \psi_{0 -}(x) \psi_{3 +}(x) \, \phi(x)^{-3}
\right. \nonumber\\
&& \left.\qquad
+
y_{45} \,\bar \psi_{4 +}(x) \psi_{5 -}(x) \, \phi(x)^{-1}
+
y_{54} \,\bar \psi_{5 -}(x) \psi_{4 +}(x) \, \phi(x)^{1}
\right. \nonumber\\
&& \left.\qquad
+
y_{40} \,\bar \psi_{4 +}(x) \psi_{0 -}(x) \, \phi(x)^{4}
+ 
y_{04} \,\bar \psi_{0 -}(x) \psi_{4 +}(x) \, \phi(x)^{-4} \,\,
\right\}
\nonumber\\
&+&
\left\{\,\,\,\,
\bar h_{35} \,\bar \psi_{3 +}(x) \,c_D\, \bar\psi_{5 -}(x) \, \phi(x)^{8}
+
h_{35} \,\psi_{3 +}(x) \,c_D\, \psi_{5 -}(x) \, \phi(x)^{-8}
\right. \nonumber\\
&& \left.\qquad
+
\bar h_{30} \,\bar \psi_{3 +}(x) \,c_D\,  \bar\psi_{0 -}(x) \, \phi(x)^{3}
+
h_{30} \,\psi_{3 +}(x) \,c_D\, \psi_{0 -}(x) \, \phi(x)^{-3}
\right. \nonumber\\
&& \left.\qquad
+
\bar h_{45} \,\bar \psi_{4 +}(x)\,c_D\, \bar \psi_{5 -}(x) \, \phi^{}(x)^{9}
+
h_{45} \,\psi_{4 +}(x)\,c_D\,\psi_{5 -}(x) \, \phi(x)^{-9}\,\,
\right. \nonumber\\
&& \left.\qquad
+
\bar h_{40} \,\bar \psi_{4 +}(x) \,c_D\, \bar \psi_{0 -}(x) \, \phi(x)^{4}
+
h_{40} \,\psi_{4 +}(x) \,c_D\, \psi_{0 -}(x) \, \phi(x)^{-4} \,\,
\right\} ,
\end{eqnarray}
\begin{eqnarray}
S_{B} &=&
\sum_{x, \mu} 
\frac{1}{2}
\left\{\,
2 
-\phi(x)^{-1} U(x,\mu) \phi(x+\hat \mu) 
-\phi(x+\hat \mu)^{-1} U(x,\mu)^{-1} \phi(x)
\right\} ,
\end{eqnarray}
using the abbreviation for the overlap Dirac operator which act on the Dirac field of charge $q$ as
$D_q = D[ U(x,\mu)^q ]$.
In the hopping parameter expansion, one can perform the path-integration of the XY-spin field first and 
formulate the fermionic effective action by the relation, 
\begin{eqnarray}
%
 \big\langle  1 \big\rangle_{M}  
&=& 
 \int 
{\cal D}[\psi_+] {\cal D}[\bar \psi_+]
{\cal D}[\psi'_-] {\cal D}[\bar \psi'_-] 
{\cal D}[\phi] \,
 {\rm e}^{-\sum_x {\cal L}(x)} \,
{\scriptstyle 
\sum_{k=0}^\infty \frac{1}{k !} \big( -\kappa S_B \big)^k }
\nonumber\\
&=&
 \int 
{\cal D}[\psi_+] {\cal D}[\bar \psi_+]{\cal D}[\psi'_-] {\cal D}[\bar \psi'_-] \,
 {\rm e}^{-S'_M} ,
\end{eqnarray}
where the fermionic effective action, $S'_M$, is defined by
the expansion w.r.t. $\kappa$ as
\begin{equation}
S'_M = \sum_{k=0}^\infty  \kappa^k S'_{\rm M}{}^{(k)} .
\end{equation}
In this respect, we note that
the path-integration measure of the chiral anti-fields,
$\bar \psi_{3+}(x)$, $\bar \psi_{4+}(x)$, $\bar \psi_{5-}(x)$
and $\bar \psi_{0-}(x)$,
can be defined as 
\begin{equation}
\prod_{x} 
d \bar \psi_{3+}(x) 
d \bar \psi_{4+}(x)
d \bar \psi_{5-}(x)
d \bar \psi_{0-}(x)
\end{equation}
and it projects out 
the local composite operators 
which include each chiral anti-fields,
$\bar \psi_{3+}(x)$, $\bar \psi_{4+}(x)$,  $\bar \psi_{5-}(x)$, 
$\bar \psi_{0-}(x)$, just once
from 
the products of the Lagrangian density
$\big\{- \sum_x {\cal L}(x) \big\}^l / l !$ $(l=4, \cdots)$.
Furthermore, 
the path-integration of the XY-spin field, $\phi(x)$, 
projects out the composite operators 
which do not include $\phi(x)$ and are neutral w.r.t. the U(1) charge.

In the leading order, 
the fermionic effective action is given by
\begin{eqnarray}
\label{eq:effective-fermionic-action-leading-order}
S'_{M}{}^{(0)} &=& \sum_{x} \, Z
\left\{  
\bar \psi_{3 +}(x) D_3 \psi_{3 +}(x)
+
\bar \psi_{4 +}(x) D_4 \psi_{4 +}(x)
+
\bar \psi_{5 -}(x) D_5 \psi_{5 -}(x)
+
\bar \psi_{0 -}(x) D_0 \psi_{0 -}(x)
\right\} 
\nonumber\\
&-&
 \sum_{x} \big\{
\,\,\,
G_{35} \,
\bar \psi_{3 +}(x) \psi_{5 -}(x) \,
\bar \psi_{5 -}(x) \psi_{3 +}(x) 
+
G_{30} \,
\bar \psi_{3 +}(x) \psi_{0 -}(x) \,
\bar \psi_{0 -}(x) \psi_{3 +}(x)
\nonumber\\
&&\qquad
+
G_{45} \,
\bar \psi_{4 +}(x) \psi_{5 -}(x) \, 
\bar \psi_{5 -}(x) \psi_{4 +}(x) 
+
G_{40} \,
\bar \psi_{4 +}(x) \psi_{0 -}(x) \, 
\bar \psi_{0 -}(x) \psi_{4 +}(x) 
\quad \big\}
\nonumber\\
&-&
 \sum_{x}
\big\{ \,
G_{3450} \,
\bar \psi_{3 +}(x) \psi_{5 -}(x) \, 
\bar \psi_{4 +}(x) \psi_{0 -}(x) \, 
\bar \psi_{5 -}(x) \psi_{4 +}(x) \,
\bar \psi_{0 -}(x) \psi_{3 +}(x) 
\,
\big\} ,
\end{eqnarray}
where the effective couplings are determined by the following 
matching conditions, 
\begin{eqnarray} 
\label{eq:matching-conditions}
(-Z)^4 &=& (-z)^4  
\nonumber\\
&-&
(-z)^2 \,\bar h_{35} \,h_{35} 
-
(-z)^2 \,\bar h_{30} \,h_{30}
-
(-z)^2 \,\bar h_{45} \,h_{45}
-
(-z)^2 \,\bar h_{40} \,h_{40}
\nonumber\\
&+&
\bar h_{35} \,h_{35} \, \bar h_{40} \,h_{40}
+
\bar h_{30} \,h_{30} \, \bar h_{45} \,h_{45}
\nonumber\\
&-& 
\bar h_{35} \,\bar h_{40} \,h_{30} \,h_{45} \,
-
\bar h_{30} \,\bar h_{45} \,h_{35} \,h_{40} \,
\\
G_{35} (-Z)^2 
&=& 
(y_{35} \,y_{53} - \bar h_{35} \,h_{35} ) \, (-z)^2 
\nonumber\\
&-&
(y_{35} \,y_{53} - \bar h_{35} \,h_{35} ) \, \bar h_{40} \,h_{40}
\nonumber\\
&-&
\bar h_{30} \,\bar h_{45} \,h_{35} \,h_{40} 
\nonumber\\
&-&
\bar h_{35} \,h_{40} \,y_{45} \,y_{03} 
 \\
G_{30} (-Z)^2 
&=& 
(y_{30} \,y_{03} - \bar h_{30} \,h_{30}) \, (-z)^2
\nonumber\\
&-& 
(y_{30} \,y_{03} - \bar h_{30} \,h_{30}) \, \bar h_{45} \,h_{45}
\nonumber\\
&-& 
\bar h_{35} \,\bar h_{40} \,h_{30} \,h_{45}
\nonumber\\
&-&
\bar h_{30} \,h_{45} \,y_{40} \,y_{53} 
  \\
G_{45} (-Z)^2 
&=& 
(y_{45} \,y_{54} - \bar h_{45} \,h_{45}) \, (-z)^2
\nonumber\\
&-&
(y_{45} \,y_{54} - \bar h_{45} \,h_{45}) \, \bar h_{30} \,h_{30} 
\nonumber\\
&-&
\bar h_{35} \,\bar h_{40} \,h_{30} \,h_{45}
\nonumber\\
&-&
h_{30} \,\bar h_{45} \,y_{35} \,y_{04} 
   \\
G_{40} (-Z)^2 
&=& 
(y_{40} \,y_{04} - \bar h_{40} \,h_{40}) \, (-z)^2
\nonumber\\
&-&
(y_{40} \,y_{04} - \bar h_{40} \,h_{40}) \, \bar h_{35} \,h_{35}
\nonumber\\
&-&
\bar h_{30} \,\bar h_{45} \,h_{35} \,h_{40} 
\nonumber\\
&-&
h_{35} \,\bar h_{40} \,y_{30} \,y_{54} 
\end{eqnarray}
and
\begin{eqnarray}
G_{3450} - G_{35} G_{40} - G_{30} G_{45} 
&=& 
( y_{35} \,y_{40} \,y_{03} \,y_{54} + 
  \bar h_{35} \,\bar h_{40} \,h_{30} \,h_{45})
\nonumber\\
&+&
( y_{53} \,y_{04} \,y_{30} \,y_{45} +
  \bar h_{30} \,\bar h_{45} \,h_{35} \,h_{40} ) 
\nonumber\\
&-&
(y_{35} \,y_{53} - \bar h_{35} \,h_{35} ) \,
(y_{40} \,y_{04} - \bar h_{40} \,h_{40}) 
\nonumber\\
&-&
(y_{30} \,y_{03} - \bar h_{30} \,h_{30}) \,
(y_{45} \,y_{54} - \bar h_{45} \,h_{45})
\nonumber\\
&+&
\bar h_{35} \,h_{40} \,y_{45} \,y_{03}
+
h_{35} \,\bar h_{40} \,y_{30} \,y_{54}
\nonumber\\
&+&
\bar h_{30} \,h_{45} \,y_{40} \,y_{53} 
+
h_{30} \,\bar h_{45} \,y_{35} \,y_{04} .
\end{eqnarray}
This result can be obtained
by noting first 
that in the limit $\kappa =0$,
$S'_M$ is obtained explicitly as
\begin{eqnarray}
S'_{M} &=& \sum_{x} \, z 
\left\{  
\bar \psi_{3 +}(x) D_3 \psi_{3 +}(x)
+
\bar \psi_{4 +}(x) D_4 \psi_{4 +}(x)
+
\bar \psi_{5 -}(x) D_5 \psi_{5 -}(x)
+
\bar \psi_{0 -}(x) D_0 \psi_{0 -}(x)
\right\} 
\nonumber\\
&-&
 \sum_{x}
\left\{\,\,\,\,
y_{35} \,y_{53} \,\bar \psi_{3 +}(x) \psi_{5 -}(x) \,
\bar \psi_{5 -}(x) \psi_{3 +}(x) 
\right. \nonumber\\
&& \left.\qquad
+
y_{30} \,y_{03} \,\bar \psi_{3 +}(x) \psi_{0 -}(x) \,
\bar \psi_{0 -}(x) \psi_{3 +}(x)
\right. \nonumber\\
&& \left.\qquad
+
y_{45} \,y_{54} \,\bar \psi_{4 +}(x) \psi_{5 -}(x) \, \bar \psi_{5 -}(x) \psi_{4 +}(x) 
\right. \nonumber\\
&& \left.\qquad
+
y_{40} \,y_{04} \,\bar \psi_{4 +}(x) \psi_{0 -}(x) \, \bar \psi_{0 -}(x) \psi_{4 +}(x) \,\,\,
\right\}
\nonumber\\
&-&
 \sum_{x}
\left\{\,\,\,\,
y_{35} \,y_{40} \,y_{03} \,y_{54} \,
\bar \psi_{3 +}(x) \psi_{5 -}(x) \, 
\bar \psi_{4 +}(x) \psi_{0 -}(x) \, 
\bar \psi_{0 -}(x) \psi_{3 +}(x) \, 
\bar \psi_{5 -}(x) \psi_{4 +}(x) \,
\right. \nonumber\\
&&
\left.\quad\quad
+
y_{53} \,y_{04} \,y_{30} \,y_{45} \,
\bar \psi_{5 -}(x) \psi_{3 +}(x) \,
\bar \psi_{0 -}(x) \psi_{4 +}(x) \,
\bar \psi_{3 +}(x) \psi_{0 -}(x) \,
\bar \psi_{4 +}(x) \psi_{5 -}(x) 
\right\}
\nonumber\\
&-&
 \sum_{x}
\left\{\,\,\,\,
\bar h_{35} \,h_{35} \,
\bar \psi_{3 +}(x) \,c_D\, \bar\psi_{5 -}(x) \,
\psi_{3 +}(x) \,c_D\, \psi_{5 -}(x) \,
\right. \nonumber\\
&& \left.\qquad
+
\bar h_{30} \,h_{30} \,
\bar \psi_{3 +}(x) \,c_D\,  \bar\psi_{0 -}(x) \,
\psi_{3 +}(x) \,c_D\, \psi_{0 -}(x) \, 
\right. \nonumber\\
&& \left.\qquad
+
\bar h_{45} \,h_{45} \,
\bar \psi_{4 +}(x)\,c_D\, \bar \psi_{5 -}(x) \,
\psi_{4 +}(x)\,c_D\,\psi_{5 -}(x) \, 
\right. \nonumber\\
&& \left.\qquad
+
\bar h_{40} \,h_{40} \,
\bar \psi_{4 +}(x) \,c_D\, \bar \psi_{0 -}(x) \, 
\psi_{4 +}(x) \,c_D\, \psi_{0 -}(x) \,
\right\}\nonumber\\
&-&
 \sum_{x}
\left\{\,\,\,\,
\bar h_{35} \,\bar h_{40} \,h_{30} \,h_{45} \,
\bar \psi_{3 +}(x) \,c_D\, \bar\psi_{5 -}(x) \, 
\bar \psi_{4 +}(x) \,c_D\, \bar \psi_{0 -}(x) \, 
\psi_{3 +}(x) \,c_D\, \psi_{0 -}(x) \, 
\psi_{4 +}(x)\,c_D\,\psi_{5 -}(x) \, 
\right. \nonumber\\
&& \left.\qquad
+
\bar h_{30} \,\bar h_{45} \,h_{35} \,h_{40} \,
\bar \psi_{3 +}(x) \,c_D\,  \bar\psi_{0 -}(x) \,
\bar \psi_{4 +}(x)\,c_D\, \bar \psi_{5 -}(x) \,
\psi_{3 +}(x) \,c_D\, \psi_{5 -}(x) \, 
\psi_{4 +}(x) \,c_D\, \psi_{0 -}(x) \,
\right\}\nonumber\\
&-&
 \sum_{x}
\left\{\,\,\,\,
\bar h_{35} \,h_{40} \,y_{45} \,y_{03} \,
\bar \psi_{3 +}(x) \,c_D\, \bar\psi_{5 -}(x) \, 
\psi_{4 +}(x) \,c_D\, \psi_{0 -}(x) \, 
\bar \psi_{4 +}(x) \psi_{5 -}(x) \, 
\bar \psi_{0 -}(x) \psi_{3 +}(x) \,
\right. \nonumber\\
&& \left.\qquad
+
h_{35} \,\bar h_{40} \,y_{30} \,y_{54} \,
\psi_{3 +}(x) \,c_D\, \psi_{5 -}(x) \, 
\bar \psi_{4 +}(x) \,c_D\, \bar \psi_{0 -}(x) \,
\bar \psi_{3 +}(x) \psi_{0 -}(x) \,
\bar \psi_{5 -}(x) \psi_{4 +}(x) \,
\right. \nonumber\\
&& \left.\qquad
+
\bar h_{30} \,h_{45} \,y_{40} \,y_{53} \,
\bar \psi_{3 +}(x) \,c_D\,  \bar\psi_{0 -}(x) \,
\psi_{4 +}(x)\,c_D\,\psi_{5 -}(x) \,
\bar \psi_{4 +}(x) \psi_{0 -}(x) \, 
\bar \psi_{5 -}(x) \psi_{3 +}(x) \, 
\right. \nonumber\\
&& \left.\qquad
+
h_{30} \,\bar h_{45} \,y_{35} \,y_{04} \,
\psi_{3 +}(x) \,c_D\, \psi_{0 -}(x) \,
\bar \psi_{4 +}(x)\,c_D\, \bar \psi_{5 -}(x) \,
\bar \psi_{3 +}(x) \psi_{5 -}(x) \,
\bar \psi_{0 -}(x) \psi_{4 +}(x) \,
\right\} .
\nonumber\\
\end{eqnarray}
We next remind the fact that the chiral fields, 
$\psi_{3+}(x)$, $\psi_{4+}(x)$, $\psi_{5-}(x)$, $\psi_{0-}(x)$,  
have two components and the bilinear operators of the kinetic, Dirac- and Majorana-type Yukawa-coupling terms 
have the following structures in the components, 
\begin{eqnarray}
\bar \psi_{q+}(x) D_{q} \psi_{q+}(x) 
&=& \bar \psi_{q+}(x) \psi_{q+}(x){}^{(2)}, \\
\bar \psi_{q'-}(x) D_{q'} \psi_{q'-}(x) 
&=& \bar \psi_{q'-}(x) \psi_{q'-}(x){}^{(1)}, 
\end{eqnarray}
\begin{eqnarray}
\bar \psi_{q+}(x) \psi_{q'-}(x) 
&=& \bar \psi_{q+}(x) \psi_{q'-}(x){}^{(2)}, \\
\bar \psi_{q'-}(x)  \psi_{q+}(x) 
&=& \bar \psi_{q'-}(x) \psi_{q+}(x){}^{(1)}, 
\end{eqnarray}
and 
\begin{eqnarray}
\psi_{q+}(x) c_D \psi_{q'-}(x) 
&=& + \psi_{q+}(x){}^{(1)} \psi_{q-}(x){}^{(2)}
- \psi_{q+}(x){}^{(2)} \psi_{q-}(x){}^{(1)} , \\
\bar \psi_{q+}(x) c_D \bar \psi_{q'-}(x) 
&=& 
- \bar \psi_{q+}(x) \bar \psi_{q' -}(x), 
\end{eqnarray}
respectively.
Then the bilinear operators of the Majorana-type Yukawa-coupling terms can be eliminated and the fermionic effective action can be rewritten with only the bilinear operators of the kinetic and Dirac-type Yukawa-coupling terms as follows.

\begin{eqnarray}
S'_{M} &=& \sum_{x} \, z 
\left\{  
\bar \psi_{3 +}(x) D_3 \psi_{3 +}(x)
+
\bar \psi_{4 +}(x) D_4 \psi_{4 +}(x)
+
\bar \psi_{5 -}(x) D_5 \psi_{5 -}(x)
+
\bar \psi_{0 -}(x) D_0 \psi_{0 -}(x)
\right\} 
\nonumber\\
&-&
 \sum_{x}
\left\{\,\,\,\,
(y_{35} \,y_{53} - \bar h_{35} \,h_{35} ) \,
\bar \psi_{3 +}(x) \psi_{5 -}(x) \,
\bar \psi_{5 -}(x) \psi_{3 +}(x) 
\right. \nonumber\\
&& \left.\qquad
+
(y_{30} \,y_{03} - \bar h_{30} \,h_{30}) \,
\bar \psi_{3 +}(x) \psi_{0 -}(x) \,
\bar \psi_{0 -}(x) \psi_{3 +}(x)
\right. \nonumber\\
&& \left.\qquad
+
(y_{45} \,y_{54} - \bar h_{45} \,h_{45}) \,
\bar \psi_{4 +}(x) \psi_{5 -}(x) \, 
\bar \psi_{5 -}(x) \psi_{4 +}(x) 
\right. \nonumber\\
&& \left.\qquad
+
(y_{40} \,y_{04} - \bar h_{40} \,h_{40}) \,
\bar \psi_{4 +}(x) \psi_{0 -}(x) \, 
\bar \psi_{0 -}(x) \psi_{4 +}(x) \,\,\,
\right\}
\nonumber\\
&-&
 \sum_{x}
\left\{\,\,\,\,
( y_{35} \,y_{40} \,y_{03} \,y_{54} 
+ \bar h_{35} \,\bar h_{40} \,h_{30} \,h_{45}) \right. \times 
\nonumber\\
&&\qquad\qquad
\bar \psi_{3 +}(x) \psi_{5 -}(x) \, 
\bar \psi_{4 +}(x) \psi_{0 -}(x) \, 
\bar \psi_{0 -}(x) \psi_{3 +}(x) \, 
\bar \psi_{5 -}(x) \psi_{4 +}(x) \,
\nonumber\\
&&
\quad\quad
+
( y_{53} \,y_{04} \,y_{30} \,y_{45} +
\bar h_{30} \,\bar h_{45} \,h_{35} \,h_{40} ) \times
\nonumber\\
&&\qquad\qquad
\left.
\bar \psi_{5 -}(x) \psi_{3 +}(x) \, 
\bar \psi_{0 -}(x) \psi_{4 +}(x) \,
\bar \psi_{3 +}(x) \psi_{0 -}(x) \,
\bar \psi_{4 +}(x) \psi_{5 -}(x) 
\quad \right\}
\nonumber\\
&-&
 \sum_{x}
\left\{
-\bar h_{35} \,h_{35} \,
\bar \psi_{3 +}(x)  D_3 \psi_{3 +}(x) \,
\bar\psi_{5 -}(x) D_5 \psi_{5 -}(x) \,
\right. \nonumber\\
&& \left.\qquad
-
\bar h_{30} \,h_{30} \, 
\bar \psi_{3 +}(x) D_3 \psi_{3 +}(x) \, 
\bar\psi_{0 -}(x) D_0 \psi_{0 -}(x) \, 
\right. \nonumber\\
&& \left.\qquad
-
\bar h_{45} \,h_{45} \,
\bar \psi_{4 +}(x) D_4 \psi_{4 +}(x) \,
\bar \psi_{5 -}(x) D_5 \psi_{5 -}(x) \, 
\right. \nonumber\\
&& \left.\qquad
-
\bar h_{40} \,h_{40} \, 
\bar \psi_{4 +}(x) D_4 \psi_{4 +}(x) \,
\bar \psi_{0 -}(x) D_0 \psi_{0 -}(x) \,
\right\}\nonumber\\
%
&-&
 \sum_{x}
\left\{
-\bar h_{35} \,\bar h_{40} \,h_{30} \,h_{45} \,
\bar \psi_{3 +}(x) D_3 \psi_{3 +}(x) \, 
\bar \psi_{0 -}(x) D_0 \psi_{0 -}(x) \, 
\bar \psi_{5 -}(x) D_5 \psi_{5 -}(x) \, 
\bar \psi_{4 +}(x) D_4 \psi_{4 +}(x) 
\right. \nonumber\\
&& \left.\qquad
-
\bar h_{30} \,\bar h_{45} \,h_{35} \,h_{40} \,
\bar \psi_{3 +}(x) D_3 \psi_{3 +}(x) \, 
\bar \psi_{5 -}(x) D_5 \psi_{5 -}(x) \, 
\bar \psi_{0 -}(x) D_0 \psi_{0 -}(x) \, 
\bar \psi_{4 +}(x) D_4 \psi_{4 +}(x) 
\right\}\nonumber\\
&-&
 \sum_{x}
\left\{
-\bar h_{35} \,\bar h_{40} \,h_{30} \,h_{45} \,
\bar \psi_{3 +}(x) D_3 \psi_{3 +}(x) \, 
\bar \psi_{0 -}(x) D_0 \psi_{0 -}(x) \, 
\bar \psi_{5 -}(x) \psi_{4 +}(x) \, 
\bar \psi_{4 +}(x) \psi_{5 -}(x)
\right. \nonumber\\
&& \left.\qquad
-\bar h_{35} \,\bar h_{40} \,h_{30} \,h_{45} \,
\bar \psi_{3 +}(x) \psi_{0 -}(x) \, 
\bar \psi_{0 -}(x) \psi_{3 +}(x) \, 
\bar \psi_{5 -}(x) D_5 \psi_{5 -}(x) \, 
\bar \psi_{4 +}(x) D_4 \psi_{4 +}(x) 
\right. \nonumber\\
&& \left.\qquad
-
\bar h_{30} \,\bar h_{45} \,h_{35} \,h_{40} \,
\bar \psi_{3 +}(x) D_3 \psi_{3 +}(x) \, 
\bar \psi_{5 -}(x) D_5 \psi_{5 -}(x) \, 
\bar \psi_{0 -}(x) \psi_{4 +}(x) \, 
\bar \psi_{4 +}(x) \psi_{0 -}(x) 
\right. \nonumber\\
&& \left.\qquad
-
\bar h_{30} \,\bar h_{45} \,h_{35} \,h_{40} \,
\bar \psi_{3 +}(x) \psi_{5 -}(x) \, 
\bar \psi_{5 -}(x) \psi_{3 +}(x) \, 
\bar \psi_{0 -}(x) D_0 \psi_{0 -}(x) \, 
\bar \psi_{4 +}(x) D_4 \psi_{4 +}(x) 
\right\}\nonumber\\
&-&
 \sum_{x}
\left\{
-
\bar h_{35} \,h_{40} \,y_{45} \,y_{03} \,
\bar \psi_{3 +}(x) \psi_{5 -}(x)  \,
\bar \psi_{5 -}(x) \psi_{3 +}(x)  \, 
\bar \psi_{4 +}(x) D_4 \psi_{4 +}(x) \, 
\bar \psi_{0 -}(x) D_0 \psi_{0 -}(x) \, 
\right. \nonumber\\
&& \left.\qquad
-
h_{35} \,\bar h_{40} \,y_{30} \,y_{54} \,
\bar \psi_{4 +}(x) \psi_{0 -}(x) \,
\bar \psi_{0 -}(x) \psi_{4 +}(x) \,
\bar \psi_{3 +}(x) D_3 \psi_{3 +}(x) \,
\bar \psi_{5 -}(x) D_5 \psi_{5 -}(x)
\right. \nonumber\\
&& \left.\qquad
-
\bar h_{30} \,h_{45} \,y_{40} \,y_{53} \,
\bar \psi_{3 +}(x) \psi_{0 -}(x) \,
\bar \psi_{0 -}(x) \psi_{3 +}(x) \,
\bar \psi_{4 +}(x) D_4 \psi_{4 +}(x) \, 
\bar \psi_{5 -}(x) D_5 \psi_{5 -}(x) 
\right. \nonumber\\
&& \left.\qquad
-
h_{30} \,\bar h_{45} \,y_{35} \,y_{04} \,
\bar \psi_{4 +}(x) \psi_{5 -}(x) \, 
\bar \psi_{5 -}(x) \psi_{4 +}(x) \,
\bar \psi_{3 +}(x) D_3 \psi_{3 +}(x)  \,
\bar \psi_{0 -}(x) D_0 \psi_{0 -}(x)
\right\}\nonumber\\
&-&
 \sum_{x}
\left\{
-
\bar h_{35} \,h_{40} \,y_{45} \,y_{03} \,
\bar \psi_{3 +}(x) \psi_{0 -}(x) \,
\bar\psi_{5 -}(x) \psi_{4 +}(x) \,  
\bar \psi_{4 +}(x) \psi_{5 -}(x) \, 
\bar \psi_{0 -}(x) \psi_{3 +}(x) \,
\right. \nonumber\\
&& \left.\qquad
-
h_{35} \,\bar h_{40} \,y_{30} \,y_{54} \,
\bar \psi_{4 +}(x) \psi_{5 -}(x) \,
\bar \psi_{0 -}(x) \psi_{3 +}(x) \,  
\bar \psi_{3 +}(x) \psi_{0 -}(x) \,
\bar \psi_{5 -}(x) \psi_{4 +}(x) \,
\right. \nonumber\\
&& \left.\qquad
-
\bar h_{30} \,h_{45} \,y_{40} \,y_{53} \,
\bar \psi_{3 +}(x) \psi_{5 -}(x) \,
\bar \psi_{0 -}(x) \psi_{4 +}(x) \,
\bar \psi_{4 +}(x) \psi_{0 -}(x) \, 
\bar \psi_{5 -}(x) \psi_{3 +}(x) \, 
\right. \nonumber\\
&& \left.\qquad
-
h_{30} \,\bar h_{45} \,y_{35} \,y_{04} \,
\bar \psi_{4 +}(x) \psi_{0 -}(x) \,
\bar \psi_{5 -}(x) \psi_{3 +}(x) \, 
\bar \psi_{3 +}(x) \psi_{5 -}(x) \,
\bar \psi_{0 -}(x) \psi_{4 +}(x) \,
\right\} . 
\end{eqnarray}
This result can be matched with $S'_M{}^{(0)}$ 
in eq.~(\ref{eq:effective-fermionic-action-leading-order})
with the given conditions eq.~(\ref{eq:matching-conditions}). 

In higher orders, 
the effective fermionic operators
${\cal O}_{f}(x)$,  ${\cal O}_{f'}(x)$ and  ${\cal O}_{T}(x)$, 
which break the fermion number symmetries 
U(1)$_f$ and U(1)$_{f'}$ 
in the mirror sector, 
are indeed generated.
${\cal O}_{f}(x)$ and ${\cal O}_{f'}(x)$ appear
 at the second order, 
\begin{eqnarray}
\Delta S'_{M}{}^{(2)} 
&=& \sum_{x} 
\big\{ 
  G_{f} \, {\cal O}_{f}(x)
+ G_{f'}\, {\cal O}_{f'}(x) 
\big\} \, + {\rm c.c.} \quad ( \subset S'_{M}{}^{(2)} ) , \\
G_{f} &=& c_{30} \, h_{30} \, y_{54}{}^3 
       +  c_{40} \, h_{40}  \, y_{53} \, y_{54}{}^2 ,\\
G_{f'} &=& c_{03} \, y_{03} \, y_{53} \, y_{54} 
       +   c_{04} \, y_{04} \, y_{53}{}^2 .
\end{eqnarray}
${\cal O}_{T}(x)$ appears at the 
eighth order,
\begin{eqnarray}
\Delta S'_{M}{}^{(8)} &=& \sum_{x} \, G_{T} \, {\cal O}_{T}(x) \, + {\rm c.c.} \quad ( \subset S'_{M}{}^{(8)} ) , \\
G_{T} 
&=&  
  c_{30,03}  \, h_{30}  \, y_{03} \, y_{53}{}^{} \,\, \, y_{54}{}^4 
+ c_{30,04}  \, h_{30}  \, y_{04} \, y_{53}{}^2 \, y_{54}{}^3 
\nonumber\\
&+&
  c_{40,03}  \, h_{40}  \, y_{03} \, y_{53}{}^2 \, y_{54}{}^3
+ c_{40,04}  \, h_{40}  \, y_{04} \, y_{53}{}^3 \, y_{54}{}^2.
\end{eqnarray}
Here $c$'s are certain numerical coefficients.

To evaluate the strength of the
fermionic operators, 
${\cal O}_{f}(x)$,  ${\cal O}_{f'}(x)$ and ${\cal O}_{T}(x)$,
we first need to renormalize the fermionic field variables 
by $Z^{-1/2}$ so that
the kinetic terms of these fields become canonical.
We then assume that
$z=0$, 
$y_{qq'}= y_{q'q} \simeq 1$ ($q=3,4, q'=5,0$), 
$h_{30}=\bar h_{30}=h_{45} = \bar h_{45} = h$,
$h_{35}=\bar h_{35}=h_{40} = \bar h_{40} = h'$,
and $h \gg h' \simeq 1$. 
In this case, 
the renormalized coupling constants, defined by
$g_{qq'} = Z^{-2} G_{qq'}$ $(q=3,4, q'=5,0)$, 
$g_{3450} = Z^{-4} G_{3450}$, 
$g_{f} = Z^{-4} G_{f}$, 
$g_{f'} = Z^{-3} G_{f'}$ and 
$g_{T} = Z^{-7} G_{T}$, 
are evaluated as
$g_{30}=g_{45} =1$, 
$g_{35}=g_{40} =- h'{}^2/ h^2$, 
$g_{3450} = 6  h'{}^2/ h^2$,
$g_{f} = c_{30}/h^3$, 
$g_{f'} = (c_{03} + c_{04})/h^3$,
and 
$g_{T} = (c_{30,03} + c_{30,04})/h^6$ 
up to the corrections of the fraction 
${\cal O}( h'{}^2/h^2, y^2/h^2)$.
And the effective fermionic 
actions reads 
\begin{eqnarray}
\label{eq:effective-fermionic-action-leading-order-large-h}
S'_{M}{}^{(0)} &=& \sum_{x} \, 
\left\{  
\bar \psi_{3 +}(x) D_3 \psi_{3 +}(x)
+
\bar \psi_{4 +}(x) D_4 \psi_{4 +}(x)
+
\bar \psi_{5 -}(x) D_5 \psi_{5 -}(x)
+
\bar \psi_{0 -}(x) D_0 \psi_{0 -}(x)
\right\} 
\nonumber\\
&-&
 \sum_{x} \big\{
\,\,\,
\bar \psi_{3 +}(x) \psi_{0 -}(x) \,
\bar \psi_{0 -}(x) \psi_{3 +}(x) 
+
\bar \psi_{4 +}(x) \psi_{5 -}(x) \, 
\bar \psi_{5 -}(x) \psi_{4 +}(x) 
\quad \big\} , 
\end{eqnarray}
and 
\begin{eqnarray}
\Delta S'_{M}{}^{(2)} 
&=& \sum_{x} \, \frac{1}{h^3}
\big\{ c_{30} \, {\cal O}_{f}(x)
      +(c_{03} + c_{04})\, {\cal O}_{f'}(x) 
\big\} \, + {\rm c.c.} \quad ( \subset S'_{M}{}^{(2)} ) , \\
\Delta S'_{M}{}^{(8)} &=& \sum_{x} \, \frac{1}{h^6} 
(c_{30,03} + c_{30,04})\, {\cal O}_{T}(x)
 \, + {\rm c.c.}
\quad ( \subset S'_{M}{}^{(8)} ) , 
\end{eqnarray}
up to the corrections of order ${\cal O}( h'{}^2/h^2, y^2/h^2)$, respectively.
We can see that
the leading effective fermionic action is 
symmetric under U(1)$_f$ and U(1)$_{f'}$,
while 
the symmetry-breaking operators 
${\cal O}_{f}(x)$ and ${\cal O}_{f'}(x)$
are suppressed by the factor $\kappa^2 / h^3$ 
and ${\cal O}_{T}(x)$ by $\kappa^8 / h^6 $.

In the limit $h \rightarrow \infty$, 
the 345 model turns out to be 
massless Thirring model of the two Dirac pairs 
$\{ \psi_{3+}(x), \psi_{0-}(x) \}$ and 
$\{ \psi_{4+}(x), \psi_{5-}(x) \}$
in the framework of the overalp/Ginsparg-Wilson fermions.
In the continuum limit, the massless Thirring model is known to be equivalent to the model of free massless bosons. When coupled to the (external) gauge field, these massless degrees of freedom can produce singular and non-local terms in the
two-point vertex function $ \tilde \Pi'_{\mu \nu} (k)$
of the (external) gauge field. 
It is suspected that the similar result holds true in the lattice model and this explains the numerical result observed in the works\cite{Chen:2012di,Giedt:2014pha}.

\subsubsection{Limit of the large Majorana-type 
Yukawa-coupling/Mass terms}

From the result of the effective fermionic action 
in the hopping parameter expansion, 
we can see that the kinetic terms are not suppressed 
in the limit, $z \rightarrow 0$ and $h, h' \rightarrow \infty$.
To make the effective kinetic terms vanish, one should choose $h$, $h'$ and $z$ so that 
they satisfy the condition,
\begin{eqnarray}
Z^4 &=& z^4 - 2 z^2 (h^2 + h'{}^2) +(h^2 - h'{}^2)^2 =0.
\end{eqnarray}
Namely, 
\begin{equation}
z=| h + h' | , \quad | h - h' | .
\end{equation}

We note that this is the common property of the mass-like terms of the Ginsparg-Wilson fermion.
For the Dirac mass term, it is usually formulated as
\begin{equation}
S_{\rm D} = \sum_{x}  \{ 
\bar \psi(x) D \psi(x) + m_D \, \bar \psi (1-D) \psi(x) 
\},
\end{equation}
because the scalar and pseudo scalar operators,
$\bar \psi (1-D) \psi(x)$ and 
$\bar \psi i \gamma_3(1-D) \psi(x) $,
have the good transformation properties
under the chiral transformation, 
$\delta \psi(x)=\gamma_3 (1-2D) \psi(x)$, 
$\delta \bar\psi(x) = \bar\psi(x) \gamma_3$.
However, this choice makes the limit of the large mass parameter
$m_D$ singular, 
because the factor $(1-D)$ projects out
the modes with the momenta 
$p_\mu = 
(\pi, 0), (0,\pi), (\pi,\pi)$.
The maximal value of the mass is given at $m_D=1$, where the kinetic term $\bar \psi D \psi$ cancels out in the action and 
the simple bilinear operator $\bar\psi(x)\psi(x)$ saturates
the path-integral measure of the Dirac field completely.
To make the limit of the large mass parameter well-defined, we should write the action as
\begin{equation}
S_{\rm D} = \sum_{x \in \Lambda}  \{ 
 z \, \bar \psi(x) D \psi(x) + m \, \bar \psi(x) \psi(x) 
\},
\end{equation}
where $z= 1-m_D$ and $m=m_D$ and should take the limit
$z/m = (1-m_D)/m_D \rightarrow 0$.

As for the Majorana mass term, it is often formulated as
\begin{eqnarray}
S_{\rm M} 
&=& \sum_{x }  \{ 
 \bar \psi(x)  D  \psi(x)
\nonumber\\
&&\qquad
 + M' \,( 
\psi_+(x)^T  c_D  \psi_-(x) 
+
\bar \psi_+(x) c_D \bar \psi_-(x)^T 
)
\} .
\end{eqnarray}
Indeed, this type of the Majorana-Yukawa couplings are used
in the formulation of the 345 model by Bhattacharya, Chen, Giedt, Poppitz and Shang in consideration.\cite{Bhattacharya:2006dc,Giedt:2007qg,Poppitz:2007tu,Poppitz:2008au,Poppitz:2009gt,Poppitz:2010at,
Chen:2012di,Giedt:2014pha} 
However, again, the limit of the large Majorana mass parameter $M'$ is singular.
In fact, in the chiral basis, the Majorana mass term has the
matrix elements as
\begin{eqnarray}
M' u^T_j  c_D v_k 
&=&
- M' \delta_{p+p',0} \, \frac{b(p')}{\omega(p')} \, \quad 
(j=\{p\}, k=\{p'\}) , \\
M' \bar u_j  c_D \bar v_k^T
&=&
- M' \delta_{x,x'} \, \quad \quad \,\,\,\qquad
(j=\{x \}, k=\{x' \}) ,
\end{eqnarray}
where $b(p)= \sum_\mu (1- \cos p_\mu) - m_0$ 
and $\omega(p)=\sqrt{{\scriptstyle \sum_\mu} \sin^2  p_\mu   
+ \big\{ {\scriptstyle \sum_\mu } (1- \cos p_\mu ) - m_0 \big\}^2}$.
And 
the first matrix is singular because
its determinant has the factor
$\prod_{p'} \{ b(p')/\omega(p')\}$ and
$b(p')$ can vanish for $ 0 < m_0 < 2$.

Instead, one can formulate the action as
\begin{eqnarray}
S_{\rm M}
&=& \sum_{x }  \{ 
 z \, \bar \psi(x)  D  \psi(x)
\nonumber\\
&&\qquad
 + M \,( 
\psi_+(x)^T i\gamma_3 c_D  \psi_-(x) 
+
\bar \psi_+(x) i\gamma_3 c_D \bar \psi_-(x)^T 
)
\} ,
\end{eqnarray}
where the matching conditions are given by $(-z)^2=1-2 M'{}^2$ and $M=M'$.
Then the limit $z/M = \sqrt{1-2 M'{}^2}/M' \rightarrow 0$ is well-defined.
In fact, in the chiral basis, the Majorana mass term has the
matrix elements as
\begin{eqnarray}
M u^T_j  i\gamma_3 c_D v_k 
&=&
i M \delta_{p+p',0} \,  \quad 
(j=\{p\}, k=\{p'\}) , \\
M \bar u_j  i\gamma_3 c_D \bar v_k^T
&=&
i M \delta_{x,x'} \, \quad \quad
(j=\{x \}, k=\{x' \}) ,
\end{eqnarray}
and the 
determinants of these matrices are both unity.
Therefore the bilinear operator
$M \,( 
\psi_+(x)^T i\gamma_3 c_D \psi_-(x) 
+
\bar \psi_+(x) i\gamma_3 c_D \bar \psi_-(x)^T 
)$ saturates
the path-integral measure of the Dirac field completely.


\section{1$^4$(-1)$^4$ axial gauge model
in the mirror-fermion approach 
}
\label{sec:1^4(-1)^4-model}

In this section, 
we consider a simpler four-flavor axial gauge model,
a 1$^4$(-1)$^4$ model, 
and 
clarify 
the effect of the symmetry-breaking operators 
like the 't Hooft vertices 
in the mirror fermion sector  
to the behavior of 
the 
correlation functions of the (external) gauge field.

\subsection{1$^4$(-1)$^4$ axial gauge model 
with Spin(6)(SU(4)) symmetry}

We consider the axial gauge model with Spin(6)(SU(4)) flavor symmetry, which is defined by the charge assignment of the U(1) gauge symmetry as
\begin{eqnarray}
Q
&=&{\rm diag}(q_1, q_2, q_3, q_4)
={\rm diag}(+1, +1, +1, +1), \\
\quad
Q'
&=&{\rm diag}(q'_1, q'_2, q'_3, q'_4)
={\rm diag}(-1, -1, -1, -1) .
\end{eqnarray}
The left- and right-handed Weyl fermions,
$\psi_-(x)$ and $\psi'_+(x)$, 
are assumed  in $\underline{4}$, the four-dimensional
irreducible spinor representation of SO(6).
The generators of 
the spinor representation of SO(6), i.e. 
Spin(6)($\cong$ SU(4)),  are defined by
\begin{equation}
\Sigma^{ab} = -\frac{i}{4} \big[ \Gamma^a, \Gamma^b \big] \quad (a,b = 1, \cdots, 6),
\end{equation}
where
$\Gamma^a$ 
are the eight-dimensional representation 
of the Clifford algebra $\Gamma^a \Gamma^b + \Gamma^b \Gamma^a = 2 \delta^{ab}$
($a, b = 1,\cdots,6$) 
specified by
\begin{eqnarray*}
\Gamma^1 &=& \sigma_1 \times \sigma_1 \times \sigma_1 ,\\ 
\Gamma^2 &=& \sigma_2 \times \sigma_1 \times \sigma_1 ,\\ 
\Gamma^3 &=& \sigma_3 \times \sigma_1 \times \sigma_1 ,\\ 
\Gamma^4 &=& I               \times \sigma_2 \times \sigma_1 ,\\ 
\Gamma^5 &=& I               \times \sigma_3 \times \sigma_1 ,\\ 
\Gamma^6 &=& I               \times I              \times \sigma_2 ,
\end{eqnarray*}
and 
the Weyl fields
$\psi_-(x)$
and $\psi'_+(x)$ 
satisfy the constraints,
\begin{eqnarray}
&& {\rm P}_+  \, \psi_-(x) = + \psi_-(x),  \quad \bar \psi_-(x) \, {\rm P}_+   = + \bar \psi_-(x) ,  \\
&& {\rm P}_+    \, \psi'_+ (x) = + \psi'_+(x), \quad \bar \psi'_+(x) \, {\rm P}_+  = + \psi'_+(x) ,
\end{eqnarray}
where
\begin{equation}
{\rm P}_\pm = \frac{1 \pm \Gamma^7}{2}, \qquad
\Gamma^7 = i \Gamma^1 \cdots \Gamma^6 .
\end{equation}
The U(1) gauge and Spin(6)(SU(4)) global symmetries
prohibit the Dirac- and Majorana-type bilinear mass terms for these fermions.

We assume accordingly that the right- and left-handed mirror fermions,
$\psi_+(x)$ and $\psi'_-(x)$, 
are in $\underline{4}$, the same 
four-dimensional
irreducible spinor representation of SO(6).
\begin{eqnarray}
&&{\rm P}_+  \, \psi_+ (x) = + \psi_+(x), \quad \bar \psi_+(x) \,   {\rm P}_+ = + \psi_+(x) , \\
&& {\rm P}_+  \, \psi'_-(x) = + \psi'_-(x),  \quad \bar \psi'_-(x) \, {\rm P}_+  = + \bar \psi'_-(x) .
\end{eqnarray}
Then, as shown in table~\ref{table:1^4(-1)^4-anomaly},
the remaining continuous 
symmetry in the mirror sector is the vector U(1) symmetry,
U(1)$_V$.

\begin{table}[hb]
\begin{center}
\begin{tabular}{|c||c|c|l|l|}
\hline
 & $+$ & $-$ & gauge anomaly & chiral anomaly \\ \hline
U(1)$_g$ & 1 & -1 & matched (gauged)& ---\\
Spin(6)/SU(4) & \underbar{4} & \underbar{4} & matched (can be gauged) 
& anomaly free \\
U(1)$_V$& 1 & 1 & not matched & anomalous \\
\hline
\end{tabular}
\end{center}
\caption{Fermionic continuous symmetries in the mirror sector
of the 1$^4$(-1)$^4$ model and their would-be gauge anomalies}
\label{table:1^4(-1)^4-anomaly}
\end{table}
\noindent
The U(1) gauge and Spin(6)(SU4)) global symmetries 
prohibit the bilinear terms, but 
allow the following quartic terms to break the U(1)$_V$ in the mirror sector,
\begin{eqnarray}
\label{eq:O_V-vertex}
{O}_V(x) 
&=& 
\frac{1}{2}\,
\psi_+(x)^T i \gamma_3 c_D 
{\rm T}^a \psi'_-(x) \,
\psi_+(x)^T i \gamma_3 c_D 
{\rm T}^a \psi'_-(x) , 
\\
\label{eq:bar-O_V-vertex}
\bar { O}_V(x)
&=&
\frac{1}{2}\,
\bar \psi_+ (x)  i \gamma_3 c_D 
{\rm T}^{a \dagger}
 {\bar\psi'_-(x)}^T \,
\bar \psi_+ (x)  i \gamma_3 c_D 
{\rm T}^{a \dagger}
 {\bar\psi'_-(x)}^T   ,
\end{eqnarray}
where ${\rm T}^a$ $(a=1, \cdots, 6)$ are defined by
${\rm T}^a = {\rm C} \Gamma^a$ $(a=1,\cdots,6)$ 
and satisfying $\{{\rm T}^a\}^T = - {\rm T}^a$.
$C$ is the charge-congugation operator 
satisfying 
${\rm C} \Gamma^a {\rm C}^{-1} = - \{ \Gamma_a \}^T$,  
${\rm C} \Gamma^{7} {\rm C}^{-1} =  - \Gamma_{7}$,
${\rm C}^T = - {\rm C}^{-1} =-  {\rm C}^\dagger = {\rm C}$.
%
The explicit representations of ${\rm T}^a$ $(a=1, \cdots, 6)$
and ${\rm C}$ 
are given as follows.
\begin{eqnarray*}
{\rm T}^1 &=&
(+1) \,
\, \sigma_3 \times \sigma_2 \times \sigma_3 , \\ 
{\rm T}^2 &=& 
(+ i) \, 
\, \,\,\, I \times \sigma_2 \times \sigma_3 , \\ 
{\rm T}^3 &=&
(- 1) \,
\, \sigma_1 \times \sigma_2 \times \sigma_3 , \\ 
{\rm T}^4 &=& 
(- i) \, 
\, \sigma_2 \times \sigma_1 \times \sigma_3 , \\ 
{\rm T}^5 &=&
(+ 1) \, 
\, \sigma_2 \times \,\, I \times \sigma_3 , \\ 
{\rm T}^6 &=&
(+ i) \,
\, \sigma_2 \times \sigma_3 \times \,\, I , \\ 
%
{\rm C} &=& (+i) \, 
\, \sigma_2 \times \sigma_3 \times \sigma_2 .
\end{eqnarray*}
The squares of 
these operators are nothing but 
the 't Hooft vertices which can be induced by the U(1) instantons in two-dimensions.
\begin{eqnarray}
\label{eq:tHooft-vertices-1^4(-1)^4}
&& O_T(x) = \frac{1}{2}\, O_V(x) O_V(x) , \qquad \quad 
\bar O_T(x) = \frac{1}{2}\, \bar O_V(x)  \bar O_V(x) . 
\end{eqnarray}

\subsection{Mirror sector of the 1$^4$(-1)$^4$ model 
with the Majorana-type Yukawa-coupling to
SO(6)-vector spin fields}

We formulate 
the mirror fermion sector of 
the 1$^4$(-1)$^4$ model
with the Majorana-type Yukawa-couplings
to the auxiliary SO(6)-vector spin fields,
$E^a(x)$, $\bar E^a(x)$ $(a=1,\cdots,6)$ 
with the unit lengths $ E^a(x)E^a(x)=1$, $\bar E^a(x) \bar E^a(x)=1 $ as follows.
\begin{eqnarray}
\label{eq:action-mirrorsector-1^4-(-1)^4-model}
S_M 
&=& \sum_x \, z \left\{  
  \bar \psi_+(x) D_{+1} \psi_+(x) 
+ \bar \psi'_-(x) D_{-1} \psi'_-(x) 
\right\}  \nonumber\\
&+&  \sum_x \, h \left\{ 
  \psi_+(x)^T i \gamma_3 c_D 
{\rm T}^a E^a(x) \psi'_-(x) 
+ \bar \psi_+ (x)  i \gamma_3 c_D 
{\rm T}^{a \dagger} \bar E^a(x) 
 {\bar\psi'_-(x)}^T   
\right\}   \nonumber\\
&+& \sum_{x, \mu} \, \kappa \, \left\{ 
E^a(x) E^a(x +\hat \mu)
+ \bar E^a(x) \bar E^a(x +\hat \mu) 
\right\}.
\end{eqnarray}
The path-integral measure of the SO(6)-vector spin fields are defined by
\begin{eqnarray}
{\cal D}[E^a]
&=&
\prod_{x} \Big[ (\pi^3)^{-1} \prod_{a=1}^6 
 d E^a(x)\delta( |E(x)|-1) \Big] , \\
{\cal D}[\bar E^a]
&=&
\prod_{x} \Big[ (\pi^3)^{-1} \prod_{a=1}^6 
 d \bar E^a(x)\delta( |\bar E(x)|-1) \Big] .
\end{eqnarray}
Note that we adopt the type of the Majorana-Yukawa coupling with
the factor $i \gamma_3 c_D$ instead of $c_D$, trying
to make the large coupling limit $z/h \rightarrow 0$ well-defined.



We then consider
the limit 
$z/h \rightarrow 0$ and $\kappa \rightarrow 0$ in
the mirror fermion sector of 
the 1$^4$(-1)$^4$ model 
defined by eq.~(\ref{eq:action-mirrorsector-1^4-(-1)^4-model}), 
where
the kinetic terms of the mirror fermion and the spin fields
are both suppressed.
In this limit,
the partition function of the mirror sector is obtained 
by performing the path-integration of the mirror fermion 
fields in the chiral basis as
\begin{eqnarray}
 \big\langle  1 \big\rangle_{M}  
 &\equiv& 
 \int 
{\cal D}[\psi_+] {\cal D}[\bar \psi_+]
{\cal D}[\psi'_-] {\cal D}[\bar \psi'_-] 
{\cal D}[E^a]{\cal D}[\bar E^a] \,
 {\rm e}^{-S_M} 
\\
&=& \int {\cal D}[E^a] \,
\det (u^T \, i\gamma_3 c_D \check {\rm T}^a E^a v') \,
\int {\cal D}[\bar E^a] \,
\det (\bar u  \, i\gamma_3 c_D \check {\rm T}^{a \dagger} \bar E^a 
\bar v'{}^T ) ,
\end{eqnarray}
where 
$(u^T \, i\gamma_3 c_D {\rm T}^a E^a v')$ and 
$(\bar u  \, i\gamma_3 c_D  {\rm T}^{a \dagger} 
\bar E^a \bar v'{}^T )$ are the complex matrices given by
\begin{eqnarray}
\label{eq:first-matrix}
(u^T \, i\gamma_3 c_D  {\rm T}^a E^a v')_{ij}
&=&
u_i^T \, i\gamma_3 c_D  {\rm T}^a E^a v'_j , \\
\label{eq:second-matrix}
(\bar u  \, i\gamma_3 c_D  {\rm T}^{a \dagger} \bar E^a 
\bar v'{}^T )_{kl} &=&
\bar u_k  \, i\gamma_3 c_D {\rm T}^{a \dagger} \bar E^a 
\bar v_l'{}^T .
\end{eqnarray}
%
The chiral basis for the anti-fields can be chosen as
$\bar u_k(x) = (0,1) \delta_{s,s'} \delta_{x,x'}$
for $k=(s', x')$ 
and  
$\bar v_l(x) = (1,0) \delta_{s s''} \delta_{x x''}$ 
for $l=(s'', x'')$. Then
the second matrix eq.~(\ref{eq:second-matrix}) is given by 
\begin{equation}
\bar u_k  \, i\gamma_3 c_D 
{\rm T}^{a \dagger} \bar E^a 
\bar v_l'{}^T = i  
\{ \check {\rm T}^{a \dagger} \}_{s' s''} \bar E^a(x') 
\delta_{x' x''},
\end{equation}
where
$\check {\rm T}^a$ $(a=1,\cdots,6)$ are $4 \times 4$ 
matrices defined as 
${\rm T}^{a'} = \check {\rm T}^{a'} \otimes \sigma_3$ 
$(a'=1,\cdots,5)$,
${\rm T}^6 = \check {\rm T}^6 \otimes I$.
And its determinant turns out to be unity,
\begin{equation}
\det (\bar u  \, i\gamma_3 c_D {\rm T}^{a \dagger} \bar E^a 
\bar v'{}^T ) =1 .
\end{equation}
Therefore the partition function is given simply by
\begin{eqnarray}
\big\langle  1 \big\rangle_{M}
&=&
\big\langle
1 
\big\rangle_{E} ,
\end{eqnarray}
where $\langle \cdots \rangle_E$
is the abbreviation for the path-integration of the spin-fields $E^a(x)$: 
\begin{equation}
\big\langle
{\cal O}
\big\rangle_{E}  \, 
\equiv
\int {\cal D}[E^a] \, 
\det (u^T \, i\gamma_3 c_D \check {\rm T}^a E^a v') 
\,\, 
{\cal O}[E^a] .
\end{equation}

We note that the above formula of the partition function
of the mirror sector makes sense in
all topological sectors of the admissible U(1) link fields,
$\mathfrak{U}[m]$. This is because
the excess (or decrease) in the number of the right-handed
basis vectors
$\{ u_j(x) \}$ due to the topologically non-trivial link
fields is always equal to that of the left-handed ones $\{ v'_j(x) \}$ thanks to the axial assignment of the U(1) charges,
and 
the matrix $(u^T \, i\gamma_3 c_D \check {\rm T}^a E^a v')$
remains to be a square matrix.

The chiral basis for the fields, on the other hand,
can be chosen so that
the basis vectors satisfy the relation
\begin{eqnarray}
u_i^T \gamma_3 c_D {\rm C} \Gamma^6 &=& {\cal C}_{ij} v'_j{}^\dagger ,
\qquad {\cal C}^\dagger = {\cal C}^{-1}, 
\end{eqnarray}
because the chiral projectors 
commute with the Gamma matrices,
\begin{equation}
 \Gamma^a \hat P_+  \Gamma^a  = \hat P_+  , \quad
 \Gamma^a \hat P'_- \Gamma^a  =  \hat P'_- \quad 
(a=1, \cdots, 6) 
\end{equation}
and 
satisfy the charge-conjugation relation,
\begin{eqnarray}
{\rm C}^{-1} ( \gamma_3 c_D)^{-1} \hat P_+[U]^T (\gamma_3 c_D ) {\rm C}
= \hat P_-[U^\ast] = \hat P'_-[U] .
\end{eqnarray}
Then the first matrix eq.~(\ref{eq:first-matrix}) is given by
\begin{eqnarray}
(u^T \, i\gamma_3 c_D {\rm T}^a E^a v') 
&=&
{\cal C} \times 
(v'{}^\dagger \,  i  \Gamma^6 \Gamma^a E^a v') 
\\
&=&
(u^T \, i \Gamma^{a}{}^T E^a \Gamma^{6}{}^T u^\ast) 
\times  {\cal C}
\end{eqnarray}
where ${\cal C} = (u^T \gamma_3 c_D {\rm C} \Gamma^6 v')$.
And its determinant can be written as
\begin{eqnarray}
\det (u^T \, i\gamma_3 c_D  {\rm T}^a E^a v') 
&=& \det {\cal C} \,  
    \det (v'{}^\dagger \,  i  \Gamma^6 \Gamma^a E^a v')
\\
&=&  \det {\cal C} \,  
   \det (u{}^\dagger \,  i  \Gamma^6 \Gamma^a E^a u) .
\end{eqnarray}

\subsection{Properties in the weak gauge-coupling limit}

We examine the properties of the 1$^4$(-1)$^4$ model
in the weak gauge-coupling limit,  where the U(1) link variables are set to unity, $U(x,\mu)=1$.

\subsubsection{Positive semi-definite mirror-fermion determinant}

In the weak gauge-coupling limit,
one can choose
the chiral basis of the fields so that the basis vectors
satisfy the relations, 
\begin{eqnarray}
&& u'_j(x) = u_j(x) , \quad v'_j(x) = v_j(x) ,
\end{eqnarray}
\begin{equation}
u_j^T(x) \gamma_3 c_D {\rm C} \Gamma^6 = v_j^\dagger(x) .
\end{equation}
And the basis vectors $\{ u_j(x) \}$ 
$(j=\{p_\mu, t\}, t=1,\cdots,4 )$ can be chosen 
explicitely as 
\begin{eqnarray}
\label{eq:chiral-basis-u-free}
u_j(x) = \frac{1}{\sqrt{L^2}} { \rm e}^{ i p x } \, 
u_\alpha (p) \, \delta_{s,t}    \qquad   ( j=\{ p, t \} ) ,
\end{eqnarray}
where $\{ u_\alpha (p) \}$ are the two-spinor eigenvectors
of the free hermitian Wilson-Dirac operator 
$H_{\rm w} = \gamma_3 (D_{\rm w} - m_0) \,(0 < m_0 < 2)$
with the negative eigenvalues
in the plane-wave basis given by
\begin{eqnarray}
\label{eq:chiral-basis-u-planewave}
u_\alpha (p) &=&
\left\{ 
\begin{array}{lcl}
\left( \begin{array}{c} 
-c(p) \\ 
( \omega(p)+ b(p) )  \end{array}\right) 
               / \sqrt{2 \omega(p) ( \omega(p)+ b(p) )}
 && ( p \not = 0 )
\\
&&\\
\left( \begin{array}{c} 1 \\ 
                          0     \end{array} \right) 
%
&& ( p = 0 )
\end{array}
\right.
\end{eqnarray}
and
\begin{eqnarray}
b(p) &=&  \sum_\mu (1- \cos p_\mu) - m_0 , \\
c(p) &=&  i \sin p_0  +  \sin p_1 , \\
\omega(p) &=& \sqrt{{\scriptstyle \sum_\mu} \sin^2  p_\mu   
+ \big\{ {\scriptstyle \sum_\mu } (1- \cos p_\mu ) - m_0 \big\}^2} .
\end{eqnarray}
The two-momentum $p_\mu$ is given by 
$p_\mu = 2\pi n_\mu / L \,\, (n_\mu \in \mathbb{Z})$ 
for the periodic boundary condition and 
$p_\mu = 2\pi (n_\mu + 1/2) / L \,\, (n_\mu \in \mathbb{Z})$
for the anti-periodic boundary condition.
The zero modes with $p_\mu=0$ in eq.~(\ref{eq:chiral-basis-u-planewave}) exist only for the periodic boundary condition.
(See the appendix~\ref{app:chiral-basis-free} for detail.)
In this basis, the matrix elements of $(u{}^\dagger \,  i  \Gamma^6 \Gamma^a E^a u)$ are given by
\begin{eqnarray}
(u{}^\dagger \,  i  \Gamma^6 \Gamma^a E^a u)_{ij}
&=&
 i \{\Gamma^{6}\Gamma^a\}_{t t'} \tilde E^a(k) \times
\left\{
\begin{array}{ll}
\frac{c(p)^\ast c(p') + (\omega+b)(p) (\omega+b)(p')}
{ L^2 \sqrt{2 \omega ( \omega+ b )}(p)
      \sqrt{2 \omega ( \omega+ b )}(p')}
\delta_{p,p'+ k}
& (p\not = 0, p' \not = 0) \\
\frac{- c(p')}{ L^2 \sqrt{2 \omega ( \omega+ b )}(p')}
\delta_{0,p'+ k}
& (p=0, p' \not = 0) \\
\frac{-  c(p)^\ast}{L^2 \sqrt{2 \omega ( \omega+ b )}(p)}
\,\,
\delta_{p,k}
& (p \not = 0, p' =0) \\
\quad 1  \,\, \delta_{0, k} & (p=0, p'=0)
\end{array}
\right.  \, , 
\nonumber\\
\end{eqnarray} 
where $ \tilde E^a(k)$ is the Fourier components of $E^a(x)$  defined by
$\tilde E^a(k) \equiv \sum_{x} \, {\rm e}^{-i k x} \,E^a(x)  $ with the constraints, $\sum_{k_\mu} \tilde E^a(k)^\ast \tilde E^a(k) = L^4$ and
$\sum_{k_\mu} \tilde E^a(k)^\ast \tilde E^a(k+ p) = 0$ $(p \not = 0)$.

Moreover, 
the matrices 
$(u{}^\dagger \,  i  \Gamma^6 \Gamma^a E^a u)$ 
and 
$(v^\dagger \,  i  \Gamma^6 \Gamma^a E^a v)$ 
are the block-diagonal parts of the matrix
\begin{eqnarray}
U &=& 
\left( 
\begin{array}{cc}
(u^\dagger \, i \Gamma^6 \Gamma^a E^a u) 
&
(u^\dagger\, i \Gamma^6  \Gamma^a E^a v)
\\
(v^\dagger \, i \Gamma^6 \Gamma^a E^a u)
&
(v^\dagger \, i \Gamma^6 \Gamma^a E^a v)
\end{array}
\right) ,
\end{eqnarray}
which is unitary $U^\dagger = U^{-1}$ and
has the unit determinant $\det U =1$. Then, 
as long as these matrices are not singular, 
they satisfy the relation
\begin{equation}
\det (u^\dagger \, i \Gamma^6 \Gamma^a E^a u) 
=\det (v^\dagger \, i \Gamma^6 \Gamma^a E^a v)^\ast .
\end{equation}

From these results,  
it follows that
\begin{eqnarray}
\det  (u^T \gamma_3 c_D {\rm C} \Gamma^6 v') &=& 
\det  (u^T \gamma_3 c_D {\rm C} \Gamma^6 v) =
1, \\
\det (u^T \, i\gamma_3 c_D  {\rm T}^a E^a v') 
&=& 
    \det (u{}^\dagger \,  i  \Gamma^6 \Gamma^a E^a u)
=  
   \det (u{}^\dagger \,  i  \Gamma^6 \Gamma^a E^a u)^\ast .
\end{eqnarray}
And 
it turns out that
the determinant  $\det (u^T \, i\gamma_3 c_D  {\rm T}^a E^a v')$ is
real for any spin-field configuration $E^a(x)$,
\begin{equation}
\det (u^T \, i\gamma_3 c_D  {\rm T}^a E^a v')
=
\det ( u^\dagger \, i \Gamma^{6} \Gamma^{a} E^{a} u )
\, \,\,  \in \, \, \mathbb{R} \, ,
\end{equation}
and, in particular, it is unity
for the constant configuration 
$E_0^a(x) = \delta^{a, 6}$,
\begin{equation}
\det (u^T \, i\gamma_3 c_D  {\rm T}^a E_0^a v') = 1 .
\end{equation}

On the other hand, 
by inspecting the matrix elements of 
$(u{}^\dagger \,  i  \Gamma^6 \Gamma^a E^a u)$,
one can see that
the zero modes with $p_\mu =0$ mix with linear-combinations of the modes 
with $p'_\mu \not = 0$ for which
$ - c(p') \delta_{0,p'+ k} \tilde E^a(k) \not = 0$,
but they decouple completely
from the modes with the
momenta $p'_\mu = \pi_\mu^{(A)}$ $(A=1,2,3)$ where
$\pi^{(1)} \equiv (\pi, 0), 
\pi^{(2)} \equiv (0,\pi), \pi^{(3)} \equiv(\pi,\pi)$.
This implies that 
the mixing of the zero modes is completely
suppressed 
for the following class of the spin configurations,
\begin{eqnarray}
E_\ast^a(x) &=& \frac{1}{V} \sum_{A=1,2,3} \, 
\cos(\pi^{(A)} x )\, 
\tilde E^a(\pi^{(A)}), \\
&& \sum_{A=1,2,3} \tilde E^a(\pi^{(A)})\tilde E^a(\pi^{(A)}) = V^2 , \\
&& \sum_{A \not = B} \tilde E^a(\pi^{(A)}) \tilde E^a(\pi^{(A)}+ \pi^{(B)}) = 0 \qquad (B =1, 2, 3).
\end{eqnarray}
For these configurations, zero eigenvalues appear in the eigenspectrum 
of $(u{}^\dagger \,  i  \Gamma^6 \Gamma^a E^a u)$
and the multiplicity of the zero eigenvalues is at least eight.

\begin{figure}[tbh]
\begin{center}
\includegraphics[width =60mm]{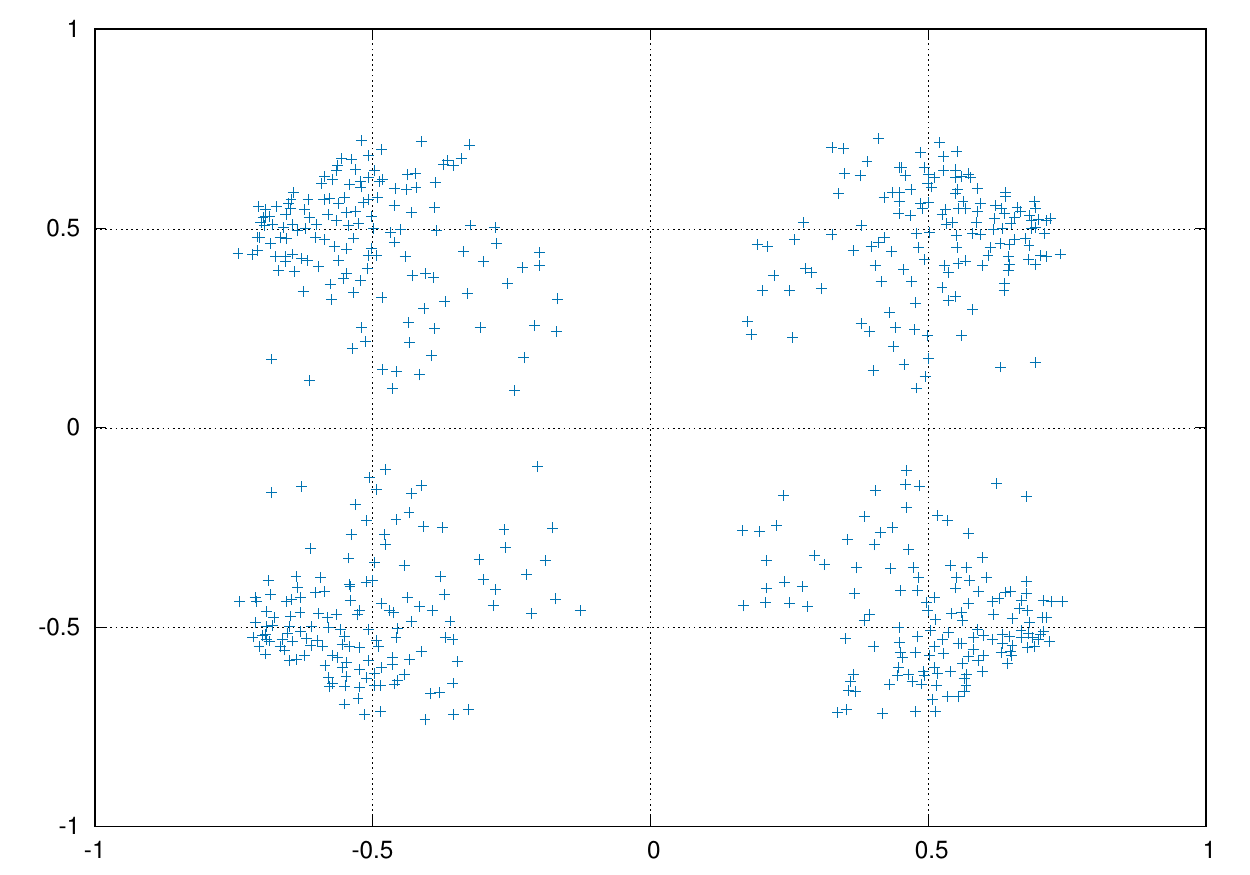} 
\hspace{2em}
\includegraphics[width =60mm]{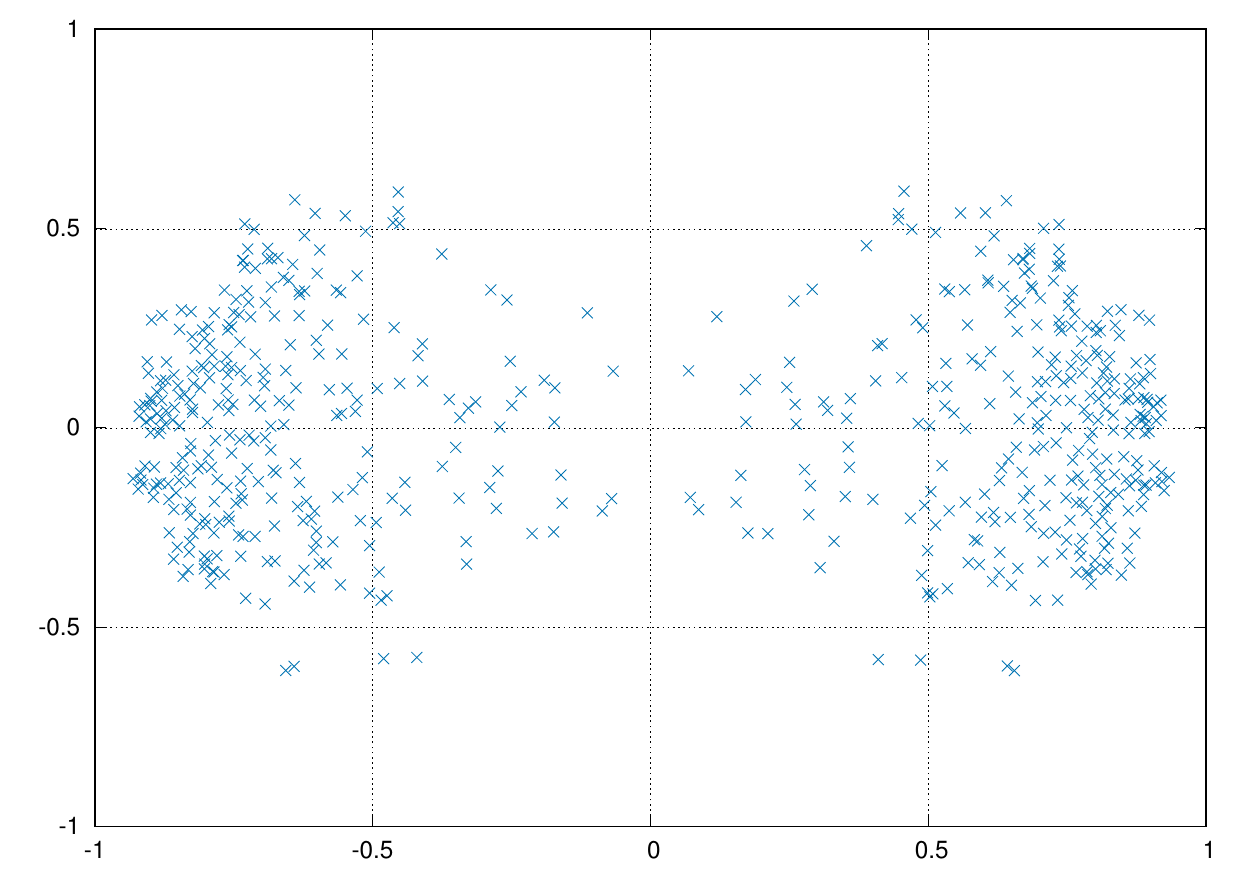} 
\end{center}
\caption{
Eigenvalue spectra of 
$\big(u^{\rm T} \,  i \gamma_3 c_D {\rm T}^a E^a v \big)$ [left] and
$\big( u^\dagger \Gamma^{6} \Gamma^{a} E^{a} u \big)$ [right]
for a randomly generated spin-field configuration $E^a(x)$.
The lattice size is $L=12$ and the periodic boundary condition is imposed on the fermion fields.}
\label{fig:eigenvalues-12x12}
\end{figure}

\begin{figure}[tbh]
\begin{center}
\includegraphics[width =60mm]{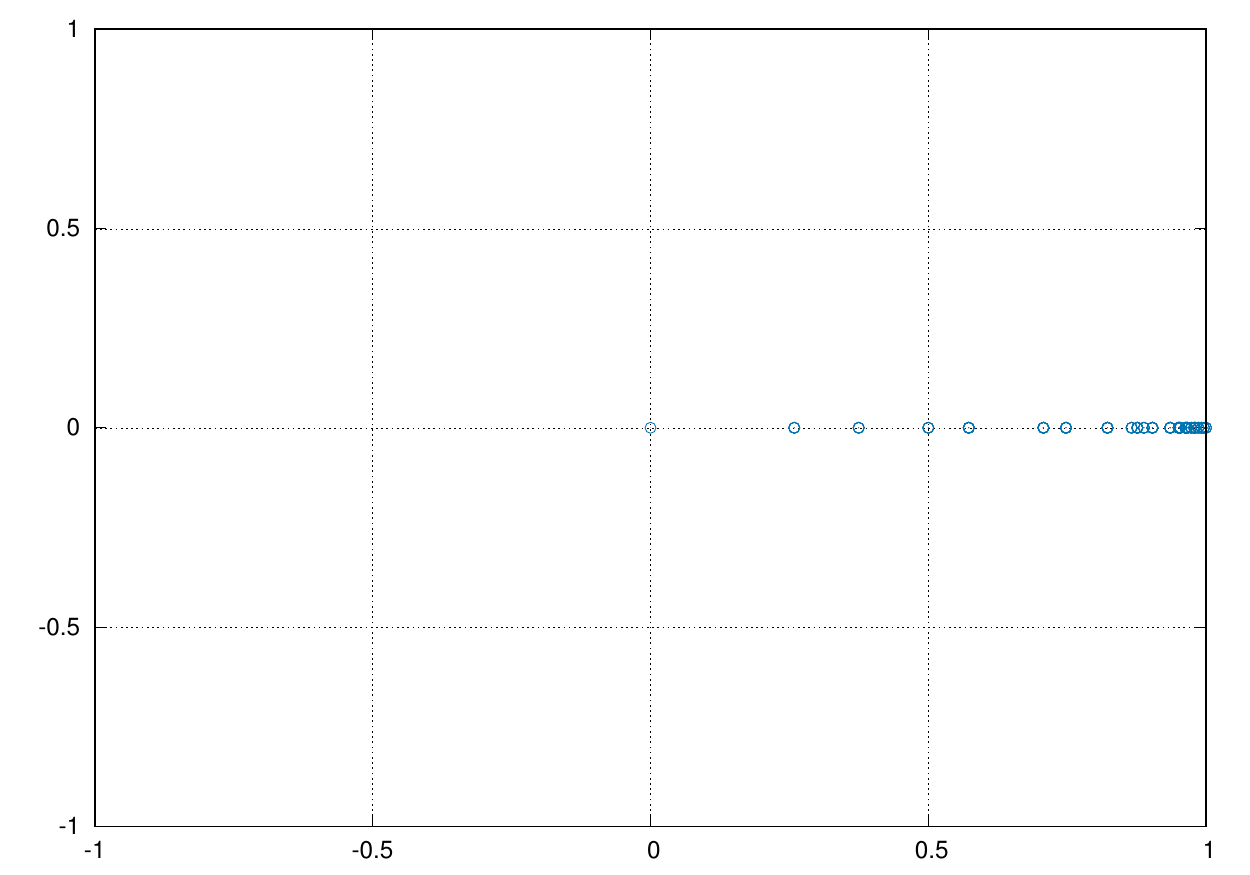} 
\end{center}
\caption{
Eigenvalue spectrum of $(\bar u D u)$.
The lattice size is $L=12$ and the periodic boundary condition is imposed on the fermion fields.}
\label{fig:eigenvalues-chiralD-12x12}
\end{figure}

It is instructive to verify the above results numerically. 
For randomly generated spin-field configurations,
we found that 
the eigenvalues of 
$\big(u^{\rm T} \,  i \gamma_3 c_D {\rm T}^a E^a v \big)$ and
$\big( u^\dagger \Gamma^{6} \Gamma^{a} E^{a} u \big)$
are all non-zero, 
and 
the determinants  
$\det ( u^{\rm T}\, i \gamma_3 c_D {\rm T}^a E^a v )$ 
and 
$\det \big( u^\dagger \Gamma^{6} \Gamma^{a} E^{a} u \big)$
are both real and positive.
We also observed that 
the eigenvalue spectra of 
$\big( u^\dagger \Gamma^{6} \Gamma^{a} E^{a} u \big)$
have the structure like
$\{ (\lambda_i, - \lambda_i^\ast) \, \vert \,  i=1,\cdots, L^2/2 \}$ approximately.
The typical examples of the eigenvalue spectra 
are shown in fig.~\ref{fig:eigenvalues-12x12} for $L=12$ with the periodic boundary condition.
For comparison, the eigenvalue spectrum of the chiral Dirac
matrix $(\bar u D u)$ is shown  in fig.~\ref{fig:eigenvalues-chiralD-12x12} for $L=12$ with the same periodic boundary condition.

\begin{figure}[tbh]
\begin{center}
\includegraphics[width =60mm]{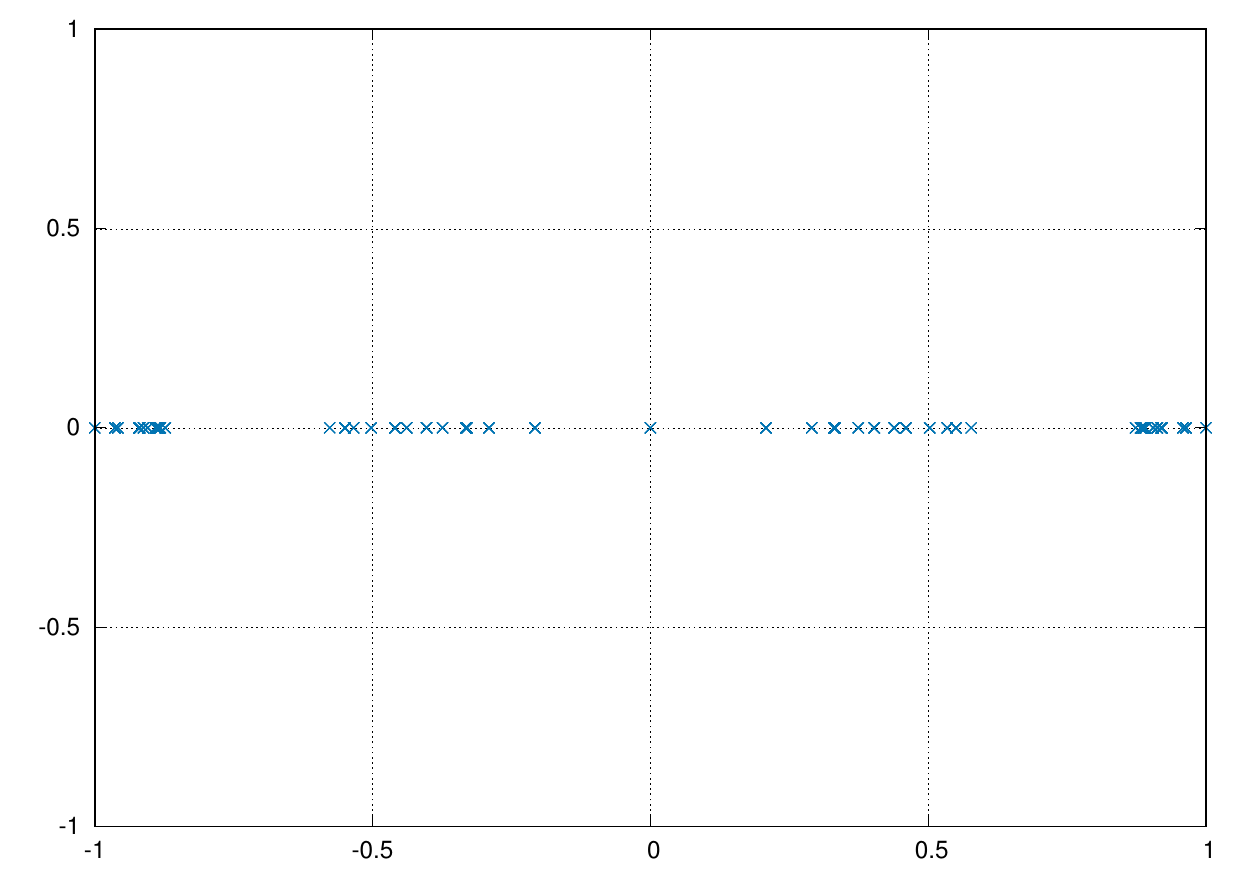} 
\hspace{2em}
\includegraphics[width =60mm]{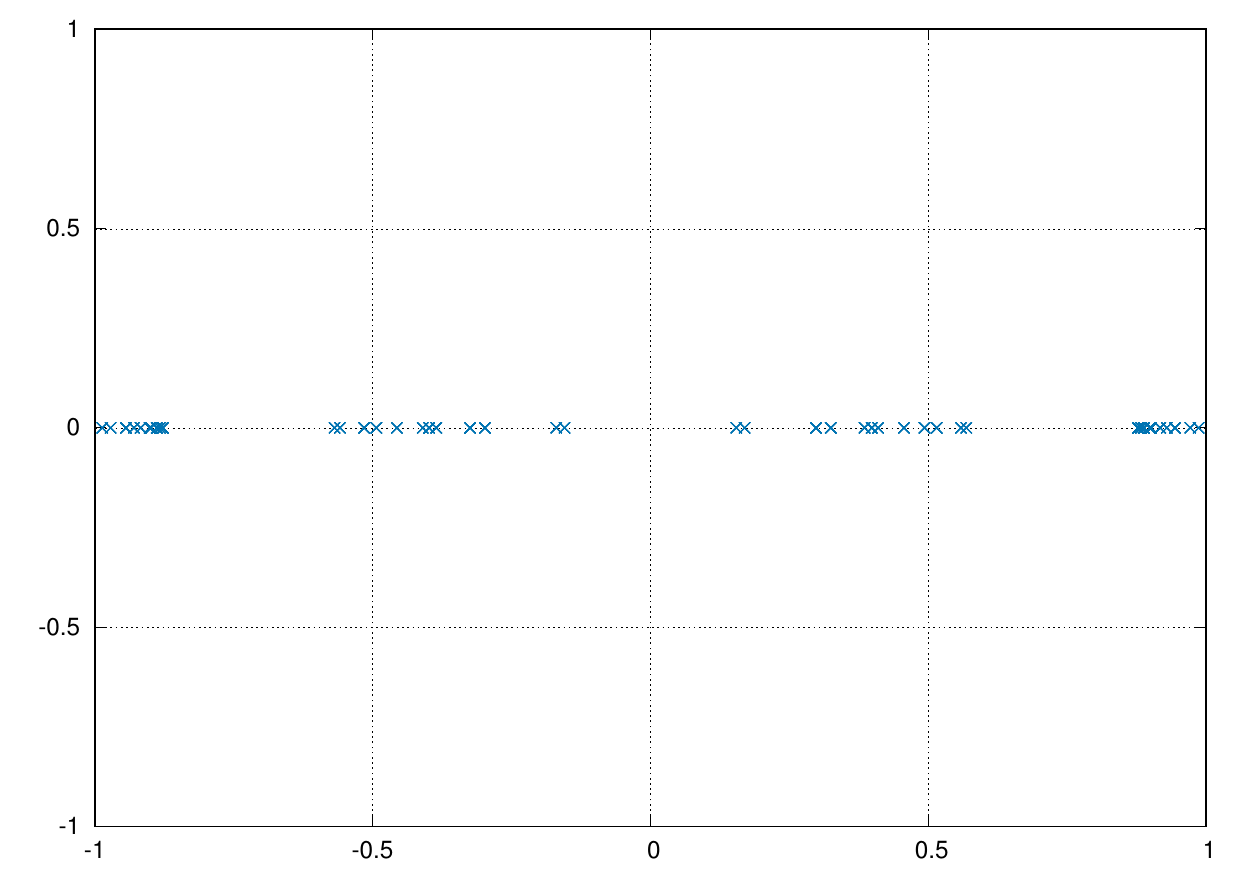} 
\end{center}
\caption{
Eigenvalue spectra of 
$\big( u^\dagger \Gamma^{6} \Gamma^{a} E^{a} u \big)$
for a spin-field configuration of the class $E^a_\ast(x)$.
with the periodic b.c. [left] 
and the anti-periodic b.c. [right].
The lattice size is $L=12$.
}
\label{fig:eigenvalues-12x12-PI}
\end{figure}

\begin{figure}[h]
\begin{center}
\includegraphics[width =60mm]{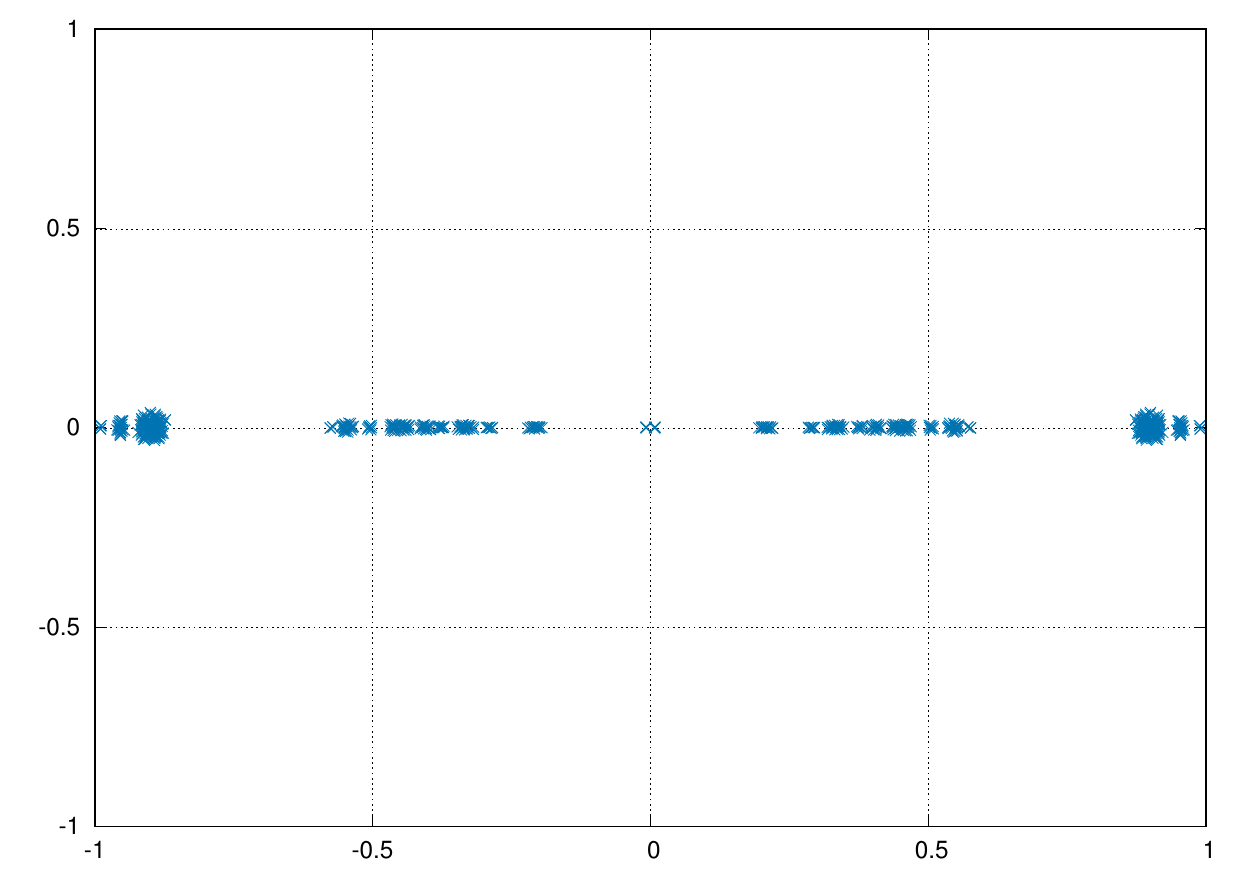} 
\hspace{2em}
\includegraphics[width =60mm]{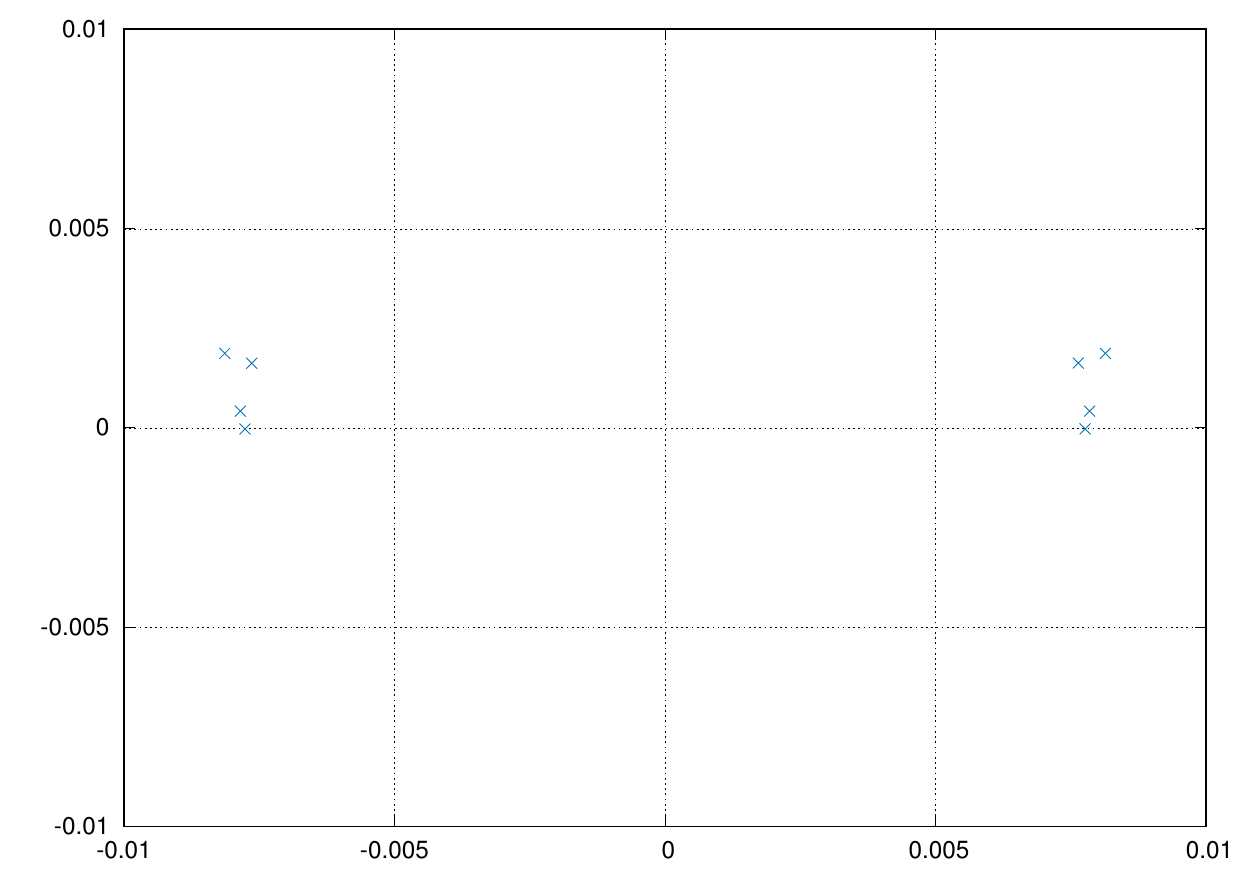} 
\end{center}
\caption{
Eigenvalue spectra of 
$\big( u^\dagger \Gamma^{6} \Gamma^{a} E^{a} u \big)$
for the spin-field configuration 
$E^a_\ast(x)$ plus
a small randomly-generated perturbation
with the periodic b.c. [left].
The range of the would-be zero eigenvalues are zoomed up in the [right] figure.
The lattice size is $L=12$.
}
\label{fig:eigenvalues-12x12-PIeps}
\end{figure}

\noindent
For the spin field configurations $E^a_\ast(x)$,
we found eight zero eigenvalues for the periodic boundary condition, but none for the anti-periodic boundary condition.
The typical examples of the eigenvalue spectra
are shown in fig.~\ref{fig:eigenvalues-12x12-PI} for $L=12$ 
with the periodic and anti-periodic boundary conditions.
We also observed that the degeneracy of the eight zero modes
are resolved by small randomly-generated perturbations to
$E^a_\ast(x)$ in the structure like
$\{ (\lambda_i, - \lambda_i^\ast) \, \vert \,  i=1,\cdots, 4 \}$ approximately, as shown in fig.~\ref{fig:eigenvalues-12x12-PIeps}, 
and  
the determinant 
$\det \big( u^\dagger \Gamma^{6} \Gamma^{a} E^{a} u \big)$
remains real and positive.

Based on the above analytical results and numerical observations,
we can argue that the determinant 
 $\det (u^T \, i\gamma_3 c_D  {\rm T}^a E^a v')$
($=\det ( u^\dagger \, i\Gamma^{6} \Gamma^{a} E^{a} u )$) 
is positive semi-definite for any spin-field configuration $E^a(x)$
in the weak gauge-coupling limit: 
\begin{eqnarray}
\det (u^T \, i\gamma_3 c_D  {\rm T}^a E^a v') 
=
\det ( u^\dagger \Gamma^{6} \Gamma^{a} E^{a} u )
\, \ge 0  \qquad ( g_0 = 0 ) .
\end{eqnarray}
We first note that
the space of the SO(6)-vector spin field configurations, 
which we denote with ${\cal V}_E$, 
is the direct product of 
multiple $S^5$
and is pathwise connected.
Then any configuration of the spin field $E^a(x)$ can be reached
from the constant configuration 
$E_0^a(x) = \delta^{a, 6}$ 
through a continuous deformation.
Since 
it is unity for the constant configuration, 
the determinant of 
$(u^\dagger \, i \Gamma^6 \Gamma^a E^a u)$
should be positive for a given configuration $E^a(x)$
as long as
there exists a path 
to $E^a(x)$ 
from $E_0^a(x) (= \delta^{a, 6})$ 
such that
the determinant never vanish along the path.
%
On the other hand, 
for the spin configurations with which the determinant is zero, 
a certain subset in the eigenvalue spectrum
of $(u^\dagger \, i \Gamma^6 \Gamma^a E^a u)$
should be zero.
Along the path which goes though 
such a spin configuration, 
the eigenvalue spectrum 
flow 
and the subset of would-be zeros pass the origin in the complex plane.
Then 
the determinant as the product of the eigenvalues,
$\det (u^\dagger \, i \Gamma^6 \Gamma^a E^a u)
={\prod_{j=1}^{4L^2}} \lambda_j$,
can change discontinuously
in its signature(phase).
Since the signature(phase) of the determinant
stays constant as far as the determinant is nonzero,
this could happen if and only if
the subspace of the configurations with the vanishing determinant, 
which we denote with ${\cal V}_{E}^0$,
can divide
the entire space of the spin configurations ${\cal V}_E$ 
into the subspaces 
which are disconnected each other.
And the divided disconnected space, ${\cal V}_E \, \backslash \, {\cal V}_{E}^0$, should be classified by the 
values of the signature(phase) of the determinant.
%
In this respect, however, one notes that 
$\pi_k(S^6) = 0 \, (k < 6)$ and 
any topological obstructions and
the associated topological terms 
are not known in the continuum limit
for the SO(6)-vector spin field $E^a(x)$ 
on the two-dimensional spacetime $S^2$ or $T^2$.
In particular, 
any topologically non-trivial configurations/defects of the SO(6)-vector spin field and the associated
fermionic massless excitations are not known
in the continuum limit.
Then it seems reasonable to assume that 
${\cal V}_{E}^0$ consists of lattice artifacts and 
in particular it is given solely by the subspace of the configurations $E^a_\ast(x)$, which we denote with ${\cal V}_{E}^\ast$. If one assumes that ${\cal V}_{E}^0 = {\cal V}_{E}^\ast$, 
the multiplicity of the zero eigenvalues are eight
and the would-be zero eigenvalues have the approximate
structure 
$\{ (\lambda_i, - \lambda_i^\ast) \, \vert \,  i=1,\cdots, 4 \}$.
Then
the signature(phase) of the determinant does not change
in passing ${\cal V}_{E}^0  ( = {\cal V}_{E}^\ast)$.
Therefore the determinant 
$\det ( u^\dagger \Gamma^{6} \Gamma^{a} E^{a} u )$
is positive semi-definite.

It then follows that the partition function of the mirror fermion sector is real and positive in the weak gauge-coupling limit:
\begin{eqnarray}
\big\langle  1 \big\rangle_{M}
&=&
\big\langle
1
\big\rangle_{E}
\nonumber\\
&=&
\int {\cal D}[E^a] \,
\det (u^T \, i\gamma_3 c_D \check{\rm T}^a E^a v') 
\nonumber\\
&=&
\int {\cal D}[E^a] \,
\det ( u^\dagger \Gamma^{6} \Gamma^{a} E^{a} u )
\qquad  \, \, > \,\,  0 \qquad (g_0 = 0) .
\end{eqnarray}

\subsubsection{Monte Carlo simulation of 
the SO(6)-vector spin field}

\begin{figure}[h]
\begin{center}
\includegraphics[width =90mm]{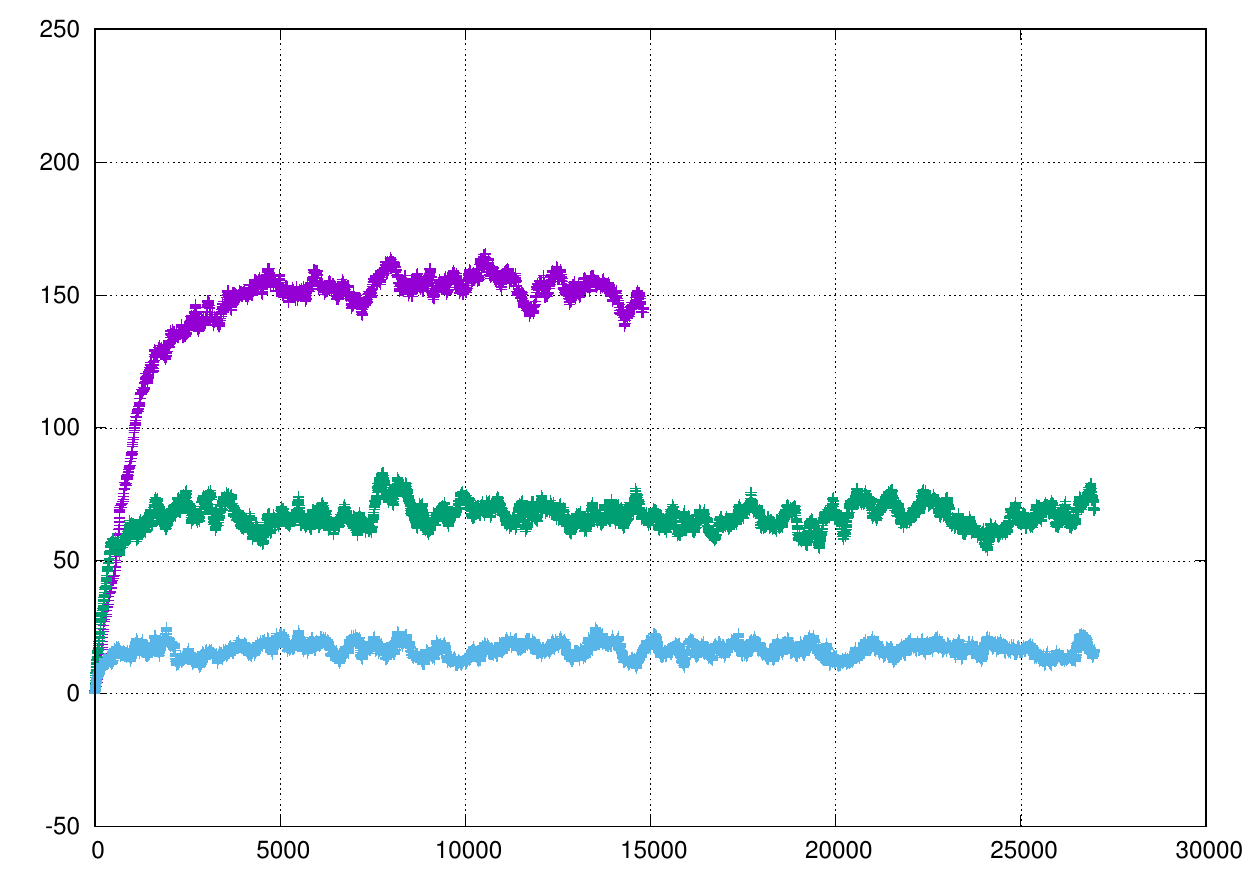} 
\end{center}
\caption{
Monte Carlo histories of the effective action $S_E[E^a]$
fot the lattice sizes $L=4, 8, 12$. The periodic boundary condition is used for the fermion fields.
}
\label{fig:histories-effective-action-L=4-12}
\end{figure}

From the positive semi-definiteness of the determinant
$\det (u^T \, i\gamma_3 c_D  {\rm T}^a E^a v')$, 
it also follows that the Monte Carlo method can be applied
to the path-integration of the SO(6)-vector spin field $E^a(x)$
in the weak gauge-coupling limit,
using the effective action 
\begin{eqnarray}
S_E[E^a] &=& 
- \ln \det (u^T \, i\gamma_3 c_D \check {\rm T}^a E^a v') 
\nonumber\\
&=& - \ln 
\det ( u^\dagger \Gamma^{6} \Gamma^{a} E^{a} u ) .
\end{eqnarray}
We have applied a hybrid Monte Carlo method to this 
spin model and have performed simulations 
for the range of lattice sizes $L=4, 8, 12$.
The examples of the histories of the effective action
$S_E[E^a]$ are shown in fig~\ref{fig:histories-effective-action-L=4-12} for the various lattice sizes.
The trajectory length is 0.05 and the average acceptance ratio is 0.5.
We have used the spin-field configurations generated by these simulations
to compute the observables of the mirror sector such as
the correlation functions of the mirror-sector fields
and 
the two-point vertex function of the (external) gauge fields.
These results are shown and discussed in the following sections.

\subsubsection{Short-ranged correlation functions}

We first examine the correlation functions of the fields of the mirror fermion sector in the weak gauge coupling limit.
We consider the following two-point correlation functions
in the channels of $\underbar{6}$
and 
$\underbar{4}$ representations of SO(6) and Spin(6). 
\begin{eqnarray}
G_E(x, y)^{ab} &\equiv&
\big\langle E^a(x) E^b(y) \big\rangle_{M} \,  /  \,
\big\langle  1 \big\rangle_{M} \\
&=&
\big\langle E^a(x) E^a(y) \big\rangle_{E} \,  /  \,
\big\langle  1 \big\rangle_{E} 
\end{eqnarray}
\begin{eqnarray}
G_{\psi' \psi E}(x, y) &=& 
\big\langle \psi'_-(x) \, \psi_+{}^T(y) i \gamma_3 c_D {\rm T}^a E^a(y)
\big\rangle_M  \,  /  \,
\big\langle  1 \big\rangle_{M} , \\
G_{\psi \psi' E}(x, y) &=& 
\big\langle \psi_+(x) \, \psi'_-{}^T(y) i \gamma_3 c_D {\rm T}^a E^a(y)
\big\rangle_M  \,  /  \,
\big\langle  1 \big\rangle_{M} . 
\end{eqnarray}
The fermionic correlation functions above satisfy
the Schwinger-Dyson equations given as follows:
\begin{eqnarray}
&&
\big\{ G_{\psi' \psi E} \,  \hat P_- \big\}(x, y) =  \hat P_-(x, y) ,
\\
&&
\big\{ G_{\psi \psi' E} \,  \hat P_+ \big\}(x, y) =  \hat P_+(x, y) .
%
\end{eqnarray}
And these equations can be solved as
\begin{eqnarray}
G_{\psi' \psi E} (x,y) &=&
\hat P_-(x, y)
+
\big\{ G_{\psi' \psi E} \,  \hat P_+ \big\}(x, y) , \\
G_{\psi \psi' E} (x,y) &=&
\hat P_+(x, y)
+
\big\{ G_{\psi \psi' E} \,  \hat P_- \big\}(x, y)  . 
\end{eqnarray}
Therefore, non-trivial parts of the fermionic correlation functions are given by
$\big\{ G_{\psi' \psi E} \,  \hat P_+ \big\}(x, y)$ and 
$\big\{ G_{\psi \psi' E} \,  \hat P_- \big\}(x, y)$,
which may be expressed explicitly in terms of the chiral basis as follows.
\begin{eqnarray}
\big\{ G_{\psi' \psi E} \,  \hat P_+ \big\}(x, y) &=&
\big\langle 
v'(x) 
(u^T 
{\cal M}_E \, v')^{-1}
(u^T
{\cal M}_E \, u) u(y)^\dagger
\big\rangle_{E}  \,  /  \,
\big\langle  1 \big\rangle_{E} , 
\\
\big\{ G_{\psi \psi' E} \,  \hat P_- \big\}(x, y) &=&
\big\langle 
u(x) 
(v'{}^T 
{\cal M}_E \, u)^{-1}
(v'{}^T
{\cal M}_E \, v') {v'(y)}^\dagger
\big\rangle_{E}  \,  /  \,
\big\langle  1 \big\rangle_{E} ,
\end{eqnarray}
where ${\cal M}_E = i \gamma_3 c_D T^a E^a$.
As to the similar correlation functions which are related to the anti-fields, 
\begin{eqnarray}
G_{\bar E}(x, y)^{ab} &\equiv&
\big\langle \bar E^a(x) \bar E^b(y) \big\rangle_{M} \,  /  \,
\big\langle  1 \big\rangle_{M}  , 
%
\end{eqnarray}%
\begin{eqnarray}
\bar G_{\bar \psi' \bar \psi \bar E}(x, y) &=& 
\big\langle 
 {\bar\psi'_-(x)}^T   
\,
\bar \psi_+ (y)  \,
i \gamma_3 c_D 
{\rm T}^{a \dagger} \bar E^a(y) 
\big\rangle_M  \,  /  \,
\big\langle  1 \big\rangle_{M} , \\
\bar G_{\bar \psi \bar \psi' \bar E}(x, y) &=& 
\big\langle 
{\bar \psi_+ (x)}^T
\,
\bar\psi'_-(y) 
\,
i \gamma_3 c_D 
{\rm T}^{a \dagger} \bar E^a(y) 
\big\rangle_M  \,  /  
\big\langle  1 \big\rangle_{M} , 
\end{eqnarray}
they are obtained exactly and have the short-ranged property as follows.
\begin{eqnarray}
G_{\bar E}(x, y)^{ab} 
&=&  \delta_{x y} \delta^{ab} , \\
%
G_{\bar \psi' \bar \psi \bar E}(x, y) &=& P_+ {}^T \delta_{xy} \delta_{s t},
\\
G_{\bar \psi \bar \psi' \bar E}(x, y) &=& {P_-}^T \delta_{x y} \delta_{s t} .
\end{eqnarray}


%
\begin{figure}[h]
\begin{center}
\includegraphics[width =90mm]{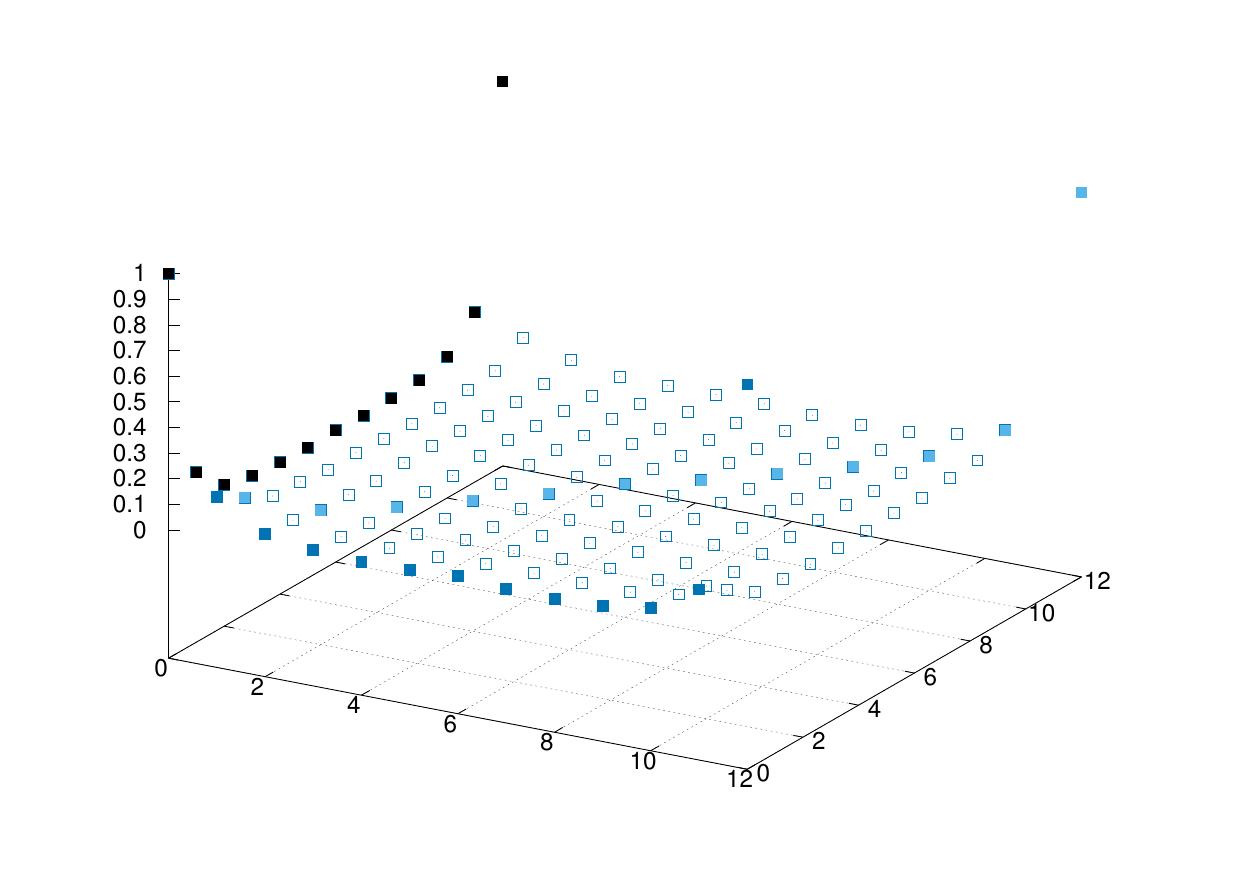} 
\\
\includegraphics[width =80mm]{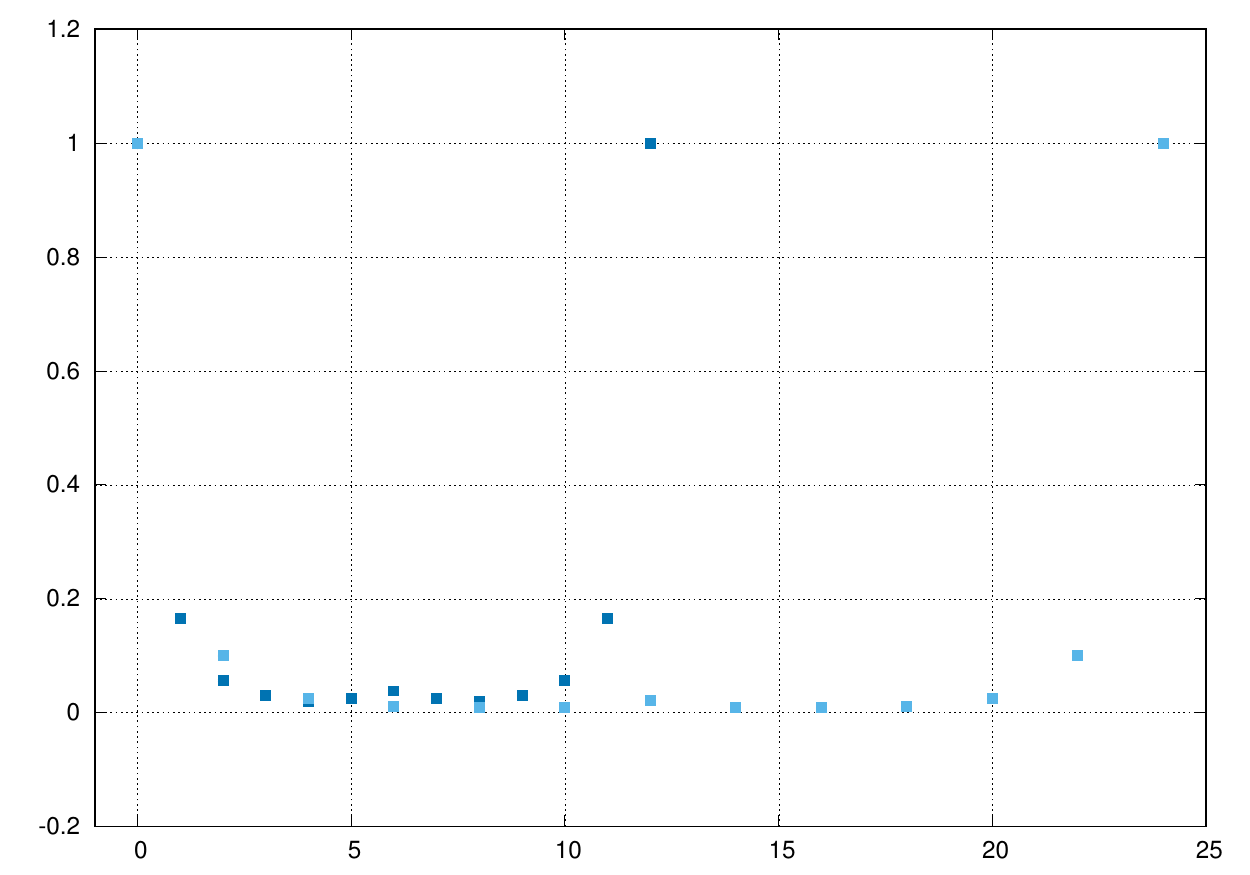} 
\\
\vspace{2em}
\includegraphics[width =80mm]{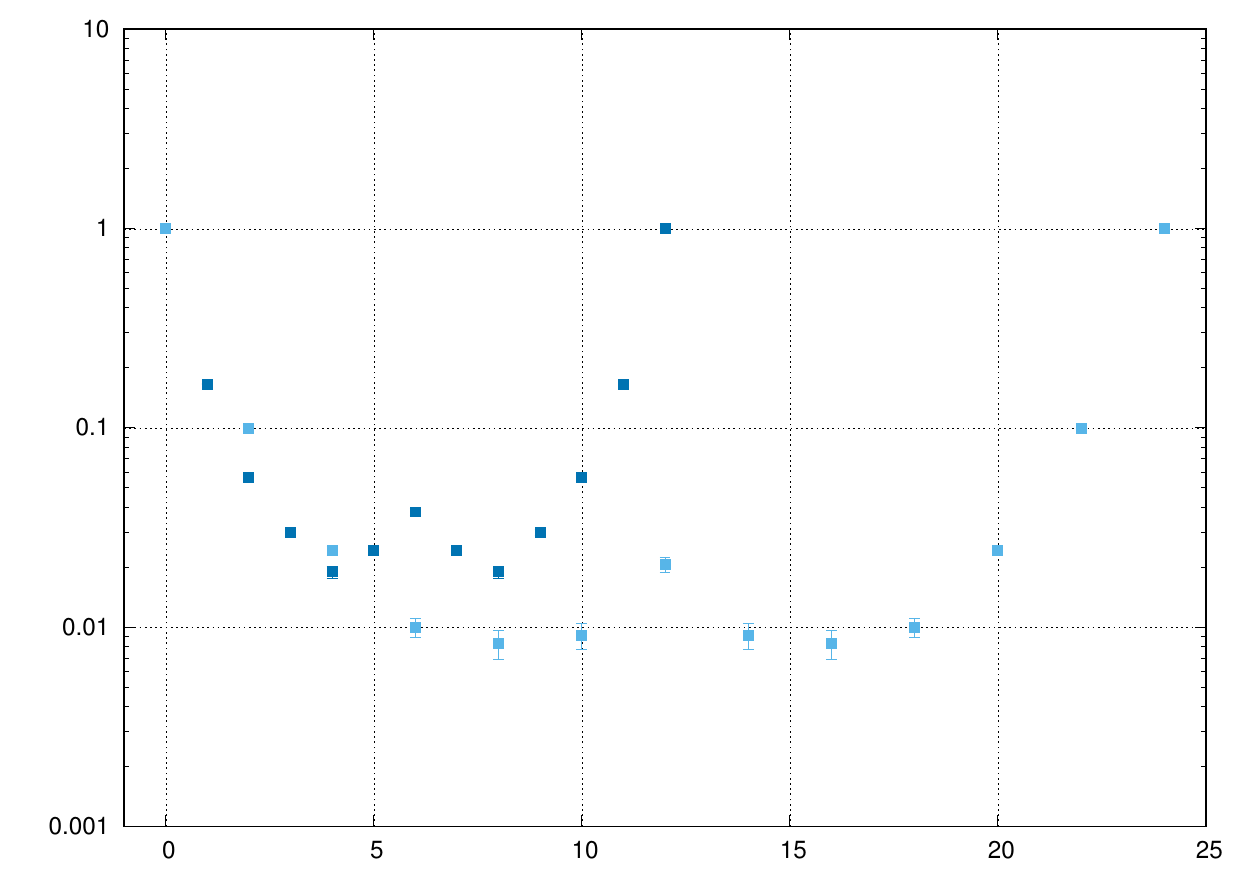} 
\end{center}
\caption{
$G_E(x)$ [left] and $ \ln \vert G_E(x) \vert $ [right] vs. 
$\vert x \vert_1 \equiv \vert x_0 \vert + \vert x_1
\vert$, where $G_E(x) \equiv \sum_{a=1}^6 G_E(x)^{aa}$.
The lattice size is $L=12$. The square-symbol(blue) plot is along
the temporal axis ($x_0=0$), while the triangle-symbol(light blue) plot is along
the diagonal axis ($x_0=x_1$).
1,100 configurations are sampled with the interval of 20 trajectories. 
The errors are simple statistical ones and the translational invariance is assumed. 
}
\label{fig:correlation-EE-12x12}
\end{figure}
In fig.~\ref{fig:correlation-EE-12x12}, we show
the numerical-simulation results of $G_E(x, y)^{ab}$ 
on the lattice with $L=12$. 
1,100 configurations are sampled with the interval of 20 trajectories. 
The errors are simple statistical ones and the translational invariance is assumed. 
We found that
the simulation results respect
the SO(6) symmetry, i.e. 
the diagonal components with $a=b$ are equal to each other and 
the off-diagonal components with $a \not = b$ are vanishing
upto the statistical errors.
We can see that
the correlation length is of order the lattice spacing
and 
the spin field $E^a(x)$ is disordered almost completely
just like $\bar E^a(x)$.

\begin{figure}[h]
\begin{center}
\includegraphics[width =90mm]{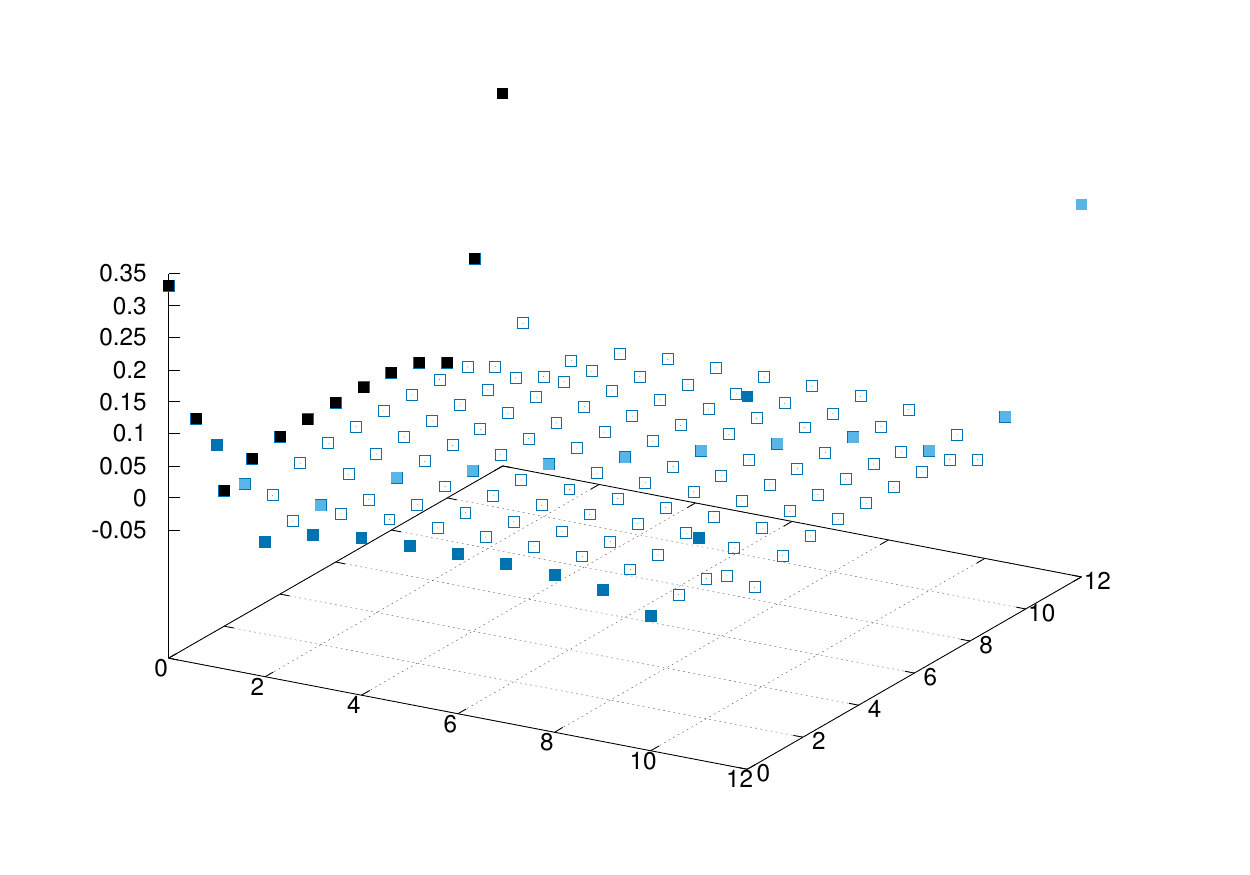} \\
\includegraphics[width =80mm]{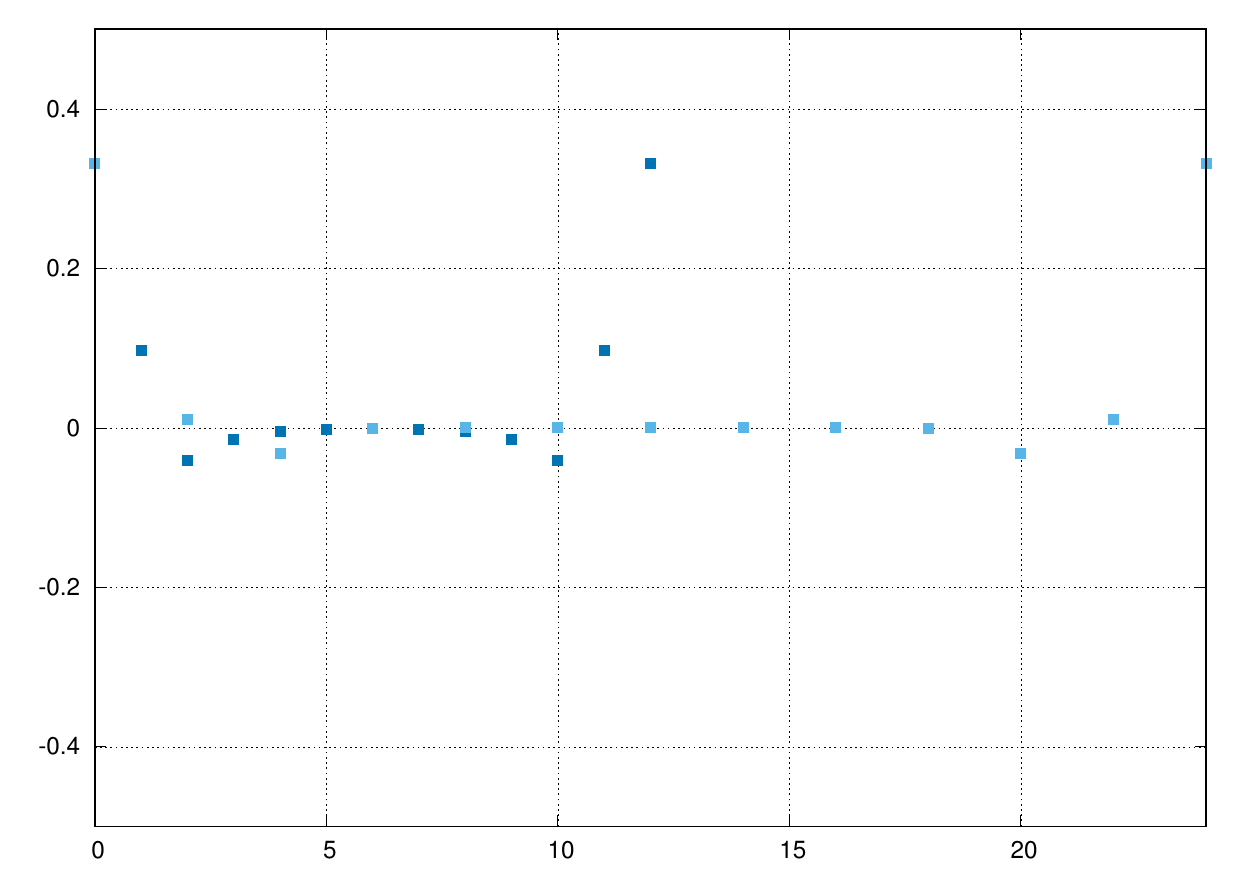} \\
\vspace{2em}
\includegraphics[width =80mm]{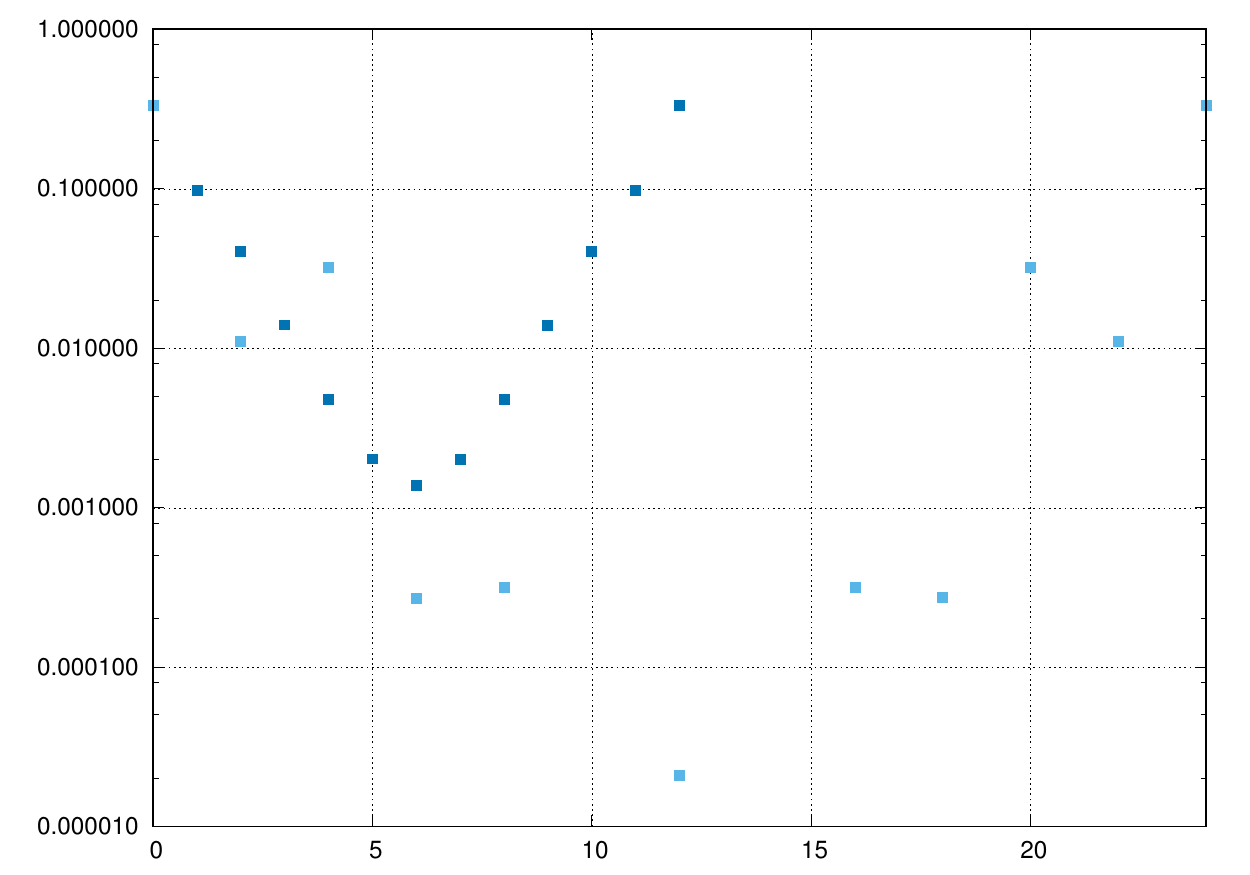} 
\end{center}
\caption{
$\sum_{s=1}^4 \{ G_{\psi' \psi E} \hat P_+\}_{00,s s}(x)$
vs. $x=(x_0, x_1)$ [top] and 
$\vert x \vert_1 \equiv \vert x_0 \vert + \vert x_1
\vert$ [middle, bottom].
The lattice size is $L=8$. 
The blue-symbol and black-symbol plots are along
the spacial axis ($x_0=0$)
and
temporal axis ($x_1 =0$), respectively,
while the lightblue-symbol plot is along
the diagonal axis ($x_0=x_1$).
1,100 configurations are sampled with the interval of 20 trajectories. 
The errors are simple statistical ones.
}
\label{fig:correlation-PPT00-12x12}
\end{figure}

\begin{figure}[h]
\begin{center}
\includegraphics[width =75mm]{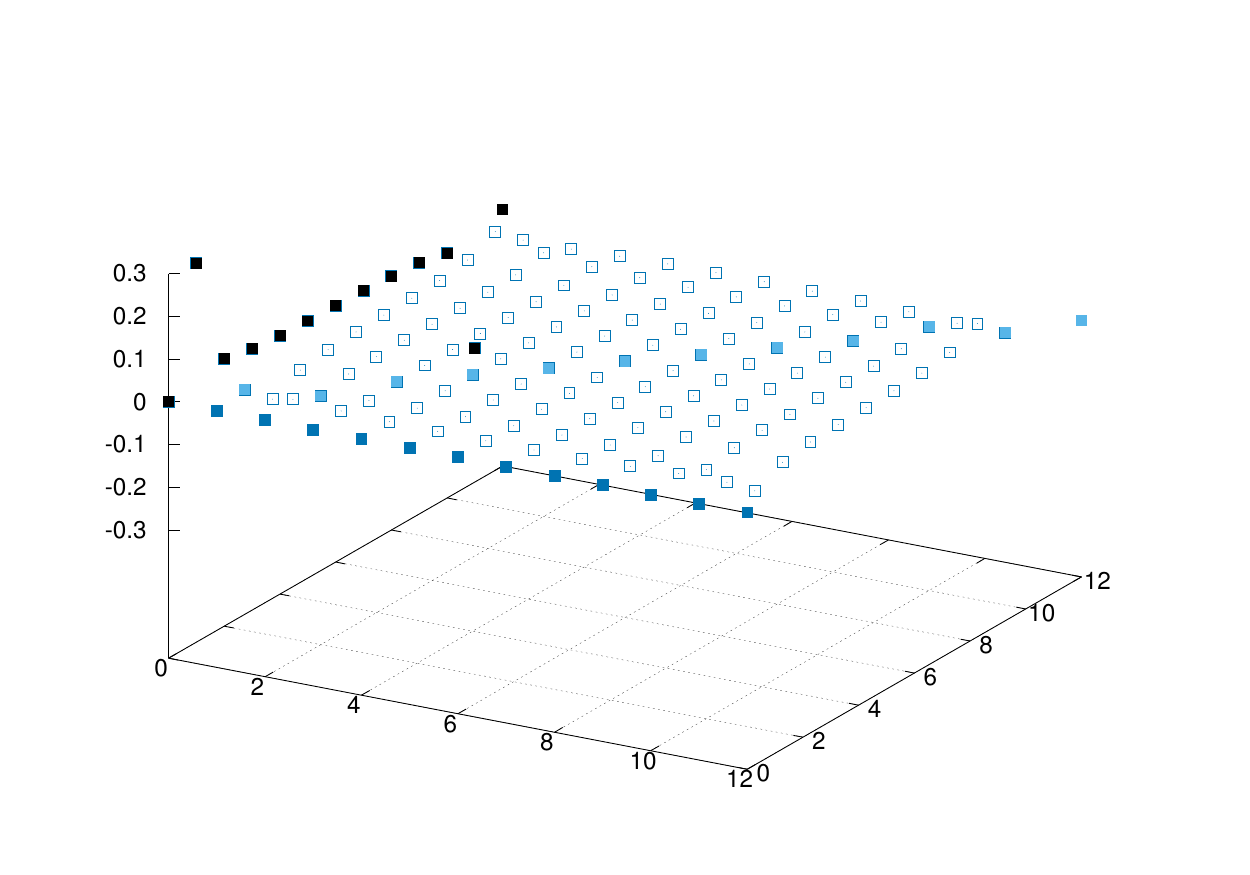} 
\includegraphics[width =75mm]{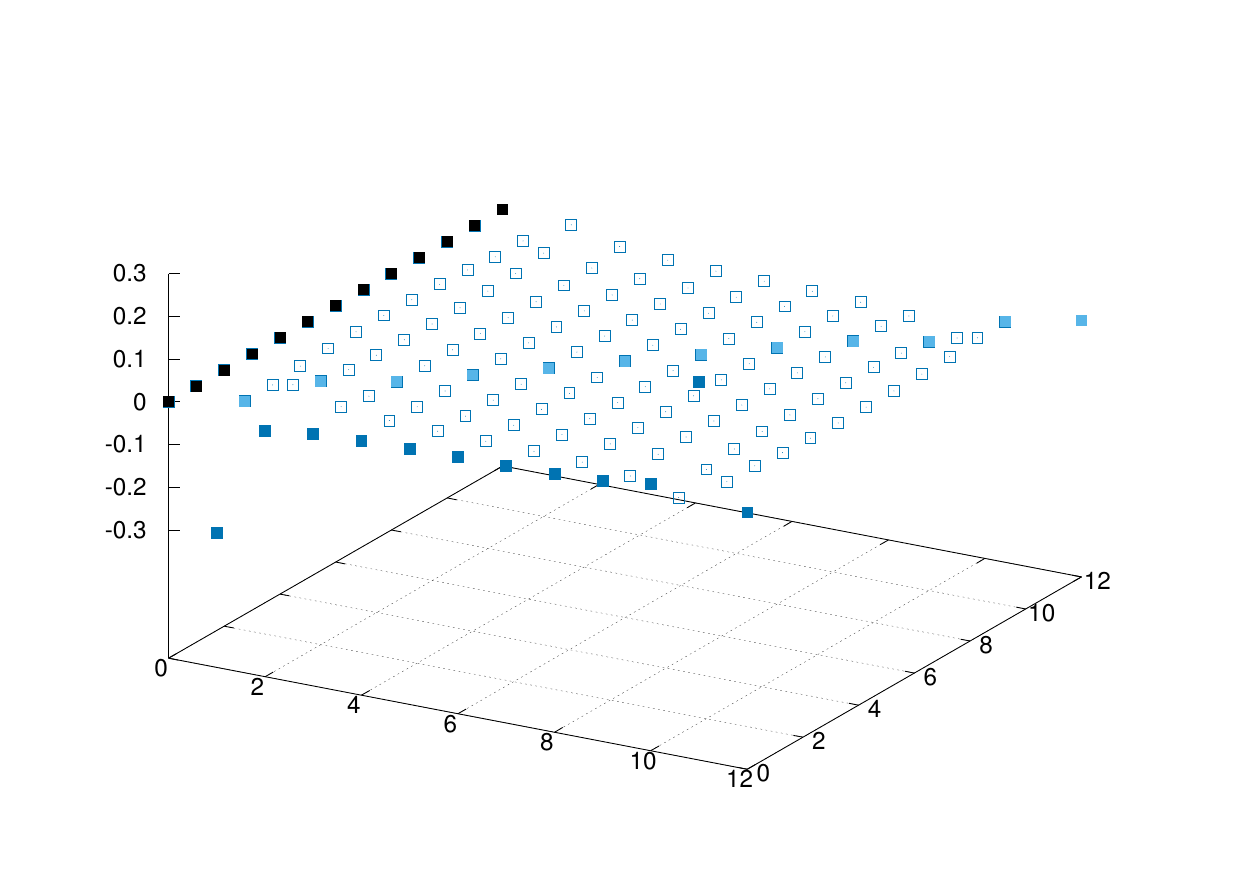}
\\
\vspace{2em}
\includegraphics[width =70mm]{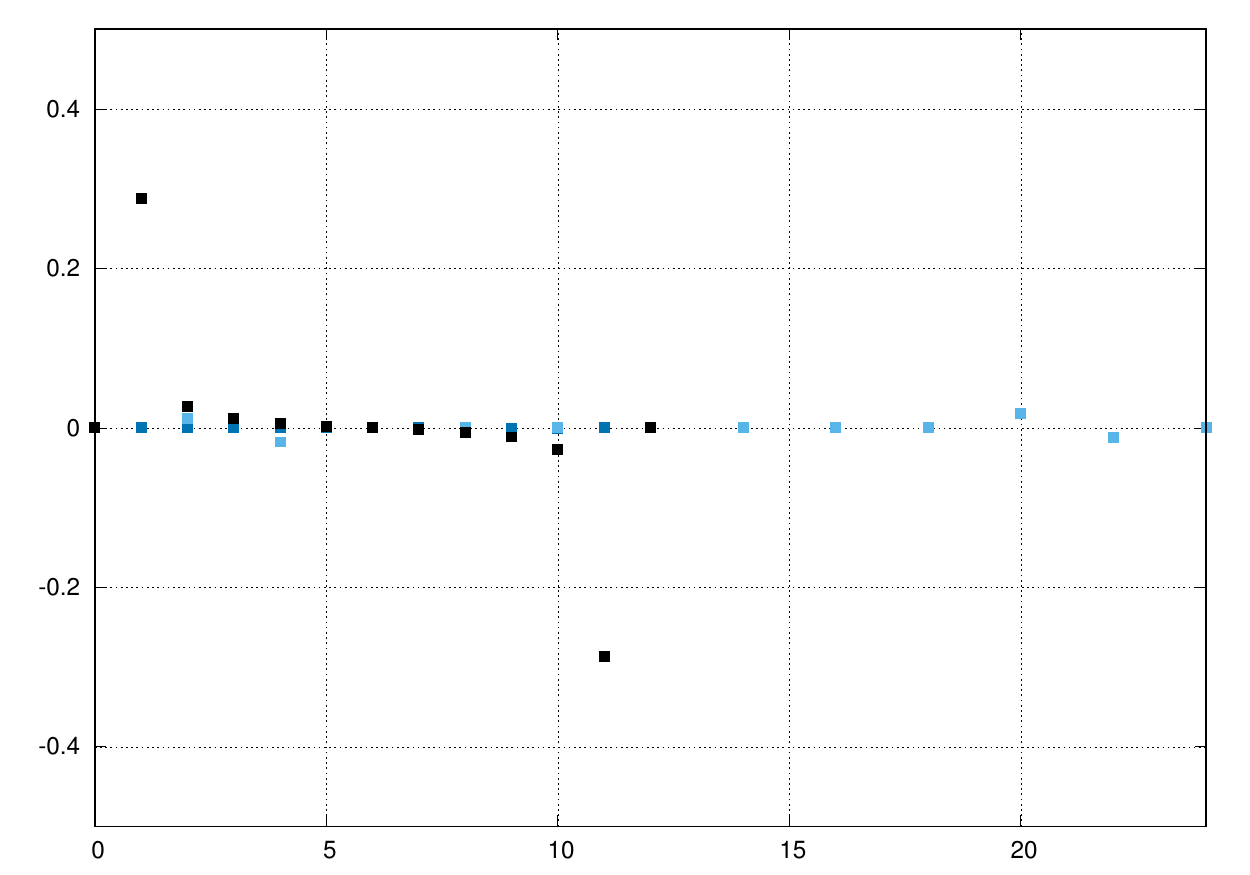} 
\hspace{1em}
\includegraphics[width =70mm]{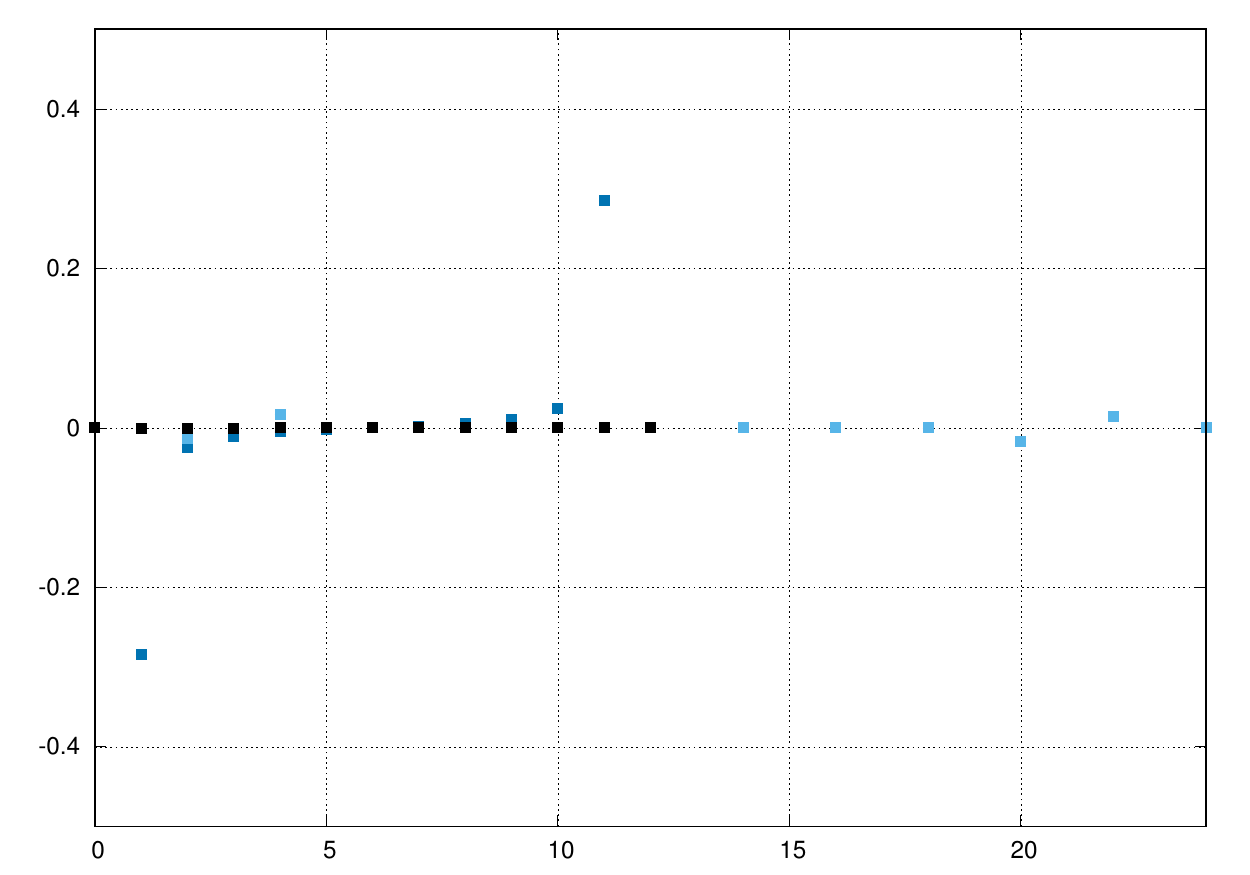} 
\\
\vspace{3em}
\includegraphics[width =70mm]{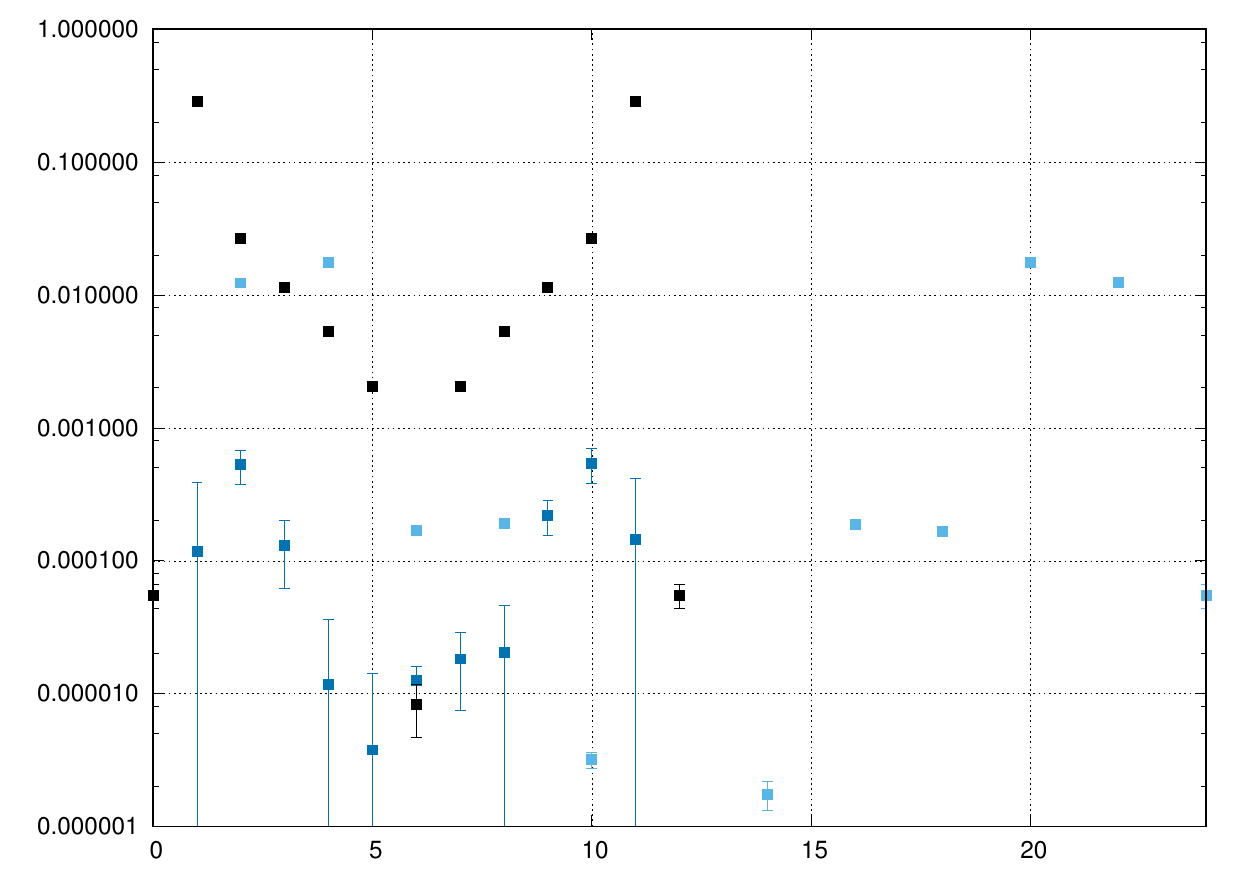} 
\hspace{1em}
\includegraphics[width =70mm]{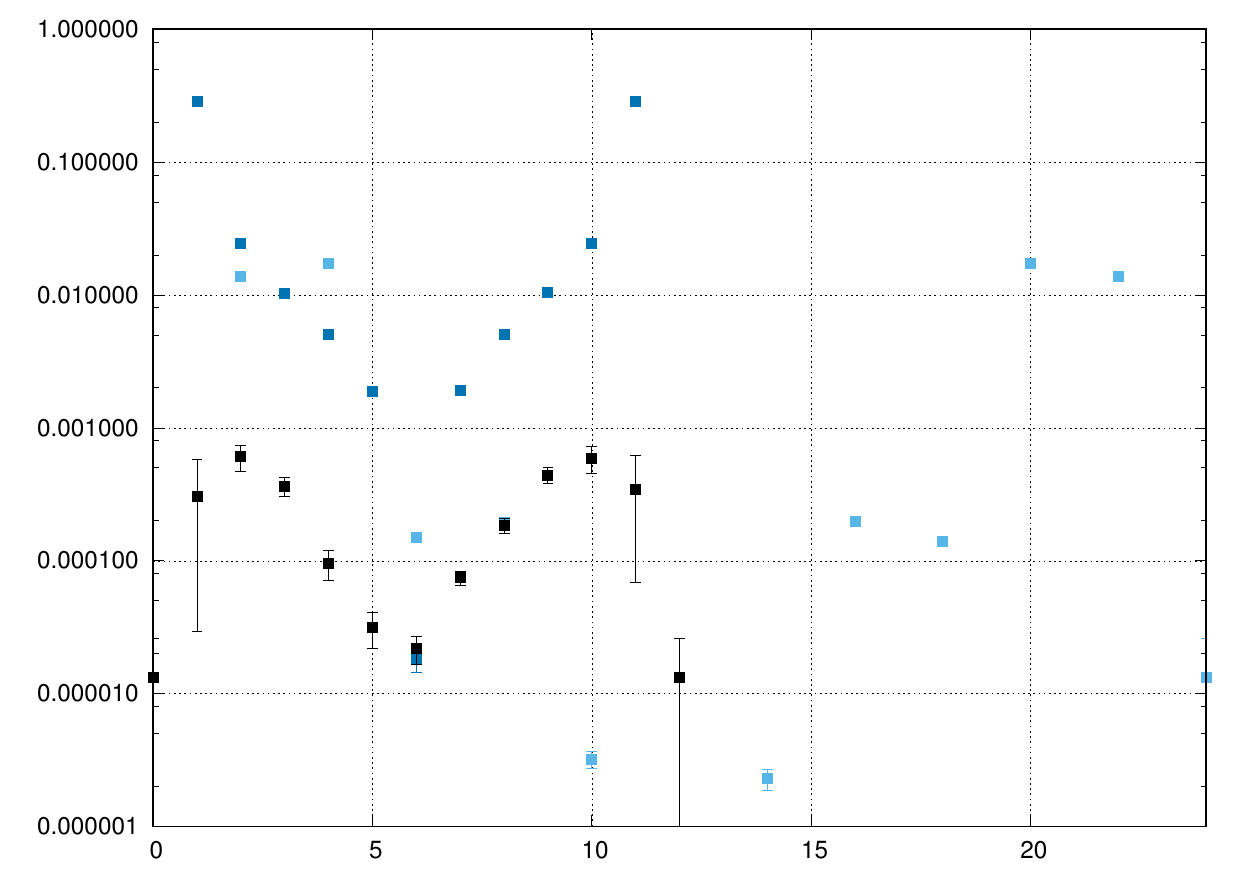} 
\end{center}
\caption{
The real [left] and imaginary [right] parts of 
$\sum_{s=1}^4 \{ G_{\psi' \psi E} \hat P_+\}_{01,s s}(x)$
vs. $x=(x_0, x_1)$ [top] and 
$\vert x \vert_1 \equiv \vert x_0 \vert + \vert x_1
\vert$ [middle, bottom].
The lattice size is $L=12$. The blue-symbol and black-symbol plots are along
the spacial axis ($x_0=0$)
and
temporal axis ($x_1 =0$), respectively,
while the light-blue-symbol plot is along
the diagonal axis ($x_0=x_1$).
1,100 configurations are sampled with the interval of 20 trajectories. 
The errors are simple statistical ones.
}
\label{fig:correlation-PPT01-12x12}
\end{figure}

\begin{figure}[h]
\begin{center}
\includegraphics[width =75mm]{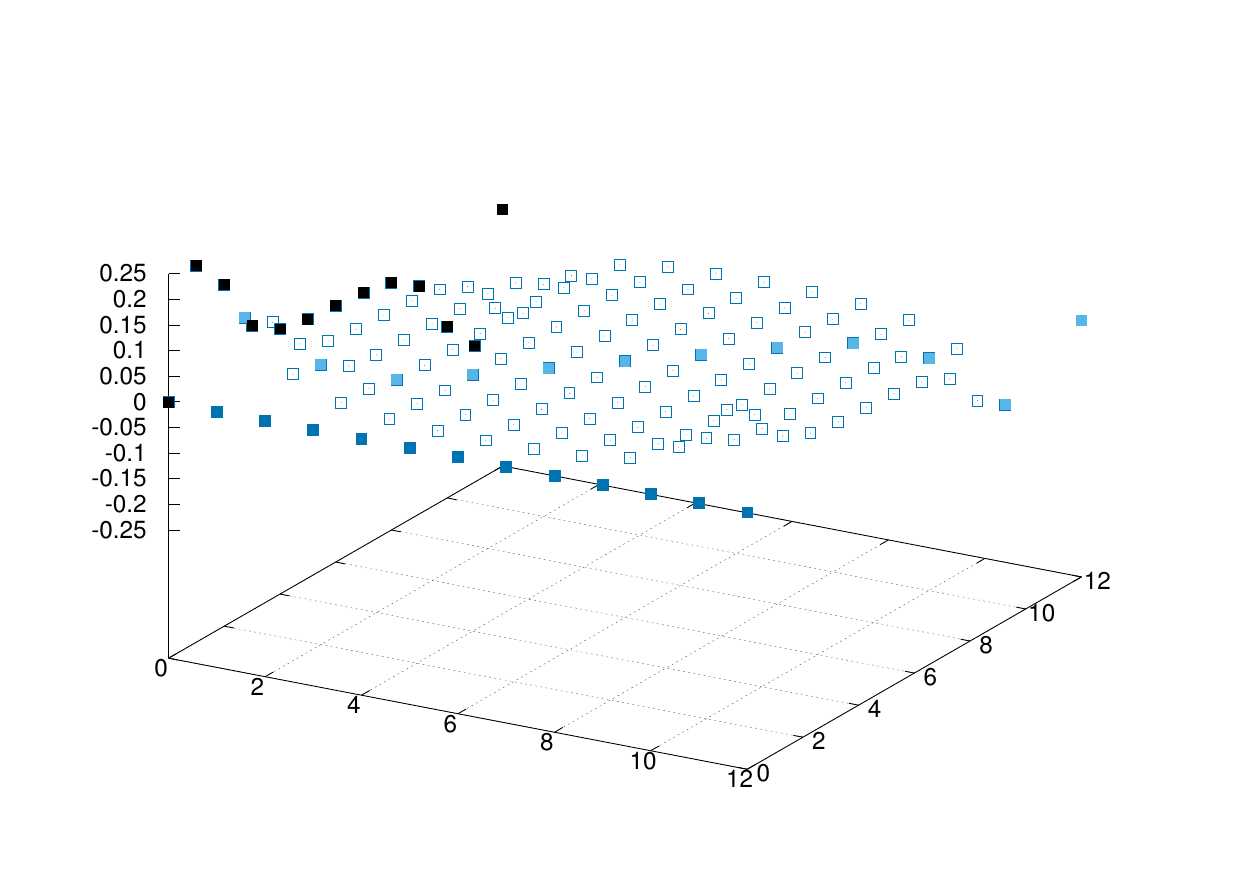} 
\includegraphics[width =75mm]{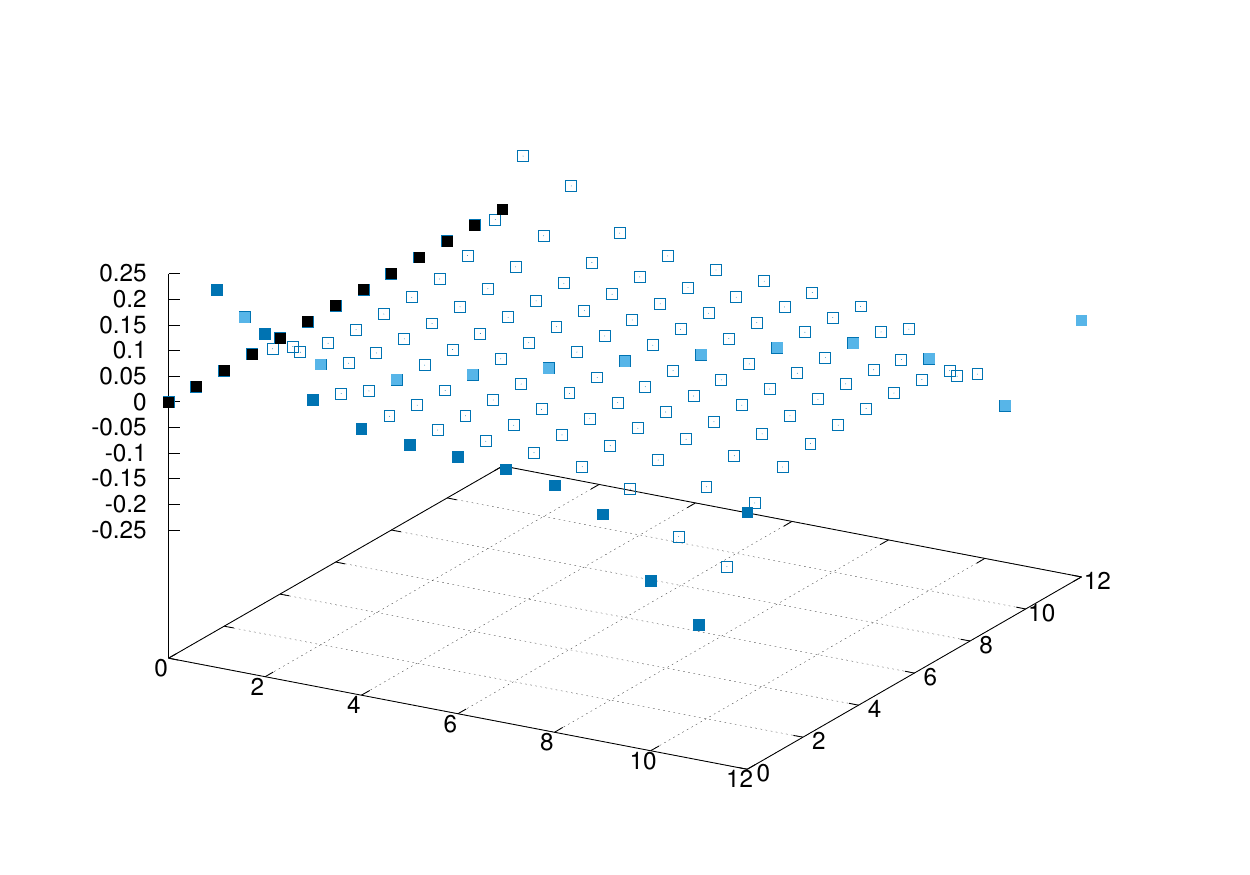}
\\
\vspace{2em}
\includegraphics[width =70mm]{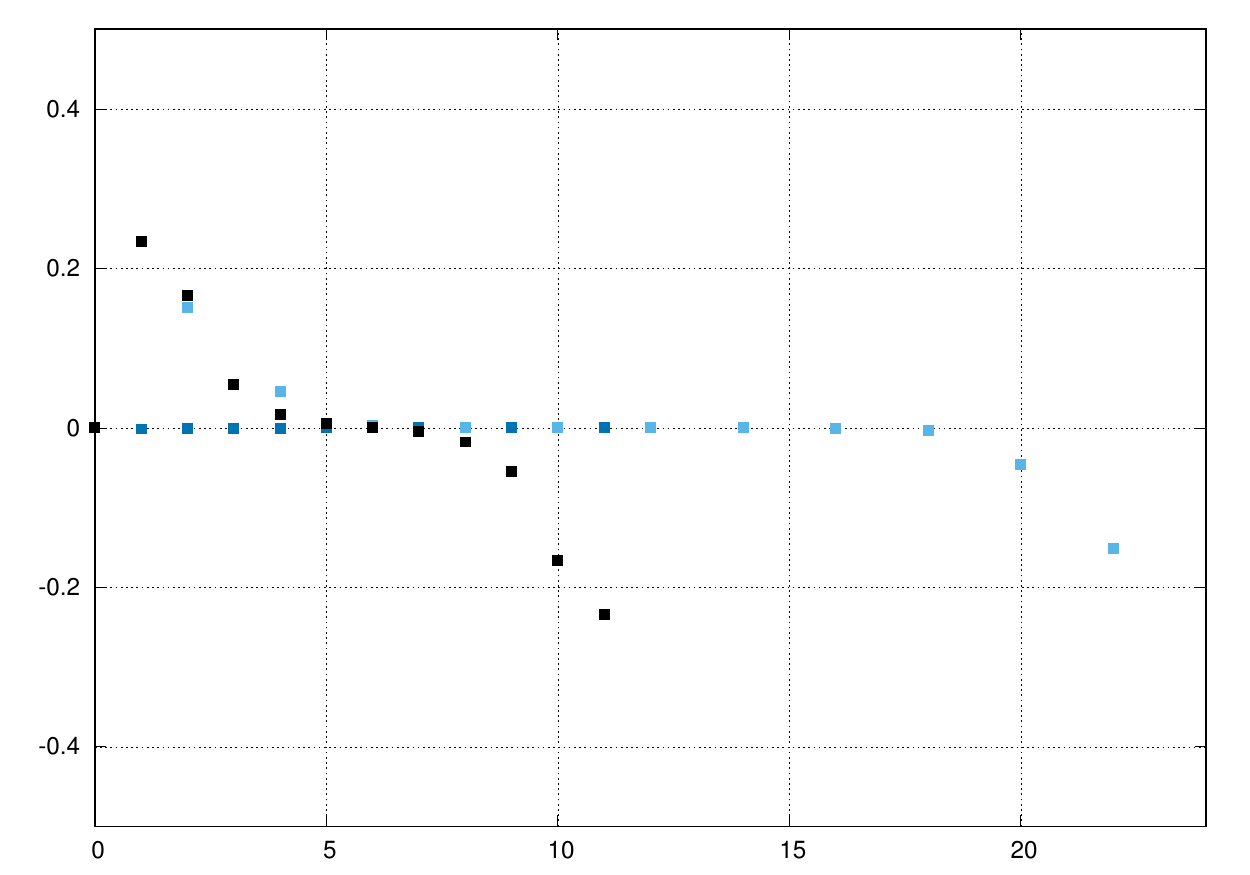} 
\hspace{1em}
\includegraphics[width =70mm]{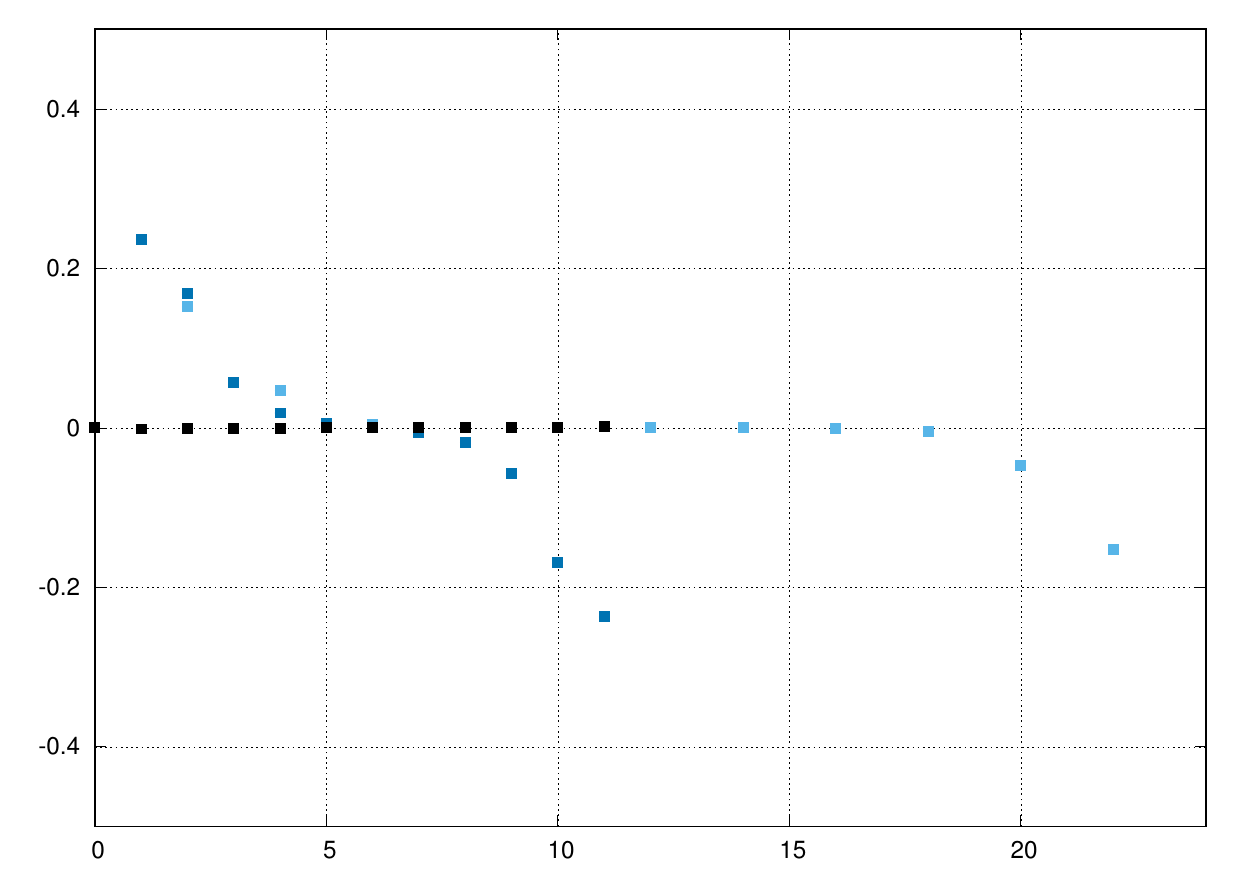} 
\\
\vspace{3em}
\includegraphics[width =70mm]{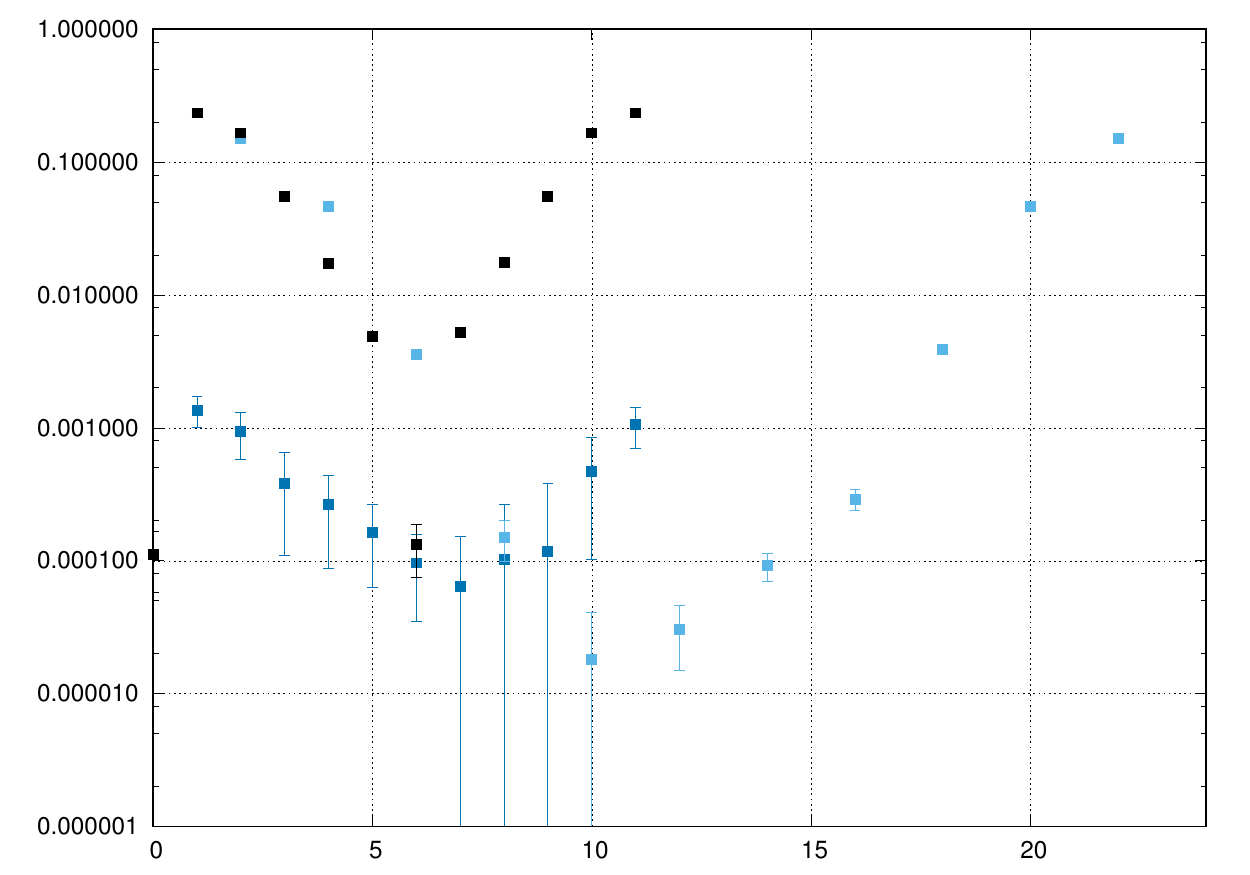} 
\hspace{1em}
\includegraphics[width =70mm]{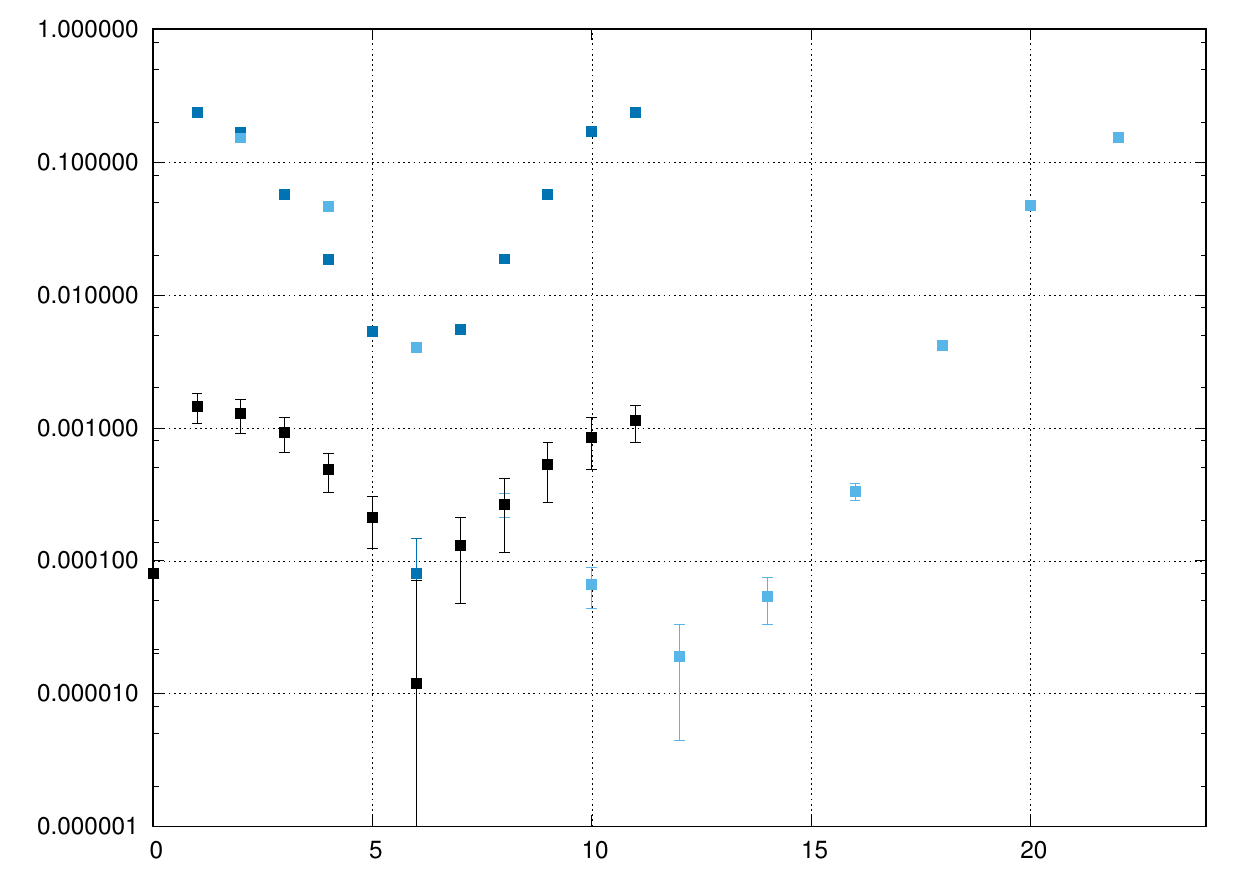} 
\end{center}
\caption{
The real [left] and imaginary [right] parts of 
$\sum_{s=1}^4 \{ G_{\psi' \psi E} \hat P_+\}_{10,s s}(x)$
vs. $x=(x_0, x_1)$ [top] and 
$\vert x \vert_1 \equiv \vert x_0 \vert + \vert x_1
\vert$ [middle, bottom].
The lattice size is $L=12$. The blue-symbol and black-symbol plots are along
the spacial axis ($x_0=0$)
and
temporal axis ($x_1 =0$), respectively,
while the light-blue-symbol plot is along
the diagonal axis ($x_0=x_1$).
1,100 configurations are sampled with the interval of 20 trajectories. 
The errors are simple statistical ones.
%
}
\label{fig:correlation-PPT10-12x12}
\end{figure}

In figs.~\ref{fig:correlation-PPT00-12x12},
\ref{fig:correlation-PPT01-12x12} and
\ref{fig:correlation-PPT10-12x12},
we show
the numerical-simulation results of $\{ G_{\psi' \psi E} \hat P_+ \}(x, y)$ for the lattice size $L=12$.
1,100 configurations are sampled with the interval of 20 trajectories. 
The errors are simple statistical ones.
Again, we found that the simulation results respect
the Spin(6) symmetry, i.e. 
the diagonal components with $s=t$ are equal to each other and 
the off-diagonal components with $s \not = t$ are vanishing within the statistical errors.
We can see that these correlation functions are short-ranged
and 
the correlation length is estimated as 
$\xi \simeq 12 / \ln 10^4 \simeq 1.30$.


From these results, we can see that 
the fields of the mirror fermion sector 
in the representations
$\underbar{6}$,
$\underbar{4}$ 
and $\underbar{4}^\ast$ of SO(6) and Spin(6)
have the short-range correlation lengths 
of order the lattice spacing.

\newpage
\subsubsection{Regular two-point vertex function of the U(1) gauge field}

We next examine 
the two-point vertex function of the U(1) gauge field in the mirror fermion sector, $ \tilde \Pi'_{\mu \nu} (k)$, 
which is defined by eq.~(\ref{eq:two-point-vertex-U1-gauge-field}),
\begin{eqnarray}
\label{eq:two-point-vertex-U1-gauge-field-second}
\frac{1}{L^2}\sum_{k} \tilde \eta_\mu(-k)\, \tilde \Pi'_{\mu \nu} (k) \, \tilde \zeta_\nu(k) &=&
\delta_\zeta 
\left[
\big\langle  - \delta_\eta S_M  \big\rangle_{M} \slash  \big\langle  1  \big\rangle_{M} 
\right]\, \Big\vert_{U(x,\mu) \rightarrow 1} .
\end{eqnarray}
In the $1^4 (-1)^4$ axial gauge model in consideration, 
it is given by 
\begin{eqnarray}
\label{eq:two-point-vertex-U1-gauge-field-1^4(-1)^4}
&&\frac{1}{L^2}\sum_{k} 
\tilde \eta_\mu(-k)\, \tilde \Pi'_{\mu \nu} (k) \, 
\tilde \zeta_\nu(k) 
\nonumber\\
&=&
\delta_\zeta
\left[
\big\langle 
{\rm Tr} \big\{
(u^T
\delta_\eta 
\{ \hat P_+{}^T 
{\cal M}_E 
\hat P'_-\} v')
\, 
( u^T 
{\cal M}_E 
\, v')^{-1}
\big\}
\big\rangle_E
\slash  \big\langle  1  \big\rangle_{E}
\right]
\Big\vert_{U(x,\mu) \rightarrow 1} 
\nonumber\\
&=&
\left[
\big\langle 
{\rm Tr} \big\{
(u^T
\delta_\zeta
\{ \hat P_+{}^T
\delta_\eta 
\{ \hat P_+{}^T 
{\cal M}_E 
\hat P'_-\}
\hat P'_- \}
 v')
\, 
( u^T 
{\cal M}_E 
\, v')^{-1}
\big\}
\big\rangle_E
\slash  \big\langle  1  \big\rangle_{E}
\right.
\nonumber\\
&&
\left.
-
\big\langle 
{\rm Tr} \big\{
(u^T
\delta_\eta 
\{ \hat P_+{}^T 
{\cal M}_E 
\hat P'_-\} v')
\, 
( u^T 
{\cal M}_E 
\, v')^{-1}
(u^T
\delta_\zeta 
\{ \hat P_+{}^T 
{\cal M}_E 
\hat P'_-\} v')
\, 
( u^T 
{\cal M}_E 
\, v')^{-1}
\big\}
\big\rangle_E
\slash  \big\langle  1  \big\rangle_{E}
\right.
\nonumber\\
&&
\left.
+
\big\langle 
{\rm Tr} \big\{
(u^T
\delta_\eta 
\{ \hat P_+{}^T 
{\cal M}_E 
\hat P'_-\} v')
\, 
( u^T 
{\cal M}_E 
\, v')^{-1}
\big\} \, \times
\right.
\nonumber\\
&&
\left.
\qquad\qquad\qquad\qquad\qquad\qquad\qquad\qquad\quad
{\rm Tr} \big\{
(u^T
\delta_\zeta 
\{ \hat P_+{}^T 
{\cal M}_E 
\hat P'_-\} v')
\, 
( u^T 
{\cal M}_E 
\, v')^{-1}
\big\}
\big\rangle_E
\slash  \big\langle  1  \big\rangle_{E}
\right.
\nonumber\\
&&
\left.
-
\big\langle 
{\rm Tr} \big\{
(u^T
\delta_\eta 
\{ \hat P_+{}^T 
{\cal M}_E 
\hat P'_-\} v')
\, 
( u^T 
{\cal M}_E 
\, v')^{-1}
\big\} 
\big\rangle_E
\slash  \big\langle  1  \big\rangle_{E}
\, \, \times
\right.
\nonumber\\
&&
\left.
\qquad\qquad\qquad\qquad\qquad\qquad
\big\langle
{\rm Tr} \big\{
(u^T
\delta_\zeta 
\{ \hat P_+{}^T 
{\cal M}_E 
\hat P'_-\} v')
\, 
( u^T 
{\cal M}_E 
\, v')^{-1}
\big\}
\big\rangle_E
\slash  \big\langle  1  \big\rangle_{E}
%
%
%
%
%
\right]
\Big\vert_{U(x,\mu) \rightarrow 1} 
\nonumber\\
&=&
\left[
{\rm Tr} \big\{
\hat P_+{}^T
\delta_\zeta \hat P_+{}^T
\delta_\eta \hat P_+{}^T 
\big\}
+
{\rm Tr} \big\{
\hat P_-{}
\delta_\eta \hat P'_-{}
\delta_\zeta \hat P'_-{}
\big\}
\right.
\nonumber\\
&&
\left.
+ \big\langle 
{\rm Tr} \big\{
(u^T
\{ \hat P_+{}^T
\delta_\zeta
\delta_\eta 
\{ \hat P_+{}^T 
{\cal M}_E 
\hat P'_-\}
\hat P'_- \}
 v')
\, 
( u^T 
{\cal M}_E 
\, v')^{-1}
\big\}
\big\rangle_E
\slash  \big\langle  1  \big\rangle_{E}
\right.
\nonumber\\
&&
\left.
-
\big\langle 
{\rm Tr} \big\{
(u^T
\delta_\eta 
\{ \hat P_+{}^T 
{\cal M}_E 
\hat P'_-\} v')
\, 
( u^T 
{\cal M}_E 
\, v')^{-1}
(u^T
\delta_\zeta 
\{ \hat P_+{}^T 
{\cal M}_E 
\hat P'_-\} v')
\, 
( u^T 
{\cal M}_E 
\, v')^{-1}
\big\}
\big\rangle_E
\slash  \big\langle  1  \big\rangle_{E}
\right.
\nonumber\\
&&
\left.
+
\big\langle 
{\rm Tr} \big\{
(u^T
\delta_\eta 
\{ \hat P_+{}^T 
{\cal M}_E 
\hat P'_-\} v')
\, 
( u^T 
{\cal M}_E 
\, v')^{-1}
\big\} \, \times
\right.
\nonumber\\
&&
\left.
\qquad\qquad\qquad\qquad\qquad\qquad\qquad\qquad\quad
{\rm Tr} \big\{
(u^T
\delta_\zeta 
\{ \hat P_+{}^T 
{\cal M}_E 
\hat P'_-\} v')
\, 
( u^T 
{\cal M}_E 
\, v')^{-1}
\big\}
\big\rangle_E
\slash  \big\langle  1  \big\rangle_{E}
\right.
\nonumber\\
&&
\left.
-
\big\langle 
{\rm Tr} \big\{
(u^T
\delta_\eta 
\{ \hat P_+{}^T 
{\cal M}_E 
\hat P'_-\} v')
\, 
( u^T 
{\cal M}_E 
\, v')^{-1}
\big\} 
\big\rangle_E
\slash  \big\langle  1  \big\rangle_{E}
\, \, \times
\right.
\nonumber\\
&&
\left.
\qquad\qquad\qquad\qquad\qquad\qquad
\big\langle
{\rm Tr} \big\{
(u^T
\delta_\zeta 
\{ \hat P_+{}^T 
{\cal M}_E 
\hat P'_-\} v')
\, 
( u^T 
{\cal M}_E 
\, v')^{-1}
\big\}
\big\rangle_E
\slash  \big\langle  1  \big\rangle_{E}
%
%
%
%
%
\right]
\Big\vert_{U(x,\mu) \rightarrow 1} .
\nonumber\\
\end{eqnarray}
It satisfies the Ward-Takahashi relation,
\begin{eqnarray}
\label{eq:two-point-vertex-U1-gauge-field-WT1}
&& 
\tilde \Pi'_{\mu \nu} (k) \, \,  2 \sin \Big( \frac{k_\nu}{2} \Big) 
= 0 ,
\end{eqnarray}
because 
\begin{equation}
\big\langle  - \delta_\eta S_M  \big\rangle_{M} \slash  \big\langle  1  \big\rangle_{M} 
= \big\langle  - \delta_\eta S_M  \big\rangle_{WM} \slash  \big\langle  1  \big\rangle_{WM}
\end{equation}
 is gauge invariant.
In the weak gauge-coupling limit in particular, 
it also satisfies the relation,
\begin{eqnarray}
\label{eq:two-point-vertex-U1-gauge-field-WT2}
&& 2 \sin \Big( \frac{k_\mu}{2} \Big) \,
\tilde \Pi'_{\mu \nu} (k) \,  
= 0 , 
\end{eqnarray}
because one can show for the gauge-variation
$\eta_\mu(x) = - \partial_\mu \omega(x)$ that
\begin{eqnarray}
\delta_\zeta 
\left[
\big\langle  - \delta_\eta S_M  \big\rangle_{M} \slash  \big\langle  1  \big\rangle_{M} 
\right]\, \Big\vert_{U(x,\mu) \rightarrow 1} 
&=&
\delta_\zeta 
\left[
{\rm Tr} \big\{ i \omega \hat P_+^T \big\}
+
{\rm Tr} \big\{ (- i \omega) \hat P'_- \big\}
\right]\, \Big\vert_{U(x,\mu) \rightarrow 1} 
\nonumber\\
&=&
\left[
{\rm Tr} \big\{ i \omega  \delta_\zeta \hat P_+ \big\}
+
{\rm Tr} \big\{ (- i \omega) \delta_\zeta  \hat P'_- \big\}
\right]\, \Big\vert_{U(x,\mu) \rightarrow 1} 
\nonumber\\
&=&
\left[
{\rm Tr} \big\{ i \omega  \delta_\zeta \hat P_+ \big\}
+
{\rm Tr} \big\{  (- i \omega) (-\delta_\zeta  \hat P_- ) \big\}
\right]\, \Big\vert_{U(x,\mu) \rightarrow 1} 
\nonumber\\
&=& 0.
\end{eqnarray}
\begin{figure}[h]
\begin{center}
\includegraphics[width =75mm]{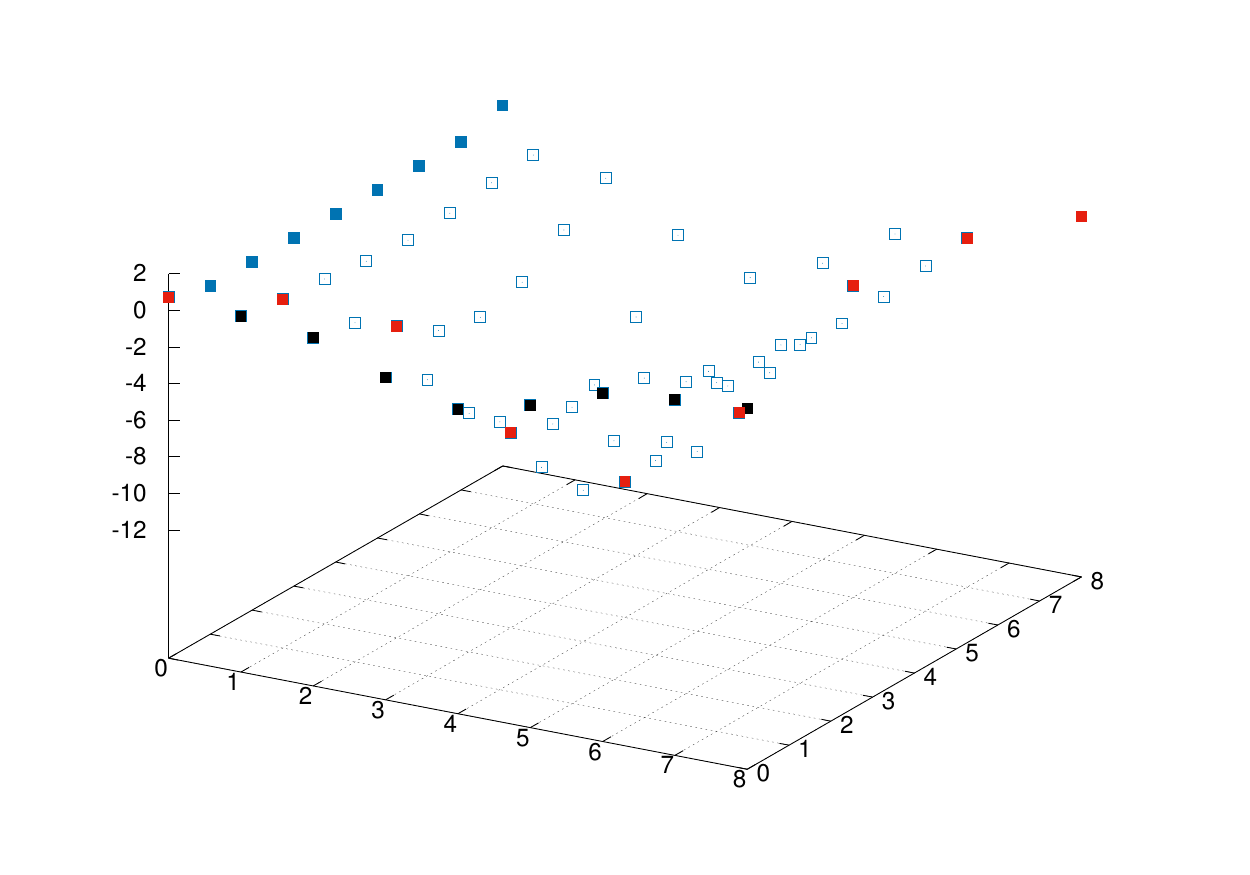} 
\includegraphics[width =75mm]{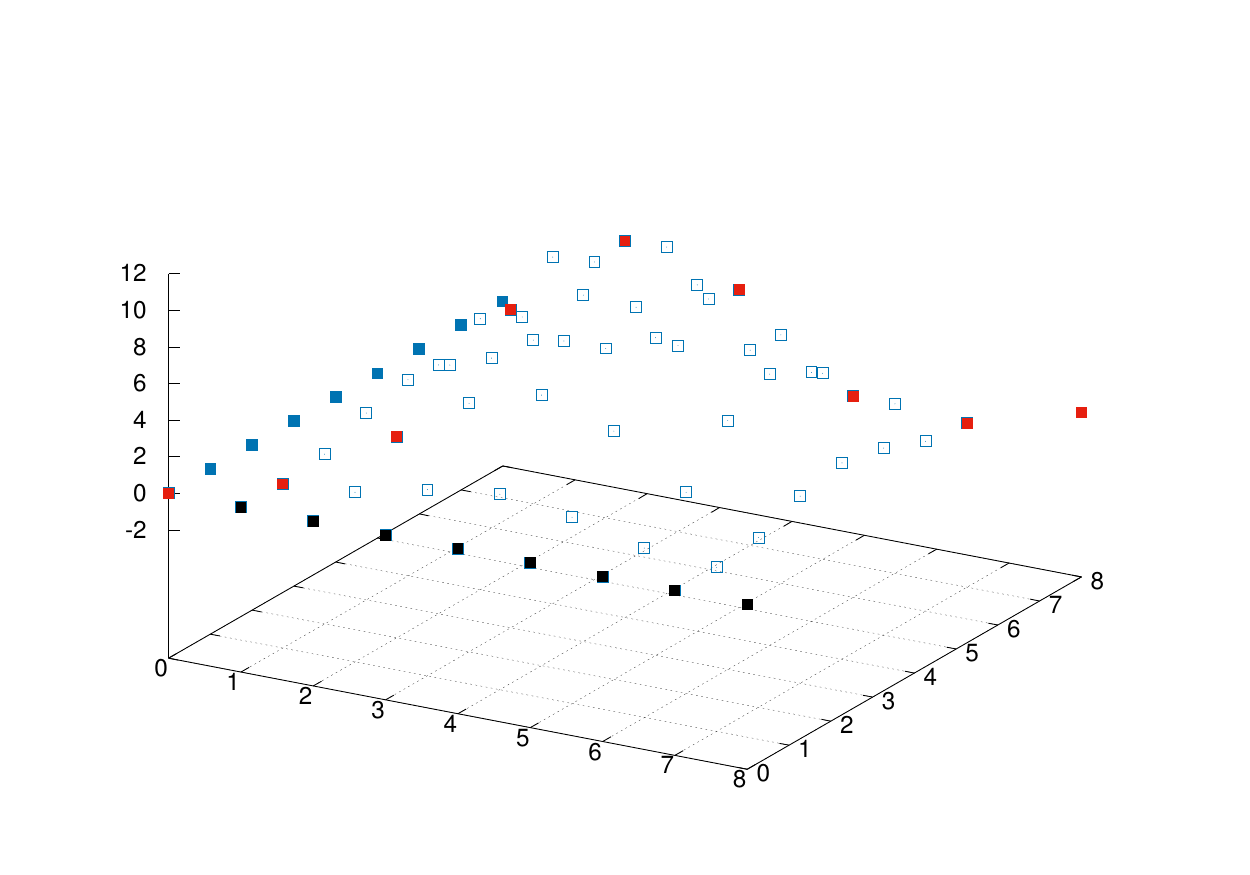}
\\
\includegraphics[width =70mm]{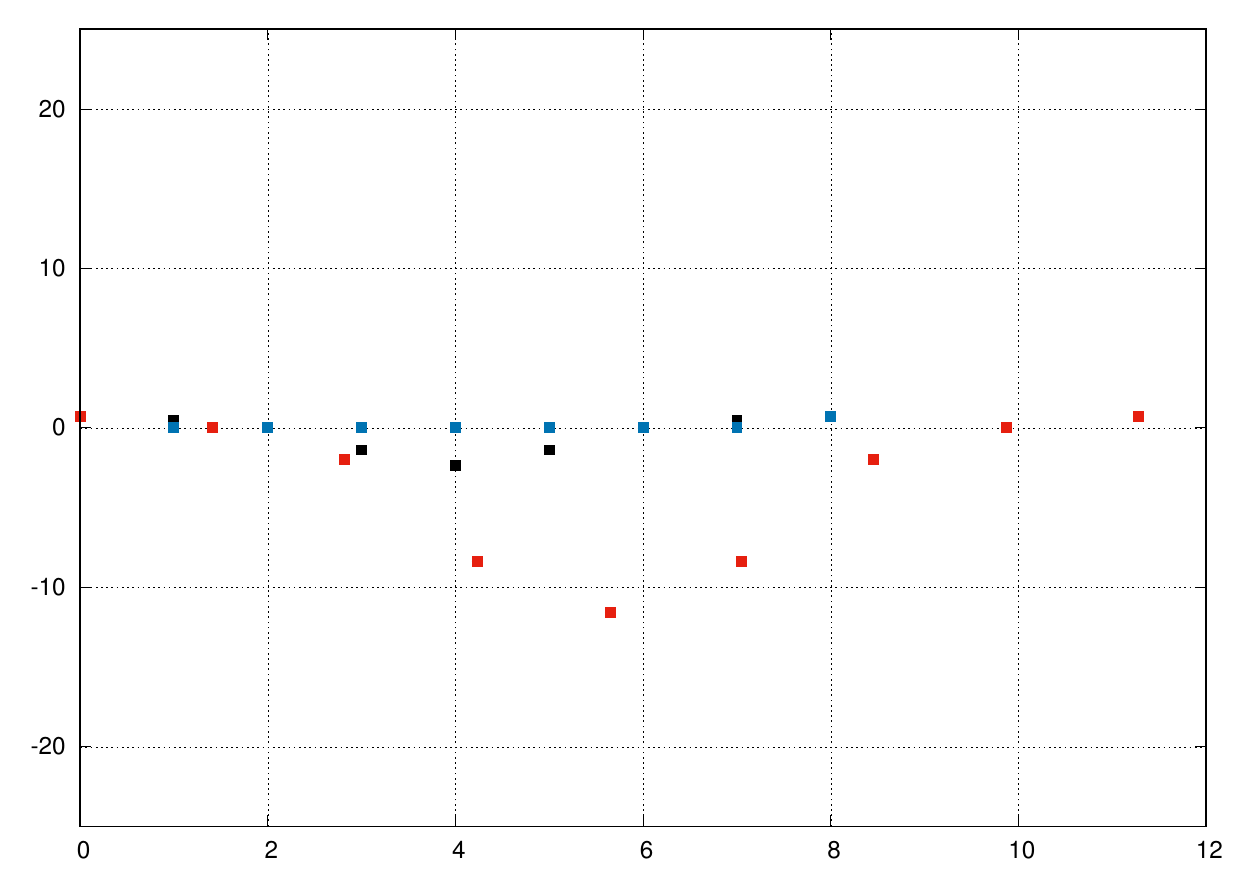} 
\hspace{1em}
\includegraphics[width =70mm]{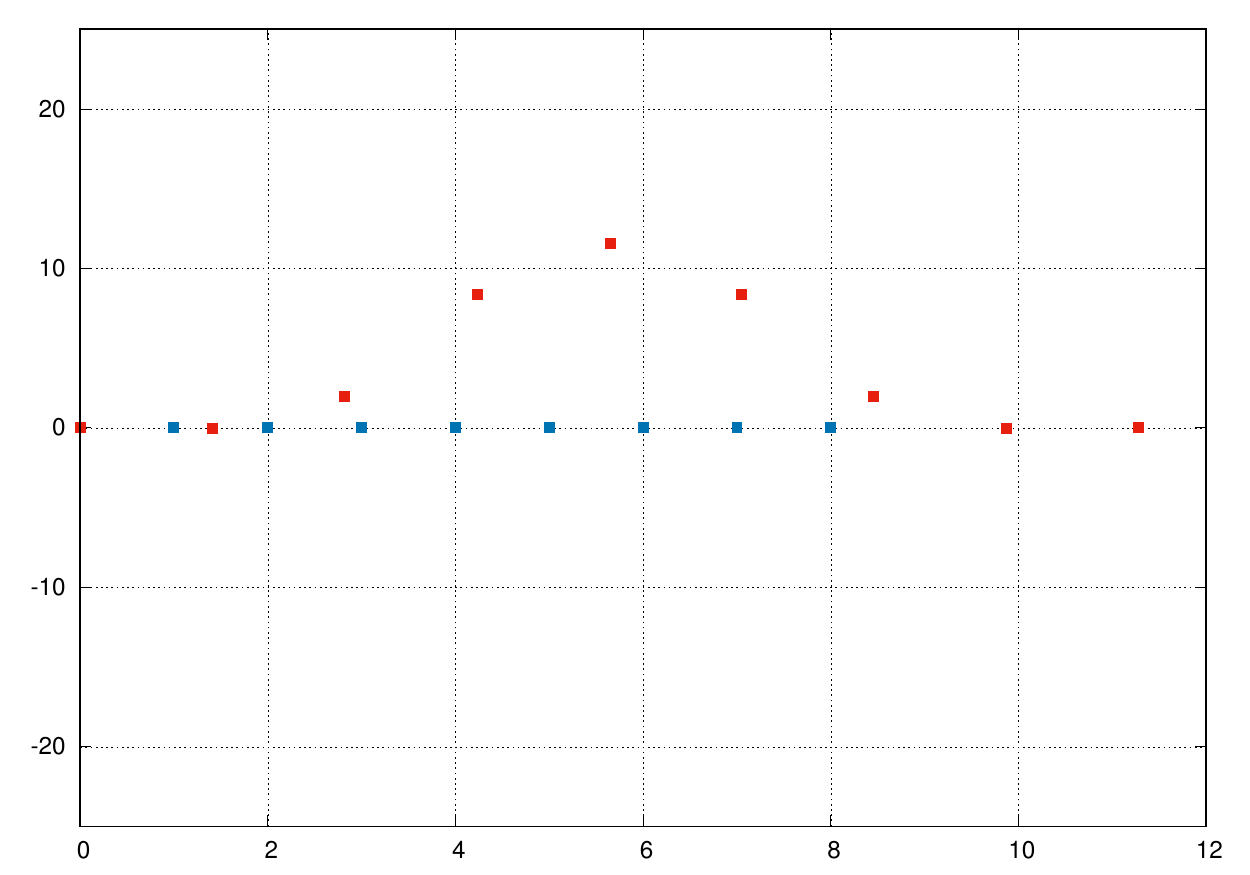} 
\\
\end{center}
\caption{
$L^2 \, \tilde  \Pi'_{00}(k)$ [left]
and 
$L^2 \, \tilde  \Pi'_{01}(k)$ [right]
 vs. 
$\vert k \vert_2 \equiv \sqrt{ k_0^2 + k_1^2 }$.
The lattice size is $L=8$. 
The periodic boundary condition is assumed
for the fermion fields.
The black-, blue-, red-symbol plots
are 
along
the spacial momentum axis ($k_0=0$),
the temporal momentum (energy) axis ($k_1 =0$) and
the diagonal momentum axis ($k_0=k_1$), respectively.
5,000 configurations are sampled with the interval of 20 trajectories. 
The errors are simple statistical ones.
}
\label{fig:correlation-AA00-8x8}
\end{figure}
\begin{figure}[h]
\begin{center}
\includegraphics[width =75mm]{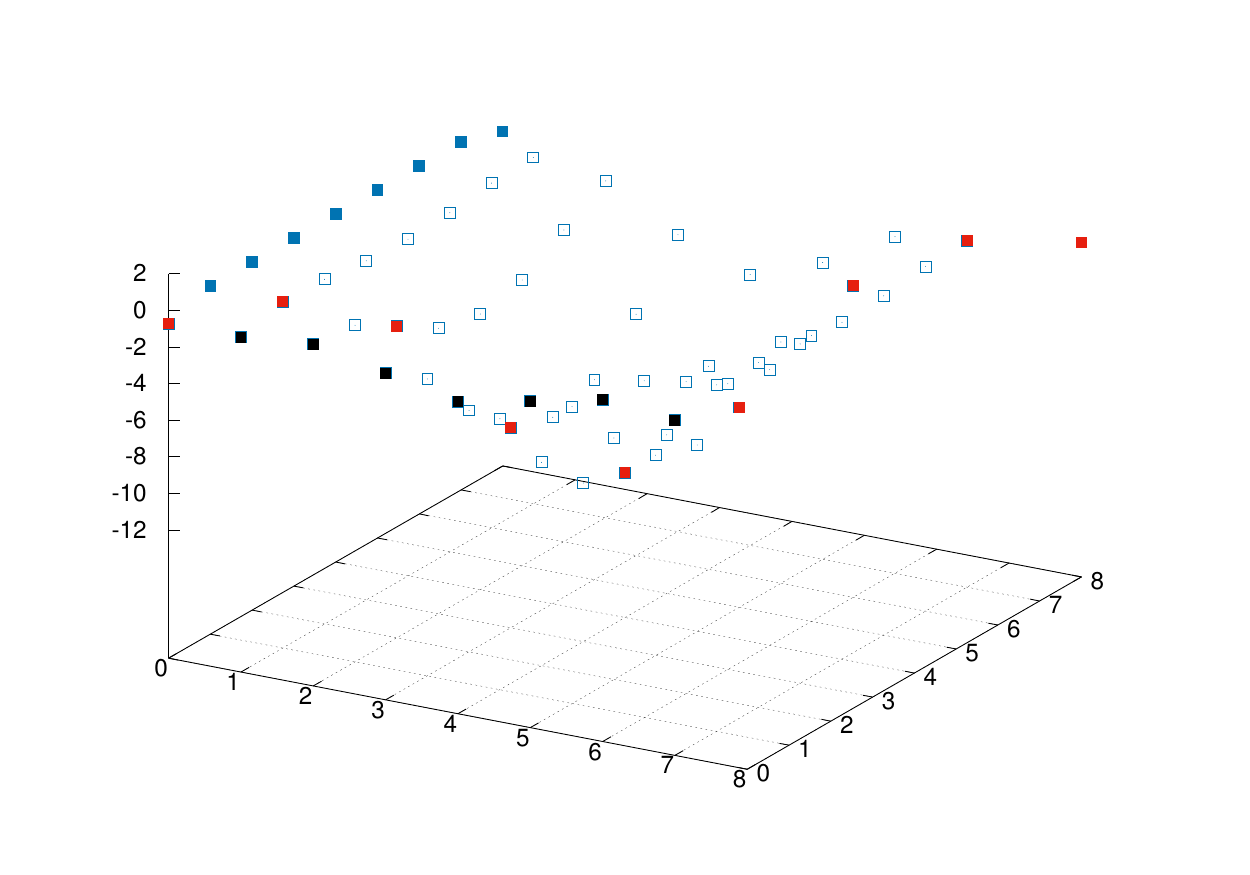} 
\includegraphics[width =75mm]{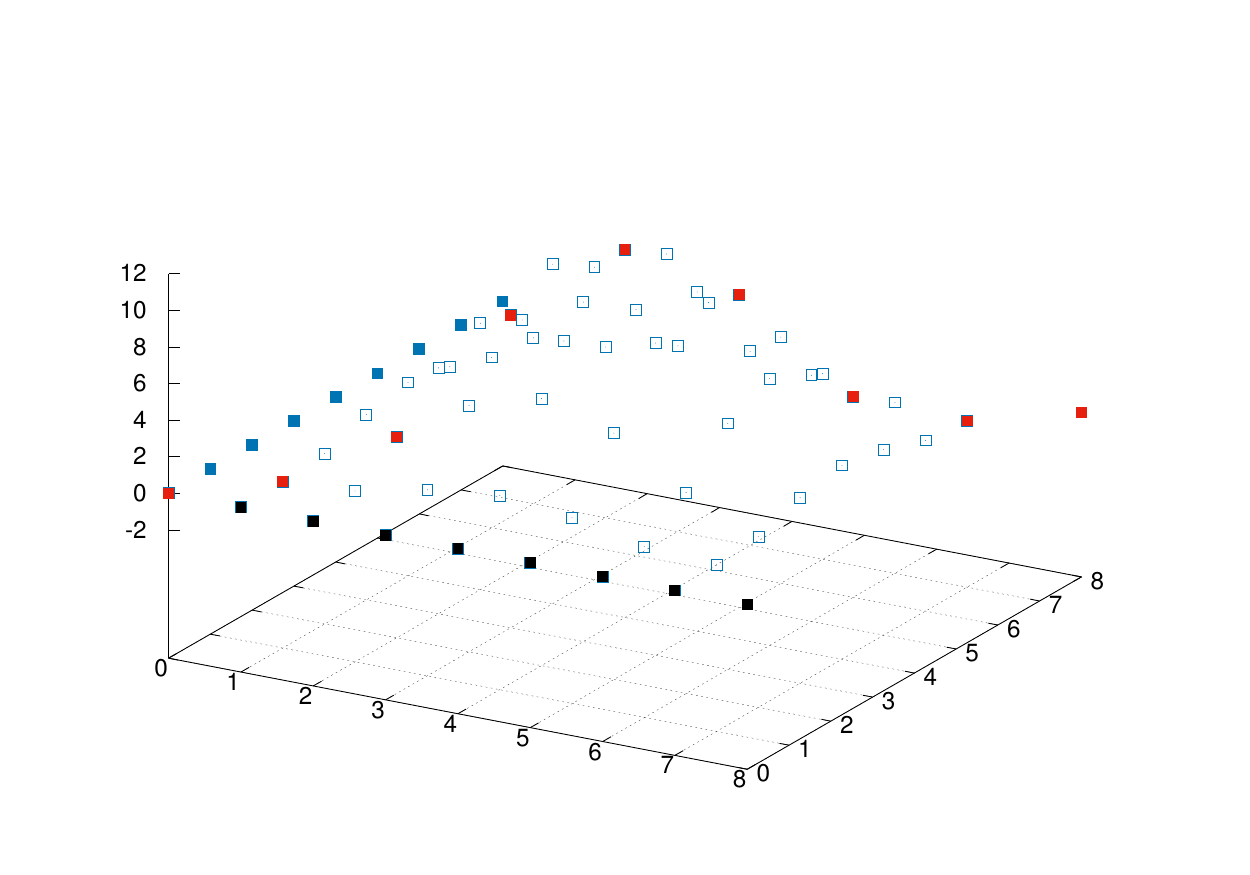}
\\
\includegraphics[width =70mm]{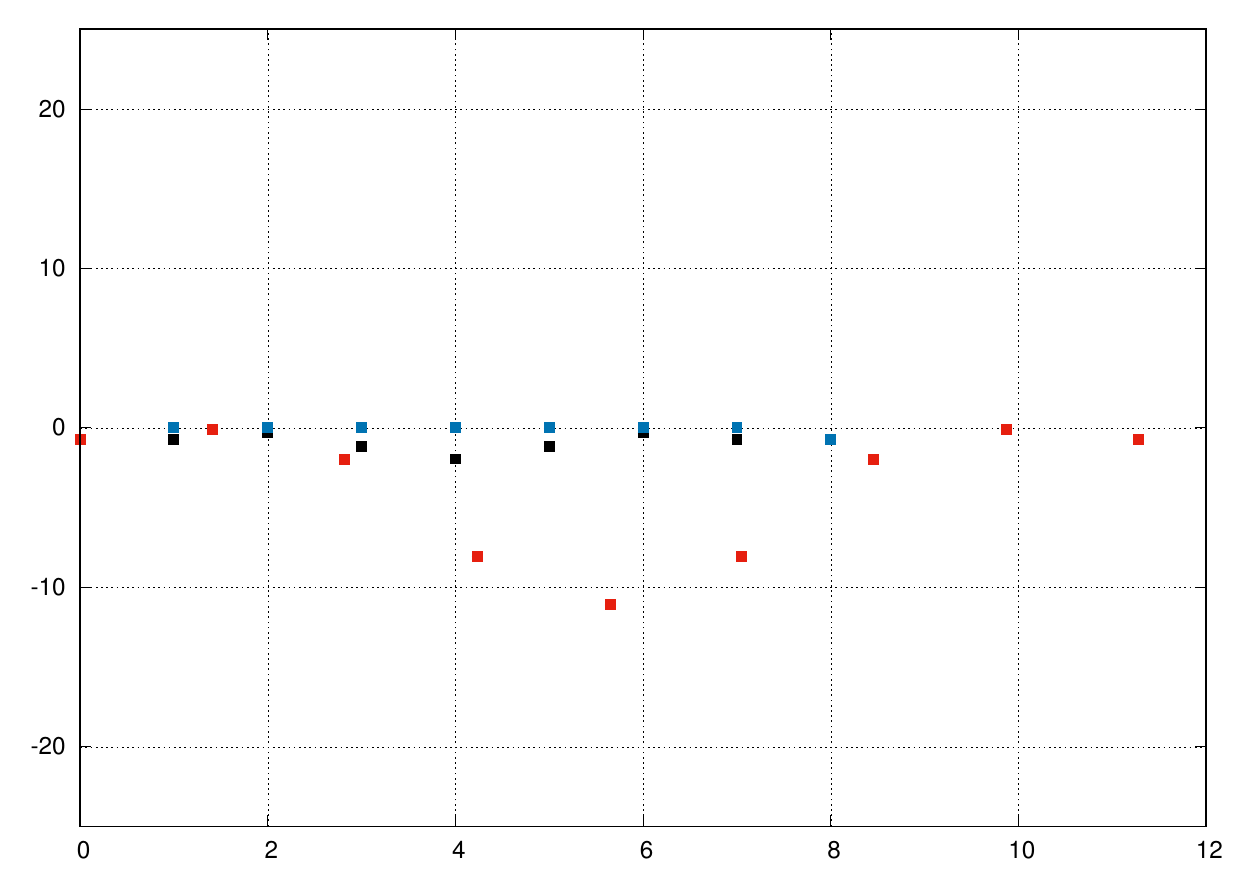} 
\hspace{1em}
\includegraphics[width =70mm]{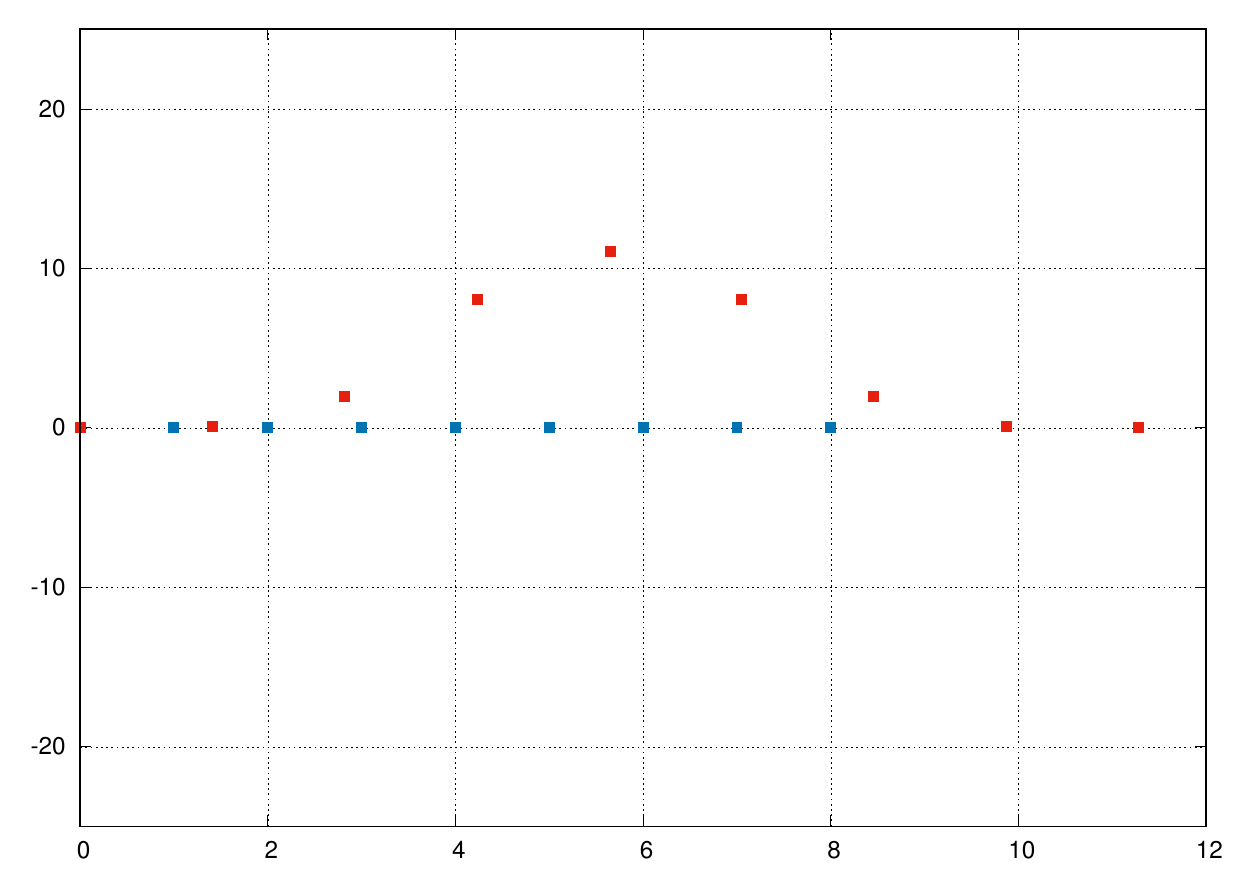} 
\\
\end{center}
\caption{
$L^2 \, \tilde  \Pi'_{00}(k)$ [left]
and 
$L^2 \, \tilde  \Pi'_{01}(k)$ [right]
 vs. 
$\vert k \vert_2 \equiv \sqrt{ k_0^2 + k_1^2 }$.
The lattice size is $L=8$. 
The anti-periodic boundary condition is assumed
for the fermion fields.
The black-, blue-, red-symbol plots
are 
along
the spacial momentum axis ($k_0=0$),
the temporal momentum axis ($k_1 =0$) and
the diagonal momentum axis ($k_0=k_1$), respectively.
5,000 configurations are sampled with the interval of 20 trajectories. 
The errors are simple statistical ones.
}
\label{fig:correlation-AA00-8x8-apbc}
\end{figure}

In figs.~\ref{fig:correlation-AA00-8x8} and
\ref{fig:correlation-AA00-8x8-apbc}, we show
the numerical-simulation results of 
$\tilde  \Pi'_{00}(k)$ 
and 
$\tilde  \Pi'_{01}(k)$ 
for the lattice size $L=8$
and for both the periodic and anti-periodic boundary conditions 
of the mirror fermion fields.
5,000 configurations are sampled with the interval of 20 trajectories. The errors are simple statistical ones.
It was verified that the Ward-Takahashi relations eqs~(\ref{eq:two-point-vertex-U1-gauge-field-WT1}) and (\ref{eq:two-point-vertex-U1-gauge-field-WT2}) 
are satisfied up to
the machine precision (double precision) of order $10^{-16}$.

The above result of
the contribution of the mirror fermions $\tilde\Pi'_{\mu\nu}(k)$ should be 
compared with that of the (target) Weyl fermions 
$\tilde\Pi_{\mu\nu}(k)$,
which is given by 
\begin{eqnarray}
\label{eq:two-point-vertex-U1-gauge-field-Weyl-sector}
\frac{1}{L^2}\sum_{k} \tilde \eta_\mu(-k)\, 
\tilde \Pi_{\mu \nu} (k) \, \tilde \zeta_\nu(k) &=&
\delta_\zeta 
\left[
 {\rm Tr}\{ P_+ \delta_\eta D D^{-1}  \}     
+ {\rm Tr}\{ P_- \delta_\eta D' {D'}^{-1}  \}  
\right]\, \Big\vert_{U(x,\mu) \rightarrow 1} 
\nonumber\\
&=&
\left[
 {\rm Tr}\{ \delta_\zeta \delta_\eta D D^{-1}  \}  
-  {\rm Tr}\{ \delta_\eta D {D}^{-1} 
               \delta_\zeta D {D}^{-1} \}  
\right]\, \Big\vert_{U(x,\mu) \rightarrow 1} .
\nonumber\\
\end{eqnarray}
It shows a singular non-local behavior due to the 
massless singularities of the Weyl fermion propagators
$D^{-1} P_-$ and $D'{}^{-1} P_+$.
For small momentum region $ \vert k \vert \ll \pi $
(in the thermodynamic limit $L = \infty$), it is given as
\begin{eqnarray}
\tilde \Pi_{\mu \nu} (k) \simeq
[4 \times 1^2 + 4 \times (-1)^2]  \, \frac{1}{2 \pi} \frac{ \delta_{\mu \nu} k^2 - k_\mu k_\nu }{ k^2 } 
\qquad (\vert k \vert \ll \pi).
\end{eqnarray}
Then it shows the non-uniform behavior
depending on how the limit 
$\vert k \vert \rightarrow 0$ is approached as follows\cite{Bhattacharya:2006dc,Giedt:2007qg,Poppitz:2007tu,Poppitz:2008au,Poppitz:2009gt,Poppitz:2010at,Chen:2012di,Giedt:2014pha}.
\begin{eqnarray}
\tilde \Pi_{0 0} (k) \simeq 
\frac{4}{\pi} \frac{ k_1^2}{ k_0^2 + k_1^2 }
& \longrightarrow &
\frac{4}{\pi} \times 
\left\{
\begin{array}{cl}
0   \qquad & k_\mu=(\vert k \vert , 0)  \\
1/2 \qquad & k_\mu=(\vert k \vert , \vert k \vert )/\sqrt{2} \\
1   \qquad & k_\mu=(0 ,\vert k \vert ) 
\end{array}
\right. ,
\end{eqnarray}
\begin{eqnarray}
\tilde \Pi_{0 1} (k) \simeq 
\frac{4}{\pi} \frac{ -k_0  k_1}{ k_0^2 + k_1^2 }
& \longrightarrow &
\frac{4}{\pi} \times 
\left\{
\begin{array}{cl}
0   \qquad & k_\mu=(\vert k \vert , 0)  \\
-1/2 \qquad & k_\mu=(\vert k \vert , \vert k \vert )/\sqrt{2} \\
0   \qquad & k_\mu=(0 ,\vert k \vert ) 
\end{array}
\right. .
\end{eqnarray}

\begin{figure}[h]
\begin{center}
\includegraphics[width =75mm]{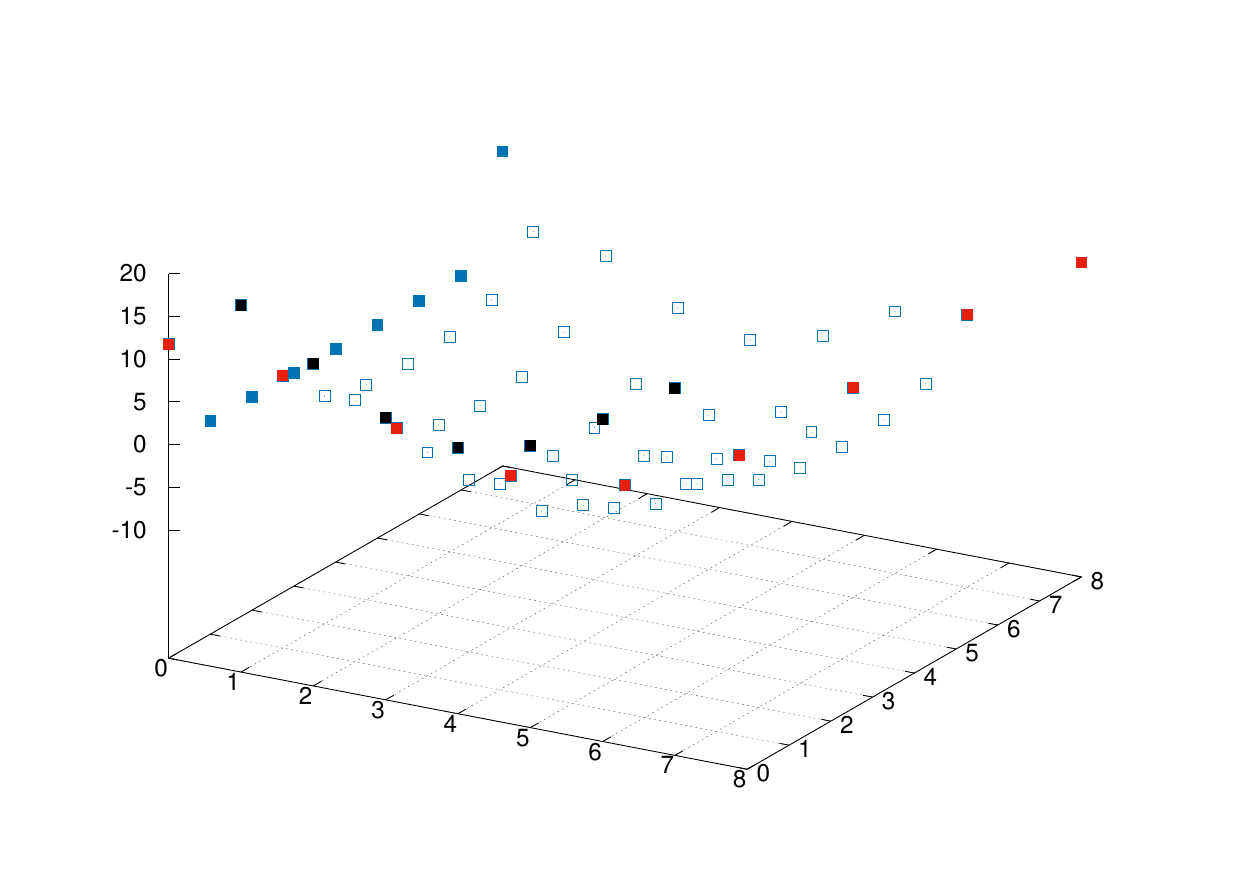} 
\includegraphics[width =75mm]{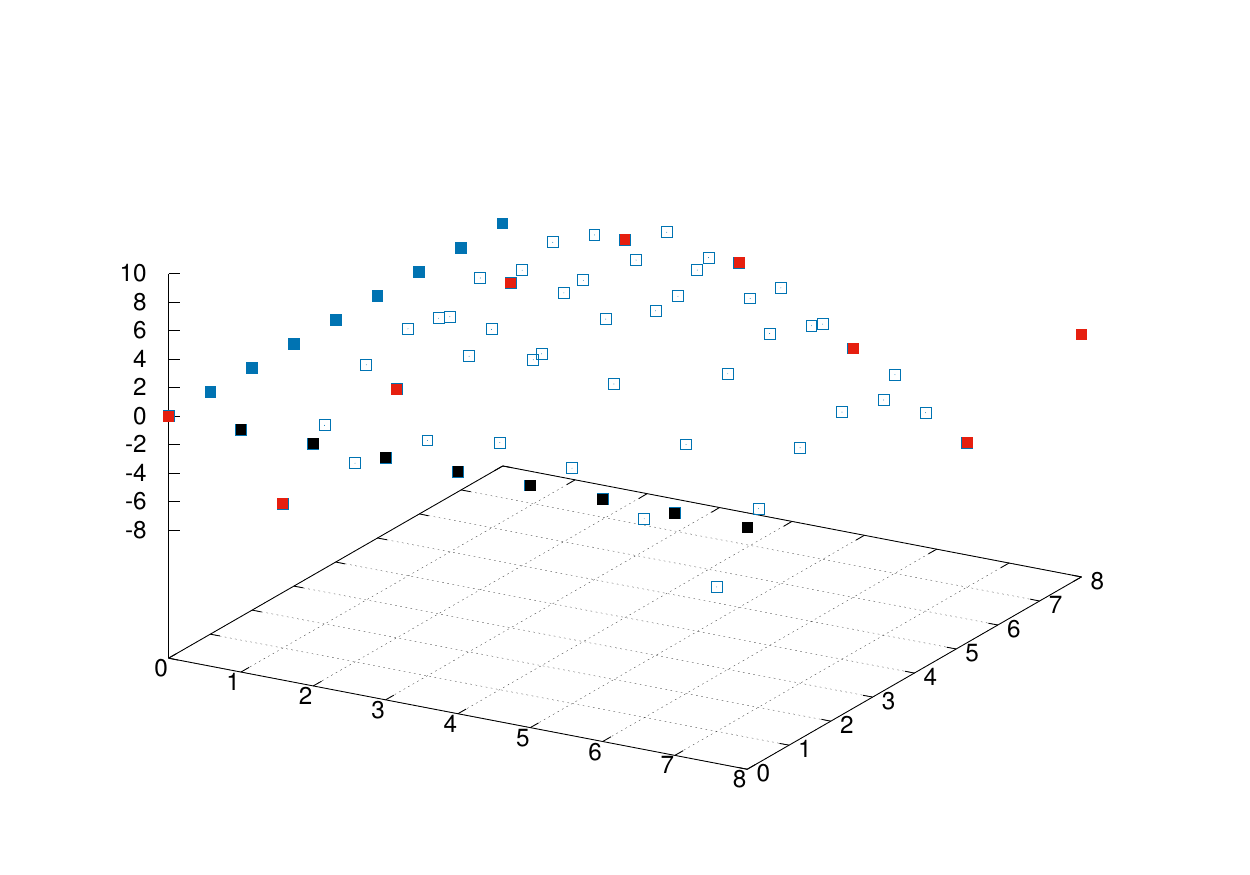}
\\
\includegraphics[width =70mm]{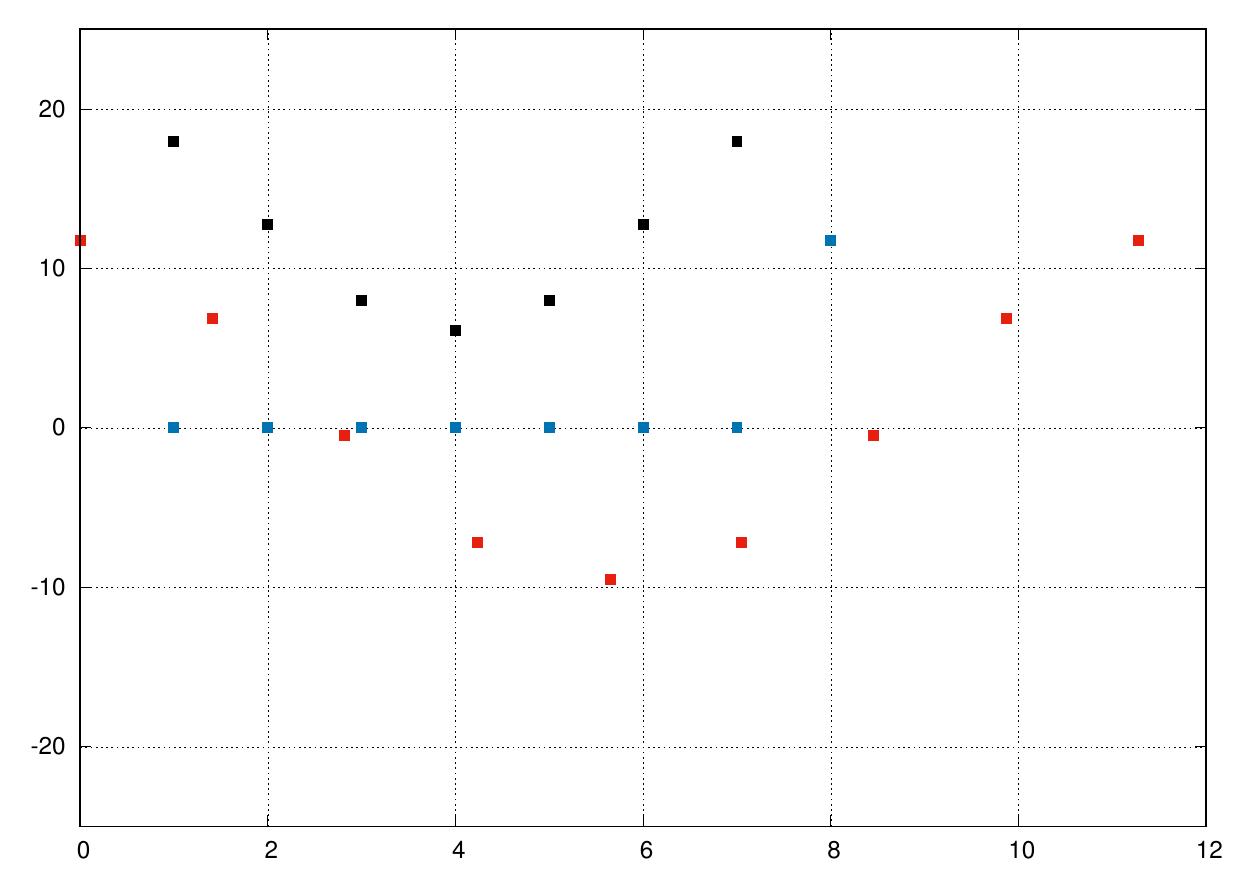} 
\hspace{1em}
\includegraphics[width =70mm]{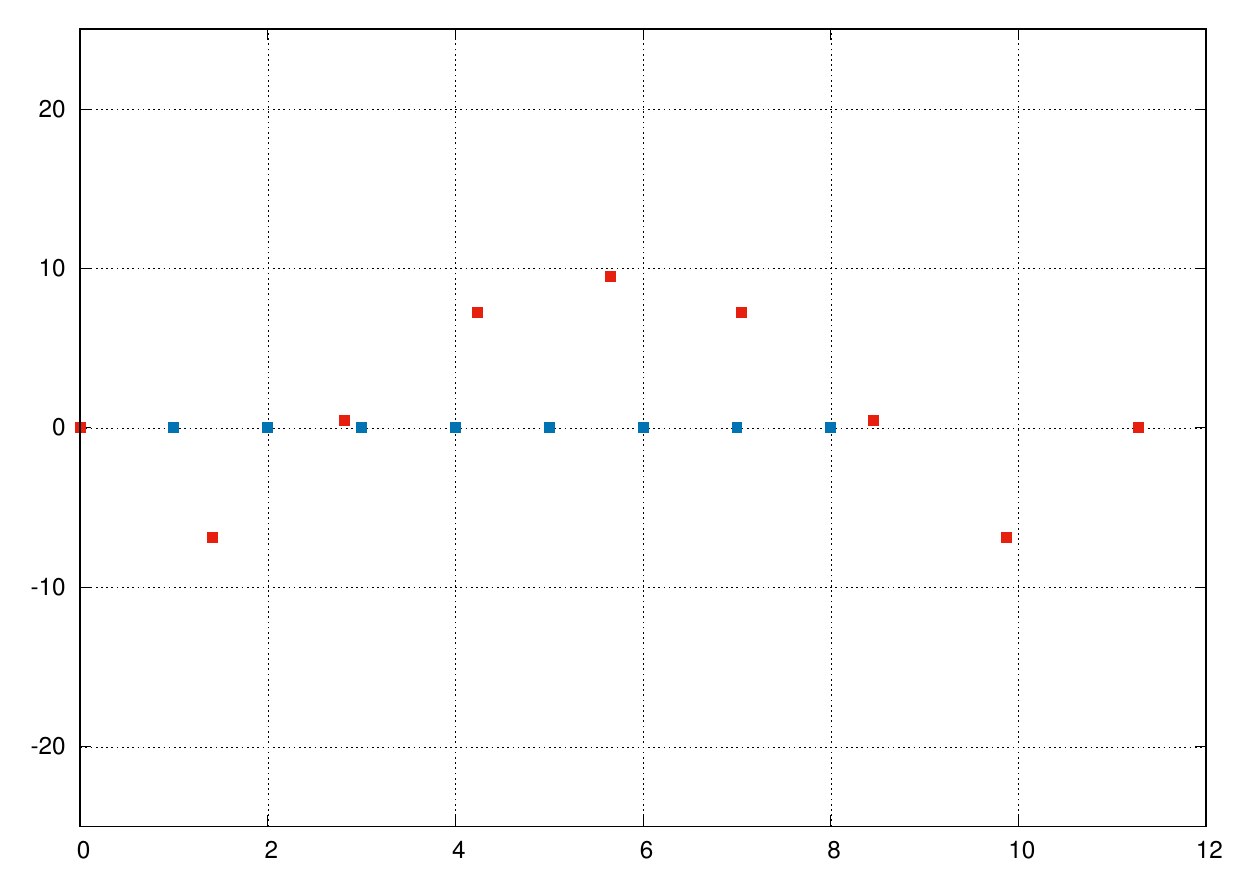} 
\\
\end{center}
\caption{
$(1/4) L^2 \, \tilde  \Pi_{00}(k)$ [left]
and 
$(1/4) L^2 \, \tilde  \Pi_{01}(k)$ [right]
 vs. 
$\vert k \vert_2 \equiv \sqrt{ k_0^2 + k_1^2 }$.
The lattice size is $L=8$. 
The anti-periodic boundary condition is assumed
for the fermion fields.
The black-, blue-, red-symbol plots
are 
along
the spacial momentum axis ($k_0=0$),
the temporal momentum axis ($k_1 =0$) and
the diagonal momentum axis ($k_0=k_1$), respectively.
}
\label{fig:correlation-AA-8x8-Weyl}
\end{figure}
This singular behavior of $\tilde\Pi_{\mu\nu}(k)$ can be verified
by numerical computation.
In fig.~\ref{fig:correlation-AA-8x8-Weyl}, we show
the numerical-computation result of 
$\tilde  \Pi_{00}(k)$ 
and 
$\tilde  \Pi_{01}(k)$ 
for the lattice size $L=8$
and 
the anti-periodic boundary conditions 
of the (target) Weyl fermion fields.
We can see rather clearly 
the non-uniform limits to $\vert k \vert = 0$.
The normalizations of 
$(1/4) L^2 \, \tilde  \Pi_{00}(k)$ 
and 
$(1/4) L^2 \, \tilde  \Pi_{01}(k)$ 
at the singularity point 
shown in fig.~\ref{fig:correlation-AA-8x8-Weyl}
are
also consistent with the result in the thermodynamic limit $L=\infty$: $L^2 \times \frac{1}{ \pi } \times \{0, \pm 1/2, 1\} \simeq 20.47 \times \{0, \pm 1/2, 1\}$.

By comparing the numerical result of
the contribution of the mirror fermions 
$\tilde\Pi'_{\mu\nu}(k)$ 
with that of the (target) Weyl fermions 
$\tilde\Pi_{\mu\nu}(k)$,
we can see that 
the mirror fermion contribution 
does not show any 
evidence of the singularities due to charged massless excitations. It behaves like a regular function of momentum $k_\mu$.
This result is consistent with the fact that 
the mirror fermions decouple by acquiring
the masses of order the inverse lattice spacing
and leave only local terms in the effective action.


\subsubsection{Regular two-point vertex function of the 
Spin(6) (SU(4)) vector field}

In the $1^4(-1)^4$ axial gauge model in consideration, 
the global Spin(6)(SU(4)) symmetry can be gauged
consistently. Then it is instructive 
to examine 
the vertex functions of the (external) Spin(6) gauge field in the mirror fermion sector, 
which can be defined in the similar manner as
those of the U(1) gauge field. Let us denote the 
link field of the (external) Spin(6) gauge field with
$V(x, \mu)$ and its variation with
$\delta_{\eta} V(x,\mu) = i \, \eta_\mu(x) V(x,\mu)$
where 
$\eta_\mu(x)=(1/2)\, \eta_\mu^{ab}(x) \Sigma^{ab}$.
Then the two-point vertex function is given by 
\begin{eqnarray}
\label{eq:two-point-vertex-Spin6-gauge-field-second}
\frac{1}{L^2}\sum_{k} \tilde \eta_\mu^{A}(-k)\, \tilde \Pi^{AB}_{\mu \nu} (k) \, \tilde \zeta_\nu^{B}(k) &=&
\delta_\zeta 
\left[
\big\langle  - \delta_\eta S_M  \big\rangle_{M} \slash  \big\langle  1  \big\rangle_{M} 
\right]\, \Big\vert_{V(x,\mu) \rightarrow 1} ,
\end{eqnarray}
where $A, B$ stand for the anti-symmetrized indices
$A=[ab], B=[cd]$.
The link field of the U(1) gauge field can be set to unity,
$U(x,\mu)=1$, from the beginning in the weak gauge-coupling limit.

The two-point vertex function satisfies the Ward-Takahashi relations,
\begin{eqnarray}
\label{eq:two-point-vertex-Spin6-gauge-field-WT1}
&& 2 \sin \Big( \frac{k_\mu}{2} \Big) \,
\tilde \Pi^{AB}_{\mu \nu} (k) \,  
= 0 ,
\\
\label{eq:two-point-vertex-Spin6-gauge-field-WT2}
&& 
\tilde \Pi^{AB}_{\mu \nu} (k) \, \,  2 \sin \Big( \frac{k_\nu}{2} \Big) 
= 0 .
\end{eqnarray}
For the gauge-variation
$\eta_\mu(x) = - D_\mu \omega(x)$ with 
$\omega(x) =\Sigma^A \omega^A(x)$, it gives 
\begin{eqnarray}
&& \delta_\zeta 
\left[
\big\langle  - \delta_\eta S_M  \big\rangle_{M} \slash  \big\langle  1  \big\rangle_{M} 
\right]\, \Big\vert_{V(x,\mu) \rightarrow 1} 
\nonumber\\
&=&
\delta_\zeta 
\left[
{\rm Tr} \big\{ i \omega^T \hat P_+^T \big\}
+
{\rm Tr} \big\{ i \omega \hat P'_- \big\}
\right.
\nonumber\\
&& \left.
\quad
+ \big\langle 
{\rm Tr} \big\{
(u^T
\{ \hat P_+{}^T 
i [ \omega^T {\cal M}_E + {\cal M}_E \, \omega ]
\hat P'_-\} v')
\, 
( u^T 
{\cal M}_E 
\, v')^{-1}
\big\}
\big\rangle_E
\slash  \big\langle  1  \big\rangle_{E}
\right]\, \Big\vert_{V(x,\mu) \rightarrow 1} 
\nonumber\\
&=&
\delta_\zeta 
\Big[
\big\langle 
{\rm Tr} \big\{
(u^T
i [ \omega^T {\cal M}_E + {\cal M}_E \, \omega ]
v')
\, 
( u^T 
{\cal M}_E 
\, v')^{-1}
\big\}
\big\rangle_E
\slash  \big\langle  1  \big\rangle_{E}
\Big]\, \Big\vert_{V(x,\mu) \rightarrow 1} .
\end{eqnarray}
And the term in the square bracket $[ \cdots ]$
vanishes identically because of the Schwinger-Dyson equation 
w.r.t. the spin field $E^a(x)$,
\begin{eqnarray}
\label{eq:SO(6)spin-SD-eq}
\big\langle 
{\rm Tr} \big\{
(u^T
i [ \omega^T {\cal M}_E + {\cal M}_E \, \omega ]
v')
\, 
( u^T 
{\cal M}_E 
\, v')^{-1}
\big\}
\big\rangle_E
\slash  \big\langle  1  \big\rangle_{E} = 0,
\end{eqnarray}
which holds true with a nontrivial external Spin(6)
vecor field, $V(x,\mu) \not = 1$.
For the gauge-variation
$\zeta_\mu(x) = - D_\mu \omega(x)$, on the other hand, 
$\big\langle  - \delta_\eta S_M  \big\rangle_{M} \slash  \big\langle  1  \big\rangle_{M} 
= \big\langle  - \delta_\eta S_M  \big\rangle_{WM} \slash  \big\langle  1  \big\rangle_{WM}$ is gauge covariant
and 
it gives
\begin{eqnarray}
&& 
\delta_\zeta 
\left[
\big\langle  - \delta_\eta S_M  \big\rangle_{M} \slash  \big\langle  1  \big\rangle_{M} 
\right]\, \Big\vert_{V(x,\mu) \rightarrow 1} 
\nonumber\\
&=&
\delta_\zeta
\left[
\big\langle 
{\rm Tr} \big\{
(u^T
\delta_\eta 
\{ \hat P_+{}^T 
{\cal M}_E 
\hat P'_-\} v')
\, 
( u^T 
{\cal M}_E 
\, v')^{-1}
\big\}
\big\rangle_E
\slash  \big\langle  1  \big\rangle_{E}
\right]
\Big\vert_{V(x,\mu) \rightarrow 1} 
\nonumber\\
&=&
\delta_\eta 
\Big[
\big\langle 
{\rm Tr} \big\{
(u^T
i [ \omega^T {\cal M}_E + {\cal M}_E \, \omega ]
v')
\, 
( u^T 
{\cal M}_E 
\, v')^{-1}
\big\}
\big\rangle_E
\slash  \big\langle  1  \big\rangle_{E}
\Big]\, \Big\vert_{V(x,\mu) \rightarrow 1} 
\nonumber\\
&& +
\left[
\big\langle 
{\rm Tr} \big\{
(u^T
\delta_{[\eta , i \omega]}
\{ \hat P_+{}^T 
{\cal M}_E 
\hat P'_-\} v')
\, 
( u^T 
{\cal M}_E 
\, v')^{-1}
\big\}
\big\rangle_E
\slash  \big\langle  1  \big\rangle_{E}
\right]
\Big\vert_{V(x,\mu) \rightarrow 1} 
\nonumber\\
\Big( &=&
\delta_\eta 
\left[
\big\langle  - \delta_\zeta S_M  \big\rangle_{M} \slash  \big\langle  1  \big\rangle_{M} 
\right]\, \Big\vert_{V(x,\mu) \rightarrow 1} 
+
\left[
\big\langle  - \delta_{[\eta, i \omega]} S_M  \big\rangle_{M} \slash  \big\langle  1  \big\rangle_{M} 
\right]\, \Big\vert_{V(x,\mu) \rightarrow 1} \Big) .
\end{eqnarray}
Again, the first term 
vanishes identically because of the Schwinger-Dyson equation 
w.r.t. the spin field $E^a(x)$, eq.~(\ref{eq:SO(6)spin-SD-eq}),
while the second term 
is the one-point
vertex function and vanishes identically.

\begin{figure}[h]
\begin{center}
\includegraphics[width =75mm]{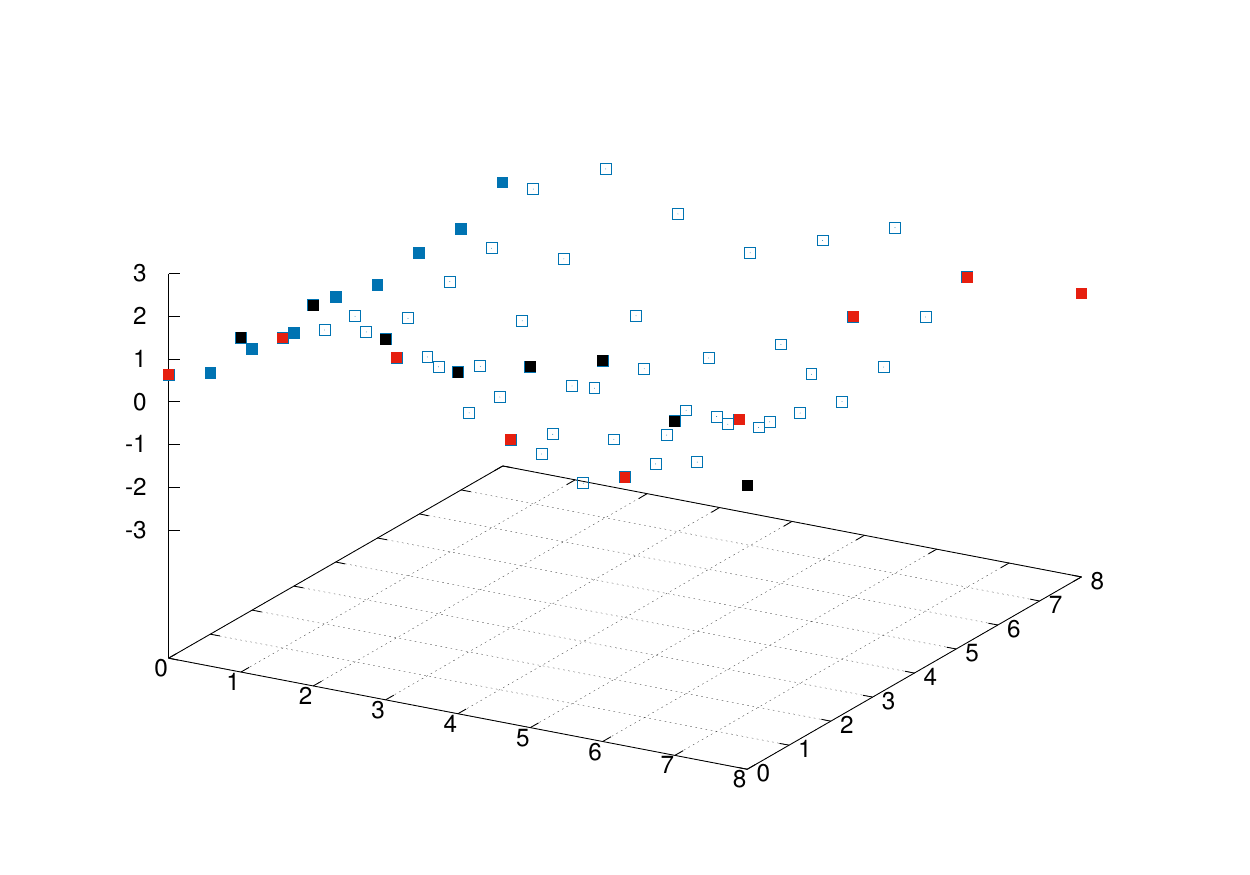} 
\includegraphics[width =75mm]{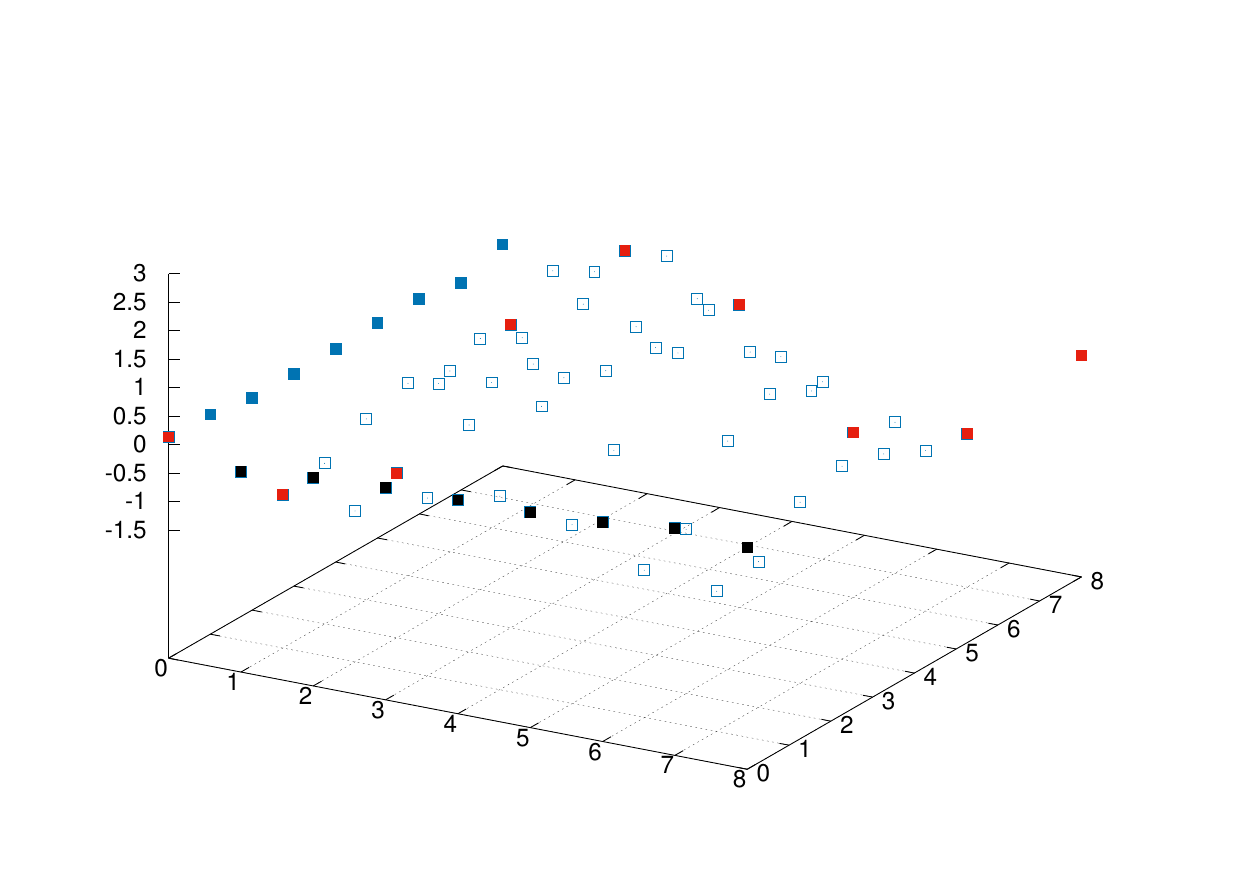}
\\
\includegraphics[width =70mm]{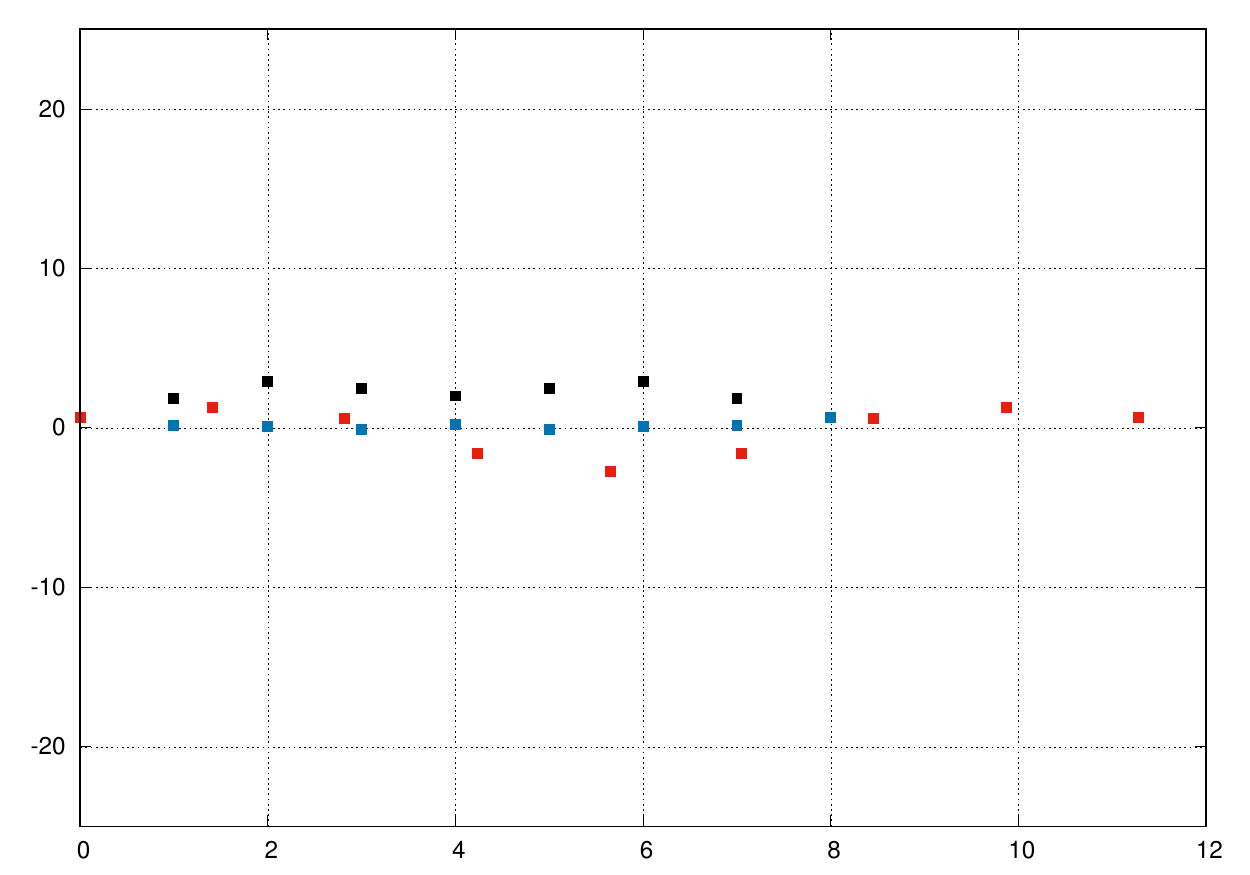} 
\hspace{1em}
\includegraphics[width =70mm]{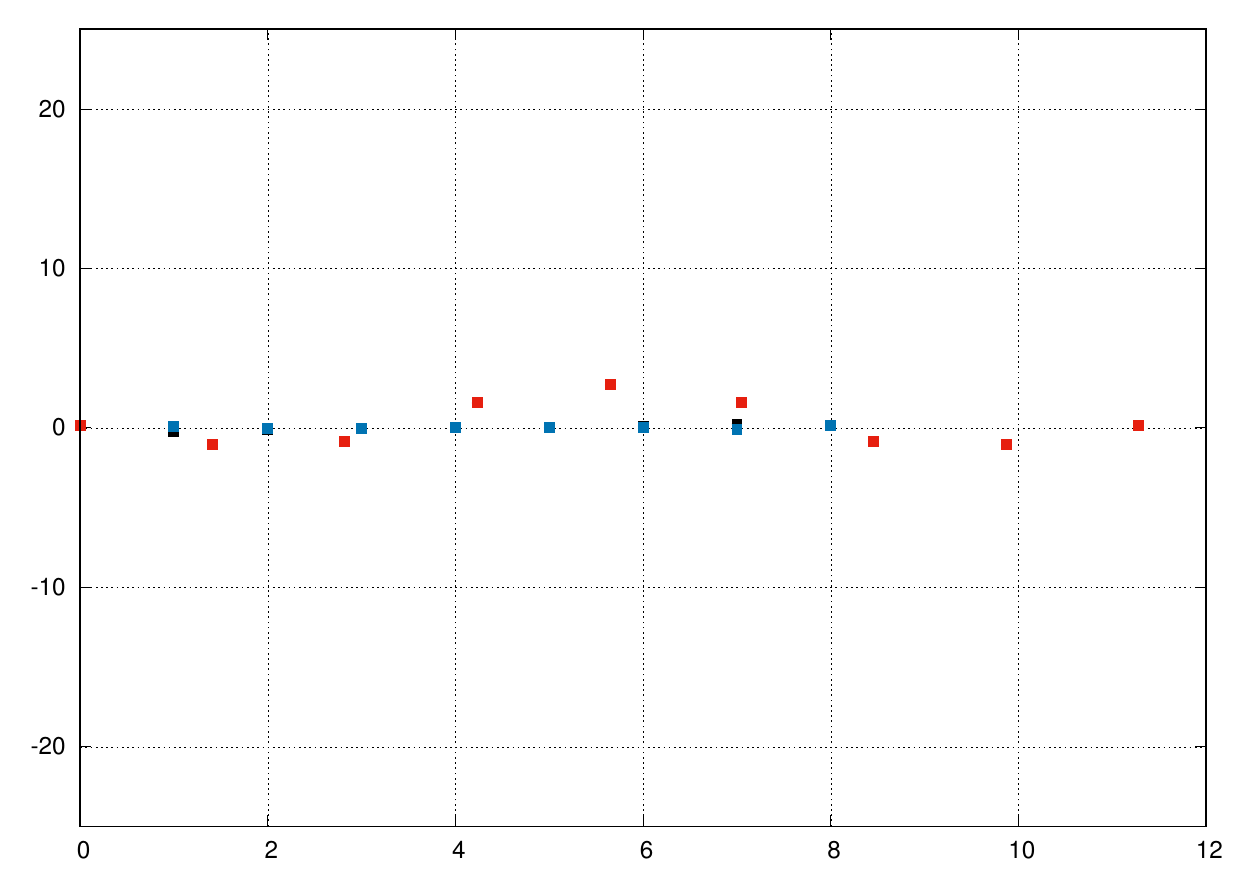} 
\\
\end{center}
\caption{
$L^2 \, \tilde  \Pi^{AA}_{00}(k)$ [left]
and 
$L^2 \, \tilde  \Pi^{AA}_{01}(k)$ [right]
 vs. 
$\vert k \vert_2 \equiv \sqrt{ k_0^2 + k_1^2 }$ where
$A=[12]$.
The lattice size is $L=8$. 
The periodic boundary condition is assumed
for the fermion fields.
The black-, blue-, red-symbol plots
are 
along
the spacial momentum axis ($k_0=0$),
the temporal momentum (energy) axis ($k_1 =0$) and
the diagonal momentum axis ($k_0=k_1$), respectively.
5,000 configurations are sampled with the interval of 20 trajectories. 
The errors are simple statistical ones.
}
\label{fig:correlation-AAS12-8x8}
\end{figure}

\begin{figure}[h]
\begin{center}
\includegraphics[width =75mm]{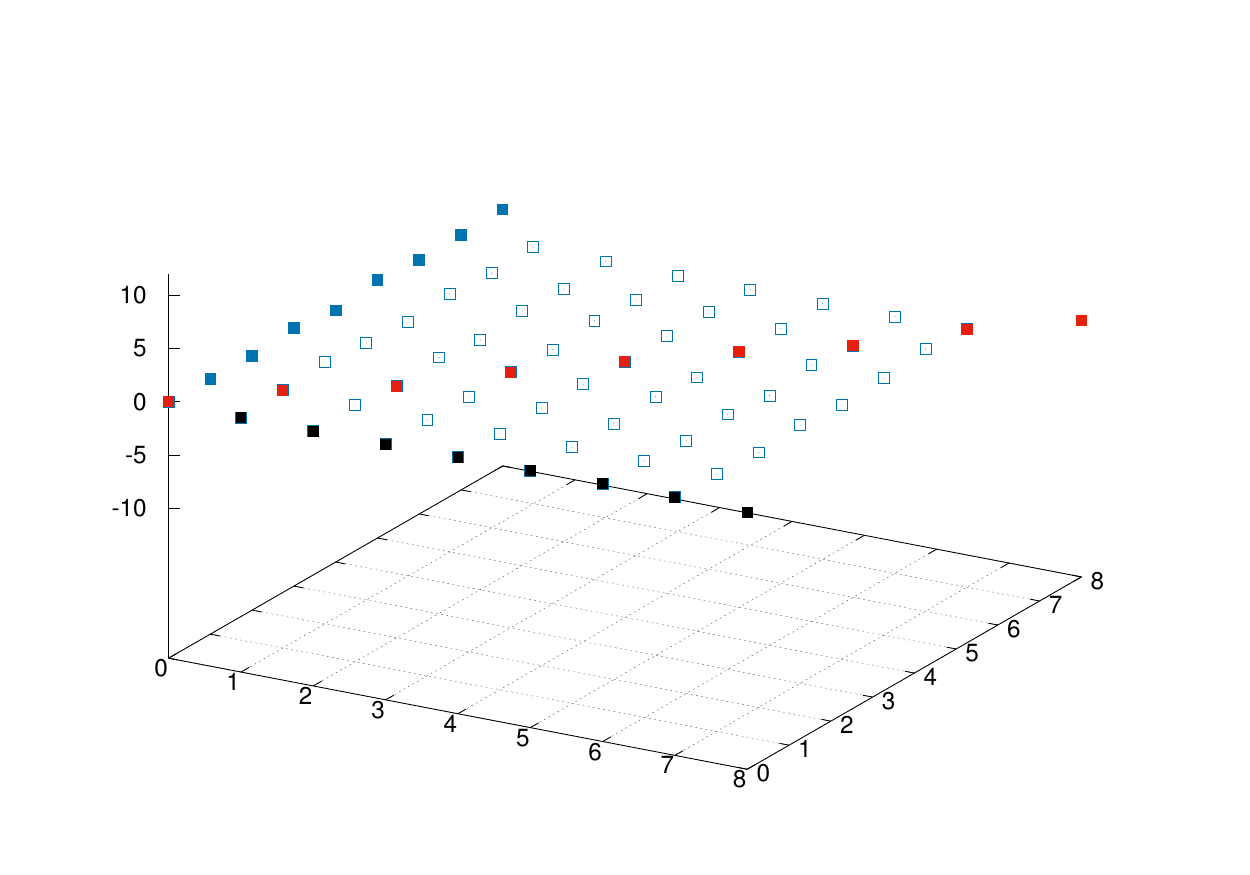} 
\includegraphics[width =75mm]{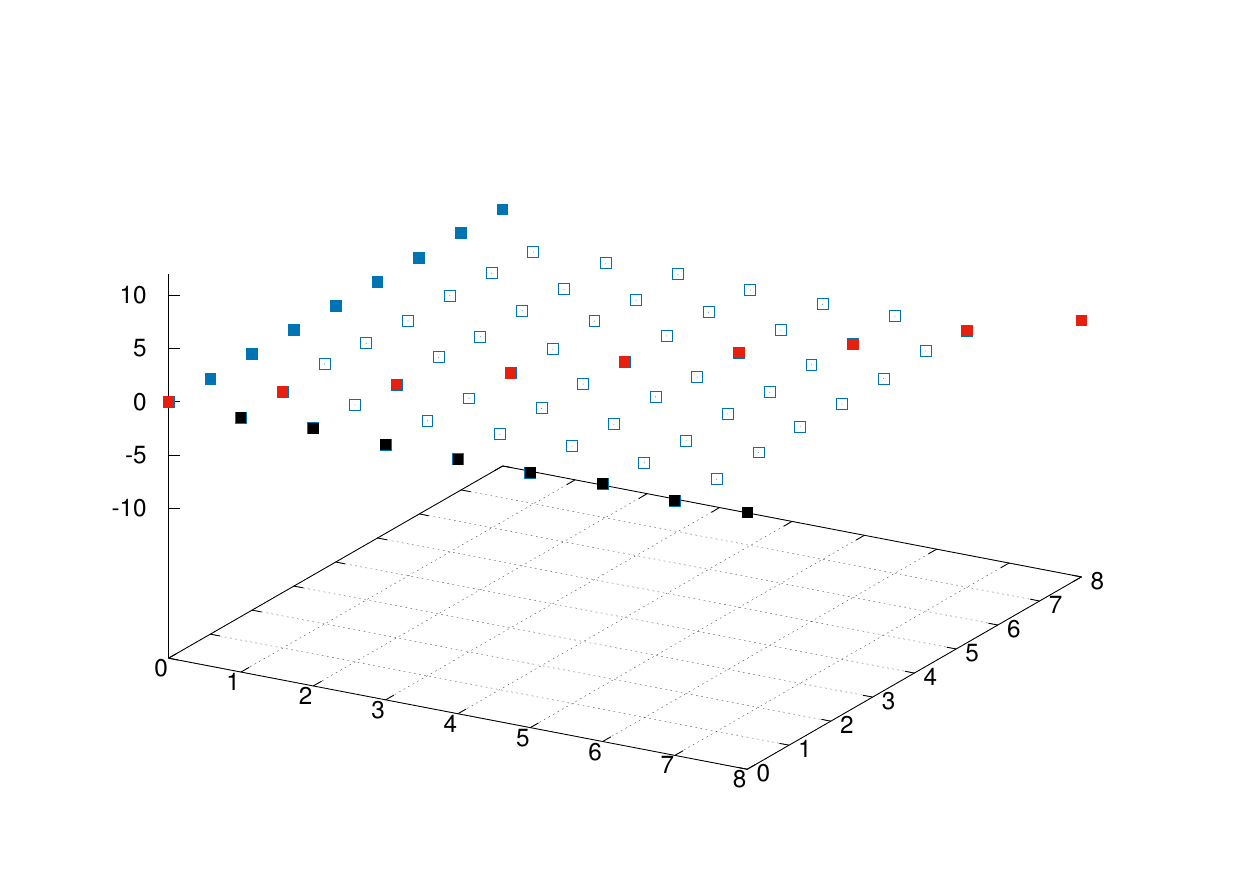}
\\
\includegraphics[width =70mm]{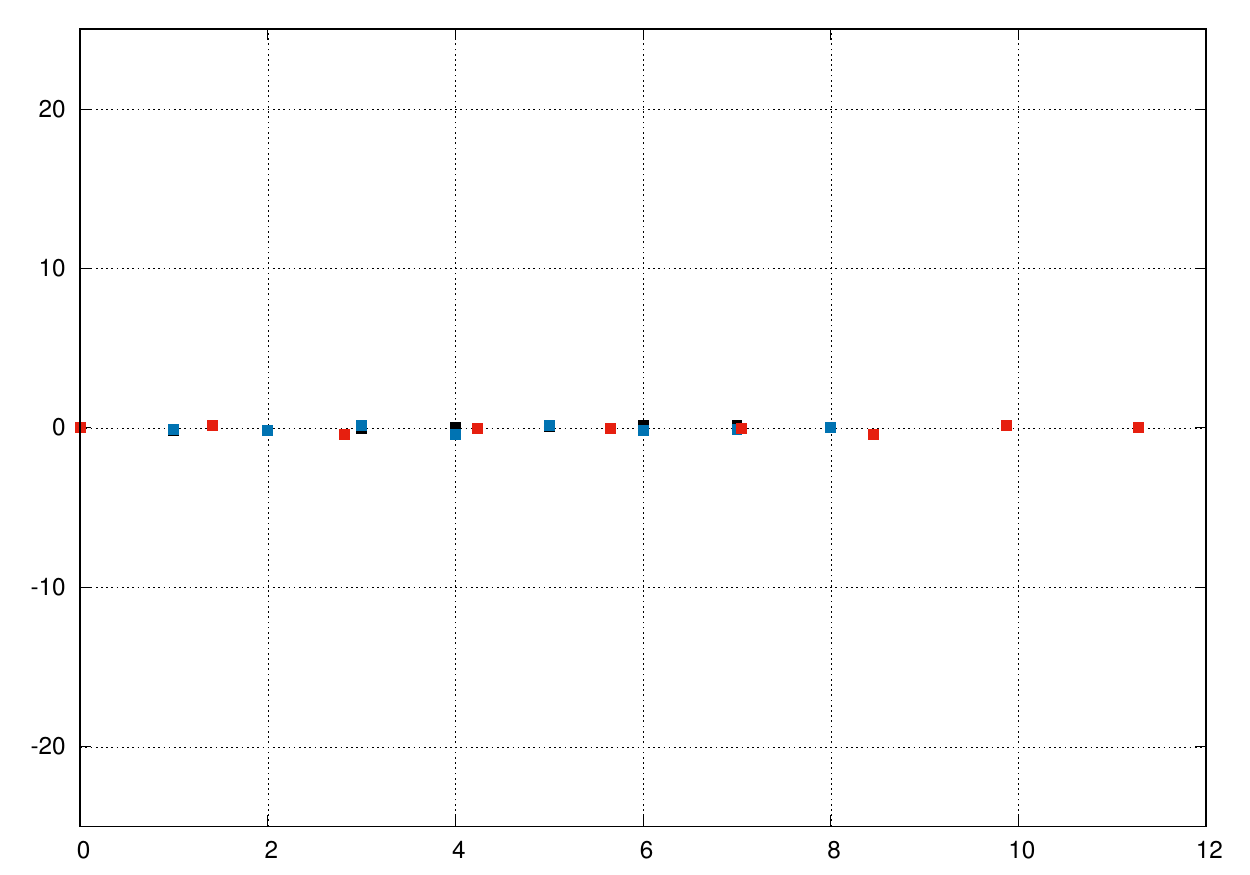} 
\hspace{1em}
\includegraphics[width =70mm]{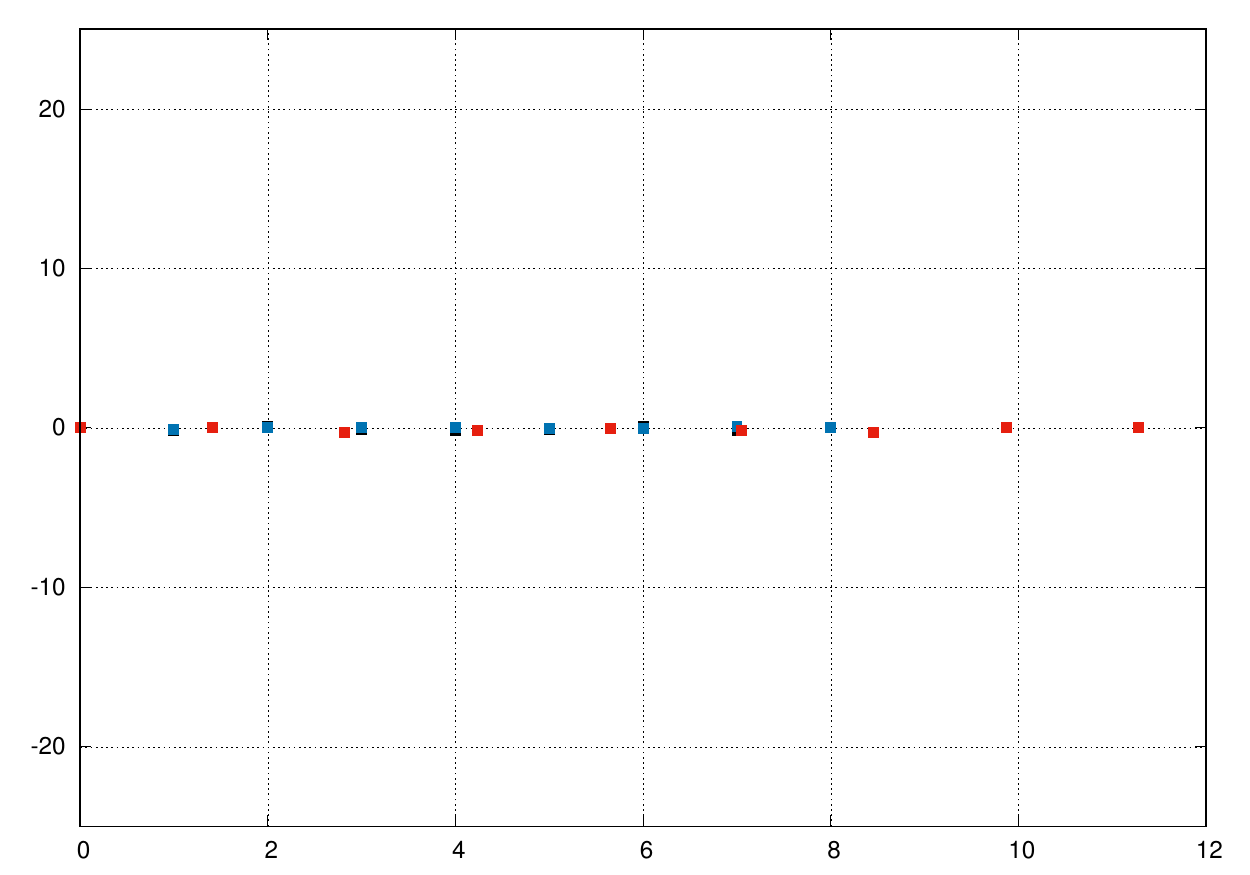} 
\\
\end{center}
\caption{
$ 2 \sin(k_\mu /2) \times  L^2 \, \tilde  \Pi^{AA}_{\mu 0}(k)$ [left]
and 
$ 2 \sin(k_\mu /2) \times L^2 \, \tilde  \Pi^{AA}_{\mu 1}(k)$ [right]
 vs. 
$\vert k \vert_2 \equiv \sqrt{ k_0^2 + k_1^2 }$ where
$A=[12]$.
The lattice size is $L=8$. 
The periodic boundary condition is assumed
for the fermion fields.
The black-, blue-, red-symbol plots
are 
along
the spacial momentum axis ($k_0=0$),
the temporal momentum (energy) axis ($k_1 =0$) and
the diagonal momentum axis ($k_0=k_1$), respectively.
5,000 configurations are sampled with the interval of 20 trajectories. 
The errors are simple statistical ones.
}
\label{fig:correlation-DAAS12-8x8}
\end{figure}

In fig.~\ref{fig:correlation-AAS12-8x8},
we show
the numerical-simulation results of 
$\tilde  \Pi^{AA}_{00}(k)$ 
and 
$\tilde  \Pi^{AA}_{01}(k)$ with $A=[12]$
for the lattice size $L=8$
and the periodic boundary condition 
of the mirror fermion fields.
5,000 configurations are sampled with the interval of 20 trajectories. The errors are simple statistical ones.
In fig.~\ref{fig:correlation-DAAS12-8x8},
one can see that the Ward-Takahashi relations eqs.~(\ref{eq:two-point-vertex-Spin6-gauge-field-WT1}) and (\ref{eq:two-point-vertex-Spin6-gauge-field-WT2}) 
are satisfied only up to the statistical error 
in this case. This is because 
the Schwinger-Dyson equation 
w.r.t. the spin field $E^a(x)$, eq.~(\ref{eq:SO(6)spin-SD-eq}),
holds true at most in the same numerical precision.

From these results, 
we can see that 
the mirror fermion contribution, 
$\tilde\Pi^{AB}_{\mu\nu}(k)$,
does not show any 
evidence of the singularities due to Spin(6)-charged massless excitations. It behaves like a regular function of momentum $k_\mu$.
This result is again consistent with the fact that 
the mirror fermions decouple by acquiring
the masses of order the inverse lattice spacing
and leave only local terms in the effective action.

\section{21(-1)$^3$ chiral gauge model -- A solution to the 
reconstruction theorem}
\label{sec:21(-1)^3-model}

In this section, 
we consider the $21(-1)^3$ chiral gauge model, 
which is  obtained from the previous $1^4(-1)^4$
axial model by modifying the gauge group
from U(1)$_A$ to a U(1) subgroup 
of U(1)$_A$ $\times$ Spin(6) (SU(4)).
We first formulate the model
in the mirror fermion approach with the Ginsparg-Wilson fermions.
We then deduce a definition of the path-integral
measure for the (target) Weyl fermions of the $21(-1)^3$ chiral gauge model,
and argue that the induced measure-term current fulfills
the requirement of the original reconstruction theorem
of the Weyl fermion measure.

\subsection{21(-1)$^3$ chiral gauge model}

We consider the $21(-1)^3$ chiral gauge model 
which is defined by the charge assignment of the U(1) gauge symmetry as
\begin{eqnarray}
Q
&=&{\rm diag}(q_1, q_2, q_3, q_4)
={\rm diag}(+2, \, \, \, 0, \, \, \, 0, \, \, \, 0), \\
\quad
Q'
&=&{\rm diag}(q'_1, q'_2, q'_3, q'_4)
={\rm diag}(+1, -1, -1, -1) .
\end{eqnarray}
We note that 
$Q$ and $Q'$ can be regarded as the 
linear combinations of the axial charge of U(1)$_A$ 
and the Cartan subalgebra of
Spin(6),  
$\{ \Sigma^{12}, \Sigma^{34},\Sigma^{56} \}$,  
in the previous four-flavor axial gauge model
as follows,
\begin{eqnarray}
Q &=& + \frac{1}{2} + \Sigma^{12} + \Sigma^{34} + \Sigma^{56} , 
\\
Q' &=& - \frac{1}{2} + \Sigma^{12} + \Sigma^{34} + \Sigma^{56} ,
\end{eqnarray}
assuming 
that in the weak gauge-coupling limit
the left- and right-handed Weyl fermions,
$\psi_-(x)$ and $\psi'_+(x)$, 
are in $\underline{4}$,
the four-dimensional
irreducible (spinor) representation of Spin(6) (SO(6)).

A comment is in order about the relation with 21$^4$ model.
In the 21$^4$ model, the anomaly matching condition for the 
flavor chiral SU(4) symmetry can be saturated by
a Majorana-Weyl field in  $\underbar{6}$ of SU(4), and it predicts the appearance of such excitation as a composite state\cite{Narayanan:1996kz,Kikukawa:1997md,Kikukawa:1997dv}.
However, it is known to be difficult to formulate the local lattice action of Majorana-Weyl fermions without species doubling (even with overlap fermions)\cite{Inagaki:2004ar}. 
Then, it seems difficult to formulate
the three neutral spectator Weyl fields, 0$^3$, 
into the Majorana-Weyl field in  $\underbar{6}$ of SU(4) to saturate the anomaly.
In the case of the 21(-1)$^3$ model, on the other hand, the anomaly matching condition for the 
flavor chiral SU(3) symmetry can be saturated by
a Weyl field in  $\underbar{3}$ of SU(3), and
the three neutral spectator Weyl fields, 0$^3$, can do the job.

\subsection{21(-1)$^3$ chiral gauge model in the mirror-fermion approach 
}

To formulate the $21(-1)^3$ model in the mirror fermion approach, we introduce that the (four-flavor) right- and left-handed mirror fermions,
$\psi_+(x)$ and $\psi'_-(x)$.
Then, as shown in table~\ref{table:21(-1)^3-anomaly},
the remaining continuous 
symmetry in the mirror sector is 
the vector flavor symmetry SU(3), 
the vector and axial U(1) symmetries U(1)$_b$ and U(1)$_a$
acting on the flavor SU(3) sector, and another vector U(1)
symmetry U(1)$_{b-3l}$.
For SU(3) and U(1)$_{b-3l}$, the would-be gauge anomalies
are matched. U(1)$_b$ and U(1)$_a$ are anomalous and
should be broken explicitly.

\begin{table}[h]
\begin{center}
\begin{tabular}{|c||c|c|c|c|l|l|}
\hline
 & $+$ & $+$ & $-$ & $-$& (mixed) gauge anomaly & chiral anomaly \\ \hline
U(1)$_g$ & 2 & 0 & 1 & -1 & matched (gauged)& ---\\
SU(3)    & $\underbar{1}$ & $\underbar{3}$ & $\underbar{1}$ & $\underbar{3}$ & matched (can be gauged) 
& anomaly free \\
U(1)$_{b}$& 0 & 1 & 0 & 1 &not matched & anomalous \\
U(1)$_{a}$& 0 & 1 & 0 & -1 &not matched & anomalous \\
U(1)$_{b-3l}$& -3 & 1 & -3 & 1 & matched (can be gauged)& anomaly free \\
\hline
\end{tabular}
\end{center}
\caption{Fermionic continuous symmetries in the mirror sector
of the 21(-1)$^3$ model and their would-be gauge anomalies}
\label{table:21(-1)^3-anomaly}
\end{table}

Then we can formulate 
the mirror fermion sector of 
the $21(-1)^3$  model in the same manner as that of the $1^4(-1)^4$ model given by
eq.~(\ref{eq:action-mirrorsector-1^4-(-1)^4-model}),
using the Majorana-type Yukawa-couplings
to the auxiliary SO(6)-vector spin fields,
$E^a(x)$, $\bar E^a(x)$ $(a=1,\cdots,6)$ 
with the unit lengths $ E^a(x)E^a(x)=1$, $\bar E^a(x) \bar E^a(x)=1 $. And we can consider 
the limit 
$z/h \rightarrow 0$ and $\kappa \rightarrow 0$ in
the mirror fermion sector,  
where
the kinetic terms of the mirror fermion and the spin fields
are both suppressed.

\begin{figure}[h]
\begin{center}
\includegraphics[width =75mm]{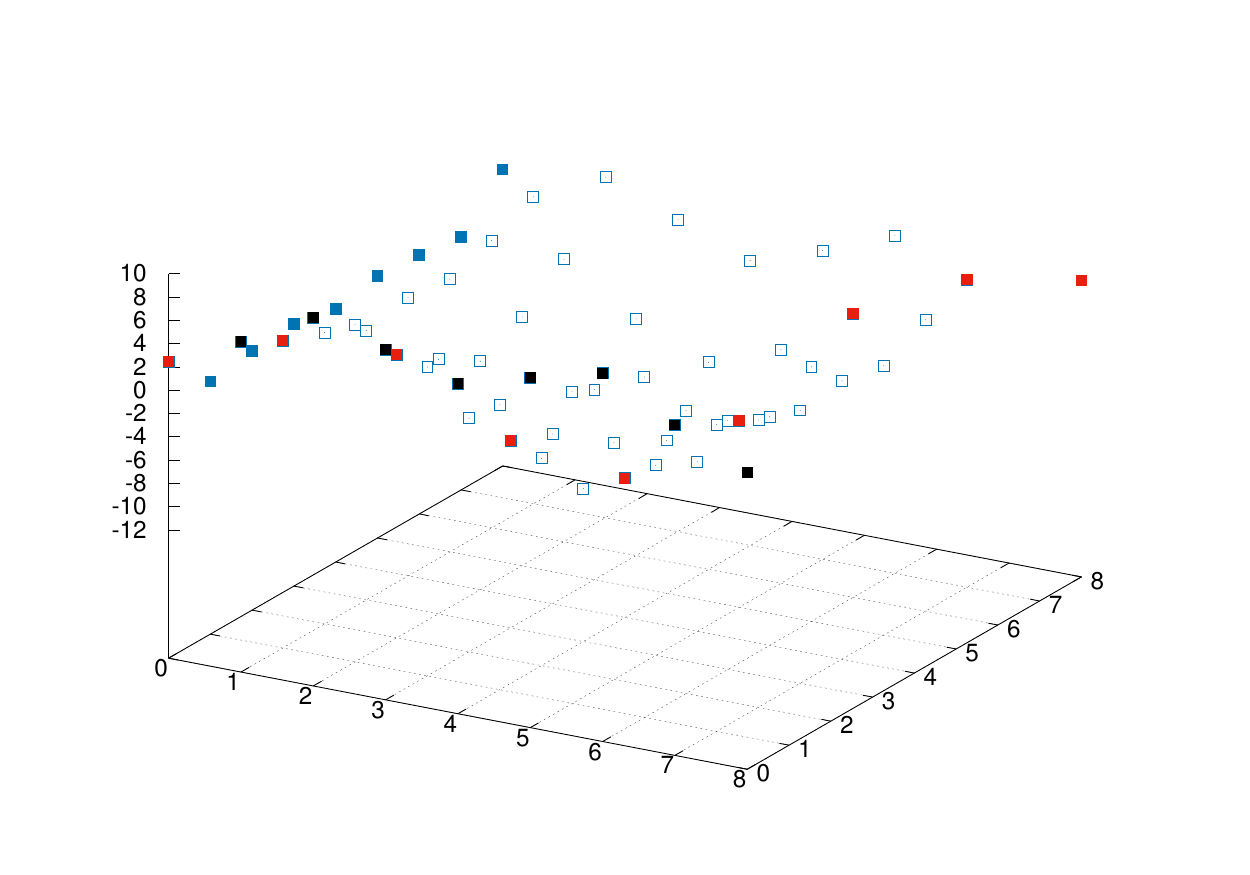} 
\includegraphics[width =75mm]{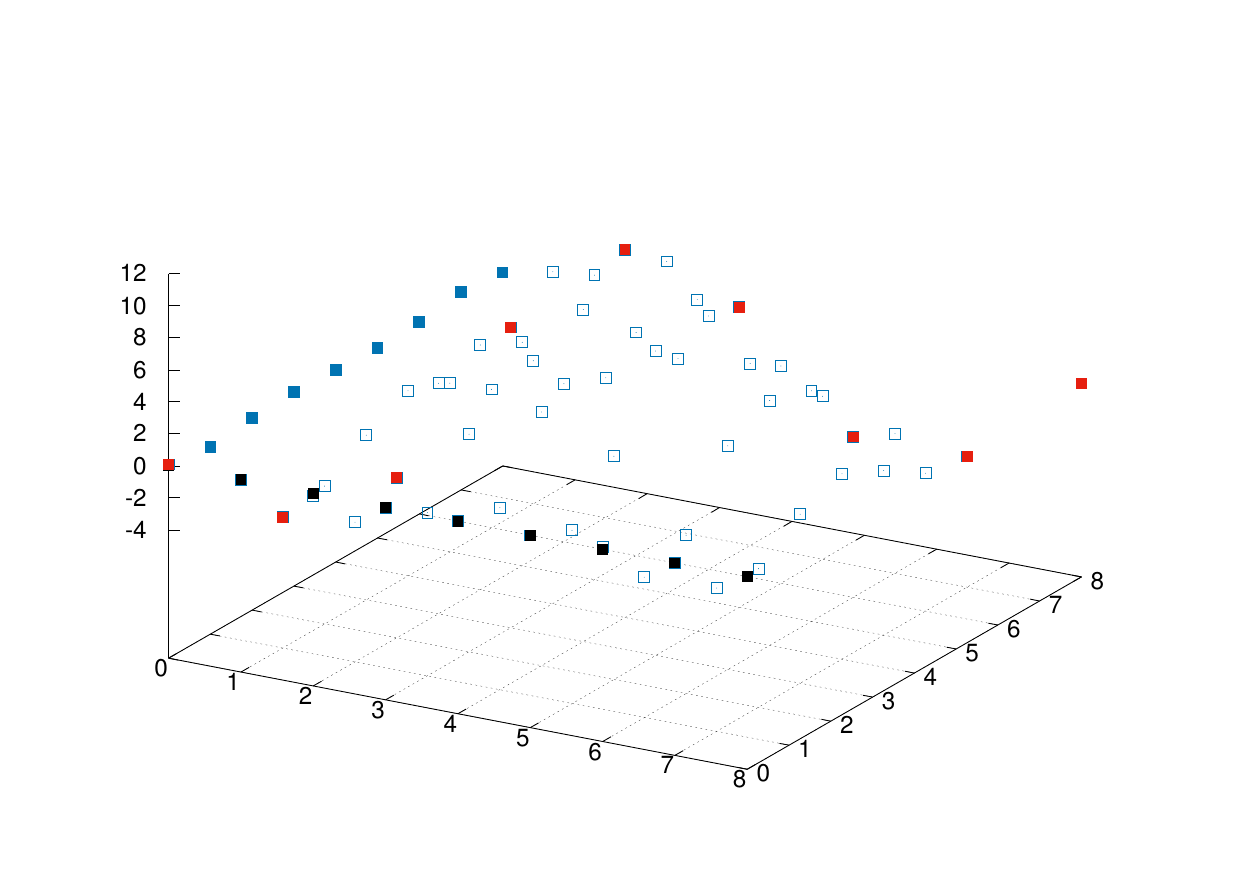}
\\
\includegraphics[width =70mm]{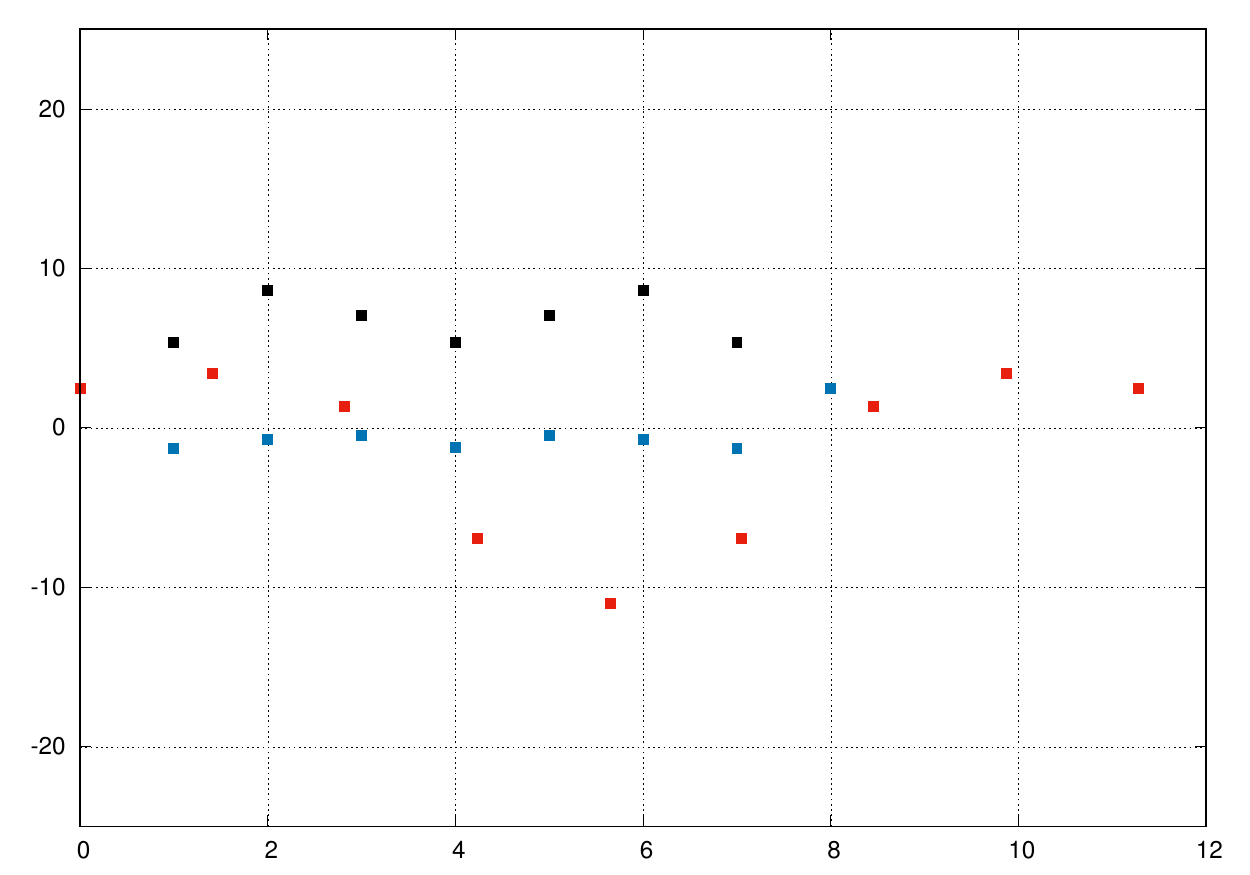} 
\hspace{1em}
\includegraphics[width =70mm]{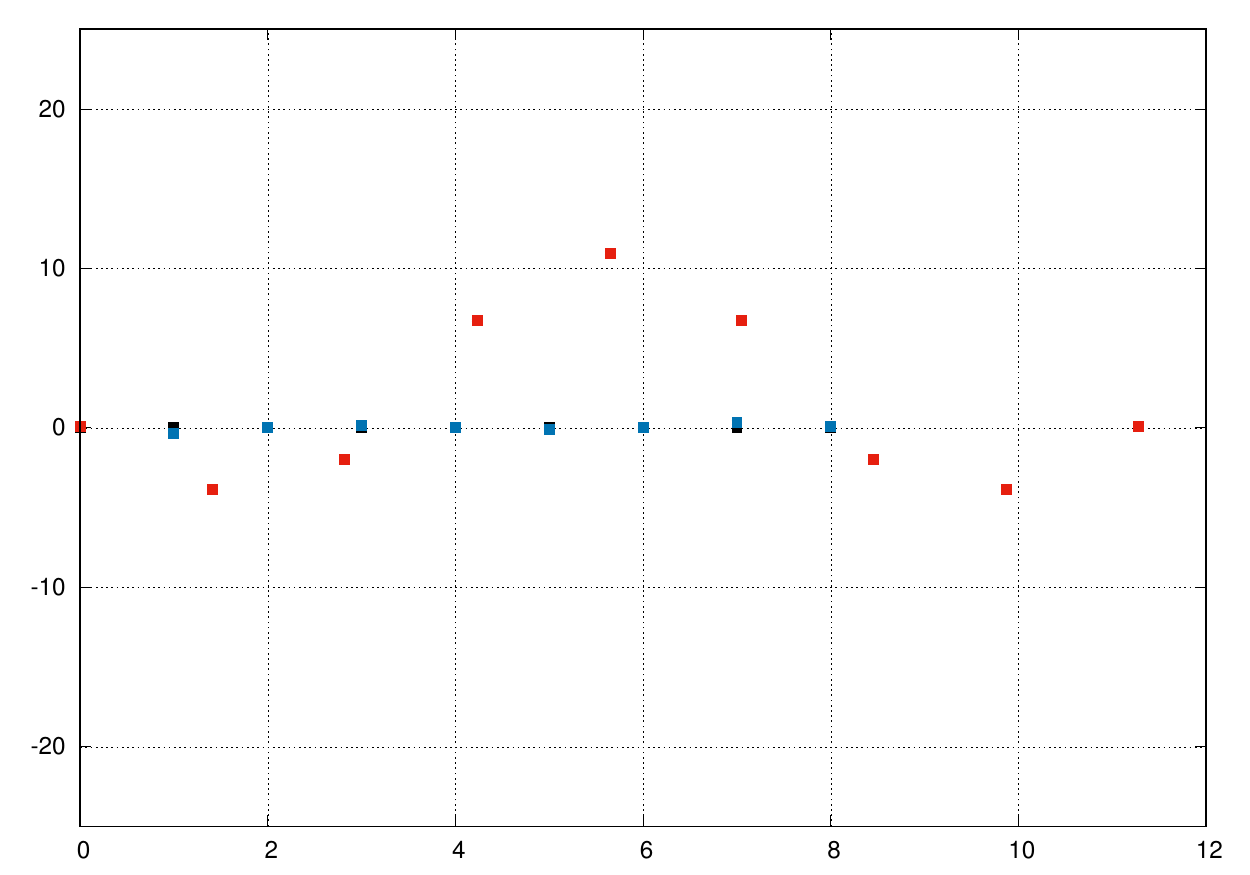} 
\\
\end{center}
\caption{
$L^2 \, \tilde  \Pi^{\hat Q \hat Q}_{00}(k)$ [left]
and 
$L^2 \, \tilde  \Pi^{\hat Q \hat Q}_{01}(k)$ [right]
 vs. 
$\vert k \vert_2 \equiv \sqrt{ k_0^2 + k_1^2 }$.
The lattice size is $L=8$. 
The periodic boundary condition is assumed
for the fermion fields.
The black-, blue-, red-symbol plots
are 
along
the spacial momentum axis ($k_0=0$),
the temporal momentum (energy) axis ($k_1 =0$) and
the diagonal momentum axis ($k_0=k_1$), respectively.
5,000 configurations are sampled with the interval of 20 trajectories. 
The errors are simple statistical ones.
}
\label{fig:correlation-AAQ-8x8}
\end{figure}

\begin{figure}[h]
\begin{center}
\includegraphics[width =75mm]{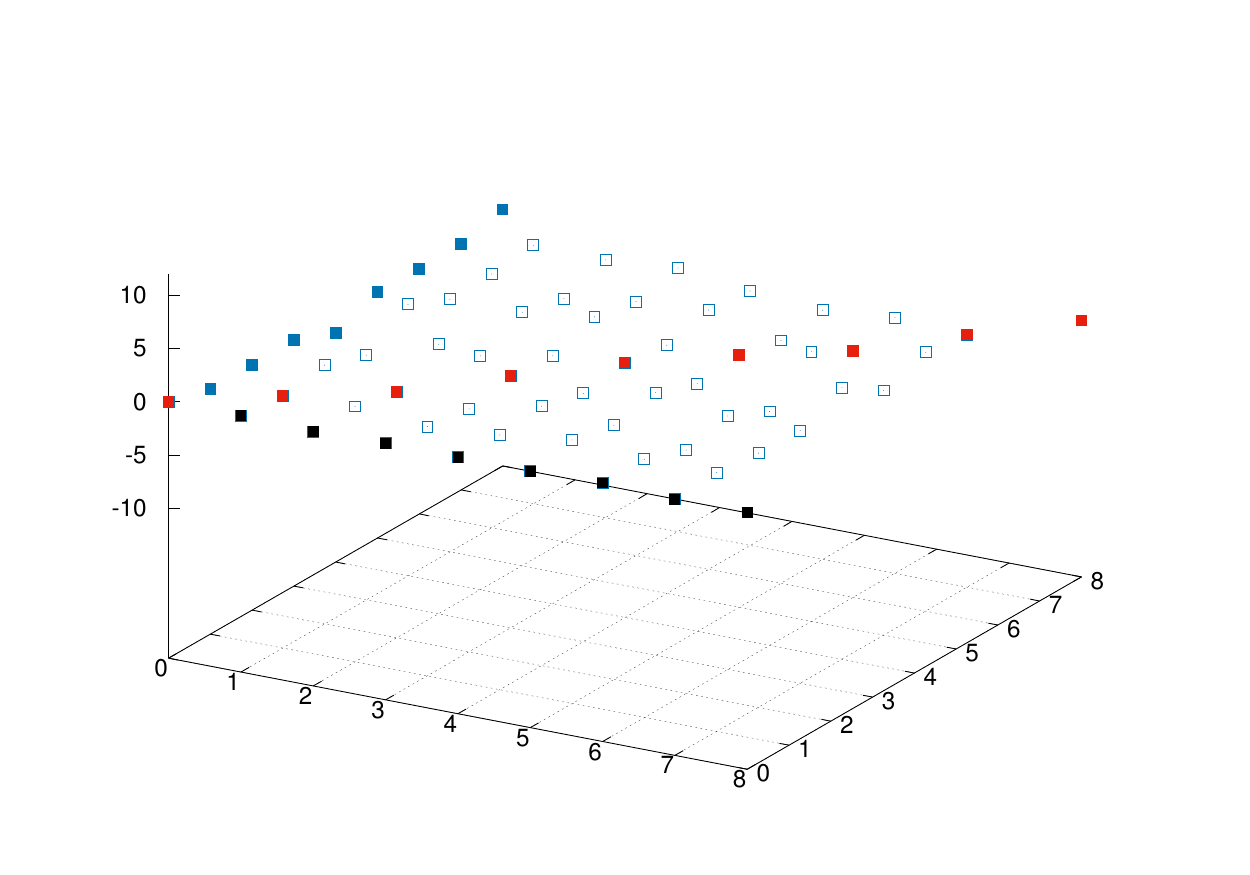} 
\includegraphics[width =75mm]{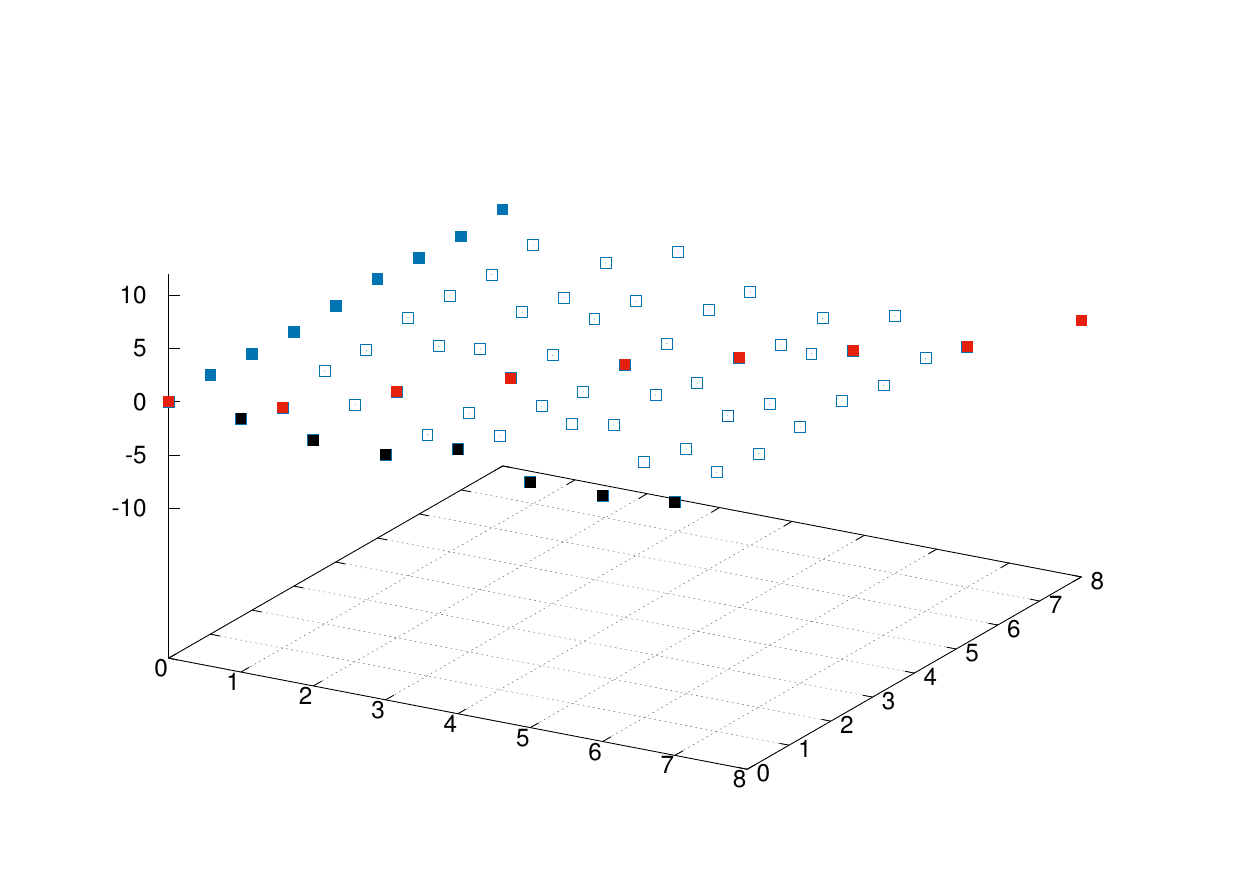}
\\
\includegraphics[width =70mm]{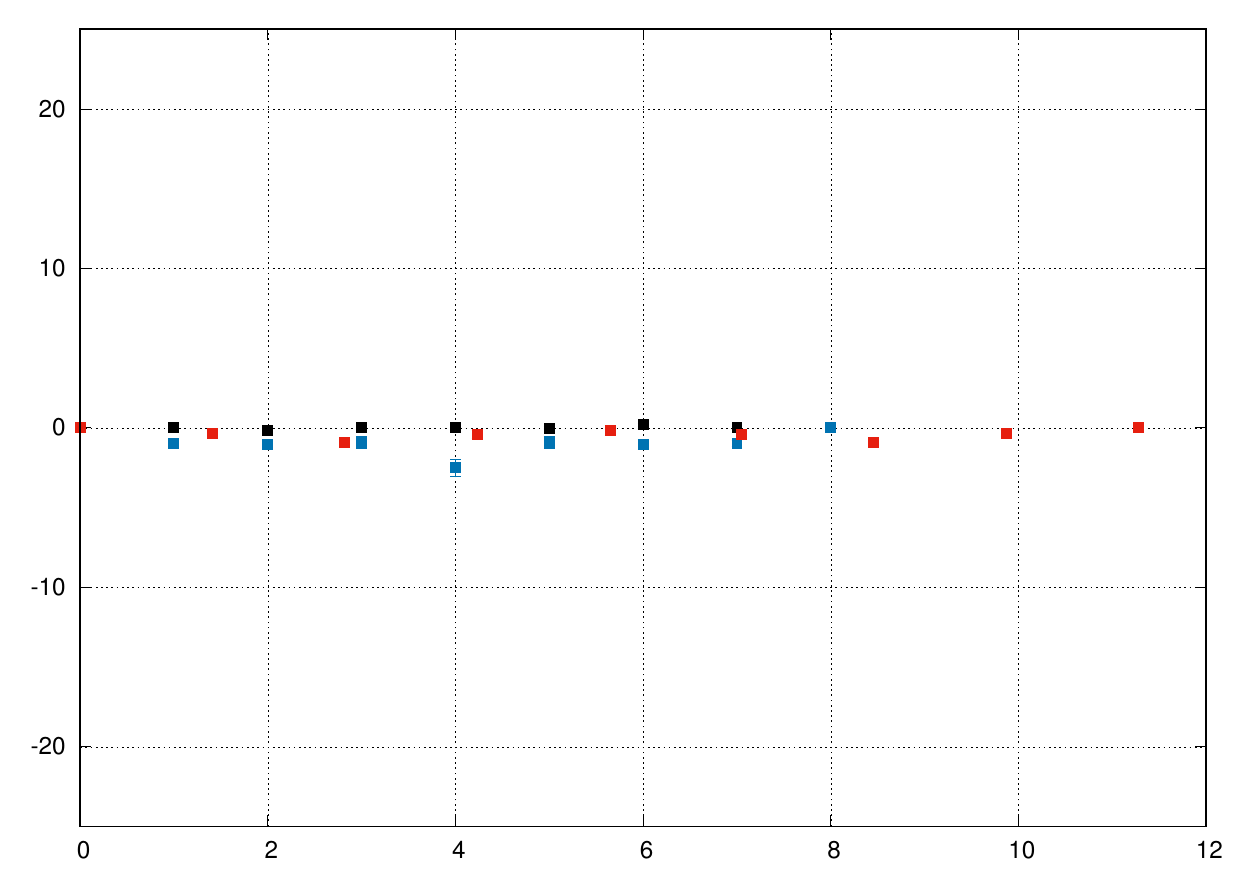} 
\hspace{1em}
\includegraphics[width =70mm]{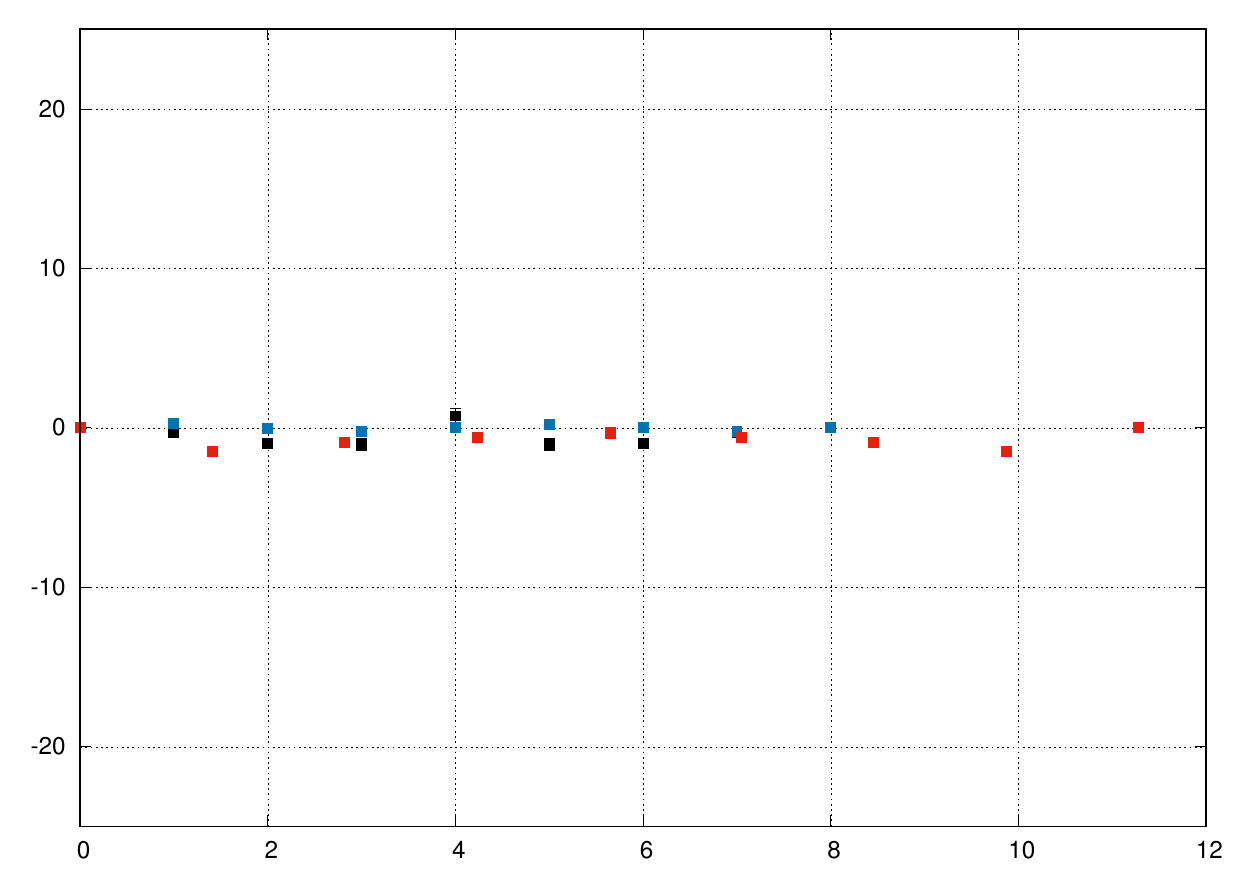} 
\\
\end{center}
\caption{
$ 2 \sin(k_\mu /2) \times  L^2 \, \tilde  \Pi^{QQ}_{\mu 0}(k)$ [left]
and 
$ 2 \sin(k_\mu /2) \times L^2 \, \tilde  \Pi^{QQ}_{\mu 1}(k)$ [right]
 vs. 
$\vert k \vert_2 \equiv \sqrt{ k_0^2 + k_1^2 }$.
The lattice size is $L=8$. 
The periodic boundary condition is assumed
for the fermion fields.
The black-, blue-, red-symbol plots
are 
along
the spacial momentum axis ($k_0=0$),
the temporal momentum (energy) axis ($k_1 =0$) and
the diagonal momentum axis ($k_0=k_1$), respectively.
5,000 configurations are sampled with the interval of 20 trajectories. 
The errors are simple statistical ones.
}
\label{fig:correlation-DAAQ-8x8}
\end{figure}

The mirror fermion sector of the $21(-1)^3$ model so defined
shares almost all the properties in the weak gauge-coupling limit
with that of the $1^4(-1)^4$ model.
The only non-trivial one is the behavior of 
the vertex functions of the U(1) gauge field.
In fig.~\ref{fig:correlation-AAQ-8x8},
we show
the numerical-simulation results of 
$\tilde  \Pi^{\hat Q \hat Q}_{00}(k)$ 
and 
$\tilde  \Pi^{\hat Q \hat Q}_{01}(k)$ 
for the lattice size $L=8$
and the periodic boundary condition 
of the mirror fermion fields.
($\hat Q$ is the abbreviation for $Q, Q' =\pm 1/2+\Sigma^{12}+\Sigma^{34}+\Sigma^{56}$ on 
$\psi_+$, $\psi'_-$, respectively.)
5,000 configurations are sampled with the interval of 20 trajectories. The errors are simple statistical ones.
In fig.~\ref{fig:correlation-DAAQ-8x8},
it is verified that the Ward-Takahashi relations 
are satisfied upto the statistical error.
These results 
should be compared with that of the (target) Weyl fermions 
$\tilde\Pi_{\mu\nu}(k)$ shown 
in fig.~\ref{fig:correlation-AA-8x8-Weyl}.
From these results, 
we can see that 
the mirror fermion contribution of the $21(-1)^3$ model, 
$\tilde\Pi^{\hat Q \hat Q}_{\mu\nu}(k)$,
does not show any 
evidence of the singularities due to charged massless excitations. It behaves like a regular function of momentum $k_\mu$.
This result is again consistent with the fact that 
the mirror fermions decouple by acquiring
the masses of order the inverse lattice spacing
and leave only local terms in the effective action.

\subsection{Weyl-field measure through the saturation of the mirror-fermion part of Dirac-field measure by the 't Hooft vertices}

Let us recall the fermion path-integral of the $21(-1)^3$ chiral gauge model in the mirror-fermion approach.
In particular, 
in the limit where
the kinetic terms of the mirror fermions and the spin fields
are both suppressed ($z/h \rightarrow 0$ and $\kappa \rightarrow 0$), it is formulated as follows.
\begin{eqnarray}
{\rm e}^{\Gamma_{\rm mirror} [ U ]} 
&=& \big\langle  1 \big\rangle_{W} 
   \big\langle  1 \big\rangle_{M} 
\nonumber\\
&=& 
 \int 
{\cal D}[\psi_-] {\cal D}[\bar \psi_-]{\cal D}[\psi'_+] {\cal D}[\bar \psi'_+] \, 
 {\rm e}^{-S_W}  \, \times 
\nonumber\\
&& \qquad
 \int 
{\cal D}[\psi_+]{\cal D}[\bar \psi_+]
{\cal D}[\psi'_-] {\cal D}[\bar \psi'_-] 
{\cal D}[E^a] {\cal D}[\bar E^a]  \,
 {\rm e}^{-S_M}  
\nonumber\\
&=&
 \int 
{\cal D}[\psi] {\cal D}[\bar \psi]
{\cal D}[\psi'] {\cal D}[\bar \psi'] 
{\cal D}[E^a] {\cal D}[\bar E^a]  \,
 {\rm e}^{-S_W - S_M } ,
\end{eqnarray}
where
\begin{eqnarray}
\label{eq:action-weylsector-21(-1)^3-model}
S_W &=& \sum_{x\in\Gamma} \bar \psi (x) P_+ D  \psi(x)  + \sum_{x\in\Gamma} \bar \psi' (x) P_- D'  \psi'(x) ,  \\
\label{eq:action-mirrorsector-21(-1)^3-model}
S_M &=& 
 \sum_{x \in \Gamma} \, \left\{ 
  \psi_+(x)^T i \gamma_3 c_D 
{\rm T}^a E^a(x) \psi'_-(x) 
+ \bar \psi_+ (x)  i \gamma_3 c_D 
{\rm T}^{a \dagger} \bar E^a(x) 
 {\bar\psi'_-(x)}^T   
\right\} .  
\end{eqnarray}
This formula can be rewritten further 
through the path-integration of 
$E^a(x)$ and $\bar E^a(x)$ using the integral,
\begin{equation}
\label{eq:integral-formula-e}
(\pi^3)^{-1} \, \int \prod_{a=1}^{6} d e^a \delta( \sqrt{e^b e^b} -1)\, {\rm e}^{e^c u^c}
= \left. 2!\, \sum_{k=0}^\infty \frac{w^k}{k! (k+2)!} 
 \, \right\vert_{w=(1/2) u^a u^a} .
\end{equation}
The result is given by
\begin{eqnarray}
{\rm e}^{\Gamma_{\rm mirror} [ U ]} 
&=& 
\int 
{\cal D}[\psi] {\cal D}[\bar \psi]
{\cal D}[\psi'] {\cal D}[\bar \psi'] \,
%
%
\prod_{x} F( O_V(x) ) \, 
\prod_{x} F( \bar O_V(x) ) \, \, 
 {\rm e}^{-S_W}  , 
\nonumber\\
\end{eqnarray}
where the function $F(\omega)$ is defined by
\begin{equation}
F(w) \equiv 2!\, \sum_{k=0}^\infty \frac{w^k}{k! (k+2)!}
= 2!\,(z/2)^{-2} I_2(z) \, 
\Big\vert_{(z/2)^2=w} 
\end{equation}
and $I_\nu(w)$ is the modified Bessel function of the first kind.
$O_V(x)$ and $\bar O_V(x)$ are given by eqs.~(\ref{eq:O_V-vertex}) and (\ref{eq:bar-O_V-vertex}), 
\begin{eqnarray}
{O}_V(x) 
&=& 
\frac{1}{2}\,
\psi_+(x)^T i \gamma_3 c_D 
{\rm T}^a \psi'_-(x) \,
\psi_+(x)^T i \gamma_3 c_D 
{\rm T}^a \psi'_-(x) , 
\\
\bar { O}_V(x)
&=&
\frac{1}{2}\,
\bar \psi_+ (x)  i \gamma_3 c_D 
{\rm T}^{a \dagger}
 {\bar\psi'_-(x)}^T \,
\bar \psi_+ (x)  i \gamma_3 c_D 
{\rm T}^{a \dagger}
 {\bar\psi'_-(x)}^T   .
\end{eqnarray}
We note that 
in this model
the counterparts of the 't Hooft vertices in the mirror sector,
which are induced by the U(1) instantons in two-dimensions,
can be written as
\begin{eqnarray}
\label{eq:tHooft-vertices-21(-1)^3}
&& O_T(x) = \frac{1}{2}\, O_V(x) O_V(x) \bar O_V(x) , 
\qquad \quad 
\bar O_T(x) = \frac{1}{2}\, \bar O_V(x) \bar O_V(x) O_V(x) . 
\end{eqnarray}
Therefore $O_V(x)$ and $\bar O_V(x)$ give 
essentially the 't Hooft vertices in the mirror sector.

The above result implies that the path-integral measure of the 
(target) Weyl fields  
$\psi_-(x)$, $\bar \psi_-(x)$ 
and 
$\psi'_+(x)$, $\bar \psi'_+(x)$ in the $21(-1)^3$ 
model
can be defined
simply through the saturation of the
mirror-fermion part of the Dirac field measure
by the suitable products of the 't Hooft vertices
in terms of the mirror-fermion fields.
Namely, one can define the path-integral measure of the 
(target) Weyl fields  
$\psi_-(x)$, $\bar \psi_-(x)$ 
and 
$\psi'_+(x)$, $\bar \psi'_+(x)$
as
\begin{eqnarray}
\label{eq:Weyl-field-measure-star}
{\cal D}_\star[\psi_-] {\cal D}_\star[\bar \psi_-]
{\cal D}_\star[\psi'_+] {\cal D}_\star[\bar \psi'_+] 
& \equiv &
\prod_{x}\prod_{s=1}^4 \prod_{\alpha=1}^2
\{ d \psi_{s \alpha}(x) d \bar \psi_{s \alpha}(x) 
d \psi'_{s \alpha}(x) d \bar \psi'_{s \alpha}(x)  \} \, 
\times 
\nonumber\\
&& \qquad\quad 
\prod_{x} F( O_V(x) ) \, 
\prod_{x} F( \bar O_V(x) ) .
\end{eqnarray}
It is manifestly gauge-invariant,  but it depends
on the gauge (link) field through the chiral projectors
$\hat P_+$ and $\hat P'_-$ which necessarily appear
in the definitions of $O_V (x)$ in terms of the mirror-fermion 
fields 
$\psi_+(x) = \hat P_+ \psi (x)$ and
$\psi'_-(x) = \hat P'_- \psi' (x)$.
It applies to all topological sectors of the admissible
U(1) link field $\mathfrak{U}[m]$,
because 
the matrix 
$\big( u^T i \gamma_3 c_D {\rm T}^a E^a v' \big)$
can change in size depending on the
topological charge, but not in shape (square).

\subsection{A solution to the measure term current
required for the reconstruction theorem}

As discussed in 
section~\ref{sec:mirror-fermion-approach-in-general}, 
the variation of the effective action 
$\Gamma_{\rm mirror} [ U ]$ with respect to the U(1) link field is given by
\begin{eqnarray}
\delta_\eta \Gamma_{\rm mirror} [ U ] 
&=&
\big\langle - \delta_\eta S_W  \big\rangle_{W} 
\slash  \big\langle  1  \big\rangle_{W}
+  
\big\langle - \delta_\eta S_M  \big\rangle_{M} 
\slash  \big\langle  1  \big\rangle_{M}
\nonumber\\
&=& {\rm Tr}\{ P_+ \delta_\eta D D^{-1}  \}     
+ {\rm Tr}\{ P_- \delta_\eta D' {D'}^{-1}  \}  
+  \big\langle  - \delta_\eta S_M  \big\rangle_{M} \slash  \big\langle  1  \big\rangle_{M} ,
\end{eqnarray}
and the contribution of the mirror sector, 
$\big\langle  - \delta_\eta S_M  \big\rangle_{M} \slash  \big\langle  1  \big\rangle_{M}$, 
should play the role of the measure term $\mathfrak{L}_\eta$. 
In the weak gauge-coupling expansion, the vertex functions are derived from this contribution as
\begin{eqnarray}
&& \big\langle - \delta_\eta S_M  \big\rangle_{M} \slash  \big\langle  1  \big\rangle_{M} 
\nonumber\\
&&=
\sum_{m=0}^\infty
\frac{1}{L^{2+2m}} \, \frac{1}{m!}\, 
\sum_{k,p_1,\cdots,p_m} \tilde \eta_\mu(-k)\,
 \tilde \Gamma'_{\mu \nu_1, \cdots, \nu_m} (k,p_1,\cdots,p_m ) \,
 \tilde A_{\nu_1}(p_1)\cdots  \tilde A_{\nu_m}(p_m) .
\nonumber\\
\end{eqnarray}

The simulation results about
the leading two-point vertex function shown in fig.~\ref{fig:correlation-AAQ-8x8}
provide a numerical evidence
that 
$\big\langle  - \delta_\eta S_M  \big\rangle_{M} \slash  \big\langle  1  \big\rangle_{M}$
is a local functional of the U(1) link field.
Then it (or its axial part) can provide 
a solution to the local current
required in the reconstruction theorem of the 
Weyl fermion measure reviewed in section~\ref{sec:review-abelian-chiral-gauge-model-two-dim}.

Thus the Weyl field measure defined by
eq.~(\ref{eq:Weyl-field-measure-star}) can provide
a solution to the gauge-invariant and local construction
of the fermion path-integral measure in the
$21(-1)^3$  chiral lattice gauge theory.

\section{The mirror-fermion models through three-dimensional domain wall fermions with boundary interaction terms}
\label{sec:DWF-to-Mirror}

The mirror-fermion models formulated with overlap fermions 
in 
sections~\ref{sec:1^4(-1)^4-model} and \ref{sec:21(-1)^3-model},  
the $1^4(-1)^4$- and $21(-1)^3$- models, 
can be also constructed through the three-dimensional vector-like domain wall fermions
by adding suitable boundary interaction terms\cite{Creutz:1996xc}.
The following action defines the explicit form of the boundary terms which precisely reproduce
the U(1)$_A$ $\times$ Spin(6)(SU(4))-invariant multi-fermion interaction in the mirror sector
without the singularity in the large-coupling limit.
\begin{eqnarray}
\label{eq:action-DW-U(1)A-Spin(6)}
S_{\rm DW} 
&=& \sum_{t=1}^{L_3} \sum_{x \in \Lambda}
\bar \psi(x,t) \big\{ [1+ a'_3 (D_{2 \rm w} - m_0)] \delta_{t t'}  - P_- \delta_{t+1, t'} -P_+ \delta_{t,t'+1} \big\} \psi(x, t') ,
\nonumber\\
S'_{\rm DW}
&=& \sum_{t=1}^{L_3} \sum_{x \in \Lambda}
\bar \psi'(x,t) \big\{ [1+ a'_3 (D'_{2 \rm w} - m_0)] \delta_{t t'}  - P_+ \delta_{t+1, t'} -P_- \delta_{t,t'+1} \big\} \psi'(x, t') ,
\nonumber\\
S_{\rm bd}
&=&
\sum_{x \in \Lambda}
(z -1) \, \bar \psi(x, L_3)P_- [1+a'_3(D_{2 \rm w} - m_0) ] \psi(x,L_3)
\nonumber\\
&+&
\sum_{x \in \Lambda}
(z -1) \, \bar \psi'(x, L_3)P_+ [1+a'_3(D'_{2 \rm w} - m_0) ] \psi'(x,L_3)
\nonumber\\
&+&
\sum_{x \in \Lambda}  
h \, 
\{ 
\psi^{\rm T}(x, L_3) 
 i \gamma_3 c_D {\rm T}^a E^a(x) 
\psi'(x, L_3)
\, + \, 
\bar \psi(x, L_3) 
P_- i \gamma_3 c_D {{\rm T}^a}^\dagger \bar E^a(x)
\bar \psi'(x, L_3)^T
\}   
\nonumber\\
&+& \sum_{x, \mu} \, \kappa \, \left\{ 
E^a(x) E^a(x +\hat \mu)
+ \bar E^a(x) \bar E^a(x +\hat \mu) 
\right\},
\end{eqnarray}
where the Dirichlet b.c. is assumed, 
\begin{eqnarray}
&&
P_+ \psi(x,0) = 0, \quad  \, \bar \psi(x,0) P_-  \, \, = 0 
\quad ; \quad
P_- \psi(x, L_3 + 1) \, \, =0, \quad \bar \psi(x, L_3 +1) P_+  \, = 0 ,
\nonumber\\
&&
P_- \psi'(x,0) = 0, \quad \bar \psi'(x,0) P_+ = 0 
\quad ; \quad
P_+ \psi'(x, L_3 + 1) =0, \quad \bar \psi'(x, L_3 +1) P_- = 0 .
\nonumber\\
\end{eqnarray}
$D_{\rm 2w}$ and $D'_{\rm 2w}$ are the two-dimensional
Wilson-Dirac operators and $a'_3 (= a_3 /a)$ is the lattice spacing of extra third dimension in the lattice unit.
The three-dimensional Dirac fields
$\psi(x,t)$ and $\psi'(x,t)$ 
are assumed  in $\underline{4}$, the four-dimensional
irreducible spinor representation of SO(6), satisfying
the constraints
\begin{eqnarray}
&& {\rm P}_+  \, \psi(x) \,= + \psi(x),  \quad \,\, \bar \psi(x) \,\, {\rm P}_+   \, = + \bar \psi(x) ,  \\
&& {\rm P}_+    \, \psi' (x) = + \psi'(x), \quad \bar \psi'(x) \, {\rm P}_+  = + \psi'(x) .
\end{eqnarray}

In the boundary action $S_{\rm bd}$, the first and second terms in the r.h.s. are introduced so that all the terms which involve the fields 
$ \bar \psi(x, L_3) P_-$ 
and 
$\bar \psi'(x, L_3) P_+$ 
in the original bulk actions of the domain wall fermions 
$S_{\rm DW}$ and $S'_{\rm DW}$
can be rescaled by the factor $z$ 
and 
made vanished in the limit $z \rightarrow 0$. Then the second part of the third term of the Majorana-Yukawa couplings is required so that it saturates the path-integral measure of 
those fields
$ \bar \psi(x, L_3) P_-$ and $ \bar \psi'(x, L_3) P_+$.
On the other hand, 
the fields $\psi(x,L_3)$ and $\psi'(x,L_3)$ are
related to the (truncated) overlap fermion fields 
$\psi(x)$ and $\psi'(x)$ by the relations 
$\psi(x,L_3)  = (- \gamma_5)(1+T^{L_3})^{-1} \psi(x)$ and
$\psi'(x,L_3) = (- \gamma_5)(1+{T'}{}^{-L_3})^{-1} \psi'(x)$,
respectively
in the usual subtraction scheme with the anti-periodic b.c.\cite{Vranas:1997da, Neuberger:1997bg}. 
They are projected
to the right- and left-handed Weyl fields 
$\psi_+(x) = \hat P_+ \psi(x)$ 
and 
$\psi'_-(x) = \hat P_- \psi'(x)$, respectively
in the limit $L_3 \rightarrow \infty$ and $a'_3 \rightarrow 0$. See \cite{Kikukawa:1999sy} for detail.

Then one can show the equality
of the fermion partition functions\cite{Vranas:1997da,Neuberger:1997bg,Kikukawa:1999sy},
\begin{eqnarray}
\big\langle  1 \big\rangle_{WM}  
&=&
 \int {\cal D}[E^a]{\cal D}[\bar E^a] \,
 \int {\cal D}[\psi] {\cal D}[\bar \psi] 
      {\cal D}[\psi'] {\cal D}[\bar \psi'] \,
{\rm e}^{-S_W-S_M} 
\nonumber\\
&=&
\lim_{a'_3 \rightarrow 0}
\lim_{L_3 \rightarrow \infty}
 \int {\cal D}[E^a]{\cal D}[\bar E^a] \,
\frac{
\int 
\prod_{x,t} 
d \bar \psi 
d \psi 
d \bar \psi' 
d \psi'(x,t) \ 
\, {\rm e}^{- S_{\rm DW}- S'_{\rm DW} - S_{\rm bd}}
\big\vert_{\rm Dir}
}
{
\int 
\prod_{x,t} 
d \bar \psi 
d \psi 
d \bar \psi' 
d \psi'(x,t) \ 
\, {\rm e}^{- S_{\rm DW}- S'_{\rm DW} \phantom{- S_{\rm bd}}}
\big\vert_{\rm AP} 
} ,
\nonumber\\
\end{eqnarray}
where $S_W$ and $S_M$ are given by 
eq.~(\ref{eq:S_W}) and 
eq.~(\ref{eq:action-mirrorsector-1^4-(-1)^4-model}),
respectively. 

From the avove equality, we can see that
the limits 
$z/h \rightarrow 0$
and 
$\kappa \rightarrow 0$ are both well-defined
in the domain wall formulation.
In this respect, we note that
if one uses
the boundary field variables 
introduced by Furmam and Shamir, 
\begin{eqnarray}
q(x) = \psi_-(x, 1) + \psi_+(x, L_3) , && \quad
\bar q(x) = \bar \psi_-(x, 1) + \bar \psi_+(x, L_3) \\
q'(x) = \psi_+(x, 1) + \psi_-(x, L_3), && \quad
\bar q'(x) = \bar \psi_+(x, 1) + \bar \psi_-(x, L_3),
\end{eqnarray}
and formulate the boundary interaction terms\cite{Creutz:1996xc} as
\begin{eqnarray}
\sum_{x \in \Lambda} 
h \,
\{ 
q^{\rm T}(x) i \gamma_3 c_D {\rm T}^a E^a(x) P_- q'(x)
+ 
\bar q(x)  P_-  i \gamma_3 C_D {{\rm T}^a}^\dagger 
\bar E^a(x) \bar q'(x)^T
\} , 
\end{eqnarray}
it is singular in the large-coupling limit $z/h \rightarrow 0$. This is because
$q(x)$ and $q'(x)$ can be related to the overlap Dirac fields as
$q(x) = (1-D) \psi(x)$ and $q'(x) = (1-D')\psi'(x)$ \cite{Kikukawa:1999sy} 
and the factors $(1-D)$ and $(1-D')$ project
out the modes with the momenta 
$p_\mu = 
(\pi, 0), (0,\pi), (\pi,\pi)$
in the interaction terms.

\section{Relations to 1D/2D Topological
Insulators/Superconductors with Gapped Boundary Phases}
\label{sec:relation-1D-2D-TI-TSC}

\subsection{Eight-flavor 1D Majorana chain with SO(7)-invariant quartic interaction \\ --- A 1+1D classical formulation on Euclidean lattice}
\label{sec:rel-8MC-SO7}

Let us recall the fact that the mirror-fermion sectors of 
the $1^4(-1)^4$ axial gauge model and the $21(-1)^3$ chiral gauge model both consist of the four-flavor right- and left-handed Weyl fields, $\psi_+(x)$ and $\psi'_-(x)$. In the weak gauge-coupling limit, these Weyl fields can be combined into four Dirac fields
$\psi(x) = \psi_+(x) + \psi'_-(x)$. Then the action of the mirror sector eq.~(\ref{eq:action-mirrorsector-1^4-(-1)^4-model}) may be written for 
$z=1$ and $\kappa = 0$ as follows
\begin{eqnarray}
S_{M} &=& \sum_x \big\{
\bar \psi(x) D \psi(x) 
+ h \big(
\psi_+^T i\gamma_3 c_D \check {\rm T}^a E^a(x) \psi_-(x) 
+
\bar \psi_+(x) i\gamma_3 c_D \check {\rm T}^a{}^\dagger \bar E^a(x) \bar \psi_-(x)^T
\big)
\big\} .
\nonumber\\
\end{eqnarray}
After the path-integration of the spin fields
$E^a(x)$ and $\bar E^a(x)$, if one keeps only 
the leading terms of 
the multi-fermion interactions, the effective action may be given by
\begin{eqnarray}
\label{eq:effective-quartic-action-1^4(-1)^4-model-mirrorsector}
S'_{M} &=& \sum_x \Big\{
\bar \psi(x) D \psi(x) 
- \frac{h^2}{6} 
\big[
\big( \psi_+(x)^T i\gamma_3 c_D \check {\rm T}^a \psi_-(x) \big)^2
+
\big( 
\bar \psi_+(x) i\gamma_3 c_D \check {\rm T}^a{}^\dagger 
\bar \psi_-(x)^T 
\big)^2
\big]
\,\,
\Big\} .
\nonumber\\
\end{eqnarray}
This model with U(1)$_A$ $\times$ Spin(6)(SO(6)) symmetry 
has a close relation with the eight-flavor 1D Majorana chain
with SO(7)-invariant quartic interaction,
which was examined by 
Fidkowski and Kitaev\cite{Fidkowski:2009dba,Fidkowski:2011},
when it is formulated 
in two-dimensional Euclidean spacetime.

The eight-flavor 1D Majorana chain
with SO(7)-invariant quartic interaction\cite{Fidkowski:2009dba,Fidkowski:2011} is defined by the following quantum
Hamiltonian.
\begin{eqnarray}
\hat H &=& \sum_{\alpha = 1}^8 \hat H_\alpha + \hat V,
\end{eqnarray}
where
\begin{eqnarray}
\hat H_\alpha &=& \frac{i}{2} 
\left(
u \sum_{l=1}^n \hat c_{2 l - 1}^\alpha \hat c_{2 l}^\alpha
+
v \sum_{l=1}^{n-1} \hat c_{2 l }^\alpha \hat c_{2 l + 1}^\alpha
\right) , 
\\
\hat V &=& \sum_{l=1}^n 
\left( \hat W_{2 l -1} + \hat W_{2 l} \right) 
\end{eqnarray}
and
\begin{eqnarray}
\hat W &=&
\phantom{+} \hat c^{1} \hat c^{2} \hat c^{3} \hat c^{4}
+\hat c^{5} \hat c^{6} \hat c^{7} \hat c^{8}
+\hat c^{1} \hat c^{2} \hat c^{5} \hat c^{6}
+\hat c^{3} \hat c^{4} \hat c^{7} \hat c^{8}
-\hat c^{2} \hat c^{3} \hat c^{6} \hat c^{7}
\nonumber\\
&&
-\hat c^{1} \hat c^{4} \hat c^{5} \hat c^{8}
+\hat c^{1} \hat c^{3} \hat c^{5} \hat c^{7}
+\hat c^{3} \hat c^{4} \hat c^{5} \hat c^{6}
+\hat c^{1} \hat c^{2} \hat c^{7} \hat c^{8}
-\hat c^{2} \hat c^{3} \hat c^{5} \hat c^{8}
\nonumber\\
&&
-\hat c^{1} \hat c^{4} \hat c^{6} \hat c^{7}
+\hat c^{2} \hat c^{4} \hat c^{6} \hat c^{8}
-\hat c^{1} \hat c^{3} \hat c^{6} \hat c^{8}
-\hat c^{2} \hat c^{4} \hat c^{5} \hat c^{7} ,
\\
&&\nonumber\\
\hat W_m &=& \hat W \big\vert_{\hat c^\alpha \rightarrow \hat c^\alpha_m} .
\end{eqnarray}
The quartic interaction $\hat W$ is invariant under the
SO(7) transformation which acts on the operators $\hat c^\alpha$
$(\alpha=1, \cdots, 8)$ 
so that they are in  $\underbar{8}$, 
the irreducible spinor representation of SO(7).\footnote{The eight-component Majorana operator $\hat c^\alpha$ can be 
regarded as a real vector or a real spinor of SO(8), 
thanks to the triality of SO(8).}
In fact, as shown by Y.-Z. You and C. Xu \cite{You:2014vea}, 
the operator $\hat W$ can be written into
the form which is manifestly SO(7)-invariant:
\begin{eqnarray}
\hat W &=& - \frac{1}{4 !} 
\Big(
\sum_{a=1}^7
\hat c^T \gamma^a \hat c \,\, \hat c^T \gamma^a \hat c - 16
\Big) ,
\end{eqnarray}
where $\{ \gamma^a \,\vert a=1, \cdots, 7 \}$ is the set of the gamma matrices for SO(7) given explicitly as
\begin{eqnarray}
\gamma^1 &=& I \otimes I \otimes \sigma_2 , \\
\gamma^2 &=& \sigma_3 \otimes \sigma_2 \otimes \sigma_3 , \\
\gamma^3 &=& I \otimes \sigma_2 \otimes \sigma_1 , \\
\gamma^4 &=& \sigma_2 \otimes I \otimes \sigma_3 , \\
\gamma^5 &=& \sigma_2 \otimes \sigma_3 \otimes \sigma_1 , \\
\gamma^6 &=& \sigma_1 \otimes \sigma_2 \otimes \sigma_3 , \\
\gamma^7 &=& \sigma_2 \otimes \sigma_1 \otimes \sigma_1 ,
\end{eqnarray}
which are pure imaginary and anti-symmetric,
$ \gamma^a{}^T = - \gamma^a $.
This Hamiltonian can be rewritten further 
using the two-component Majorana field operator, 
$\hat \psi_l^\alpha 
= ( \hat c_{2l}^\alpha , \hat c_{2 l -1}^\alpha )^T$
as 
\begin{eqnarray}
\hat H_\alpha &=&
\sum_{l=1}^n 
\frac{z}{2} \, 
\hat \psi_l^\alpha {}^T \, 
\Big\{
 \alpha \frac{1}{2i} (\nabla - \nabla^\dagger) 
+\beta \frac{1}{2} \nabla \nabla^\dagger 
+ \beta m_0 
\Big\}
\hat \psi_l^\alpha , \\
\hat V &=& - \frac{1}{4!} 
\sum_{l=1}^n 
 \big\{ 
\sum_{a=1}^7
\big(
\hat \psi_l^T 
\tilde P_+
\gamma^a \hat \psi_l 
\big)^2
+
\sum_{a=1}^7
\big(
\hat \psi_l^T 
\tilde P_- \gamma^a \hat \psi_l 
\big)^2
-32
\big\} ,
\end{eqnarray}
where $\alpha = -\sigma_1$, $\beta = \sigma_2$,
$\tilde P_\pm = (1 \mp i \beta \alpha)/2 = (1 \pm \sigma_3)/2$
and
the matching condition of the couplings are given by
$v=2z$ and $u/v= 1+ m_0 $.
We note that the Majorana condition for the eight-flavor Majorana field $\psi^\alpha(x)$ is formulated 
in general by 
${\bar \psi(x)} =  \psi(x)^\dagger \gamma^0 =
 \psi(x)^T c_D C $. The choice of the 
representation in the above case is understood as follows:
$\gamma^0 = \beta =\sigma_2$, $c_D = i \sigma_2$, and $C= -i$.
The time-reversal transformation acts on $\hat \psi^\alpha_l$
as $\hat T \hat \psi^\alpha_l \hat T^{-1} 
= \sigma_3 \hat \psi^\alpha_l
= \gamma_3 \hat \psi^\alpha_l$ and $\hat H^\alpha$, $\hat V$ are both invariant.

The 1D quantum lattice model of the eight-flavor Majorana chain defined with $\hat H$ may be formulated
as a 1+1D classical lattice model in the Euclidean metric
within the framework of the path-integral quantization\cite{Ohnuki:1978jv,Creutz:1986ky}.
The action can be chosen as
\begin{eqnarray}
S_{8MC} 
&=& \sum_x 
\Big\{
\frac{z}{2} \,\psi_M(x)^T c_D {\rm C} \big(  D_{\rm w} + m_0 \big)  \psi_M(x)
\nonumber\\
&& \qquad
-
\frac{1}{4!}
\sum_{a=1}^7
\big(
\psi_M(x)^T 
c_D 
\big(\frac{\gamma_0 - i \gamma_3}{2} \big)
{\rm C} \Gamma^a \psi_M(x)
\big)^2
\nonumber\\
&& \qquad\qquad
-
\frac{1}{4!}
\sum_{a=1}^7
\big(
\psi_M(x)^T 
c_D 
\big(\frac{\gamma_0 + i \gamma_3}{2} \big)
{\rm C} \Gamma^a \psi_M(x)
\big)^2
\Big\} ,
\end{eqnarray}
where $D_{\rm w}$ is the two-dimensional massless Wilson-Dirac operator,
$D_{\rm w} = \sum_\mu \big\{ \gamma_\mu (\nabla_\mu - \nabla_\mu^\dagger)/2 + \nabla_\mu \nabla^\dagger_\mu /2\big\}$.
$\psi_M(x)$ is the Grassmann-number field, obeying the 
constraint 
$\bar \psi_M(x) 
= \psi_M(x) c_D {\rm C}$,
if it is taken as complex.
One nay assume generic representations for the Dirac- and SO(7)-
gamma matrices. 
(Our choice of
the representation of the Dirac gamma matrices
in the Euclidean metric 
is specified 
as $\gamma_0 = \sigma_1, \gamma_1 = \sigma_2$, $\gamma_3=\sigma_3$.)
The action is invariant under
the parity and charge conjugation transformations, 
${\cal P} : \psi_M(x) \rightarrow i \gamma_0 
\psi_M( x_{\cal P} )$ where $x_{\cal P} =(x_0, - x_1)$ and 
${\cal C} : \psi_M(x) \rightarrow \psi_M(x)$.
We note that the SO(7)-invariant quartic interaction terms
possess the Z$_2$ symmetry under the discrete chiral
transformation, $Z_2: \psi_M(x) \rightarrow \gamma_3 \psi_M(x)$,
but it is broken by the mass and Wilson terms.
We also note that the 
quartic interaction terms do not respect the covariance w.r.t. 
2 dim. (hyper-cubic) rotation in the case of Euclidean metric
nor 
Lorentz transformation in the case of Minkowski metric
by the terms with $\gamma_0$.

The eight-flavor Majorana field in $\underbar{8}$ of SO(7),
$\psi_M(x)$, 
can be composed into
the four-flavor Dirac pairs of left- and right-handed Weyl fields
in the $\underbar{4}$ of SO(6), 
$\psi(x) = \psi_+(x) + \psi_-(x)$. 
In the representation of the SO(7) gamma matrices
specified in section~\ref{sec:1^4(-1)^4-model}, 
we have
${\rm C} = \check {\rm C} \otimes \sigma_2$,
${\rm C}\Gamma^a = {\rm T}^{a'} 
= \check {\rm T}^{a'} \otimes \sigma_3$ ($a'=1, \cdots, 5$), 
${\rm C} \Gamma^6 = {\rm T}^6 = \check {\rm T}^6 \otimes I$,
${\rm C} \Gamma^7 = {\rm T}^7 = \check {\rm C} \otimes ( i \sigma_1 )$
and 
$ \big( \check {\rm T}^{a'} \big)^\dagger 
= \check {\rm C}^T \check {\rm T}^{a'} \check {\rm C}$ 
($a'=1, \cdots, 5$), $\big( \check {\rm T}^6 \big)^\dagger
= - \check {\rm T}^6$.
Therefore
the Majorana field $\psi_M(x)$ can be parametrized
as 
\begin{equation}
\psi_M(x) = 
\left(
\begin{array}{c}
\psi(x) \\
 -i c_D \check{\rm C} \bar\psi(x)^T 
\end{array}
\right) .
\end{equation}
%
And the bilinear operators in the quartic interaction can be rewritten as
\begin{eqnarray}
\big(
\psi_M(x)^T 
c_D ( \mp i \gamma_3) 
{\rm C} \Gamma^a \psi_M(x)
\big) 
&=&
\pm  \,  \big\{
\psi(x) i \gamma_3 
c_D \check {\rm T}^a \psi(x)
-
\bar \psi(x) i \gamma_3 
c_D \check {\rm T}^a{}^\dagger \bar \psi(x)^T
\big\} , 
\nonumber\\
&=&
\pm 2 \,  \big\{
\psi_+(x) i \gamma_3 
c_D \check {\rm T}^a \psi_-(x)
-
\bar \psi_+(x) i \gamma_3 
c_D \check {\rm T}^a{}^\dagger \bar \psi_-(x)^T
\big\}  
\nonumber\\
&&\\
\big(
\psi_M(x)^T 
c_D ( \mp i \gamma_3) 
{\rm C} \Gamma^7 \psi_M(x)
\big) 
&=&
\mp 2 \,  \bar \psi(x) i \gamma_3 \psi(x)
\nonumber\\
&=&
\mp 2 \, 
\big\{
 \bar \psi_-(x) i \gamma_3 \psi_+(x)
+
 \bar \psi_+(x) i \gamma_3 \psi_-(x)
\big\}
\\
&&\nonumber\\
\big(
\psi_M(x)^T 
c_D (\gamma_0 ) 
{\rm C} \Gamma^a \psi_M(x)
\big) 
&=&
\psi(x) 
c_D \gamma_0 \check {\rm T}^a \psi(x)
-
\bar \psi(x) 
c_D \gamma_0 \check {\rm T}^a{}^\dagger \bar \psi(x)^T
\nonumber\\
&=&
\psi_+(x) 
c_D \gamma_0 \check {\rm T}^a \psi_+(x)
-
\bar \psi_+(x) 
c_D \gamma_0 \check {\rm T}^a{}^\dagger \bar \psi_+(x)^T
\nonumber\\
&&
+
\psi_-(x) 
c_D \gamma_0 \check {\rm T}^a \psi_-(x)
-
\bar \psi_-(x) 
c_D \gamma_0 \check {\rm T}^a{}^\dagger \bar \psi_-(x)^T ,
\\
&&\nonumber\\
\big(
\psi_M(x)^T 
c_D (\gamma_0 ) 
{\rm C} \Gamma^7 \psi_M(x)
\big) 
&=&
+ 2 \, \bar \psi(x) \gamma_0 \psi(x)
\nonumber\\
&=&
+ 2 \, \big\{
\bar \psi_+(x) \gamma_0 \psi_+(x)
+
\bar \psi_-(x) \gamma_0 \psi_-(x)
\big\} ,
\end{eqnarray}
for $a=1,\cdots, 6$.
The four quartic terms, which are obtained 
with the above bilinear operators squared, 
compose the original SO(7)-invariant interaction terms. 
U(1)$_V$ is broken by the first and third ones.
U(1)$_A$ is broken by the second and third ones, but
its Z$_2$ subgroup is preserved.

One can reduce the SO(7) symmetry of the model to SO(6) by restricting
the summations of the group index in the quartic interaction
$\hat V$ to $a=1, \cdots, 6$ 
without affecting the non-degenerate gapped ground state\cite{Fidkowski:2009dba, You:2014vea}. Then, 
the action of 1+1D lattice model of
the eight-flavor Majorana chain with the reduced
SO(6) symmetry can be given by
\begin{eqnarray}
\label{eq:8MC-W-SO6}
S_{8MC/SO(6)} 
&=& 
\sum_x 
\Big\{
z \, \bar \psi(x) \big(  D_{\rm w} + m_0 \big)  \psi(x)
\nonumber\\
&& \qquad
-
\frac{1}{12}
\sum_{a=1}^6
\big(
\psi_+(x) i \gamma_3 
c_D \check {\rm T}^a \psi_-(x)
-
\bar \psi_+(x) i \gamma_3 
c_D \check {\rm T}^a{}^\dagger \bar \psi_-(x)^T
\big)^2
\nonumber\\
&& \qquad
-
\frac{1}{48}
\sum_{a=1}^6
\big(
\psi_+(x) 
c_D \gamma_0 \check {\rm T}^a \psi_+(x)
+
\bar \psi_+(x) 
c_D \gamma_0 \check {\rm T}^a{}^\dagger \bar \psi_+(x)^T
\nonumber\\
&& \qquad\qquad\qquad
-
\psi_-(x) 
c_D \gamma_0 \check {\rm T}^a \psi_-(x)
-
\bar \psi_-(x) 
c_D \gamma_0 \check {\rm T}^a{}^\dagger \bar \psi_-(x)^T 
\big)^2 
\,\,\,
\Big\} .
\nonumber\\
%
%
\end{eqnarray}
This action is invariant under
the parity and charge conjugation transformations, 
${\cal P} : \psi(x) \rightarrow i \gamma_0 \psi( x_{\cal P} )$,
$\bar \psi(x) \rightarrow - i \bar\psi(x_{\cal P}) \gamma_0$ 
where $x_{\cal P} =(x_0, - x_1)$ and 
${\cal C} : \psi(x) \rightarrow c_D \check {\rm C} \bar \psi(x)^T$,
$\bar \psi(x) \rightarrow - \psi(x)^T c_D{}^T \check {\rm C}^T$.

In this model with the reduced SO(6) symmetry, 
the axial U(1)$_A$ symmetry
is broken by the bilinear mass- and Wilson-terms
and also by the quartic, but non-covariant interaction term,
although the Z$_2$ subgroup of U(1)$_A$ is intact by the latter.
One can expect that the Z$_2$ symmetry is restored 
in the chiral limit
$m_0 \rightarrow m_c = \pm 0 + \delta m(1/z)$ 
at least in the weak quartic-coupling region $1/z^2 \ll 1$\footnote{This point deserves further studies, because this question is related to Aoki phase.}.
If the non-covariant quartic terms are irrelevant in this region
of the couplings, 
then one can further expect the
restration of the covariance and  the full U(1)$_A$ symmetry.
To make this chiral limit clear and manifest, one can use the Ginsparg-Wilson fermion instead of the Wilson fermion, 
making the replacements,
$D_{\rm w} \rightarrow D$ and 
$\psi_\pm(x) = P_\pm \psi(x) \rightarrow \hat P_\pm \psi(x)$.
Neglecting the non-covariant terms of the quartic interaction,
we obtain
\begin{eqnarray}
\label{eq:8MC-OV-SO6}
S'_{8MC/SO(6)} 
&=&
\sum_x 
\Big\{
z \, \bar \psi(x) \big(  D + m_0 \big)  \psi(x)
\nonumber\\
&& \qquad
-
\frac{1}{12}
\sum_{a=1}^6
\big(
\psi_+(x) i \gamma_3 c_D \check {\rm T}^a \psi_-(x) 
-
\bar \psi_+(x) i \gamma_3 c_D \check {\rm T}^a{}^\dagger 
\bar \psi_-(x)^T
\big)^2
%
\, \, \Big\} .
\nonumber\\
\end{eqnarray}
Omitting further the cross term of the quartic interaction
in eq.~(\ref{eq:8MC-OV-SO6}), 
which is irrelevant in breaking the U(1)$_V$ symmetry,  
we end up with the action 
eq.~(\ref{eq:effective-quartic-action-1^4(-1)^4-model-mirrorsector}) with the matching condition
$h^2 = 1/ 2 z^2$.\footnote{In the original work
by Fidkowski and Kitaev\cite{Fidkowski:2009dba}, they simply neglected the non-cross terms of the quartic interaction 
(as well as the non-covariant terms) 
in order to respect the vector U(1)$_V$ symmetry for their continuum-theory analysis
of the phase transition at the chiral limit 
$m_0 \rightarrow \pm 0$. In this case, 
the effective model is given by
the Gross-Neveu-type model with SO(7) 
symmetry.}
%
In this case, however,  we note that the charge conjugation invariance
is not manifest, but the chiral projection operators are
interchanged as follows.\cite{Suzuki:2000ku, Fujikawa:2002vj, Fujikawa:2002up}
\begin{eqnarray}
\psi_\pm(x) = \hat P_\pm \psi(x) 
&\rightarrow& \psi_\pm = P_\pm \psi(x) , \\
\bar \psi_\pm (x) = \bar \psi(x) P_\mp 
&\rightarrow&
\bar\psi_\pm = \bar \psi \{ \gamma_3 \hat P_\mp \gamma_3 \}(x) .
\end{eqnarray}

Thus our four-flavor axial model with U(1)$_A$ $\times$ Spin(6)(SU(4)) symmetry can be regarded as an effective model
for the chiral limit of 
the eight-flavor 1D Majorana chain with the reduced SO(6) symmetry\cite{Fidkowski:2009dba,You:2014vea}.
The rigorous result
about 
the mass gap of the eight-flavor 1D Majorana chain
with the SO(7)-invariant quartic interaction
by Fidkowski and Kitaev\cite{Fidkowski:2009dba} 
and its extension to the model with the reduced SO(6) symmetry
by Y.-Z. You and C. Xu \cite{You:2014vea}
therefore 
suggest 
strongly 
that
the four-flavor axial model with U(1)$_A$ $\times$ Spin(6)(SU(4)) symmetry is indeed gapped.
On the other hand, 
our numerical-simulation results that
the correlation lengths of the mirror-sector fields are
of order multiple lattice spacings
provide a numerical evidence for the mass gap
of the eight-flavor 1D Majorana chain
based on the framework of  1+1D Euclidean 
path-integral quantization.
%
What is actually new in our numerical-simulation results
is the finding that the two-point vertex functions of the U(1)$_A$ and Spin(6)(SU(4)) gauge fields are regular and local,
which implies that
the gapped eight-flavor 1D Majorana chain with the reduced SO(6) symmetry is indeed robust against the couplings of (external) gauge fields to its 
continuous symmetries.

\subsection{Eight-flavor 2D Topological Superconductor
with Gapped Boundary Phase 
\\ --- A description of gapped boundary phase in terms of overlap fermions}

The 2+1D domain wall fermion (using Wilson fermions)
is nothing but
a classical Euclidean formulation of the 2D Topological 
Insulator (Chern Insulator/IQHE without time-reversal symmetry:
class A in 2D classified by $\mathbb{Z}$)\cite{Qi:2006xx,Creutz:1994ny,Qi:2008ew}.
Then  
our result in section~\ref{sec:DWF-to-Mirror}
provides the explicit procedure to bridge
between the two constructions
for 1+1D chiral gauge theories,
the 2+1D classical construction of
the domain wall fermion with boundary interactions
to decouple the mirror-modes\cite{Creutz:1996xc}
and 
the 2D quantum Hamiltonian construction
of TI/TSC with gapped boundary phases\cite{Wen:2013ppa,Wang:2013yta,You:2014oaa,You:2014vea,DeMarco:2017gcb}.
The 1+1D mirror-fermion model in terms of overlap fermions
is derived precisely as a low-energy effective {\em local} lattice
theory of the domain wall fermion, 
and the lattice theory can describe directly the gapless/gapped boundary phases of the TI/TSC.

To illustrate the above point, 
it is instructive to consider the eight-flavor 
2D 
chiral p-wave 
TSC 
with the time-reversal and Z$_2$ symmetries (class D'/DIII+R in 2D classified by 
$\mathbb{Z}_8$($\leftarrow\mathbb{Z}$))\cite{Qi:2013dsa,Yao-Ryu:2013,Ryu-Zhang:2012,Gu-Levin:2014}.
In this class of TSC, the edge modes are 
1+1D Majorana fermions those are protected 
from acquiring mass terms
by
the discrete chiral symmetry, $\psi_M(x) \rightarrow \gamma_3 \psi_M(x)$ in the continuum limit.
When these Majorana fermions are described by
the lattice theory of overlap fermions, however, 
the asymmetric assignment of the chiral operators $\hat \gamma_3$
and $\gamma_3$ to the fields and the anti-fields 
as
$\psi(x) \rightarrow \hat \gamma_3 \psi(x)$ and
$\bar \psi(x) \rightarrow \bar \psi(x) ( - \gamma_3 )$, 
respectively
contradicts the Majorana condition 
$\bar \psi_M(x) = \psi_M(x)^T c_D {\rm C}$.
Then 
one needs to 
understand the eight-flavor Majorana field in $\underbar{8}$ of SO(7) in terms of the 
four-flavor Dirac field in $\underbar{4}$ of SO(6) through
the relation
\begin{equation}
\psi_M(x) = 
\left(
\begin{array}{c}
\psi(x) \\
 -i c_D \check {\rm C} \bar\psi(x)^T 
\end{array}
\right) ,
\end{equation}
and the discrete chiral transformation should be defined as
\begin{eqnarray}
\psi_M(x)  &\longrightarrow& 
( \hat \gamma_3 {\rm P_+ } + \gamma_3 {\rm P_-} ) \, 
\,
\psi_M(x) ,
\end{eqnarray}
where ${\rm P_\pm} = ( 1 \pm \Gamma_7 )/2 $.
This means that 
the SO(7) symmetry must be reduced to SO(6),  while 
the Majorana field are protected from acquiring the mass terms
by the discrete chiral symmetry based on the Ginsparg-Wilson relation.

With this understanding, 
we can see that
the domain wall fermion model 
with U(1)$_A$ $\times$ Spin(6)(SU4)) symmetry
discussed in section~\ref{sec:DWF-to-Mirror} 
is indeed defining 
a 2+1D classical Euclidean lattice model of the 2D quantum TSC.
In fact, the action given by 
eq.~(\ref{eq:action-DW-U(1)A-Spin(6)})
can be rewritten as follows:
\begin{eqnarray}
\label{eq:action-DW-SO(7)}
S_{\rm DW} 
&=& \sum_{t=1}^{L_3} \sum_{x \in \Lambda}
\frac{1}{2}\,
\tilde \psi_M(x,t)^T \tilde c_D {\rm C} 
\big\{ [1+ a'_3 (\tilde
D_{2 \rm w} - m_0)] \delta_{t t'}  
- \tilde P_- \delta_{t+1, t'} -\tilde P_+ \delta_{t,t'+1} \big\} \tilde \psi_M(x, t') ,
\nonumber\\
S_{\rm bd}
&=&
\sum_{x \in \Lambda}
\frac{1}{2}\,
(z -1) \,  \tilde \psi_M(x, L_3)^T \tilde c_D {\rm C} 
\tilde P_-
 [1+a'_3( \tilde D_{2 \rm w} - m_0) ] \tilde \psi_M(x,L_3)
\nonumber\\
&+&
\sum_{x \in \Lambda}  \,
\frac{h}{2} \, \,
\tilde \psi_M(x, L_3)^T
i \tilde \gamma_5  
\tilde \gamma_3  
\tilde c_D ({\rm P}_+ + \tilde P_- {\rm P}_- ) {\rm T}^a E^a(x) 
\tilde \psi_M(x, L_3) .
\end{eqnarray}
Here 
$\tilde \psi_M(x)$ is the eight-flavor four-component Majorana field given in terms of the four-flavor four-component
Dirac field as 
\begin{eqnarray}
&&
\tilde \psi_M(x) = 
\left(
\begin{array}{c}
\tilde \psi (x) \\
- i \tilde c_D \check {\rm C} \bar{\tilde \psi}(x)
\end{array}
\right) \, ; 
\qquad
\tilde \psi(x) = 
\left(
\begin{array}{c}
 \psi (x) \\
 \psi'(x) 
\end{array}
\right),  \quad
\bar{\tilde \psi}(x) = 
\left(
 \bar\psi (x), \bar\psi'(x) 
\right) .
\end{eqnarray}
Dirac gamma matrices are defined as 
$\tilde \gamma_0 = \sigma_1 \otimes \sigma_3$,
$\tilde \gamma_1 = \sigma_2 \otimes \sigma_3$,
$\tilde \gamma_3 = \sigma_3 \otimes \sigma_3$,
$\tilde \gamma_4 = I \otimes \sigma_2$, 
$\tilde \gamma_5 = I \otimes \sigma_1$.
The charge conjugation operator $\tilde c_D$ is given as
$\tilde c_D = i \sigma_2 \otimes \sigma_3$.
The 2+1D Dirac operator is defined with 1+1D Wilson-Dirac operator
$\tilde D_{\rm 2 w} = \sum_{\mu=0}^1\big\{ 
\tilde \gamma_\mu (\nabla_\mu - \nabla_\mu^\dagger)/2 + \nabla_\mu \nabla_\mu^\dagger /2 \big\}$ and 
$\tilde P_\pm = (1 \pm \tilde \gamma_3)/2$,
and $a'_3 (= a_3 /a)$ is the lattice spacing of the third dimension in the lattice unit.
The Dirichlet b.c. is imposed as 
\begin{eqnarray}
&&
\tilde P_+ \tilde \psi_M(x,0) = 0, 
\quad ; \quad
\tilde P_- \tilde \psi_M(x, L_3 + 1) =0 .
\end{eqnarray}
The action is invariant under 
the parity and charge conjugation transformations, 
${\cal P'} : 
\tilde \psi_M(x) \rightarrow \tilde \gamma_5 \tilde \gamma_1
\tilde \psi_M( x_{\cal P'} )  ; 
E^a(x) \rightarrow - E^a(x_{\cal P'})$ 
where $x_{\cal P'} =(x_0, - x_1,  x_2)$ 
and ${\cal C} : \tilde \psi_M(x) \rightarrow \tilde \psi_M(x)$.
It is also invariant under the Z$_2$ transformation, $Z_2 : \tilde \psi_M(x) \rightarrow i \tilde \gamma_4 \tilde \gamma_5 \tilde \psi_M(x) = (I \otimes \sigma_3) \tilde \psi_M(x)$, but not invariant
under the discrete chiral transformation,
$\tilde \psi_M(x) \rightarrow \tilde \gamma_3 \tilde \psi_M(x)= (\sigma_3 \otimes \sigma_3) \tilde \psi_M(x)$
by the mass and Wilson terms.

As for the 1+1D mirror-fermion model in terms of overlap fermions, 
which describes
the gapless/gapped boundary phases,
the action can be rewritten as 
%
\begin{eqnarray}
S_{W} 
&=& \sum_x 
\Big\{
\frac{1}{2} \,\psi'_M(x)^T c_D {\rm C}  D \psi'_M(x)
\Big\} ,
\\
\label{eq:2D-TSC-S_M}
S_{M} 
&=& \sum_x 
\Big\{
\frac{z}{2} \,\psi_M(x)^T c_D {\rm C}  D \psi_M(x)
+ \frac{h}{2} \, 
\psi_M^T \, 
X_E
\psi_M \, 
\Big\} ,
\end{eqnarray}
where 
\begin{eqnarray}
X_E &=&
%
\hat P_+^T {\rm P}_+ i\gamma_3 c_D {\rm T}^a E^a(x) {\rm P}_+ \hat P_-
+
\hat P_-^T {\rm P}_+ i\gamma_3 c_D {\rm T}^a E^a(x) {\rm P}_+ \hat P_+
\nonumber\\
&+& 
 P_+^T {\rm P}_- i\gamma_3 c_D {\rm T}^a E^a(x) {\rm P}_-  P_-
+
 P_-^T {\rm P}_- i\gamma_3 c_D {\rm T}^a E^a(x) {\rm P}_-  P_+ .
\end{eqnarray}
Here $\psi'_M(x)$ and $\psi_M(x)$ are 
the eight-flavor two-component Majorana fields given in terms of the four-flavor two-component 
Dirac fields $\psi_\vartriangleleft (x)$ and 
$\psi_\vartriangleright(x)$  as 
\begin{eqnarray}
&&
\psi'_M(x) = 
\left(
\begin{array}{c}
\psi_\vartriangleleft(x) \\
- i \tilde c_D \check {\rm C} \bar{\psi}_\vartriangleleft(x)
\end{array}
\right), 
\qquad
\psi_M(x) = 
\left(
\begin{array}{c}
\psi_\vartriangleright (x) \\
- i \tilde c_D \check {\rm C} \bar{\psi}_\vartriangleright(x)
\end{array}
\right),  \\
&&
\psi_\vartriangleleft(x) 
= \hat P_+ \psi'(x) + \hat P_- \psi(x) ,
\quad
\bar{\psi}_\vartriangleleft(x) 
= \bar \psi'(x) P_- + \bar \psi(x) P_+ .
\\
&&
\psi_\vartriangleright(x) 
= \hat P_+ \psi(x) + \hat P_- \psi'(x) ,
\quad
\bar{\psi}_\vartriangleright(x) 
= \bar \psi(x) P_- + \bar \psi'(x) P_+ .
\end{eqnarray}
Note that 
$\psi_\vartriangleleft (x)$ and 
$\psi_\vartriangleright(x)$ consist of the fields of the boundaries
at $t=0$ and $t=L_3$, respectively.
$S_W$ and $S_M$ stand for the actions 
of the boundary phases at $t=0$ and
$t=L_3$, respectively. 
The precise relation between the bulk TSC(domain wall fermion)
and the boundary phases(overlap fermions) are given by the 
following identity\cite{Vranas:1997da,Neuberger:1997bg,Kikukawa:1999sy}\cite{Hotta:1997my}.
\begin{eqnarray}
&&
\lim_{a'_3 \rightarrow 0}
\lim_{L_3 \rightarrow \infty}
 \int {\cal D}[E^a]\,
\prod_{x,t} 
d \tilde \psi_M(x,t) \, {\rm e}^{- S_{\rm DW}- S_{\rm bd}}
\big\vert_{\rm Dir}
\nonumber\\
&&=
\lim_{a'_3 \rightarrow 0}
\lim_{L_3 \rightarrow \infty}
\int 
\prod_{x,t} 
d \tilde \psi_M(x,t) 
\, {\rm e}^{- S_{\rm DW}} 
\big\vert_{\rm AP} 
\,
 \int {\cal D}[\psi'_M] \,  {\rm e}^{-S_W} 
\,
 \int  {\cal D}[E^a] {\cal D}[\psi_M]\, {\rm e}^{-S_M} .
\nonumber\\
\end{eqnarray}
In terms of the functional pfaffians, it is given by
\begin{eqnarray}
&& 
\lim_{a'_3 \rightarrow 0}
\lim_{L_3 \rightarrow \infty}
 \int {\cal D}[E^a]\,
{\rm pf} \big( 
\tilde c_D {\rm C} [ \tilde{D}'_{\rm 3w} - m_0 
+ h i \tilde \gamma_3 \Gamma^a E^a \delta_{t,L_3}\delta_{t',L_3}]
\big)_{\rm Dir.}
\nonumber\\
&&=
\lim_{a'_3 \rightarrow 0}
\lim_{L_3 \rightarrow \infty}
{\rm pf} \big( \tilde c_D {\rm C} [ \tilde D_{\rm 3w} - m_0 ] \big)_{\rm AP} \, \, \, 
{\rm pf} \big( c_D {\rm C} D \big) \, 
\int {\cal D}[E^a]\,
{\rm pf} \Big( 
z c_D {\rm C} D +  h X_E
\Big) ,
\nonumber\\
\end{eqnarray}
where
$ \{ \tilde{D}'_{\rm 3w} \}_{t t'} 
= \{ \tilde{D}_{\rm 3w} \}_{t t'}+ \delta_{t,L_3}\delta_{t',L_3}(z-1) \tilde P_- [1+ a_3' ( \tilde D_{\rm 2w} -m_0) ]$.

The boundary phase at $t=L_3$, which is supposed to be gapped,
is now described by the 1+1D lattice model of
the eight-flavor two-component  overlap Majorana field
$\psi_M(x)$ 
with the action 
$S_M$ of eq.~(\ref{eq:2D-TSC-S_M}). 
As argued in section~\ref{sec:1^4(-1)^4-model},
the limit of the large Majorana-Yukawa coupling, 
$z/h \rightarrow 0$, is well-defined in this formulation. 
Then
the pfaffian
factorizes as
\begin{eqnarray}
{\rm pf} \Big( X_E \Big) &=&
\det \big( u^T \, i \gamma_3 c_D \check {\rm T}^a E^a v \big) \,
\det \big( \bar u \, i \gamma_3 c_D {\check {\rm T}}^a{}^\dagger E^a 
\bar v^T \big) , 
\end{eqnarray}
where
the first determinant is positive semi-definite and the second
one is unity. Therefore 
the partition function of the boundary phase
is 
positive-definite in this limit:
\begin{eqnarray}
\big\langle  1 \big\rangle_{M}  
&=&
\int {\cal D}[E^a]\,
{\rm pf} \Big( 
z c_D {\rm C}  D +  h X_E
\Big) 
\nonumber\\
& \rightarrow &
\int  {\cal D}[E^a] \, 
\det \big( u^T i \gamma_3 c_D \check {\rm T}^a E^a v \big) \, > \, 0 \qquad (z/h \rightarrow 0) .
\end{eqnarray}
And the numerical-simulation results 
in section~\ref{sec:1^4(-1)^4-model}
show
that
the correlation lengths of the mirror-sector fields are
of order multiple lattice spacings
and that the two-point vertex functions of the U(1)$_A$ and Spin(6)(SU(4)) gauge fields are regular and local.
These results 
provide a numerical evidence 
in the framework of 2+1D path-integral quantization
that 
%
the boundary phase
of the eight-flavor 2D chiral p-wave TSC
with the time-reversal and Z$_2$ 
symmetries(class D'/DIII+R in 2D)
is indeed gapped
by the SO(6)-invariant multi-fermion interaction,
shown originally
in \cite{Qi:2013dsa,Yao-Ryu:2013,Ryu-Zhang:2012,Gu-Levin:2014}.


The above connection should hold true in lower and higher dimensions.
It is straightforward to extend the above discussion
to the case of
the eight-flavor 1D TSC with time-reversal symmetry
(class BDI in 1D classified by 
$\mathbb{Z}_8$($\leftarrow\mathbb{Z}$))\cite{Fidkowski:2009dba}
through dimensional reduction.
It would be also useful
to examine
the Hamiltonian constructions
of 3+1D chiral gauge theories
based on the 4D TI/TSC with the proposed gapped boundary phases\cite{Wen:2013ppa,You:2014oaa,You:2014vea}
from the point of view of the 3+1D/4+1D Euclidean 
construction based on
the overlap/domain wall fermions.
These topics will be discussed elsewhere\cite{Kikukawa:2017ngf,Kikukawa:2017c}.

\section{Discussions}
\label{sec:summary-discussion}

In this paper, we 
addressed
the basic question how to
decouple the mirror Ginsparg-Wilson fermions in lattice models for two-dimensional abelian chiral gauge theories.
After we investigated why the mirror-fermion approach seems to fail for the $345$-model with Dirac- and Majorana-Yukawa couplings to XY-spin field\cite{Bhattacharya:2006dc,Giedt:2007qg,Poppitz:2007tu,Poppitz:2008au,Poppitz:2009gt,Poppitz:2010at,Chen:2012di,Giedt:2014pha}, 
we proposed the two mirror-fermion models
with U(1)$_A$$\times$Spin(6)(SU(4))-invariant
Majorana-Yukawa couplings to SO(6)-vector spin field:  
$1^4(-1)^4$ axial gauge model and $21(-1)^3$ chiral gauge model.
These models are well-defined and simplified in the limit
of the large Majorana-Yukawa couplings.
We examined their properties
in the weak gauge-coupling limit through Monte Carlo simulations
and provided numerical evidences that the mirror-fermions are indeed decoupled in these models.
For the $21(-1)^3$ chiral gauge model, we deduced
a definition of the (target) Weyl-field
measure,
in which
the mirror-fermion part of the Dirac-field measure
is just saturated 
by the suitable products of the 't Hooft vertices
in terms of the mirror-fermion fields.
Based on the results of Monte Carlo simulations,
we argued that the induced fermion measure term satisfies the required {\em locality} property 
and provides a solution to the reconstruction theorem
of the Weyl field measure 
in the framework of the Ginsparg-Wilson relation\cite{Luscher:1998du}. 

As to the properties of these models in the weak gauge-coupling limit, 
we have argued that the functional determinant
of the mirror-fermions, 
$ \det \big( u^T i \gamma_3 c_D {\rm T}^a E^a v' \big)$, 
is positive semi-definite 
based on the analytical and numerical results.
We have also argued that the induced measure term 
of the mirror sector,
$ \big\langle - \delta_\eta S_M  \big\rangle_{M} \slash  \big\langle  1  \big\rangle_{M} $, is a local functional of the (extenal) U(1) and Spin(6) link fields by computing the two-point vertex functions numerically. 
It is highly desirable to 
establish these properties rigorously, 
if possible. The verification by numerical simulations should be also extended for larger lattice sizes with
higher statistics.

The deduced definition of the Weyl-field measure
of the $21(-1)^3$ chiral-gauge model seems generic, 
although
another constraint on the charge assignment,
${\rm Tr} \, Q + {\rm Tr} \, Q' = 0$,  is actually required
so that the matrix 
$\big( u^T i \gamma_3 c_D {\rm T}^a E^a v' \big)$
remains square in all topological sectors $\mathfrak{U}[m]$.\footnote{This constraint does not seem to contradict nor
to be related with the (local) 
gravitational anomaly because there 
is no mixed anomaly in two-dimensions.}
We will discuss the application of this definition to SO(10) chiral gauge theory in four-dimensions elsewhere\cite{Kikukawa:2017ngf}.

We find it interesting that
the gauge/global symmetries of the $21(-1)^3$ chiral gauge model
shown in table~\ref{table:21(-1)^3-anomaly} mimics the standard model, where the chiral U(1) gauge interaction plays the role
of SU(2)$_L$ $\times$ U(1)$_Y$ Electroweak gauge interaction.
By introducing an Abelian Higgs field and its Yukawa couplings to
``quarks'' and ``leptons'', 
the model may be used as a toy model to
study/simulate the baryon-number non-conservation in the standard model (cf. \cite{Grigoriev:1989ub,Grigoriev:1989je,Ambjorn:1990pu,Ambjorn:1990wn,Moore:1998ge,Moore:1998swa,Moore:1999fs,Moore:2000mx,GarciaBellido:1999sv,GarciaBellido:2003wd,Tranberg:2003gi,Tranberg:2006dg,DOnofrio:2014rug}).

It is known that a chiral lattice gauge theory is a difficult case for numerical simulations
because the effective action induced by 
Weyl fermions has a non-zero imaginary part.
But, 
in view of the recent studies of the 
simulation methods based on 
the complex Langevin dynamics\cite{
Parisi:1984cs, 
Klauder:1983zm, 
Klauder:1983sp, 
Ambjorn:1985iw, 
Ambjorn:1986fz, 
Berges:2006xc,
Berges:2007nr, 
Aarts:2008rr, 
Aarts:2008wh, 
Aarts:2009dg,
Aarts:2009uq, 
Aarts:2010aq, 
Aarts:2011ax, 
Aarts:2011zn, 
Seiler:2012wz,
Pawlowski:2013pje, 
Pawlowski:2013gag, 
Aarts:2013uxa, 
Sexty:2013ica, 
Aarts:2013fpa,
Giudice:2013eva,
Mollgaard:2013qra,
Sexty:2014zya,
Hayata:2014kra, 
Splittorff:2014zca, 
Aarts:2015oka,
Fodor:2015doa,
Salcedo:2015jxd,
Hayata:2015lzj, 
Li:2016srv,
Aarts:2016qrv,
Abe:2016hpd,
Ito:2016efb,
Salcedo:2016kyy, 
Aarts:2017vrv, 
Fujii:2017oti} 
and the complexified path-integration on Lefschetz thimbles\cite{
Pham:1983, 
Kaminski:1994, 
Howls:1997, 
Witten:2010cx,
Cristoforetti:2012su, 
Cristoforetti:2012uv, 
Cristoforetti:2013wha, 
Mukherjee:2013aga, 
Fujii:2013sra, 
Cherman:2014ofa,
Cristoforetti:2014gsa, 
Mukherjee:2014hsa, 
Aarts:2014nxa, 
Tanizaki:2014xba, 
Nishimura:2014kla, 
Tanizaki:2014tua, 
Kanazawa:2014qma, 
Behtash:2015kna, 
Tanizaki:2015pua, 
DiRenzo:2015foa, 
Behtash:2015kva,
Fukushima:2015qza, 
Tanizaki:2015rda, 
Fujii:2015bua, 
Fujii:2015vha,
Behtash:2015zha, 
Alexandru:2015xva, 
Behtash:2015loa,
Scorzato:2015qts, 
Alexandru:2015sua, 
Alexandru:2016lsn, 
Alexandru:2016gsd, 
Alexandru:2016san, 
Fujimori:2016ljw,
Alexandru:2016ejd, 
Tanizaki:2016xcu, 
Fujimori:2017oab,
Fukuma:2017fjq, 
Alexandru:2017oyw,  
Nishimura:2017vav,
Mori:2017zyl,
Fujimori:2017osz,
Tanizaki:2017yow}, 
one may consider to apply these methods to chiral lattice gauge theories. In particular, it seems feasible to
apply the generalized Lefschetz thimble 
method\cite{Alexandru:2015sua,Fukuma:2017fjq,Alexandru:2017oyw}, which uses
the integration contours followed from the holomorphic gradient flow with various flow-times and the exchange Monte Carlo (parallel tempering) algorithm, 
to the two-dimensional abelian
chiral lattice gauge theories discussed in this paper.
Analytical study of the Lefschetz-thimble structures
of these chiral lattice gauge theories would be also 
interesting and useful.\cite{Tanizaki:2016xcu}
The tensor renormalization group method\cite{Levin:2006jai, Gu:2010yh, Xie:2012, Shimizu:2012zza, Shimizu:2012wfa, Gu:2013gba, Liu:2013nsa, Yu:2013sbi,Shimizu:2014fsa, Shimizu:2014uva, 
Unmuth-Yockey:2014iga, Takeda:2014vwa, Kawauchi:2016xng,Sakai:2017jwp} is another option. It seems also feasible to
apply the method to 
these two-dimensional theories.

The recent proposal by Grabowska and Kaplan\cite{Grabowska:2015qpk,Grabowska:2016bis,Kaplan:2016dgz,Fukaya:2016ofi,Okumura:2016dsr,Makino:2016auf,Makino:2017pbq,Hamada:2017tny}
is ``orthogonal'' to
the mirror-fermion approach with Ginsparg-Wilson fermions
discussed in this paper.
It is based on the original domain wall fermion by Kaplan\cite{Kaplan:1992bt}, but
coupled to the
``five-dimensional'' link field
which is obtained from the dynamical four-dimensional link field
at the target wall by the gradient flow toward the mirror wall.
This choice of the ``five-dimensional'' link field 
makes possible a chiral gauge coupling for the target and mirror walls, while keeping the system four-dimensional and gauge-invariant.\footnote{
In the weak gauge-coupling region of the topologically trivial sector, the condition that the form factor of the mirror-modes
is soft enough to suppress the (transverse) gauge-coupling
is given by 
$\sum_\mu 4 \sin^2( p_\mu / 2)  \, t \gg 1$ for all possible
momenta above a certain IR cutoff. If one assumes
$ \vert p_\mu \vert \ge \pi/L$, the condition
reads 
$ \sqrt{8t} \gg \sqrt{8/\pi^2} L$, 
which apparently contradicts
the other condition 
$1 \ll  \sqrt{8t} \ll L$ for that 
local composite operators of the flowed five-dimensional link field 
is local w.r.t. the original four-dimensional link field.
This implies that the imaginary part of the
effective action, which can be written with the local operators of the five-dimensional link field, actually contains
the {\em non-local} operators w.r.t. the dynamical four-dimensional
link field. This is one reason
why the gauge-invariance is maintained in this formulation
even when any anomalous set of chiral-modes appear
in the target wall. (The other reason is that the mirror-modes are never decoupled from the gauge degrees of freedom of the dynamical four-dimensional link field.)

The fate of the non-local terms in
anomaly-free cases is not clarified yet. But the studies
of chiral anomaly in such cases\cite{Okumura:2016dsr,Hamada:2017tny} revealed that 
one needs to subtract certain non-local/five-dimensional
contributions to obtain the correct local expression
of the anomaly term
in the anomalous conservation law.

In this respect, we point out the following fact.
If one indeed tries to obtain the non-local/five-dimensional
counter terms to subtract the above non-local/five-dimensional
contributions (nonperturbatively as it should be), one actually ends up with solving 
the known local cohomology problem, which was first formulated
by L\"uscher for Ginsparg-Wilson fermions in the 4-dim. lattice plus 2-dim. continuum space\cite{Luscher:1999un,Luscher:1998kn} and then extended
by the present author for domain wall fermion
in the 5-dim. lattice plus 1-dim. continuum 
space\cite{Kikukawa:2001mw}.
And if one would include the non-local/five-dimensional counter terms so obtained to the original formulation, one can show that
the resulted four-dimensional model is {\em local} and
does not actually depend on how
the dynamical four-dimensional link field is extrapolated
to the extra dimension\cite{Kikukawa:2001mw}. 
Then there is no particular reason to
choose the method of gradient flow for this purpose.
}
It is ``orthogonal'' in the sense that
the authors do not try to decouple the massless-modes at the mirror wall, but interpret them as physical degrees of freedom with very soft form factor caused by the gradient flow, 
and that the authors do not try (do not need) to break explicitly the continuous global symmetries with ``would-be gauge anomalies'' in the mirror-wall sector,
which would be required if one would try to decouple
the mirror-modes as claimed by Eichten and Preskill
and by the other and present authors\cite{Eichten:1985ft,Bhattacharya:2006dc,Wang:2013yta}.

\appendix
\section{chiral basis in the free theory}
\label{app:chiral-basis-free}

Dirac gamma matrices:
\begin{eqnarray}
&& \gamma_0 = \sigma_1, \quad \gamma_1 = \sigma_2, \quad \gamma_3 = \sigma_3 , \\
&& \quad
\{ \gamma_\mu, \gamma_\nu \} = 2 \delta_{\mu \nu}, \quad 
\gamma_\mu^\dagger =\gamma_\mu,  \quad 
\gamma_3 = -i \gamma_0 \gamma_1 ,  \\
&& \quad
\gamma_0^{\rm T} = \gamma_0, \, 
\gamma_1^{\rm T} = - \gamma_1, \, 
\gamma_3^{\rm T} = \gamma_3, \\
&& 
c_D = i \gamma_1 = i \sigma_2, \\
&& \quad
c_D \gamma_\mu c_D^{-1} = - \gamma_\mu^{\rm T},  \,
c_D \gamma_3 c_D^{-1} = -  \gamma_3 \, ; \, 
c_D^{\rm T} = c_D^{-1}= c_D^\dagger = -c_D .
\end{eqnarray}

\noindent
Kernels of chiral operators
$\hat \gamma_3 = \gamma_3(1-2D)$, $\gamma_3$:
\begin{eqnarray}
\hat \gamma_3(x,y) &=& 
-\int^{+\pi}_{-\pi} \frac{d^2 p}{(2\pi)^2} \, {\rm e}^{i p (x-y)}
\, \frac{1}{\omega(p)}
\left( 
\begin{array}{cc}
b(p) & c(p) \cr
c(p)^\dagger & -b(p) \cr
\end{array}
\right),
\\
 \gamma_3(x,y) &=& 
+\int^{+\pi}_{-\pi} \frac{d^2 p}{(2\pi)^2} \, {\rm e}^{i p (x-y)}
\, 
\left( 
\begin{array}{cc}
1 & 0 \cr
0 & -1 \cr
\end{array}
\right),
\end{eqnarray}
where $c(p) = i \sin(p_1) + \sin(p_2)$, 
$b(p)= 2-\cos(p_1)-\cos(p_2) - m_0$, $\omega(p)=\sqrt{c^\dagger c +b^2}(p)$.

\vspace{1em}
\noindent
Orthonormal chiral bases:

\noindent 
The case with $j=p$

\begin{eqnarray}
u_j(x) &=& 
{\rm e}^{i p x} \, 
u(p)\, ; \quad 
u(0)=
\left( \begin{array}{c} 1 \cr 0\end{array} \right) , \,\,\,
u(p)=\frac{1}{\sqrt{2 \omega ( \omega + b)}}
\left( \begin{array}{c} - c \cr  \omega+ b \end{array} \right)(p)
\quad (p \not = 0) ,
\nonumber\\
&&\\
v_j(x) &=& 
{\rm e}^{i p x} \, 
v(p)\, ; \quad 
v(0)=
\left( \begin{array}{c} 0 \cr 1 \end{array} 
\right), \,\,\,
v(p)=\frac{1}{\sqrt{2 \omega ( \omega + b)}}
\left( \begin{array}{c} \omega +  b \cr c^\dagger \end{array} 
\right)(p)
\quad (p \not = 0) ,
\nonumber\\
&&\\
\bar u_j(x) &=& 
{\rm e}^{-i p x} \, \bar u(p)\, 
; \quad
\bar u(p)=
\left( \begin{array}{cc} 0 & \, 1 \end{array} \right) , \\
\bar v_j(x) &=& 
{\rm e}^{-i p x } \, \bar v(p)\, 
; \quad 
v(p)=
\left( \begin{array}{cc } 1 & \, 0 \end{array} \right) .
\end{eqnarray}

\noindent
The case with $j=x$
\begin{eqnarray}
u_j(x) &=& 
\int^{+\pi}_{-\pi} \frac{d^2 p}{(2\pi)^2} \, 
{\rm e}^{i p (x-x_j)} \, 
u(p)\, ; \quad 
u(0)=
\left( \begin{array}{c} 1 \cr 0\end{array} \right) , \,\,\,
u(p)=\frac{1}{\sqrt{2 \omega ( \omega + b)}}
\left( \begin{array}{c} - c \cr  \omega+ b \end{array} \right)(p)
\quad (p \not = 0) ,
\nonumber\\
&&\\
v_j(x) &=& 
\int^{+\pi}_{-\pi} \frac{d^2 p}{(2\pi)^2} \, 
{\rm e}^{i p (x-x_j)} \, 
v(p)\, ; \quad 
v(0)=
\left( \begin{array}{c} 0 \cr 1 \end{array} 
\right), \,\,\,
v(p)=\frac{1}{\sqrt{2 \omega ( \omega + b)}}
\left( \begin{array}{c} \omega +  b \cr c^\dagger \end{array} 
\right)(p)
\quad (p \not = 0) ,
\nonumber\\
&&\\
\bar u_j(x) &=& 
\int^{+\pi}_{-\pi} \frac{d^2 p}{(2\pi)^2} \, 
{\rm e}^{-i p (x-x_j)}  \, \bar u(p)\,
; \quad
\bar u(p)=
\left( \begin{array}{cc} 0 & \, 1 \end{array} \right) , \\
\bar v_j(x) &=& 
\int^{+\pi}_{-\pi} \frac{d^2 p}{(2\pi)^2} \, 
{\rm e}^{-i p (x-x_j)} \, \bar v(p)\, 
; \quad 
\bar v(p)=
\left( \begin{array}{cc } 1 & \, 0 \end{array} \right) .
\end{eqnarray}

\noindent
Majorana-mass-type inner products:
\begin{eqnarray}
u(-p)^T c_D v(p) &=& \frac{1}{2\omega(\omega+b)}
 [ c c^\dagger - (\omega+b)^2   ](p)
= - \frac{b}{\omega}(p), \\
u(-p)^T \gamma_3 c_D v(p) &=& \frac{1}{2\omega(\omega+b)}
 [ (\omega+b)^2 + c c^\dagger](p)
= 1
\end{eqnarray}

\section{SO(7) spinor}


\noindent
The Clifford algebra in 7 dimensions ($a=1,\cdots,7$):

\begin{eqnarray}
&&
\Gamma^a \Gamma^b + \Gamma^b \Gamma^a = 2 \delta^{ab}, \quad 
{\Gamma^a}^\dagger = \Gamma^a, \quad 
\Gamma^{7}= i \Gamma^1 \Gamma^2 \cdots \Gamma^{6}, \\
&&
{\rm C} \Gamma^a {\rm C}^{-1} = - \{ \Gamma^a \}^T, \,  \,   
{\rm C}^T = - {\rm C}^{-1} =-  {\rm C}^\dagger = {\rm C}
\end{eqnarray}
\begin{eqnarray}
\Gamma^1 &=& \sigma_1 \times \sigma_1 \times \sigma_1 ,\\ 
\Gamma^2 &=& \sigma_2 \times \sigma_1 \times \sigma_1 ,\\ 
\Gamma^3 &=& \sigma_3 \times \sigma_1 \times \sigma_1 ,\\ 
\Gamma^4 &=& I               \times \sigma_2 \times \sigma_1 ,\\ 
\Gamma^5 &=& I               \times \sigma_3 \times \sigma_1 ,\\ 
\Gamma^6 &=& I               \times I              \times \sigma_2 ,\\ 
\Gamma^7 &=& I               \times I              \times \sigma_3 ,\\ 
{\rm C} &=& i \sigma_2 \times \sigma_3 \times \sigma_2 . 
\end{eqnarray}

\noindent
{\rm T} matrices:
\begin{equation}
{\rm T}^a = {\rm C} \Gamma^a  , \qquad
%
\{{\rm T}^a\}^T = - {\rm T}^a
\end{equation}

\begin{eqnarray}
{\rm T}^1 &=&
 i (-i)(+i) (-i) \, 
 \sigma_3 \times \sigma_2 \times \sigma_3 
\,\,= \, \check{\rm T}^1 \times \sigma_3 ,\\
{\rm T}^2 &=& 
i (+1)(+i) (-i)\, 
 I \times \sigma_2 \times \sigma_3 
\,\,\,\, = \, \check{\rm T}^2 \times \sigma_3 , \\ 
{\rm T}^3 &=&
i (+i)(+i) (-i)\, 
 \sigma_1 \times \sigma_2 \times \sigma_3 
\,\, = \, \check{\rm T}^3 \times \sigma_3 , \\ 
{\rm T}^4 &=& 
 i (+1)(-i) (-i)\, 
 \sigma_2 \times \sigma_1 \times \sigma_3 
\, = \, \check{\rm T}^4 \times \sigma_3 , \\ 
{\rm T}^5 &=&
 i (+1)(+1) (-i)\, 
 \sigma_2 \times I \times \sigma_3 
\,\,\, = \, \check{\rm T}^5 \times \sigma_3 , \\ 
{\rm T}^6 &=&
i (+1)(+1) (+1)\, 
 \sigma_2 \times \sigma_3 \times I 
\,\,\, =  \, \check{\rm T}^6 \times I , \\ 
{\rm T}^7 &=&
i (+1)(+1) (+i) \, 
\sigma_2 \times \sigma_3 \times \sigma_1 
\, =  + i \, \check{\rm C} \, \times \sigma_1 . 
\end{eqnarray}

\newpage
\noindent
The Clifford algebra in 5 dimensions ($a=1,\cdots,5$):

\begin{eqnarray}
\check \Gamma^1 &=& \sigma_1 \times \sigma_1 , \\
\check \Gamma^2 &=& \sigma_2 \times \sigma_1 , \\
\check \Gamma^3 &=& \sigma_3 \times \sigma_1 , \\
\check \Gamma^4 &=& I               \times \sigma_2 , \\
\check \Gamma^5 &=& I               \times \sigma_3 , \\
\check {\rm C} &=& i \sigma_2 \times \sigma_3 . 
\end{eqnarray}

\noindent
Reduced {\rm T} matrices:
\begin{eqnarray}
\check {\rm T}^a &=& - i \check{\rm C} \check\Gamma^a  , 
\quad
\{\check{\rm T}^a\}^T = - \check{\rm T}^a , \\
\check {\rm T}^6 &=& \check{\rm C}   , \quad\quad
\{\check{\rm T}^6\}^T = - \check{\rm T}^6 .
\end{eqnarray}

\acknowledgments

The authors would like to thank M.~Sato, H.~Fujii, M.~Kato, Y.~Okawa, T.~Okuda for enlightening discussions.
This work is supported in part by JSPS KAKENHI Grant Numbers 24540253, 16K05313 and 25287049.

\end{document}